\documentclass[cernpreprint, atlasdraft=false, UKenglish, coverpage=false, texmf, orcidlogo]{atlasdoc}
\usepackage{atlaspackage}
\usepackage{atlasbiblatex}

\usepackage[export]{adjustbox}

\usepackage{atlasphysics}

\graphicspath{{logos/}{figures/}}

\usepackage{multirow}
\AtlasTitle{ATLAS searches for additional scalars and exotic Higgs boson decays with the LHC Run~2 dataset}
\AtlasAbstract{%
This report reviews the published results of searches for possible additional scalar particles and exotic decays of the Higgs boson performed by the ATLAS Collaboration using up to 140~fb$^{-1}$ of 13~TeV proton--proton collision data collected during Run~2 of the Large Hadron Collider.
Key results are examined, and observed excesses, while never statistically compelling, are noted.
Constraints are placed on parameters of several models which extend the Standard Model, for example by adding one or more singlet or doublet fields, or offering exotic Higgs boson decay channels.
Summaries of new searches as well as extensions of previous searches are discussed.
These new results have a wider reach or attain stronger exclusion limits.
New experimental techniques that were developed for these searches are highlighted.
Search channels which have not yet been examined are also listed, as these provide insight into possible future areas of exploration.
}

\author{The ATLAS Collaboration}

\AtlasRefCode{HDBS-2023-15}

\PreprintIdNumber{CERN-EP-2024-094}

\AtlasDate{\today}

\AtlasJournal{Physics Reports}

\AtlasJournalRef{Phys. Rep. 1116 (2025) 184-260}
\AtlasDOI{https://doi.org/10.1016/j.physrep.2024.09.002}

\hypersetup{pdftitle={ATLAS document},pdfauthor={The ATLAS Collaboration}}

\begin{document}

\maketitle

\tableofcontents

\section{Introduction}
\label{sec:intro}

The discovery of a new particle in 2012 at a mass of about
125~\GeV by ATLAS~\cite{HIGG-2012-27} and CMS~\cite{CMS-HIG-12-028}
marked a fundamental milestone in high energy physics. Since then, a multitude of measurements have confirmed
that the properties of this new particle are consistent with the Higgs boson as predicted by the Standard Model (SM)~\cite{HIGG-2021-23,CMS-HIG-22-001}. A huge number of searches for additional Higgs bosons or other phenomena indicating new physics beyond the SM (BSM physics)  have all yielded null results, thereby constraining the phase space of possible models considerably. However, the parameter space to which these searches are sensitive is limited by the available data and the analysis tools used. Extending the coverage, and the presence of small excesses in a number of searches, call for more data and continued effort.

This report reviews the landscape and results of searches for new BSM Higgs bosons or exotic Higgs boson decays conducted by ATLAS~\cite{PERF-2007-01} during the \RunTwo of the Large Hadron Collider (LHC) at CERN, using up to 140~\ifb of 13~\TeV proton--proton ($pp$) collision data. First, a brief overview of potential models and their motivation is given. Next, the experimental signatures are listed, including those used in searches for additional neutral or charged Higgs bosons, for decays of heavy Higgs bosons to final states with multiple scalars (at least one of which can be the SM Higgs boson), or for non-standard, exotic decays of the 125~\GeV Higgs boson. Then, upper limits set on cross-sections are presented, constraints placed on model parameters are discussed, and small excesses -- if present -- are highlighted. Finally, a summary of the small excesses, a list of uncovered signatures, and a discussion of current limitations and promising techniques for future BSM Higgs boson searches are provided.

With regard to the notation used for Higgs particles in this report: usually $h$ denotes a light Higgs boson $(m_h<125~\GeV)$, the SM Higgs boson is called $h_{125}$, and $H$ is a heavy Higgs boson. However, if the paper under discussion has a different convention, that one is used.


\section{Theoretical overview and motivation}
\label{sec:theory}

A complete description of particle physics theory is beyond the scope of this experimental review. This section introduces only the main concepts that are needed to understand the experimental searches and the motivation behind them.

In the Standard Model (SM) of particle physics~\cite{Weinberg:1967tq,Glashow:1961tr,Salam:1968rm,glashow70,tHooft:1972tcz}, the fundamental forces (the weak force, strong force and electromagnetic force) are mediated by gauge bosons ($W$ and $Z$ bosons, gluons and photons), while matter is built from fermions (the leptons and quarks). After spontaneous electroweak symmetry breaking (EWSB), i.e.\ the Higgs mechanism~\cite{PhysRevLett.13.321,osti_4009504,PhysRevLett.13.585}, the $W$ and $Z$ gauge bosons acquire their masses by  absorbing the Goldstone bosons while the photon remains massless. The Higgs mechanism postulates that a scalar field permeates the universe and also predicts one observable scalar, the Higgs boson. The fermions also acquire their masses by interacting with this scalar field. The Higgs boson interacts more strongly with massive particles. Gluon--gluon fusion production, mediated by a top-quark loop, dominates the Higgs boson production cross-section at the LHC. The cross-sections, branching fractions and total width of the Higgs boson are all determined in the SM. Only the Higgs boson mass was an unknown parameter until its discovery at about 125~\GeV (the most precise mass measurement from ATLAS is reported in Ref.~\cite{HIGG-2022-20}).

Despite its tremendous success and confirmation in experiments, the SM cannot be the final, complete theory. One of the deficiencies of the SM is the lack of a candidate that could constitute dark matter. The observation of neutrino oscillations established that neutrinos have mass, which is inconsistent with the SM. Gravity is a fundamental force that is not described by the SM. The asymmetry between baryons and anti-baryons in the universe is not explained by the SM. Hence, searches for new physics are extremely well motivated and any significant observation which is inconsistent with the SM would fundamentally advance our understanding of the universe.

Additional Higgs bosons are predicted by many BSM theories. One Higgs boson has already been found at the LHC, and this strongly motivates continuing searches for other scalars. The BSM Higgs programme complements precision measurements of the SM Higgs boson and other SM particles, as well as searches for new physics such as \emph{exotic} particles and supersymmetry (SUSY). Many of the BSM scenarios presume the existence of SUSY, which is a theory that postulates a symmetry between bosons (which have integer spin) and fermions (which have half-integer spin). It was developed in the 1970s~\cite{Golfand:1971iw,Wess:1974tw,Ramond:1971gb}, numerous theoretical reviews are available, too, e.g.~\cite{MARTIN_1998}. This symmetry then predicts a partner particle (a \emph{superpartner}) for each SM particle, with the same coupling but with the other spin type (for example, the superpartner of an electron would be called a selectron and it would be a boson). However, since no superpartners to the SM particles have been observed, it is assumed that supersymmetric particles are much heavier than their SM counterparts. The details of the breaking of SUSY, which leads to this mass difference, are described by a large number of theoretical parameters that also influence the production and decay of the superpartners. The important point here, however, is that SUSY requires an extended Higgs sector. It is therefore natural to consider models that postulate both, i.e.\ supersymmetry and additional Higgs bosons. Some SUSY particles are also suitable candidates for dark matter.

The observation of the SM Higgs boson constrains possible BSM theories considerably. Any relevant BSM scenario must predict a scalar at a mass of about 125~\GeV, and this scalar should have \enquote{SM-like} couplings, about the same as the observed Higgs boson's couplings to SM particles, which is also called \emph{alignment}. Precision studies of the spin, production, and decay rates of the 125~\GeV scalar~\cite{HIGG-2021-23} showed agreement with the SM prediction and further constrained BSM parameters. However, given the small (4.1~\MeV) width of the SM Higgs boson, even very small contributions from new physics (e.g.\ through a small coupling of the Higgs boson to a dark sector~\cite{Lagouri:2825287} or hidden sectors~\cite{Gopalakrishna_2008}) could lead to sizeable decay rates into BSM particles~\cite{Curtin_2014}. This makes searches for exotic decays of the 125~\GeV boson particularly relevant. %

Additional experimental constraints on BSM theories come from the searches for SUSY that resulted in lower limits on gluino and squark masses~\cite{ATL-PHYS-PUB-2023-025}. Flavour-changing processes such as $b\to s\gamma$ and $B\to\mu^+\mu^-$ decays also constrain BSM hypotheses; for example, they lead to a lower limit on the $H^+$ mass in some types of two-Higgs-doublet models (2HDM, see below)~\cite{Mahmoudi_2010}. LEP and SLD precision measurements of $Z$-boson decay also lead to constraints on the 2HDM~\cite{Lebedev_2000}. Electroweak precision measurements, such as those measuring the top-quark and $W$-boson masses, further constrain the 2HDM~\cite{Haller_2018}. Limits on the Higgs boson self-coupling obtained from ATLAS and CMS searches for Higgs boson pair production can also restrict the allowed parameter space in the 2HDM~\cite{Arco_2022}.

In the SM description, the scalar fields are arranged in a single complex SU(2)$_\text{L}$ doublet, which is the minimal representation that establishes mass terms for the $W$ and $Z$ bosons, as well as massive fermions, after EWSB and leads to exactly one Higgs boson, which has been discovered. The following list briefly describes the field content and Higgs boson spectrum of some of the most-used BSM theories:

\begin{itemize}

\item The simplest extension of the SM is achieved by adding a new singlet to the SM. The singlet may be real or complex. The singlet fields are allowed to mix with the SM fields and this leads to the presence of two CP-even (in the case of a real singlet) or three CP-even (in the case of a complex singlet) neutral Higgs bosons. Depending on the model parameters, one of the scalars may be a dark matter candidate which does not interact with SM particles, and one of the CP-even scalars can have a mass and other properties similar to those of the observed scalar. Benchmarks have been proposed, for example in Ref.~\cite{Costa_2016}.

\item A similar model relevant for BSM Higgs boson searches is the two-real-singlet model (TRSM)~\cite{Robens_2020}, which extends the SM by adding two real singlets, leading to three observable neutral CP-even Higgs bosons ($h_1$, $h_2$, $h_3$), which couple to each other and to the SM particles, and one of them can have a mass of 125~\GeV and properties similar to those of the observed scalar. The Higgs bosons are allowed to decay into each other, depending on the mass hierarchy.

\item A popular and widely investigated extension of the SM features the addition of a second Higgs doublet, leading to the two-Higgs-doublet model (2HDM). The 2HDM in its general form has a very rich phenomenology; this is reviewed, for example, in Ref.~\cite{Branco_2012}. Benchmarks for various realizations of the 2HDM are discussed in Ref.~\cite{deFlorian:2016spz}. Parameters are often chosen to avoid flavour-changing neutral currents (FCNC) at tree level, which can be achieved by imposing a $\mathcal{Z}_2$ symmetry. CP-violation is possible but is not considered here. The CP-conserving 2HDM predicts five observable Higgs bosons: a CP-even and often light Higgs boson $h$, a (typically) heavier CP-even scalar $H$, a CP-odd scalar $A$ and two charged Higgs bosons $H^{\pm}$. This model has seven free parameters: the masses of $h$, $H$, $A$ and $H^{\pm}$, an angle $\alpha$ that describes the mixing between the CP-even states, the ratio of the vacuum expectation values (vev) of the two doublets ($\tan\beta=\textrm{vev}_1/\textrm{vev}_2$), and a parameter $m_{12}$ which controls the soft breaking term of the $\mathcal{Z}_2$ symmetry in the potential. Many of the ATLAS results discussed in later sections consider a SUSY-inspired scenario in which several simplifications are made: $m_H=m_A=m_{H^{\pm}}$, $m_{12}^2=m_A^2\sin\beta\cos\beta$, and the $h$ is identified as the 125~\GeV Higgs boson (which is then denoted by $h_{125}$). Once the Higgs boson masses are set, the properties of the Higgs sector can be described as a function of three remaining parameters, which are typically chosen to be $m_A$, $\tan\beta$ and $\sin\left(\alpha-\beta\right)$. The CP-conserving 2HDM is categorized into four types (type-I, type-II, lepton-specific, and flipped) that are used for interpretations in ATLAS searches. They differ in the way the Higgs fields couple to the SM particles, listed in Table~\ref{tab:theory:2hdm}.%

The general 2HDM (g2HDM) without $\mathcal{Z}_2$ symmetry features FCNC but can lead to a SM-like 125~\GeV Higgs boson when all heavy Higgs boson quartic couplings are $\mathcal{O}(1)$~\cite{Hou_2018}. Therefore, this model is also relevant, and is considered for ATLAS searches.

\item A special case of the type-II 2HDM is the minimal supersymmetric extension of the SM, or MSSM. The more than 100 SUSY parameters are fixed in benchmark scenarios~\cite{Bagnaschi_2019} that explore various phenomenological aspects, such as different SUSY mass scales or the couplings of the additional Higgs bosons. The SUSY corrections directly impact the mass of the Higgs bosons. Once the SUSY parameters are chosen, the Higgs sector in the MSSM depends only on two free parameters: $m_A$ and $\tan\beta$. The $A$ boson, one of the CP-even bosons and the charged Higgs bosons are almost mass-degenerate; small remaining mass differences  depend on the value of $\tan\beta$. Among the proposed MSSM scenarios, the hMSSM~\cite{Djouadi_2013} is an even more simplified model that approximates the MSSM sufficiently well for low values of $\tan\beta$. Here the idea is that the SUSY corrections are fully determined and thus fixed by the observation of $h$ at 125~\GeV. The $h$ is therefore identified as the $h_{125}$, and its fixed mass value can be understood as an input to the scenario rather than an output.

\item Another possible class of models explored at the LHC are those that postulate a 2HDM that is extended by another singlet. In the case of supersymmetry and a complex singlet, this model is referred to as the next-to-minimal supersymmetric extension of the SM (NMSSM)~\cite{Ellwanger_2010}. The NMSSM predicts seven Higgs bosons: three CP-even and two CP-odd scalars, and two charged Higgs bosons. The model parameters can be tuned such that one of the CP-even Higgs bosons has a mass of approximately 125~\GeV and properties similar to those of the observed scalar. The generalized version of the NMSSM (without the SUSY condition) is referred to as 2HDM+S~\cite{Baum_2018}, where $S$ is a complex singlet. One of the scalars in the 2HDM+S can be a dark-matter candidate that does not couple to SM particles~\cite{baum2019benchmark}. If $S$ is a real singlet, the model is called N2HDM and is discussed, for example, in Ref.~\cite{Muhlleitner2017}. It is also possible that the additional scalar is CP-odd, and such models are referred to as 2HDM+$a$~\cite{Bauer_2017}.

\item The three-Higgs-doublet model (3HDM)~\cite{Keus_2014} extends the SM by adding two more doublets. It leads to five neutral Higgs bosons and two pairs of charged Higgs bosons $H^{\pm}_1$ and $H^{\pm}_2$. In particular, it allows the possibility that one of these charged Higgs bosons is light, without violating experimental constraints from low-energy processes such as $b\rightarrow s \gamma$~\cite{Akeroyd_2017}. Five different types of this model are used to explore its phenomenology, and the charged Higgs boson couplings depend on three parameters ($X_1$, $Y_1$, $Z_1$).

\item Another very well-motivated model is the Georgi--Machacek (GM) model~\cite{GEORGI1985463,CHANOWITZ1985105}, which extends the SM by adding two triplets -- or even higher multiplets in its most general form. In the GM model that is explored in ATLAS searches, there are ten observable Higgs bosons ($h$, $H$, $H_3$, $H_3^{\pm}$, $H_5$, $H_5^{\pm}$, $H_5^{\pm\pm}$), among which is a doubly charged Higgs boson candidate, and $h$ can have a mass and properties similar to those of the observed scalar. Benchmarks for the five-plet of scalars were proposed in Ref.~\cite{Ismail_2021}, and results are typically presented as a function of the coupling parameter $\sin\theta_H$, which quantifies how much of the $W$- and $Z$-boson masses are generated by the non-SM Higgs fields.

\item Besides the GM, other models also predict the existence of doubly charged Higgs bosons: the type-II seesaw mechanism~\cite{PhysRevD.22.2227,Fileviez_P_rez_2008} is an attractive model since it can predict massive neutrinos. This model also has an extended Higgs sector, leading to seven scalars: $h$, $H$, $A$, $H^{\pm}$ and $H^{\pm\pm}$. Specific choices of model parameters allow the $h$ to have a mass and properties similar to those of the observed scalar. Also relevant for $H^{\pm\pm}$ searches are the left-right symmetric model (LRSM)~\cite{PhysRevD.11.566}, which extends the SM with two triplets, and the Zee--Babu neutrino mass model ~\cite{ZEE1985141,BABU1988132}, which introduces two complex singlets.

\end{itemize}

Providing predictions for these models is one of the tasks of the LHC Higgs Working Group (LHCHWG)~\cite{Dittmaier:2011ti,Dittmaier:2012vm,Heinemeyer:2013tqa,deFlorian:2016spz}. This group connects theoreticians with experimentalists from the LHC experiments -- mostly but not exclusively from ATLAS and CMS. The state-of-the art calculations provided by the theoreticians allow the data to be compared with BSM expectations, and upper limits to be set on model parameters. By using the same model predictions, results from different collaborations can be compared directly.

\begin{table}[h!]
\begin{center}
\caption{The four types of the CP-conserving 2HDM and the couplings of quarks and charged leptons to the scalar doublets $\Phi_1$ and $\Phi_2$ (with vacuum expectation values $\textrm{vev}_1$ and $\textrm{vev}_2$, respectively).}
\label{tab:theory:2hdm}
\resizebox{\textwidth}{!}{
\begin{tabular}{c|c|c|c}
2HDM Type & Up-type quarks couple to & Down-type quarks couple to & Charged leptons couple to \\
\hline
Type-I          & $\Phi_2$ & $\Phi_2$ & $\Phi_2$ \\
Type-II         & $\Phi_2$ & $\Phi_1$ & $\Phi_1$ \\
Lepton-specific & $\Phi_2$ & $\Phi_2$ & $\Phi_1$ \\
Flipped         & $\Phi_2$ & $\Phi_1$ & $\Phi_2$ \\
\end{tabular}
}
\end{center}
\end{table}


\section{Experimental signatures}
\label{sec:experimalsignatures}

When examining searches for BSM Higgs bosons, a useful way to group the experimental signatures is by the mass of the new scalar resonance. If the mass of the additional Higgs boson is above 125~\GeV, the search is called \emph{high mass}, and the final-state particles are often highly energetic and have high transverse momentum (\pt), whereas if the mass of the new scalar is below that value, the search is called \emph{low mass} and the decay products often have lower momentum. Low mass searches are usually limited by the trigger, while high mass searches often have less background and low, model-dependent cross-sections. Not every search can be easily categorized in this way, but it is often useful guidance. Figure~\ref{fig:highmass_lowmass} illustrates this classification scheme.

High mass searches include those for heavy neutral or charged scalars that decay either into fermions or into vector bosons. Neutral scalars can also be CP-odd, in which case they are called pseudoscalars. However, no distinction is made between the two CP hypotheses in most BSM searches. Charged Higgs bosons can also be light, but their signatures are discussed alongside the heavy charged scalars since they have some commonalities. A special class of heavy scalar searches are those where a heavy neutral scalar decays into a pair of other scalars. This pair can be two SM-like Higgs bosons ($h_{125}h_{125}$), two new scalars with the same mass ($SS$), or two different kinds of scalars from an asymmetric decay (for example $Sh_{125}$). The common element here is the heavy scalar resonance that decays; these searches are therefore also classified as high mass ones.
Low mass searches comprise those for new light scalars (or pseudoscalars), $a$. These new light Higgs bosons are produced either directly in $pp$ collisions or via the decay of the SM-like Higgs boson, which results in either $aa$ or $Za$ final states.
Another type of search is when the 125~\GeV Higgs boson decays in ways that are suppressed or not allowed in the SM. An observation or enhancement of such a branching fraction could therefore indicate new physics. These searches are referred to as \emph{$h_{125}$ rare decays} (when the decay is allowed but very improbable in the SM) or \emph{$h_{125}$ exotic decays} (if the decay is not allowed in the SM). A special class of Higgs boson decays are those with invisible final states, which can be inferred from measurements of missing transverse momentum in the detector.

The searches are designed to enrich one or several signal regions in events from the hypothesized BSM process consistent with the expected signal signature. The aim is not only to quantify a potential excess of signal events over background, but also to constrain the cross-section for the hypothesized process as well as the parameters of relevant models that influence the production or decay of the BSM particles. The model-independent and model-dependent results are reported in Section~\ref{sec:results}.

A complex aspect of high mass searches concerns assumptions about the decay width of the new scalar. In many searches a narrow decay width, of the order of a few \MeV, is assumed in the signal model used by the event generator, although such narrow high-mass (order of \TeV) resonances lack theoretical motivation. However, detector resolution effects, which are taken into account via simulation, will lead to a much wider mass distribution being reconstructed in the analysis. This resolution is typically about a few percent of the scalar's mass, and is therefore much larger than the decay width generated from the signal model. In many cases, the mass of the scalar particle cannot be fully reconstructed because the final state has undetectable particles such as neutrinos, which degrades the signal resolution. Also, in some analyses the final discriminant, which is used to separate a potential signal from the background, is not the mass but the output score of a multivariate algorithm, which leads to even worse signal resolution. Therefore, analyses that simulate a narrow-width signal are still able to constrain models that predict signal widths of the order of \GeV to tens of \GeV.

Most searches presented in this report use the full \RunTwo dataset, which has an integrated luminosity of 139 or 140~\ifb, the exact value depending on the luminosity calibration at the time of publication~\cite{DAPR-2021-01}, and also on the triggers~\cite{TRIG-2016-01} used. Some analyses were performed with a partial \RunTwo dataset.

\begin{figure}[tb!]
\centering
\includegraphics[width=0.8\textwidth]{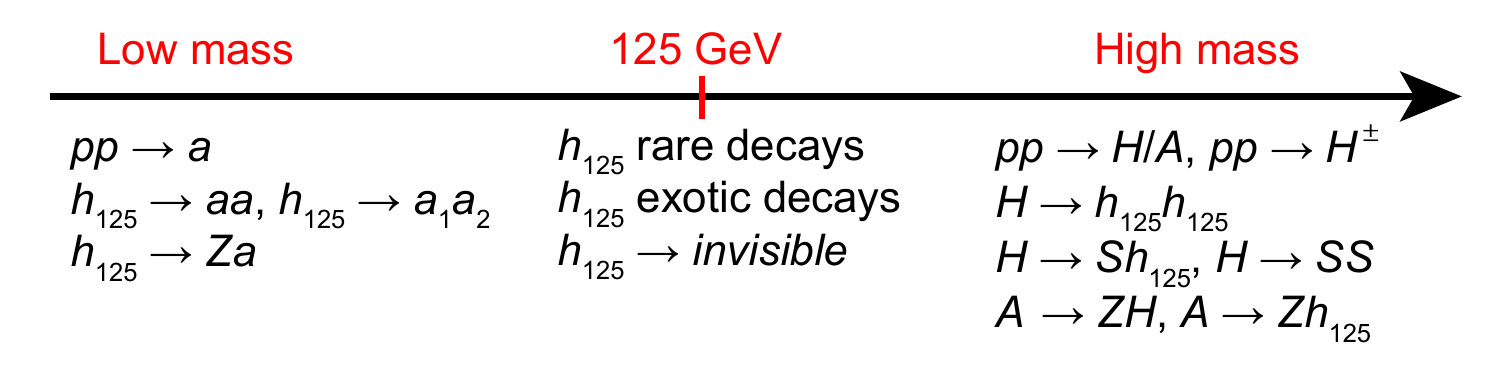}
\caption{An illustration of the landscape of BSM Higgs boson searches, for which the 125~\GeV Higgs boson ($h_{125}$) marks the border between low mass and high mass searches. Low-mass scalars are denoted by $a$, heavy Higgs bosons are labelled with an $H$ (or $A$ if they are CP-odd), and other scalars are denoted by an $S$. New scalars can be produced either directly in $pp$ collisions, or through the decay of other Higgs bosons (or, in general, through the decay of other particles). }
\label{fig:highmass_lowmass}
\end{figure}


\subsection{Searches for neutral heavy Higgs bosons}
\label{sec:neutralhiggs}

\subsubsection{Heavy Higgs bosons decaying into fermions}
\label{sec:neutralhiggs:fermions}

ATLAS has conducted searches for heavy Higgs bosons decaying into fermions in the following final states: $\tau^+\tau^-$~\cite{HDBS-2018-46}, $\mu^+\mu^-$~\cite{HIGG-2017-10}, $b\bar{b}$~\cite{HIGG-2016-32} and $t\bar{t}$~\cite{EXOT-2020-25,EXOT-2019-26}. These decays are predicted in models such as the 2HDM and MSSM. Due to the mass degeneracy of the $H$ and $A$ bosons, the production cross-sections of the two particles are summed (unless the mass difference becomes sizeable compared to the mass resolution). The Higgs boson couplings to these fermions increase with their mass, as in the case of the SM-like Higgs boson. Therefore, the largest branching fractions are predicted for decays into a top-quark pair, provided that the mass of the heavy Higgs boson exceeds twice the top-quark mass ($m_t$), about 350~\GeV. If the Higgs boson is lighter than that, then its most likely fermionic decay is the one into $b\bar{b}$, followed by $\tau^+\tau^-$. The decay into $\mu^+\mu^-$ has an even lower probability; however, this channel has excellent mass resolution. The exact branching fraction values are model-dependent. If the SUSY mass scale is low, Higgs bosons may decay into SUSY particles, which would decrease the probabilities of decays into fermion pairs. ATLAS also explored the decay of a heavy neutral Higgs boson into $t\bar{q}$~\cite{HDBS-2020-03} in the context of the g2HDM, which introduces additional Yukawa couplings.

The most common production modes considered are gluon--gluon fusion (ggF) and $b$-associated production. The $t\bar{t}H$ production mode was also explored in the case of heavy Higgs boson decay into $t\bar{t}$. For the SM Higgs boson, ggF production has the largest cross-section. In BSM theories, this cross-section depends on the model's parameters. In a type-II 2HDM, the coupling of the BSM Higgs boson to top quarks is larger at low values of $\tan\beta$, while $b$-associated production becomes significant at high values of $\tan\beta$. This is due to the enhanced coupling to quarks with charge $-1/3$ which occurs for type-II (or flipped) models. In the SM, $b$-associated production has only a very small cross-section and is not exploited. In searches motivated by type-II models, $b$-tagging can be used to enrich the signal regions in $b$-associated production. In the SM, $t\bar{t}H$ production has a small cross-section, but in BSM scenarios this process can be enhanced at low $\tan\beta$ values. Figure~\ref{fig:neutralhiggs:fermions:feynmans} displays some example Feynman diagrams for these production modes.

\begin{figure}[tb!]
\centering
\subfloat[]{
\includegraphics[width=0.4\textwidth,valign=c]{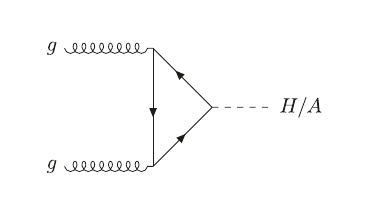}
}
\subfloat[]{
\includegraphics[width=0.4\textwidth,valign=c]{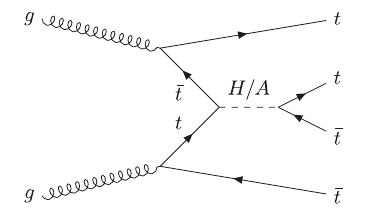}
}\\
\subfloat[]{
\includegraphics[width=0.4\textwidth,valign=c]{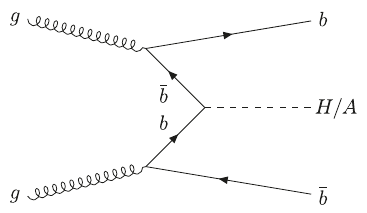}
}
\subfloat[]{
\includegraphics[width=0.4\textwidth,valign=c]{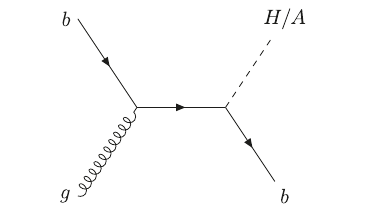}
}
\caption{\label{fig:neutralhiggs:fermions:feynmans}Illustrative Feynman diagrams for the production modes explored in ATLAS searches for heavy Higgs bosons decaying into fermions: (a) gluon--gluon fusion, (b) $t\bar{t}H$, and (c--d) $b$-associated production.}
\end{figure}

\paragraph{$H\to\tau^+\tau^-$}\mbox{}\\
ATLAS considers two $\tau$-lepton decay channels for this search~\cite{HDBS-2018-46}: either both $\tau$-leptons decay hadronically (\tauhad{}\tauhad channel), or one of the $\tau$-leptons decays to an electron or muon (\taulep{}\tauhad channel). The hadronically decaying $\tau$-leptons are identified with a boosted decision tree (BDT) using both calorimetric shower shape and tracking information~\cite{ATLAS-CONF-2017-029}, and their \pt is required to exceed 25 (65)~\GeV in the \taulep{}\tauhad (\tauhad{}\tauhad) channel. Jets\footnote{Standard jets are reconstructed either from topological clusters~\cite{PERF-2014-07} in the calorimeter or from particle flow objects~\cite{PERF-2015-09} using the anti-$k_t$ algorithm~\cite{Cacciari:2008gp,Fastjet} with a radius parameter R=0.4. They are calibrated to the particle level using a combination of simulation and in-situ measurements in data~\cite{JETM-2018-05}.} that originate from $b$-quarks are identified using a multivariate algorithm called \emph{mv2c10}~\cite{FTAG-2018-01}. Events are selected with either a single-lepton trigger~\cite{TRIG-2018-05,TRIG-2018-01} or a single-$\tau$ trigger~\cite{ATLAS-CONF-2017-061}. Events in both the \taulep{}\tauhad and \tauhad{}\tauhad channels are separated into $b$-tag or $b$-veto categories, giving a total of four signal-enriched categories. This is useful because the signal and background compositions depend on whether or not a $b$-tag is required: the ggF ($b$-associated) signal is enriched in the $b$-veto ($b$-tag) categories, and the $t\bar{t}$ background accumulates in the $b$-tagged categories. The $\tau^+\tau^-$ mass reconstruction is challenging because of the presence of neutrinos that escape detection and leave a momentum imbalance called the missing transverse momentum~\cite{JETM-2020-03}, with magnitude \met. Therefore, the transverse mass $m_\text{T}$ is used instead of the invariant mass as the final discriminant. The mass range from 200 to 2500~\GeV is explored for a heavy Higgs boson in this search; no distinction is made between a scalar or a pseudoscalar hypothesis.

\paragraph{$H\to\mu^+\mu^-$}\mbox{}\\
In the search for $H\to\mu^+\mu^-$~\cite{HIGG-2017-10}, events are selected using muon triggers. They must contain two muon candidates with \pt greater than 30~\GeV that are reconstructed using both tracking and muon spectrometer information~\cite{MUON-2018-03}. The signal is produced through either ggF or $b$-associated production, so events are grouped into $b$-veto and $b$-tag categories. The resonance mass is reconstructed from the two muons, and the mass resolution varies from 5\% to 14\% of the Higgs boson mass. The search is carried out for Higgs boson mass hypotheses between 200 and 1000~\GeV. The search is sensitive to scalars and pseudoscalars, and the analysis does not distinguish between them.

\paragraph{$H\to b\bar{b}$}\mbox{}\\
The search for $b$-associated production of $H\to b\bar{b}$~\cite{HIGG-2016-32} selects events with a combination of various $b$-jet triggers, requires one or more $b$-jets to exceed different minimum \pt values, and uses a $b$-tagging working point with 60\%--72\% efficiency depending on the trigger type and data-taking period~\cite{TRIG-2018-08}. The \emph{mv2c20} tagger is used to identify $b$-jets both online and offline~\cite{FTAG-2018-01}. Events in the signal region must contain at least three $b$-jets with a \pt of at least 20~\GeV, with the \pt requirements for the leading and sub-leading $b$-jets raised to 160 and 60~\GeV, respectively. Events are then assigned to regions with three, four or five $b$-jets. The final discriminant is the $b\bar{b}$ invariant mass, whose calculation includes a transformation to achieve better mass resolution, which then amounts to 10\%--15\%. This transformation is done with a principal component analysis (PCA) in three dimensions, using the invariant mass and the \pt of each of the two leading $b$-jets. The $m_{b\bar{b}}$ resolution deteriorates for higher \pt values, due to the presence of final-state radiation. This dependence is reduced by the PCA rotation. Higgs boson masses between 450 and 1400~\GeV are explored; scalar and pseudoscalar hypotheses are not distinguished.

\paragraph{$H\to t\bar{t}$}\mbox{}\\
If the BSM Higgs boson is heavier than twice the top-quark mass, i.e.\ more than about 350~\GeV, then the decay into $t\bar{t}$ is dominant in many models, especially if the Higgs boson is produced via ggF, which implies that the coupling to top quarks (via loop contributions) exists. Since the signal process and the largest background process, i.e.\ non-resonant SM $t\bar{t}$ production, lead to the same final state, a large destructive interference effect occurs. This distorts the signal-plus-background mass distribution shape from a peak to a broad excess below the true Higgs boson mass followed by a dip structure around the heavy Higgs boson's mass. The details of this interference are model-dependent, and this large effect needs to be taken into account when analysing this channel.\\
ATLAS searched for this signature in 13~\TeV data in final states with one or two electrons or muons in the $H$ mass range of 400--1400~\GeV, with the scalar and pseudoscalar hypotheses considered separately~\cite{EXOT-2020-25}. The events are selected using single-lepton triggers. Four signal-enriched categories are constructed: a two-lepton category and three categories with one lepton (here lepton means electron~\cite{EGAM-2018-01} or muon). The hadronic top-quark decay in the one-lepton channel is reconstructed either by using a reclustered jet with a variable radius parameter of $R = 0.4{-}1.5$~\cite{ATLAS-CONF-2017-062, ATL-PHYS-PUB-2016-013} if the top quark's \pt is large enough to collimate its decay products into one such jet (the \enquote{merged} case), or by using three standard jets in the \enquote{resolved} case. Standard jets are also used to identify $b$-quarks; at least one $b$-tagged jet is required. Leptons must have $\pt > 28$~\GeV, and the event must have $\met > 20$~\GeV. The final discriminant is the reconstructed $t\bar{t}$ mass. Angular variables sensitive to $t\bar{t}$ spin correlations are used in both the one- and two-lepton channels to increase the sensitivity of the analysis.

\paragraph{$t\bar{t}H\to 4t$}\mbox{}\\
This search considers $t\bar{t}H$ production of a heavy Higgs boson that then decays into a pair of top quarks, leading to a four-top-quark final state~\cite{EXOT-2019-26}. Events are selected using single-lepton or dilepton triggers. Events with two leptons (electrons or muons) with $\pt > 28$~\GeV and the same charge, or three leptons without any charge requirements, are retained in the analysis. The selection requires the presence of six jets (reconstructed from tracks in the inner detector and energy deposits in the calorimeters using a particle-flow algorithm~\cite{PERF-2015-09}), and two of these jets must be $b$-tagged. The \emph{DL1r} $b$-tagging algorithm~\cite{FTAG-2019-07} is used; it is based on a deep neural network and outperforms the previously mentioned \emph{mv2} taggers. The final discriminant in the signal region is the output score of a parameterized BDT. The Higgs boson mass range considered is 400--1000~\GeV. Scalar and pseudoscalar hypotheses are treated in the same way.

\paragraph{$H\to t\bar{t}/t\bar{q}$}\mbox{}\\
This search~\cite{HDBS-2020-03} targets the g2HDM that predicts new Yukawa couplings of the Higgs boson to up-type quarks ($\rho_{tt}$, $\rho_{tc}$ and $\rho_{tu}$). The dominant signal diagrams are displayed in Figure~\ref{fig:neutralhiggs:feynman:g2hdm}. The search selects a variety of channels where the heavy Higgs boson is produced in association with top quarks and then decays into a top quark and another up-type quark, leading to final states with multiple leptons and $b$-jets, which are grouped into 17 signal-enriched categories. Events are selected using single-lepton or dilepton triggers. The final discriminant in each signal region is the output score of a deep neural network. The mass range between 200 and 1500~\GeV is explored for a heavy Higgs boson.

\begin{figure}[tb!]
\centering
\subfloat[]{
\includegraphics[width=0.19\textwidth,valign=c]{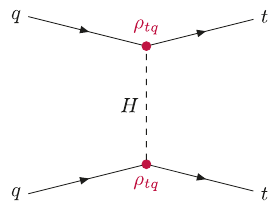}
}
\subfloat[]{
\includegraphics[width=0.19\textwidth,valign=c]{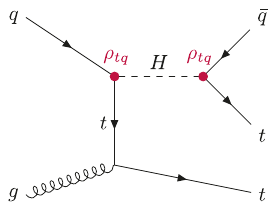}
}
\subfloat[]{
\includegraphics[width=0.19\textwidth,valign=c]{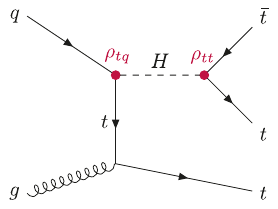}
}
\subfloat[]{
\includegraphics[width=0.19\textwidth,valign=c]{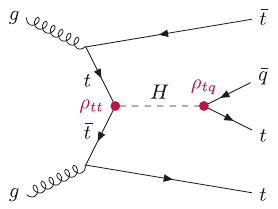}
}
\subfloat[]{
\includegraphics[width=0.19\textwidth,valign=c]{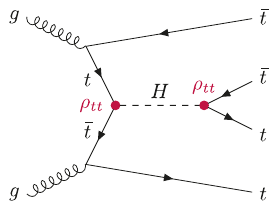}
}
\caption{\label{fig:neutralhiggs:feynman:g2hdm}Illustrative Feynman diagrams for the signal processes in the search for $H\to t\bar{t}/t\bar{q}$. The couplings $\rho_{tq}$ are introduced in the specific BSM scenario that is considered here, the g2HDM.}
\end{figure}

\subsubsection{Heavy Higgs bosons decaying into bosons}
\label{sec:neutralhiggs:bosons}

Heavy Higgs bosons decaying into gauge bosons have been investigated in the $W^+W^-$~\cite{ATLAS-CONF-2022-066}, $ZZ$~\cite{HIGG-2018-09}, $\gamma\gamma$~\cite{HIGG-2018-27,HIGG-2023-12,HIGG-2019-23} and $Z\gamma$~\cite{HIGG-2018-44,HDBS-2019-10} final states. These decay modes are disfavoured for pseudoscalar Higgs bosons because CP conservation forbids the decay of a pseudoscalar (CP-odd) to boson pairs (CP-even). In models where the heavy Higgs boson couples to bosons, production through vector-boson fusion (VBF) will occur, as depicted in Figure~\ref{fig:neutralhiggs:bosons:feynman}. Therefore, these decay channels are typically explored for ggF and VBF production modes. Since these production processes also occur in the SM, the BSM searches for bosonic decays share many similarities with the SM Higgs boson analyses.

\begin{figure}[tb!]
\centering
\includegraphics[width=0.4\textwidth]{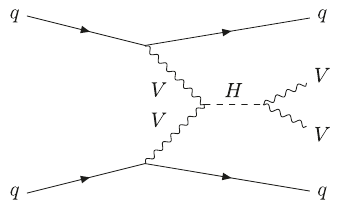}
\caption{Illustrative Feynman diagram for the vector-boson-fusion production mode, which occurs in models where the Higgs boson couples to bosons.}
\label{fig:neutralhiggs:bosons:feynman}
\end{figure}

\paragraph{$H\to W^+W^-$}\mbox{}\\
The search for $H\to W^+W^-$~\cite{ATLAS-CONF-2022-066} considers only leptonic $W$-boson decays into an electron or muon and a neutrino, and the events are selected using single-electron or single-muon triggers. The two leptons must each have $\pt > 25$~\GeV, and events with additional leptons with $\pt > 15$~\GeV are vetoed. Events with $b$-jets are rejected to suppress the $t\bar{t}$ background. The analysis uses three signal-enriched categories: two categories are optimized for VBF production and require either exactly one jet or at least two jets, and the third category is enriched in ggF production and contains events not assigned to the VBF categories. The final discriminant is the transverse mass of the dilepton and \met system. Higgs boson masses in the range 300--4000~\GeV are explored.

\paragraph{$H\to ZZ$}\mbox{}\\
The search for $H\to ZZ$~\cite{HIGG-2018-09} considers the decays of $ZZ$ into either $\ell^+\ell^-\ell'^+\ell'^-$ or $\ell^+\ell^-\nu\bar{\nu}$, where $\ell$ denotes an electron or muon. The choice to consider $Z$ decays into neutrinos is well motivated since it has a large branching fraction of about 20\%. Events are selected using single-lepton triggers. This analysis searches for Higgs bosons with a mass between 200 and 2000~\GeV.\\
In the $4\ell$ channel, the electrons (muons) must have a \pt exceeding 7 (5)~\GeV; this requirement is raised to 20~\GeV for the leading lepton. The leptons are grouped into two pairs, and each invariant mass must be close to the $Z$-boson mass (91~\GeV). Events with at least two jets are classified as VBF-like if the output score of a classifier trained to distinguish this process from the SM $ZZ$ background exceeds a certain value. An analogous classifier is used to select ggF signal events. The final discriminant is the $4\ell$ invariant mass.\\
In the $\ell^+\ell^-\nu\bar{\nu}$ channel, leptons must have a \pt above 20~\GeV; for the leading lepton this is raised to 30~\GeV. Due to the presence of neutrinos, the \met is required to exceed 120~\GeV. If events contain at least two jets that meet some kinematic selection criteria for being consistent with a typical VBF signature, they enter the VBF category. Events not meeting these criteria enter the ggF category. The final discriminant is the transverse mass calculated from the \pt of the dilepton system, the mass of the $Z$ boson, and the \met.

\paragraph{$H\to \gamma\gamma$}\mbox{}\\
Several searches for $H\to \gamma\gamma$ were carried out, targeting different Higgs boson mass ranges (high mass~\cite{HIGG-2018-27}, intermediate mass~\cite{HIGG-2023-12}, and low mass~\cite{HIGG-2019-23}). In all these searches, events are selected using a diphoton trigger~\cite{TRIG-2018-05}. Photon candidates~\cite{EGAM-2018-01} must have a \pt exceeding 25 (20--22)~\GeV in the high mass (low and intermediate mass) search, and must also meet stringent identification and isolation criteria. The final discriminant is the diphoton mass. Due to the excellent detector resolution (around 1.3\% at 125~\GeV), if the new resonance has a sizeable width, the resolution of the diphoton mass peak will be dominated by that width. Therefore, the analysis also tests non-narrow signal hypotheses in this channel. The Higgs boson mass range explored in the high mass search is 160--3000~\GeV, in the intermediate mass search it is 66--110~\GeV, and in the low mass search it is 10--70~\GeV.\\
The intermediate mass search has to deal with a difficulty caused by some events having both electrons from the $Z\to ee$ decay misidentified as photons, which can produce a peak in the $\gamma\gamma$ mass spectrum. To reduce the misidentification rate, a gradient BDT is used at object level to reject photon candidates that are likely to be fakes. The high mass search is performed inclusively, and the intermediate mass search is split into categories according to the conversion status of the selected photons.\\
The low mass search is difficult because selecting photons with transverse energies close to the trigger thresholds sculpts the $m_{\gamma\gamma}$ distribution at its lower edge.

\paragraph{$X\to Z\gamma$}\mbox{}\\
The search for a generic narrow-width scalar $X$ with $X\to Z\gamma$ was carried out separately for the leptonic~\cite{HIGG-2018-44} and hadronic~\cite{HDBS-2019-10} decays of the $Z$ boson. Two channels were considered for the leptonic decays, with the $Z$ boson decaying into either an electron pair or a muon pair. Events are selected using electron, muon or single-photon triggers. The leptons must have $\pt > 10$~\GeV, and photons must have $\pt > 15$~\GeV. Events must contain a pair of same-flavour leptons with opposite charge and an invariant mass within 15~\GeV of the $Z$ mass. A dedicated multivariate analysis (MVA) procedure was developed to improve the identification of closely spaced electrons from highly boosted $Z$ bosons. The final discriminant is the three-body invariant mass of the lepton pair and the photon. The Higgs boson mass range covered in this search is 220--3400~\GeV.\\
For hadronic $Z$-boson decays, a high-\pt single-photon trigger was used, and the minimum \pt requirement for photon candidates is 200~\GeV. Events are also required to have a large-radius jet with $\pt > 200$~\GeV. Large-radius jets~\cite{ATL-PHYS-PUB-2017-015} are built from tracks and calibrated clusters of energy in the calorimeter and have a radius parameter of R=1.0. Three analysis categories are defined, based on the identification of $b$-jets (for $Z\to b\bar{b}$), the value of $D_2$ (a variable that exploits jet substructure~\cite{Larkoski_2013}) and the mass of the jet. The final discriminant is the invariant mass of the jet and photon, and the search covers the range 1.0--6.8~\TeV.

\subsubsection{Higgs-to-Higgs decays}
\label{sec:neutralhiggs:higgstohiggs}

Another class of searches conducted by ATLAS comprises two types of cascade decays of heavy Higgs bosons. In the first, a heavy pseudoscalar Higgs boson decays into a $Z$ boson and either the 125~\GeV Higgs boson or a new scalar of different mass. In the second, discussed further in Section~\ref{sec:dihiggs}, a heavy Higgs boson decays into a pair of other scalars.\\
If the decays involve a $Z$ boson, then CP conservation requires the CP states of the heavy and lighter Higgs bosons to be different, meaning one is a CP-odd pseudoscalar and the other is a CP-even scalar. ATLAS assumes that the heavier one is the CP-odd boson, but this is not strictly necessary -- it can be the other way around, but $m_A>m_H$ is favoured for a strong first-order phase transition to occur in the early universe~\cite{Dorsch_2014}. ATLAS searched for this kind of signature in the following channels: $A\to Zh_{125}$ with $h_{125}\to b\bar{b}$ and $Z\to\nu\bar{\nu}/\ell^+\ell^-$~\cite{HDBS-2020-19}, $A\to ZH$ with $H\to b\bar{b}/W^+W^-$ and $Z\to\ell^+\ell^-$~\cite{HDBS-2018-13}, and $A\to ZH$ with $H\to b\bar{b}$ and $Z\to\nu\bar{\nu}$ or $H\to t\bar{t}$ and $Z\to\ell^+\ell^-$~\cite{HDBS-2021-02}. These kinds of decays were investigated for the ggF and $b$-associated production modes. Another Higgs-to-Higgs search channel features the decay of a heavy scalar $H$ into a pair of 125 GeV
Higgs bosons decaying to 4b~\cite{HDBS-2019-31}. The $H$ is produced either via Higgsstrahlung or the decay of an even heavier pseudoscalar $A$.

\paragraph{$A\to Zh_{125}$}\mbox{}\\
In the search for $A\to Zh_{125}$ ~\cite{HDBS-2020-19} the $h_{125}$ is assumed to be a SM-like Higgs boson that decays into $b\bar{b}$, and $A$ is a heavier pseudoscalar. Events where the $Z$ boson decays into neutrinos or charged leptons (electrons or muons) are selected using an \met trigger or lepton triggers respectively. If the $A$ is heavy compared to the $h_{125}$, the $h_{125}$ will be boosted and the emitted $b$-jets will have a small opening angle, such that they can be reconstructed as a single large-radius jet. In that case, a $b$-tagging algorithm is applied to track-jets of variable, \pt-dependent radius~\cite{ATL-PHYS-PUB-2017-010} that are associated with the large-radius jet. This approach increases the background rejection rate and improves the search's sensitivity by up to a factor of four compared to a previous analysis using 36~\ifb of \RunTwo data~\cite{EXOT-2016-10}. If the mass difference between the $A$ and $h_{125}$ bosons is moderate, the $b$-jets are reconstructed separately as small-radius jets. Four categories are defined: events contain either no lepton or exactly two leptons, and the $h_{125}$ is reconstructed as one or two jets. In the categories without a lepton, a large \met of at least 150~\GeV is required. The final discriminant is the transverse mass calculated from the Higgs boson candidate and the \met for the zero-lepton case, or the invariant mass built from the decay products of the $Z$ and $h_{125}$ bosons for the two-lepton category. The mass resolution is better for the latter case, varying from 2\% to 9\% of the mass of the $A$ boson. The investigated mass range for the heavy pseudoscalar is 220--2000~\GeV.

\paragraph{$A\to ZH$}\mbox{}\\
In this channel the $H$ is assumed to be a new Higgs boson with a mass that is not consistent with 125~\GeV, and the $A$ is a heavy pseudoscalar. One publication~\cite{HDBS-2018-13} explores the leptonic $Z$-boson decays into electrons or muons ($Z\to\ell^+\ell^-$) together with the $H\to b\bar{b}$ or $H\to W^+W^-$ decay modes. In another publication~\cite{HDBS-2021-02} the following final states are investigated: $H\to b\bar{b}$ and $Z\to\nu\bar{\nu}$, or $H\to t\bar{t}$ and $Z\to\ell^+\ell^-$.\\
For the first analysis~\cite{HDBS-2018-13}, with $H\to b\bar{b}/W^+W^-$ and $Z\to\ell^+\ell^-$, the investigated $H$-boson mass range is 130--700~\GeV and the $A$ mass range is 230--800~\GeV. The $H \to b\bar{b}$ and $H\to W^+W^-$decay modes are complementary -- some models favour the decay into fermions, and other models predict the decay into bosons to be dominant. Gluon-gluon fusion production is considered for both decays of the $H$ boson. The $b$-associated production mode is only studied for the $H\to b\bar{b}$ decay, which is relevant only for models that favour a coupling of $H$ to fermions. Events are selected with single-electron or single-muon triggers. The dilepton mass must be consistent with the $Z$-boson mass. For $H\to b\bar{b}$ decay, events are assigned to categories with either two or at least three $b$-jets, where the extra $b$-jets come from $b$-associated production. For $H\to W^+W^-$ decay, only fully hadronic decays of the $W$ bosons into quarks are considered, resulting in a final state with four jets. In each case, the invariant mass of the decay products of the $H$ boson must be consistent with the hypothesized $H$ mass (allowing for the detector resolution). The final discriminant is either $m_{\ell\ell bb}$ or $m_{\ell\ell 4q}$, depending on the $H$-boson decay mode. The $A$ mass can be reconstructed with a resolution of 1\%--16\%, depending on the category and the masses of the two heavy Higgs bosons.\\
The second analysis~\cite{HDBS-2021-02} considers two channels: $H\to b\bar{b}$ and $Z\to\nu\bar{\nu}$ for an $A$ ($H$) mass range of 350--1200~\GeV (130--800~\GeV), or $H\to t\bar{t}$ and $Z\to\ell^+\ell^-$ for an $A$ ($H$) mass range of 450--1200~\GeV (350--800~\GeV). Both channels are explored for ggF production, and $b$-associated production is studied for the $b\bar{b}+\nu\bar{\nu}$ final state. Events are selected using either single-lepton or \met triggers~\cite{TRIG-2019-01}. For the $\ell\ell+t\bar{t}$ case, one of the top quarks is required to decay leptonically, leading to a final state with three leptons, two of which must have an invariant mass consistent with the $Z$-boson mass. Both top quarks are reconstructed, using a $W$-boson mass constraint for the leptonically decaying top quark. The final discriminant is the mass difference ($\Delta m$) between the reconstructed $H$ and $A$ bosons, with a resolution ranging from 3\% to 20\% of $\Delta m$. For the other case, $\nu\bar{\nu}+b\bar{b}$, no leptons are detected and instead the \met is required to exceed 150~\GeV. The events are categorized according to the number of $b$-jets, which is either exactly two or at least three, the latter case targeting $b$-associated production. The final discriminant is the transverse mass calculated from the $b$-jets and \met, and the resolution corresponds to 8\%--27\% of the $A$-boson mass.

\paragraph{$VH$ or $A\to ZH$ with $H\to h_{125}h_{125}\to\bbbar\bbbar$}\mbox{}\\
Another Higgs-to-Higgs decay search involves the decay of a heavy scalar $H$ boson into a pair of 125~\GeV Higgs bosons and uses the $b\bar{b}b\bar{b}$ final state~\cite{HDBS-2019-31}. The $H$ is produced either in association with an off-shell vector boson $V$ ($V$ = $W$ or $Z$) (similar to \enquote{Higgsstrahlung}) or in $A\to ZH$ decay, where the $A$ boson is produced via ggF. Feynman diagrams illustrating these processes at leading order are displayed in Figure~\ref{fig:neutral:HiggstoHiggs:feynman_VHH}.\\
For $VH$ production, the explored $H$ mass range is 260--1000~\GeV. In the case of $A\to ZH$, the explored mass ranges are $360 \leq m_A \leq 800$~\GeV and $260  \leq m_H \leq  400$~\GeV. Potential signal events are characterized by a leptonically decaying $V$ boson and four $b$-jets. Three leptonic channels (0L, 1L, 2L) correspond to $Z\to\nu\nu$, $W\to\ell\nu$ and $Z\to\ell\ell$. The \pt of electrons and muons must exceed 7~\GeV. The selected events contain at least four $b$-tagged jets with $\pt > 20$~\GeV, which are paired to form the two $h_{125}$ candidates. Events with identified hadronically decaying $\tau$-leptons are vetoed to reduce backgrounds. The \met also forms part of the signal region definition, and its allowed values depend on the channel. Further criteria include several multivariate BDT discriminants to enhance the signal purity in each channel. The final discriminant is the reconstructed mass of the Higgs boson pair.

\begin{figure}[tb!]
\centering
\subfloat[]{
\includegraphics[width=0.4\textwidth,valign=c]{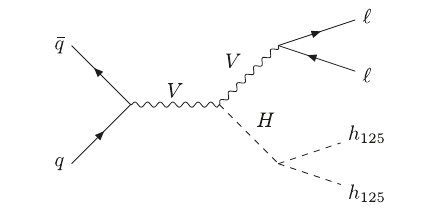}
}
\subfloat[]{
\includegraphics[width=0.4\textwidth,valign=c]{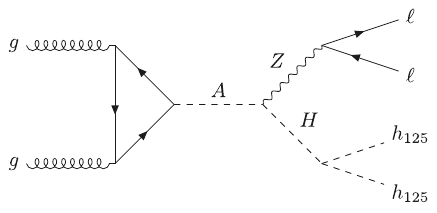}
}
\caption{\label{fig:neutral:HiggstoHiggs:feynman_VHH}Illustrative Feynman diagrams for resonant Higgs pair production in association with a vector boson $V$, as predicted in some BSM scenarios. The heavy scalar $H$ resonance, which decays into $h_{125}h_{125}$, originates from (a) an off-shell vector boson or (b) the decay of a neutral heavy pseudoscalar $A$.}
\end{figure}


\subsection{Searches for charged Higgs bosons}
\label{sec:chargedhiggs}

\subsubsection{Charged Higgs bosons decaying into fermions}
\label{sec:chargedhiggs:single_fermions}

ATLAS performed searches for charged Higgs bosons ($H^+$)\footnote{For simplicity and better readability, only the positive charge is indicated, but both charges are always implied.} in various decay channels. The fermionic final states explored in \RunTwo are $\tau^+\nu$~\cite{HIGG-2016-11}, $tb$~\cite{HDBS-2018-51} and $cb$~\cite{HDBS-2019-24}. The decays into $\tau^+\nu$ and $tb$ are predicted in the 2HDM, while the $cb$ channel is relevant in the 3HDM.\\
Charged Higgs bosons can span a broad mass range and are thus classified as light, intermediate, or heavy. Light charged Higgs bosons have a mass below 160~\GeV and can therefore be produced in top-quark decays, i.e.\ $t\to H^+ b$. Top-quark pair production has a huge cross-section, so even a tiny branching fraction to $H^+b$ can lead to a sizeable event rate. Charged Higgs bosons with a mass larger than that of the top-quark (called \emph{heavy $H^+$} in this context) are produced in $gg$ or $gb$ processes via the $tbH^+$ coupling with additional $b$- and $t$-quarks in the final state (details of these calculations and how overlap is handled can be found in Ref.~\cite{deFlorian:2016spz} and references therein). For charged Higgs bosons with a mass close to that of the top-quark, which defines the intermediate mass range, the \ttbar and $gg$ or $gb$ production processes interfere, leading to a complicated situation~\cite{Degrande_2017}. These production processes are displayed in Figure~\ref{fig:chargedhiggs:single_fermions:feynmans}.

\begin{figure}[tb!]
\centering
\subfloat[]{
\includegraphics[width=0.4\textwidth,valign=c]{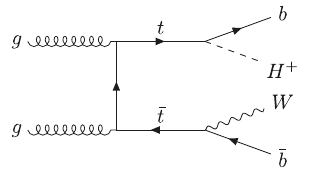}
}
\subfloat[]{
\includegraphics[width=0.4\textwidth,valign=c]{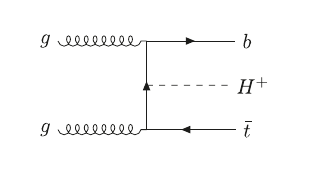}
}
\caption{\label{fig:chargedhiggs:single_fermions:feynmans}Illustrative Feynman diagrams for the production modes explored in ATLAS searches for charged Higgs bosons:\ (a) light $H^+$ production in top-quark decays, (b) $t$- and $b$-associated production of heavy $H^+$. In the intermediate mass range, both production modes contribute.}
\end{figure}

The decay of the charged Higgs boson is also governed by its mass. In an aligned 2HDM, if the $H^+$ mass is larger than the sum of the top-quark and $b$-quark masses, then its decay into $tb$ is dominant, followed by decay into $\tau^+\nu$. As in the case of neutral Higgs bosons, the Yukawa couplings to higher-mass particles are larger. If the charged Higgs boson is lighter than the top quark, then the decay into $\tau^+\nu$ is dominant. The $\tau^+\nu$ channel is therefore relevant over the entire mass range and has been explored even in the difficult intermediate region. %

\paragraph{$H^+\to \tau^+\nu$}\mbox{}\\
The search for $H^+\to \tau^+\nu$~\cite{HIGG-2016-11} is conducted over a wide mass range of 90--2000~\GeV, and includes production through top-quark decays, the intermediate region, and $t$-associated production. Hadronic $\tau$ decays ($\tauhad$) are selected, and the $W$-boson (from the decay of the top quark that is produced in association) decays either to jets or leptonically, which leads to three categories: $\tau$+jets, $\tau+e$ or $\tau+\mu$. In the case of the $\tau$+jets category, the $\tau$ candidate must have a \pt of at least 40~\GeV, and the event must have $\met > 150$~\GeV and at least three jets, one of which is $b$-tagged. These events are selected with an \met trigger. In the case of the $\tau+\ell$ categories, the events are selected with single-lepton triggers, the \tauhad must have $\pt > 30$~\GeV, there must be at least one $b$-tagged jet, and the \tauhad and the lepton must have opposite charge. The final discriminant in each category is the output score of a BDT, trained separately for five regions across the $H^+$ mass range.

\paragraph{$H^+\to tb$}\mbox{}\\
The search for $H^+\to tb$~\cite{HDBS-2018-51} considers $H^+$ production in association with, as well as decay into, $t$ and $b$, leading to final states with many jets and $b$-jets. Events are required to contain exactly one lepton (electron or muon, with a \pt above 27~\GeV), arising from one of the $t$-quark decays, which also motivates the usage of single-lepton triggers. The events must also have at least five jets with $\pt > 25$~\GeV, and at least three of them must be $b$-tagged by the \emph{mv2c10} algorithm with a working point corresponding to 70\% $b$-tagging efficiency. The events are then grouped into four signal-enriched categories:\footnote{The notation $n$j$m$b means that the event has \emph{n} jets of which \emph{m} are $b$-tagged, and $\geq$\emph{n}j$\geq$\emph{m}b means the event has at least \emph{n} jets of which at least \emph{m} are $b$-tagged.} 5j3b, 5j$\geq$4b, $\geq$6j3b and $\geq$6j$\geq$4b. Since both the production and the decay involve top quarks and $b$-jets, the modelling of $t\bar{t}$+jets backgrounds is the main challenge in this analysis. Data are used to correct and control both the shape and normalization of these background contributions. A sophisticated NN algorithm is trained separately in each category to discriminate between signal and background; this works very well when the $H^+$ is heavy and signal events are kinematically quite different from typical background events, but at lower masses the discrimination becomes weaker. The final discriminant is the output score of the neural network. The $H^+$ mass range explored is 200--2000~\GeV. Above an $H^+$ mass of about 1~\TeV the jets tend to merge, resulting in a loss of acceptance which could be recovered in a dedicated boosted analysis with large-radius jets.

\paragraph{$H^+\to cb$}\mbox{}\\
The search for $H^+\to cb$~\cite{HDBS-2019-24} focuses on $H^+$ production in $t\bar{t}$ decays, in a mass range of 60--160~\GeV. One top quark decays into $H^+b$, and the other top quark is required to decay into $Wb$ and then to a lepton, which is used to select such events via single-lepton triggers. The leptons must have a \pt greater than 27~\GeV. To enter the signal regions, the events must also contain at least four jets, and at least three are required to be $b$-tagged. The fourth jet is assumed to come from the hadronization of the $c$-quark, although no dedicated $c$-tagger is used. The $b$-tagging uses the \emph{DL1r} algorithm and four working points corresponding to different $b$-tagging efficiencies, which is also called pseudo-continuous $b$-tagging. The events are eventually grouped into six signal-enriched categories: 4j3b, 5j3b, 6j3b, 4j4b, 5j$\geq$4b and 6j$\geq$4b. The final discriminant is the output score of a neural network in regions with exactly three $b$-jets, and the total event yield is used in regions with four or more $b$-jets.

\subsubsection{Charged Higgs bosons decaying into bosons}
\label{sec:chargedhiggs:single_bosons}

The charged Higgs boson can also decay into bosons. This was investigated by ATLAS in the $Wa$~\cite{HDBS-2020-12} and $WZ$~\cite{HDBS-2018-19} final states. The decay into $Wa$ requires the presence of another scalar ($a$), which is assumed to be light in that ATLAS search. Such a signature is typical for 2HDM+$a$ scenarios. More generally, the decay of the charged Higgs boson to final states with a $W$ boson and a neutral Higgs boson can have sizeable branching fractions in large phase-space regions of the 2HDM.\\
The $WZ$ decay channel was explored in the GM model, which predicts a fermiophobic charged Higgs boson with large couplings to vector bosons. This enables the production of the charged Higgs boson through VBF, and its subsequent decay into vector bosons, displayed in Figure~\ref{fig:chargedhiggs:single_bosons:feynmans}. In the 2HDM, this process is strongly suppressed.

\begin{figure}[tb!]
\centering
\includegraphics[width=0.4\textwidth]{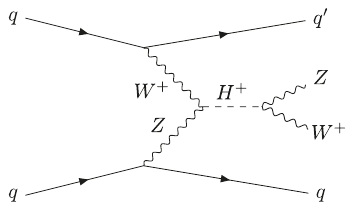}
\caption{\label{fig:chargedhiggs:single_bosons:feynmans}Illustrative Feynman diagram for the VBF production of a charged Higgs boson with subsequent decay into vector bosons.}
\end{figure}

\paragraph{$H^+\to W^+a$}\mbox{}\\
The search for $H^+\to W^+a$~\cite{HDBS-2020-12} considers light $H^+$ mass hypotheses in the range of 120--160~\GeV and also assumes the existence of a new light scalar $a$ that has a mass of 15--72~\GeV and decays into $\mu^+\mu^-$. This is a specific choice made in the analysis; the $a$-boson could decay differently, but the dimuon channel is favoured because it allows the mass to be reconstructed with very good resolution. The $H^+$ is produced in $t\bar{t}$ decays. At least one of the top quarks is required to decay to a lepton, which can then be used for triggering. The minimum \pt for the electron (muon) is 27 (10)~\GeV. Furthermore, the selected events must contain two opposite-charge muons that have an invariant mass consistent with that of the hypothesized $a$-boson. This leads to two signal-enriched categories: $e\mu\mu$ or $\mu\mu\mu$. Events are also required to contain at least three jets, one of them $b$-tagged. The final discriminant is the invariant mass of the two muons used to reconstruct the $a$-boson.

\paragraph{$H^+\to W^+Z$}\mbox{}\\
The search for $H^+\to W^+Z$~\cite{HDBS-2018-19} is tailored to the GM model, in which the charged Higgs boson of the custodial five-plet ($H_5^+$) is fermiophobic and therefore has enhanced couplings to vector bosons. The $W$ and $Z$ bosons are assumed to decay leptonically. Events are selected using single-lepton triggers, and they are required to contain three leptons. The \pt of each lepton must be at least 25~\GeV, and this requirement is raised to 27~\GeV for the trigger-matched lepton. A $W$ and a $Z$ candidate are reconstructed from the leptons and the \met (since the leptonic decay of the $W$ boson leads to a neutrino). To enrich the selection further in the signal process, events must contain two jets that are consistent with a typical VBF signature (large invariant mass of the dijet system and a high output score from a neural network that was trained on this signature). There is only one signal region, but two control regions are included in the fit to constrain the main backgrounds ($ZZ$ and QCD-produced $WZ$) using data. The final discriminant is the invariant mass of the $WZ$ system. The $H^+$ mass range explored in the search is 200--1000~\GeV.

\subsubsection{Doubly charged Higgs bosons}
\label{sec:chargedhiggs:double}

Some non-minimal BSM scenarios with an extended Higgs sector predict the existence of doubly charged Higgs bosons. ATLAS conducted searches for $H^{++}$ in two channels: $H^{++}\to W^+W^+$~\cite{HDBS-2019-06,STDM-2018-32} and $H^{++}\to\ell\ell$~\cite{EXOT-2018-34}. These decays complement each other nicely, as either one or the other usually has a large branching fraction. The Drell--Yan process would produce $H^{++}$ bosons in pairs; another possibility is production of one $H^{++}$ in association with an $H^+$, and a single $H^{++}$ could also be created via vector-boson fusion. These production modes are displayed in Figure~\ref{fig:chargedhiggs:doubly:feynmans}.

\begin{figure}[tb!]
\centering
\subfloat[]{
\includegraphics[width=0.32\textwidth,valign=b]{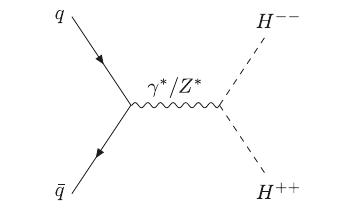}
}
\subfloat[]{
\includegraphics[width=0.32\textwidth,valign=b]{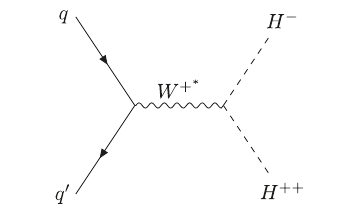}
}
\subfloat[]{
\includegraphics[width=0.32\textwidth,valign=b]{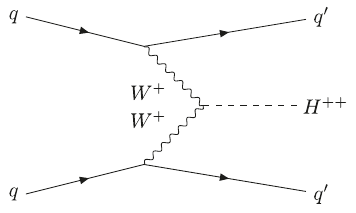}
}
\caption{\label{fig:chargedhiggs:doubly:feynmans}Illustrative Feynman diagrams for the production modes explored in ATLAS searches for doubly charged Higgs bosons:\ (a) pair production via the Drell--Yan process, (b) associated production with an $H^+$, and (c) production via $W$-boson fusion.}
\end{figure}

\paragraph{$H^{++}\to W^+W^+$}\mbox{}\\
A search for $H^{++}\to W^+W^+$~\cite{HDBS-2019-06} motivated by type-II seesaw models considered pair and associated production. In the case of pair production, both $H^{++}$ are assumed to decay into two $W$ bosons, while for associated production the $H^{++}$ is assumed to decay into $W^+W^+$, and the $H^+$ into $W^+Z$. Events are selected using single-lepton triggers. Six signal-enriched categories are defined: $2\ell$SS (with same-sign charge), which is further split into $ee$, $\mu\mu$ and $e\mu$; $3\ell$, which is split into subcategories with or without a same-flavour opposite-charge lepton pair; and $4\ell$, which has too few events to be split. The leptons must each have $\pt > 20$~\GeV, and the lepton that fired the trigger must have $\pt > 26$~\GeV. Requirements that depend on the $H^{++}$ mass are placed on the \met to further enrich the selections in potential signal events. The backgrounds are validated in regions with \met selections looser than those used to define the signal regions. The final discriminant is the event yield in each category. The $H^{++}$ mass range explored is 200--600~\GeV.\\
In a different analysis the results of a SM measurement of vector-boson scattering are reinterpreted to set limits on the VBF production of $H^{++}\to W^+W^+$ in the mass range of 200--3000~\GeV~\cite{STDM-2018-32}. This analysis selects events that have two same-sign leptons with $\pt > 27$~\GeV, $\met > 30$~\GeV, and two high-\pt jets consistent with VBF production. The final fit is performed on a 2D discriminant of the dijet mass versus the transverse mass of the dilepton and \met system.

\paragraph{$H^{++}\to \ell^+\ell^+$}\mbox{}\\
The search for $H^{++}\to \ell^+\ell^+$~\cite{EXOT-2018-34} investigates pair production of $H^{++}$, with each $H^{++}$ decaying into a pair of leptons ($e$, $\mu$ or $\tau$) with the same charge. The same-charge requirement gives a striking signature that is not expected in most other processes. Lepton-flavour-violating decays are allowed and considered. Events are selected with two-lepton triggers. Final states with two, three or four light leptons (electrons or muons) are selected, meaning the only $\tau$ decays included are leptonic. Five signal-enriched categories are constructed: $ee$, $e\mu$, $\mu\mu$, $3\ell$ and $4\ell$. The $2\ell$ and $3\ell$ categories must contain one same-sign lepton pair, and the $4\ell$ category must contain two such pairs. The leading same-sign lepton pair must have an invariant mass of at least 300~\GeV. This pair's mass is the final discriminant in the $2\ell$ and $3\ell$ regions, whereas the $4\ell$ region has far fewer events and the final fit uses the event yield instead. The mass range explored is 400--1300~\GeV.


\subsection{Searches for additional scalars decaying into Higgs boson pairs}
\label{sec:dihiggs}
Enhanced non-resonant or resonant Higgs boson pair production, $HH$, is predicted in many BSM theories, whereas in the SM, the Higgs boson pair-production cross-section is too low to be currently observable at the LHC~\cite{HIGG-2016-20, Glover:1988}. It is several orders of magnitude smaller than the single-Higgs-boson production cross-section~\cite{deFlorian:2016spz}. Searches for BSM physics in the di-Higgs sector are nevertheless very relevant.

Modifications to the non-resonant $HH$ cross-section occur in BSM scenarios with new, light, coloured scalars~\cite{Kribs_2012}, in composite Higgs models~\cite{Groeber_2011}, in theoretical scenarios with couplings between pairs of top quarks and pairs of Higgs bosons~\cite{Contino_2012}, and in models with a modified coupling of the Higgs boson to the top quark. A description of this is beyond the scope of this review.

Sources of $HH$ resonances include heavy Higgs bosons from extended Higgs sectors such as those in 2HDMs~\cite{Branco_2012}, the MSSM~\cite{Bagnaschi_2019, Djouadi_2013}, twin Higgs models~\cite{Chacko_2006}, and composite Higgs models~\cite{Groeber_2011,Mrazek_2011}. Heavy resonances that decay into pairs of Higgs bosons also include spin-0 radions and spin-2 gravitons from the Randall-Sundrum model~\cite{Randall_1999, Tang:2012pv, Cheung_2001}, and stoponium states in supersymmetric models~\cite{Kumar_2014}. The results from searches for spin-2 resonances decaying into $HH$ are not covered here.

The rich diversity of decay channels and final states presented here is important not only because it enhances the discovery potential, but also because it helps us understand the sensitivity of various decay channels for observing Higgs boson pair production, and improves the projections in studies for future high-luminosity and high-energy colliders. Now that experience has been accumulated regarding detector performance and cut-based analyses, there is increased usage of modern machine-learning techniques, which enhance the performance of many aspects of the searches.

For consistency with the cited publications, the 125-GeV Higgs boson is now denoted $H$ in the context of Di-Higgs searches.

\subsubsection{Resonant $HH$}
\label{sec:dihiggs:fermions}
This subsection covers ATLAS searches for resonant Higgs boson pair production (via a new scalar) in the ggF $HH\to\bbbar\bbbar$, VBF $HH\to\bbbar\bbbar$, $HH\to\bbbar\tautau$, $HH\to\bbbar\tautau$ (boosted) and $HH\to\bbbar\gamma\gamma$ topologies. Both the ggF and VBF production modes are used here, and are therefore displayed in Figure~\ref{fig:dihiggs:ggf:VBF}, where $X$ is a generic
spin-0 boson, also sometimes referred to as an additional scalar.
Then follows a full \RunTwo combination of the channels with these final states, based only on the ggF production mode because it is the strongest.

\begin{figure}[tb!]
\centering
\subfloat[]{
\includegraphics[width=0.4\textwidth,valign=c]{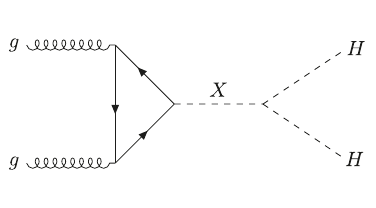}
}
\subfloat[]{
\includegraphics[width=0.4\textwidth,valign=c]{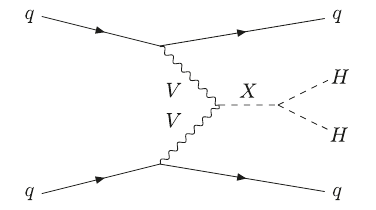}
}
\caption{\label{fig:dihiggs:ggf:VBF}Illustrative Feynman diagrams for the production modes explored in ATLAS searches
for resonant Higgs boson pairs:\ (a) gluon--gluon fusion (ggF) and (b) vector-boson fusion (VBF). $X$ is a generic
spin-0 boson, also sometimes referred to as an additional scalar.}
\end{figure}

\paragraph{ggF $HH\to\bbbar\bbbar$}\mbox{}\\
This is a search for a new boson which would be revealed by resonant pair production of SM Higgs bosons via gluon--gluon fusion, using the $\bbbar\bbbar$ final state and 126--139~\ifb of data~\cite{HDBS-2018-41}. The analysis uses two separate complementary channels, resolved and boosted, to target the 250~\GeV to 1.5~\TeV and  900~\GeV to 5~\TeV mass ranges, respectively. In the former the four $b$-quark jets are reconstructed individually, and in the latter they are paired in two large-radius jets. Large-radius jets are not $b$-tagged directly; instead, track-jets are matched to the large-radius jets and are then subjected to $b$-tagging algorithms. The four-momenta of the two SM Higgs boson candidates are rescaled so that each conforms to a mass of 125~\GeV, and the final discriminant is the \enquote{corrected $m(HH)$}, the invariant mass obtained from the sum of these rescaled four-momenta. This search improves on previous versions, with machine-learning enhancements in several areas, an extended search range and better $b$-tagging.

\paragraph{VBF $HH\to\bbbar\bbbar$}\mbox{}\\
This is a generic inclusive search for a resonance with a mass $m_X$ of 260--1000~\GeV, using 126~\ifb of data. It is based on VBF Higgs boson pair production, characterized by a large rapidity gap between the two jets from the quarks radiating the vector bosons, and uses the dominant $\bbbar\bbbar$ decay mode of the two Higgs bosons~\cite{HDBS-2018-18}. The analysis has two signal classes. Firstly, a broad resonance (width $\approx$ 10\%--20\%) is considered, corresponding to a heavy scalar of a type-II 2HDM~\cite{Branco_2012}. Secondly, a narrow resonance with a fixed generated width of 4~\MeV is used as a signal benchmark. Events with four central $b$-tagged jets with $\pt>40$~\GeV and at least two forward jets with $\pt>30$~\GeV are selected. The $b$-jet pairing optimizes the formation of dijets compatible with Higgs boson decays, taking into account the expected correlation of the four-$b$-jet invariant mass with the Lorentz boost of the Higgs bosons and thus with the angular separation of the two $b$-jets in each dijet. A machine-learning approach corrects the $b$-jet energies for effects not considered in the default calibration. As a result, the $H \to \bbbar$ mass peak is closer to 125~\GeV, and the resolution for a simulated 600~\GeV resonance improves by about 25\%. Several additional requirements suppress sources of dijet background. The final discriminant is the four-$b$-jet invariant mass.

\paragraph{$HH\to\bbbar\tautau$}\mbox{}\\
This analysis searches for resonant $HH$ production, using 139~\ifb of data, in final states with two opposite-sign $\tau$-leptons and two $b$-jets, with at least one \tauhad candidate~\cite{HDBS-2018-40}. The search targets a narrow resonance with $m_X$ in the range 251--1600~\GeV. To be selected, events must contain signatures consistent with the visible part of a \tauhad decay (\tauhadvis{}) as well as \met appropriate for the neutrinos from the decays of the $\tau$-leptons and $b$-hadrons. The analysis is split into three categories based on the trigger and the $\tau$-lepton decay mode. Events in the \tauhad{}\tauhad category are selected using a combination of single-\tauhadvis triggers, with a \pt threshold of at least 80~\GeV, and di-\tauhadvis triggers, with a \pt requirement of 35 (25)~\GeV for the leading (subleading) \tauhadvis. Events in the \taulep{}\tauhad categories were recorded using a combination of single-lepton triggers (which defines one category) or lepton-plus-\tauhadvis triggers (used for the other category). The lepton's \pt requirements depend on the lepton's flavour and the data-taking period and are in the range of 14--26~\GeV. Multivariate discriminants are used to reject the backgrounds; the most important variables entering these discriminants are the reconstructed Higgs boson masses and the di-Higgs mass. The final fits are performed on the output scores of these discriminants.

\paragraph{$HH\to\bbbar\tautau$ (boosted)}\mbox{}\\
This search uses 139~\ifb of data and it investigates resonant $HH$ production in final states with two opposite-sign hadronically decaying $\tau$-leptons and two $b$-jets. In this incarnation of the search, a new technique for reconstructing and identifying $\tau$ pairs with a large boost is exploited~\cite{HDBS-2019-22}. The Higgs boson pair is produced by ggF, and the search focuses on a \enquote{heavy}, narrow, scalar resonance in the mass range 1--3~\TeV. Reference~\cite{HDBS-2019-22} includes a dedicated benchmark of the new di-$\tau$ tagger as well as its performance in terms of rejection rate for light-quark- and gluon-initiated jets misidentified as di-$\tau$ objects using multivariate techniques. Events are selected if they satisfy a rather high trigger \et threshold (at least 360~\GeV) for a large-radius jet. For orthogonality with $HH\to\bbbar\bbbar$, events with more than one large-radius $b$-tagged jet are vetoed. The discriminating variable is the Higgs boson pair mass.

\paragraph{$HH\to\bbbar\gamma\gamma$}\mbox{}\\
Searches are performed for resonant Higgs boson pair production via ggF and VBF in the $\bbbar\gamma\gamma$ final state with 139~\ifb of data~\cite{HDBS-2018-34}. The probed range for a narrow resonance is $251 \leq m_X \leq 1000$~\GeV. The analysis employs multivariate methods for the photon energy calibration, the $b$-tagging of jets, and the rejection of background processes. In the context of the resonant search, non-resonant $HH$ production is considered as a background. Events are selected if they have exactly two $b$-jets, at least two photons, and the two leading photons have an invariant mass within a window around the SM Higgs boson mass. Other criteria reject possible backgrounds. The photon identification is particularly stringent and the latest high-performance $b$-jet tagger \emph{DL1r} is used~\cite{FTAG-2018-01}. Two BDTs are trained to discriminate between the di-Higgs signal and either the non-resonant backgrounds or the resonant single-Higgs-boson backgrounds, and the combined BDT score is used to define two signal-enriched categories. The reconstructed resonance mass, $m_{bb\gamma\gamma}$, is used in the BDT training. The final discriminant is the diphoton invariant mass $m_{\gamma\gamma}$.

\paragraph{$\text{Combination for } HH$}\mbox{}\\
A combination of searches for resonant Higgs boson pair production was performed recently~\cite{HDBS-2023-17}, using the $\bbbar\bbbar$, $\bbbar\tautau$ and $\bbbar\gamma\gamma$ final states and up to 139~\ifb of data. The gluon--gluon fusion production mode is considered because it is typically the dominant production mechanism for heavy Higgs bosons. The tested mass range for the new scalar resonance is $251~\GeV \leq m_X \leq 5~\TeV$. Compared to other channels, the three included decay modes provide better sensitivity to Higgs boson pair production because they either have large branching fractions or are easy to discriminate from background processes. The previous combination of searches for resonant Higgs boson pair production~\cite{HDBS-2018-58} covered six channels but it was performed for up to 36.1~\ifb of data.

\subsubsection{Resonant $HH/SH/SS$ decaying into $W$ bosons}
\label{sec:dihiggs:bosons}
This section covers ATLAS searches for resonant Higgs boson or heavy scalar pair production in the following topologies: $HH/SS\to WW^* WW^*$, $HH\to\bbbar WW^*$, $HH\to\gamma\gamma WW^*$ and $SH\to W^+W^-\tautau$.

\paragraph{$HH/SS\to WW^* WW^*$}\mbox{}\\
This search uses 36~\ifb of data and covers either resonant Higgs boson pair production or resonant production of a pair of heavy scalar particles ($S$)~\cite{HIGG-2016-24}. The final states considered have either two same-sign leptons, three leptons or four leptons, which defines three categories:
\begin{align*}
WW^* WW^* &\to \ell\nu + \ell\nu + 4q,\\
WW^* WW^* &\to \ell\nu + \ell\nu + \ell\nu + 2q,\\
WW^* WW^* &\to \ell\nu + \ell\nu + \ell\nu + \ell\nu.
\end{align*}
The search range is $135 \leq m_S \leq 340$~\GeV for the pair of new scalars and $260 \leq m_X \leq 500$~\GeV for the scalar progenitor of the pair of Higgs bosons or new scalars. The production modes considered are ggF, VBF and production in association with a $W$ or $Z$ boson or a top-quark pair. All scalars are assumed to be on-shell and the heavy resonance decays mostly into the pair of scalars for the models considered. Electrons are identified using medium (tight) criteria~\cite{EGAM-2018-01} for the four-lepton channel (two- and three-lepton channels) and must have $\pt > 10$~\GeV. Muons are identified using medium (tight) criteria~\cite{MUON-2018-03} for the four-lepton channel (two- and three-lepton channels) and must have $\pt > 10$~\GeV. Jets are required to have $\pt > 25$~\GeV and events containing $b$-jets are vetoed to make this search orthogonal to similar searches with final states involving at least one $H\to b\bar{b}$ decay. A final selection uses a multivariate procedure which includes several kinematic variables depending on the category. These selections optimize the discriminating power for the signal across a set of benchmark mass points for both $X$ and $S$.

\paragraph{$HH\to\bbbar WW^*$}\mbox{}\\
Searches are performed for resonant Higgs boson pair production in the $\bbbar WW^*$ decay mode and the $\bbbar\ell\nu qq$ final state with 36~\ifb of data~\cite{HIGG-2016-27}. The probed mass range for the scalar hypotheses is $500 \leq m_X \leq 3000$~\GeV. The $H\to WW^*$ branching fraction is the second largest after $H\to\bbbar$, so the $\bbbar WW^*$ final state is important provided that the signal can be well separated from the dominant $\ttbar$ background. The resonant search uses two complementary procedures to reconstruct the Higgs boson that decays into $\bbbar$, these being \enquote{resolved} and \enquote{boosted}, where for the latter case, a single large-radius jet represents the $\bbbar$ pair. The detector signature is one charged lepton $(e/\mu)$, large \met, and four or more jets (exactly two of which are $b$-tagged and at least two are non-$b$-tagged). For events to be selected, electrons must be isolated and satisfy tight identification requirements, both electrons and muons must have $\pt>27$~\GeV, resolved jets must have $\pt > 20$~\GeV and boosted jets must have $\pt > 250$~\GeV. The dominant $\ttbar$  backgrounds are constrained by using several data control regions. The final discriminant is the reconstructed resonance mass $m_{HH}$.

\paragraph{$HH\to\gamma\gamma WW^*$}\mbox{}\\
This search for resonant Higgs boson pair production probes the $\gamma\gamma WW^*$ channel and the $\gamma\gamma\ell\nu jj$ final state, using 36~\ifb of data~\cite{HIGG-2016-20}. The mass range searched for a narrow-width resonance is $260 \leq m_X \leq 500$~\GeV. This search benefits from the large $H\to WW^*$ branching fraction, the clear signature from the photon pair and lepton, and good resolution for the diphoton invariant mass $m_{\gamma\gamma}$. The selection includes $p_{\text{T}}^{\gamma\gamma} > 100$~\GeV, except for $m_X < 400$~\GeV, and $m_{\gamma\gamma}$ must lie within a window around the Higgs boson mass. A selection on \met was found to not add sensitivity. The final discriminant is $m_{\gamma\gamma}$.

\paragraph{$X\to SH\to  VV\tautau$}\mbox{}\\
This search targets a new heavy scalar particle $X$ decaying into a SM Higgs boson and a new singlet scalar particle $S$, where the Higgs boson decays into $\tau$-leptons,  $H\to\tautau$, and the new scalar decays into vector bosons, $S\to VV$ (with $V$ being $W$ or $Z$)~\cite{HDBS-2022-44}. One of these vector bosons decays to one or two light leptons and the other decays hadronically or into neutrinos. The search uses 140~\ifb of data. Figure~\ref{fig:dihiggs:XSH} displays a representative Feynman diagram. The mass range for the progenitor scalar is  $500 \leq m_X \leq 1500$~\GeV while that of the new scalar (with decay branching fractions as expected in the SM) is $200 \leq m_S \leq 500$~\GeV. Events are selected if they have two hadronically decaying $\tau$-leptons, and one or two light leptons $(\ell = e,\mu)$ from the vector bosons. The signal purity is enhanced by using multivariate techniques based on BDTs exploiting the different kinematic distributions of the signal and background. The final fit is performed using both the reconstructed $m_X$ and $m_S$ in a 2D discriminant.

\begin{figure}[tb!]
\centering
\includegraphics[width=0.4\textwidth,valign=c]{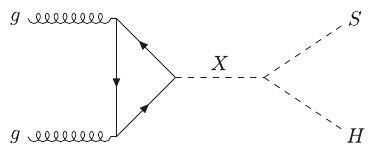}
\caption{\label{fig:dihiggs:XSH}Illustrative Feynman diagram that contributes to $X\to SH$ production via the gluon--gluon fusion process.}
\end{figure}

\paragraph{$X\to SH\to \bbbar\gamma\gamma$}\mbox{}\\
This search considers an extended BSM Higgs sector and targets a new heavy scalar $X$ decaying into a Higgs boson and a new lighter scalar $S$, given by $X\to S(\bbbar)H(\gamma\gamma)$, and uses 140~\ifb of data~\cite{HDBS-2021-17}. The search ranges are $170 \leq m_X \leq 1000$~\GeV and $15 \leq m_S \leq 500$~\GeV. The signature includes a diphoton, with a reconstructed mass corresponding to a SM $h_{125}$ Higgs boson, and two $b$-tagged jets. The new boson's width is assumed to be narrow. The final discriminant uses parameterized neural networks which enhance the signal purity and provide sensitivity across the explored region of the $(m_X, m_S)$ plane.


\subsection{Searches for additional scalars and exotic decays of the Higgs boson}
\label{sec:exoticdecays}
Many of the theoretical models proposed in Section~\ref{sec:theory} motivate the search for exotic Higgs boson decays, reviewed for  example in Ref.~\cite{Cepeda_2022}. These include models addressing the baryon asymmetry of the universe~\cite{Mukaida_2022}, the naturalness problem~\cite{craig2022naturalness}, the nature of dark matter~\cite{Arcadi_2020} or the anomalous magnetic moment of the muon~\cite{Iguro_2019}. The Higgs boson has a particularly narrow width, so its branching fraction to BSM particles via exotic decays could be sizeable. The next subsection discusses analyses to constrain the branching fraction for Higgs boson decays into invisible particles, and the two following subsections explore the suite of ATLAS \RunTwo searches probing exotic and rare decays of the Higgs boson.

\subsubsection{Exotic decays of the Higgs boson to invisible final states}
\label{sec:ex-higgs:inv}
Many BSM theories predict invisible decays of the Higgs boson.
The SM Higgs boson can decay invisibly to neutrino final states, but the branching fraction is very small (0.1\%),
leading to good sensitivity for this invisible BSM channel.
As dark matter (DM) particles must be massive, they could couple strongly to the Higgs boson, making it an
important portal for their discovery.
Measuring the branching fraction for Higgs boson decays into invisible
particles, $\mathcal{B}(H \to \text{invisible})$, can lead to constraints on the dark sector.
This section also considers the exotic semi-visible Higgs boson decay
via a SM photon and a dark photon, which can be massless or massive.
Although the three $H \to \text{invisible}$ searches described here have a progenitor Higgs boson at $
m_H = 125$~\GeV, they can also constrain other possible BSM scalar bosons.
Searching for decays of the Higgs boson to
(semi-) invisible final states is a powerful way to probe BSM physics.

\paragraph{$ZH$, $H\to$ invisible}\mbox{}\\
Using Higgs boson production in association with a $Z$ boson, a search was performed~\cite{HIGG-2018-26} for an excess of events in the invisible decay channel of the Higgs boson. In this search, events must have two leptons ($2e$ or $2\mu$) with $p^{\ell}_{\text{T}} >20, 30$~\GeV that form a $Z$ boson, have $\met >90~\GeV$, and satisfy other criteria. The final discriminant for the invisible decays of the Higgs boson is based on the output of a BDT. For the model-dependent searches (simplified DM models and the 2HDM+$a$) the final discriminant is the transverse mass calculated from the \met and the di-lepton system.

\paragraph{VBF, $H\to$ invisible}\mbox{}\\
This search probes the SM Higgs boson as well as possible additional BSM scalar bosons, produced mostly via vector-boson fusion, that are all envisaged to subsequently decay into invisible particles~\cite{EXOT-2020-11}. The scalar boson mass range is $50~\GeV \leq m_X \leq 2~\TeV$. This is an improved version of previous similar analyses~\cite{EXOT-2016-37}, not only with more data, but also with a refined analysis strategy, finer binning in the dijet mass, new binning in dijet azimuthal separation as well as in jet multiplicity, and it includes new methods to improve the evaluation of the backgrounds. The selection for the VBF production mode was improved to require events to have two energetic jets which exhibit a large rapidity separation and a large dijet invariant mass, and to contain sizeable \met. Events with additional jets are included only if the additional jets were shown to arise from final-state radiation. The event selection in the signal region relies on a number of topological and kinematical requirements that are used to define 16 signal-enriched regions (SRs). To constrain the dominant backgrounds, which are $Z\to\ell\ell$ and $W\to\ell\nu$, 16 corresponding control regions are constructed to be kinematically similar to the SRs but have a different requirement on \met in order to enrich these regions in background events. The signal and control regions are then subjected to maximum-likelihood fits to extract limits on the branching fraction of $H\to$ \text{invisible}.

\paragraph{Combination of searches for $H\to$ invisible}\mbox{}\\
A statistical combination of six searches for $H\to$ invisible decays across several production modes of the SM Higgs boson yields the most sensitive direct constraint on its invisible decays yet obtained by ATLAS~\cite{HIGG-2021-05}. The five Higgs boson production modes are VBF\,+\,\met,  $Z\to\ell\ell + \met$, $\ttbar + \met$, VBF\,+\,\met\,+\,$\gamma$, and jet\,+\,\met, all using the full \RunTwo dataset (139~\ifb). The $H\to$ invisible combination from  \RunOne is also included. The statistical combination of the analyses is performed by constructing the product of their respective likelihood functions and maximizing the resulting profile likelihood ratio, with careful treatment of possible correlations between uncertainties.

\paragraph{$ZH$, $H\to\gamma\gamma_d$}\mbox{}\\
This is a search for a dark photon~\cite{HDBS-2019-13} using 139~\ifb of data. The dark photon is the boson of the $U(1)_D$ gauge group of the dark sector. This channel is often referred to as the \emph{vector portal} because there is kinetic mixing between the Abelian gauge field of the SM and that of the dark sector. The mixing enables the exotic decay of the SM Higgs boson, $H\to\gamma\gamma_d$, which is probed in this analysis, assuming associated production of $ZH$ with $Z\to\ell^+\ell^-$ ($\ell = e,\mu$). The signature consists of a photon with an energy $E_{\gamma} =m_H /2$ in the Higgs boson's centre-of-mass frame and a similar amount of \met which originates from the escaping $\gamma_d$. The leptons are used for triggering and provide a $Z$-boson mass constraint. The dominant reducible background processes are estimated using data-driven techniques. A BDT technique is adopted to enhance the sensitivity of the search. The transverse mass $m_{\text{T}}$ of the $\gamma{-}\met$ system has a distinctive kinematic edge at the Higgs boson mass and is used as the final discriminant.

\subsubsection{Exotic decays of the Higgs boson or a heavy scalar into (pseudo)scalars or vector bosons}
\label{sec:ex-higgs:aa-XX}
This section covers ATLAS searches for exotic decays of the Higgs boson into (pseudo)scalars or vector bosons in the intermediate state, including the searches for $H\to XX/ZX \to 4\ell$, $H\to aa \to \bbbar\mumu$, $H\to aa \to (\bbbar)(\bbbar)$, $H \to aa \to \gamma\gamma jj$, $H \to aa \to 4\gamma $ and $\Phi \to SS \to \text{LLP}$, where $SS \to \text{LLP}$ denotes decays of scalar long-lived particles (LLP). Searches for $H \to Za \to\ellell + \text{jet}$, $H \to Za \to\ellell + \gamma\gamma$ and $H \to \chi_1\chi_2$ are also covered in this section.

\paragraph{$H\to XX/ZX \to 4\ell$}\mbox{}\\
This search targets a new spin-0 or spin-1 boson from the exotic decay of a SM Higgs boson into four leptons $(\ell = e,\mu)$~\cite{HDBS-2018-55}. The intermediate state contains two on-shell, promptly decaying bosons $H\to XX/ZX \to 4\ell$, where the new boson $X$ is in the mass range $1 \leq m_X \leq 60~\GeV$. The $X$ boson can be the dark vector boson $Z_d$ of the Hidden Abelian Higgs Model (HAHM)~\cite{Gopalakrishna_2008,Curtin_2014} or the scalar/pseudoscalar of the 2HDM+S or 2HDM+$a$, respectively~\cite{Curtin_2014}. Figure~\ref{fig:zd-diagrams} shows the relevant Feynman diagrams for the three analyses in Ref.~\cite{HDBS-2018-55}, where $ZZ_d$ probes the hypercharge portal, $Z_dZ_d$ explores the Higgs portal for the dark vector boson, and $aa/ss$ couple to SM particles through mixing with the SM Higgs field in models with an extended Higgs sector. In this last case, the selected final-state leptons are exclusively muons. The three analyses share a common event preselection but differ in the subsequent steps of selecting the candidate final-state leptons, assigning them to quadruplets, selecting one of those quadruplets, and placing further requirements on the selected quadruplet. In the $Z_dZ_d$ case, the analysis is somewhat inspired by the related SM Higgs boson discovery channel~\cite{HIGG-2016-22}, with some important differences, e.g.\ the similar invariant masses of dilepton pairs that each form a $Z_d$ must not be compatible with the SM $Z$ boson. The final discriminant is the reconstructed $Z_d$ mass in the range $15  \leq m_X \leq 60~\GeV$. The $4\mu$-final-state analysis extends the search region to cover $1 \leq m_X \leq 15$~\GeV, again with subtly different event selections. The $ZX$ analysis looks for an excess in what would be the off-shell boson's mass distribution in the ATLAS SM $H\to ZZ^*\to 4\ell$ analysis~\cite{HIGG-2016-22}; however, there are differences in the dilepton cuts, alternative pairing requirements and quadruplet ranking, among others.

\begin{figure}[tb!]
\begin{center}
\subfloat[]{\includegraphics[width=0.32\textwidth]{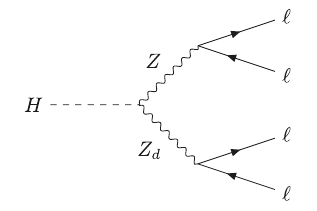}\label{fig:zx-diag}}
\subfloat[]{\includegraphics[width=0.32\textwidth]{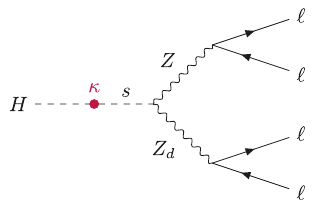}\label{fig:zd-diag}}
\subfloat[]{\includegraphics[width=0.32\textwidth]{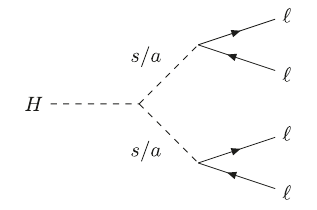}\label{fig:aa-diag}}
\caption{Illustrative Feynman diagrams for exotic decays of the Higgs~boson into four leptons induced by intermediate dark vector bosons via
(a)~the hypercharge portal where $\epsilon$ measures the hypercharge kinetic mixing and
(b)~the Higgs portal, where $s$ is a dark Higgs boson and $\kappa$ measures the Higgs boson coupling,
and (c)~is the Higgs~boson decay via scalars $s$ or pseudoscalars $a$.}
\label{fig:zd-diagrams}
\end{center}
\end{figure}

\paragraph{$H\to aa \to \bbbar\mumu$}\mbox{}\\
ATLAS has searched for the exotic decay of the Higgs boson into a pair of light new BSM pseudoscalar mediator particles, $aa$, where one of them decays into a $b$-quark pair and the other one onto a muon pair~\cite{HDBS-2021-03}. This search uses the full luminosity of 139~\ifb, and also introduces machine-learning techniques to suppress the backgrounds and enhance the search sensitivity. The final discriminant is the dimuon invariant mass, and its very clean spectrum is searched for a narrow resonance with a mass between 16 and 62~\GeV. The pseudoscalar couplings are assumed to be proportional to mass, so the rare but clean $a\to\mumu$ decay is balanced by the more probable $a\to\bbbar$ decay. This analysis has particular sensitivity  to scenarios where there are enhanced lepton couplings~\cite{Curtin2015}, in which case $\mathcal{B}(a\to\mumu)$ may also be large. The LLP version of this search is described in Section~9.1.1 of a companion report~\cite{EXOT-2023-14}.

\paragraph{$H\to aa \to (\bbbar)(\bbbar)$}\mbox{}\\
This search probes BSM exotic decays of the Higgs boson into a pair of new spin-0 particles which then each decay into a pair of $b$-quarks~\cite{HDBS-2018-47}. The mass range explored for the possible new boson is $15  \leq m_a \leq 30~\GeV$, where a large boost can lead to collimated decay products. In this case, a specific analysis strategy must be used because the jets from the hadronization of separate $b$-quarks within a pair may merge into a single jet. A multivariate technique considers several jet characteristics in order to perform the di-$b$-quark selection. The search considers the production of a Higgs boson in association with a $Z$ boson, and uses 36~\ifb of data. It complements the previous search in the same final state performed with the same dataset in the range $20  \leq m_a \leq 60~\GeV$, where the decay products are usually well separated~\cite{HIGG-2017-05}. An additional feature of the new search is a novel strategy to treat the collimated $a \to \bbbar$ decays. The events must contain the dilepton decay ($\ell^+\ell^-, \ell = e/\mu$) of the $Z$ boson and the two merged $a \to \bbbar$ decays, all satisfying minimal kinematic requirements, and mass compatibility requirements are imposed on the $a$-bosons. The final discriminant is ${m_{\bbbar}}$, which allows $m_a$ to be estimated.

\paragraph{$H \to aa \to 4\gamma $}\mbox{}\\
This is a search for the exotic Higgs boson decay into two axion-like particles (ALPs) where each ALP decays into two photons~\cite{HDBS-2019-19}. The search is sensitive to recently proposed models that could explain the tension between theory and experiment for the anomalous magnetic moment of the muon. The probed mass range is from 100~\MeV to 62~\GeV and ALP--photon couplings $C_{a\gamma\gamma}$ are in the range $10^{-5}~\TeV{}^{-1} \leq C_{a\gamma\gamma}/\Lambda \leq 1~\TeV{}^{-1}$ (where $\Lambda$ is the new physics scale, assumed to be in the \TeV range). The signatures may therefore include significantly displaced $a\to\gamma\gamma$ decay vertices and highly collinear photons. The search uses 140~\ifb of data. An ALP mass of $m_a =3.5$~\GeV marks the transition from a boosted event topology to a resolved event topology. Neural network classifiers were trained to distinguish between single-photon and collimated-photon signatures in the boosted region. The prompt-decay and long-lived scenarios are separated at a coupling of $C_{a\gamma\gamma}/\Lambda \geq 0.1~\TeV{}^{-1}$, while small couplings $C_{a\gamma\gamma}/\Lambda \leq 10^{-5}~\TeV{}^{-1}$ imply that the ALP escapes the detector. A dedicated set of search strategies were developed to handle the various scenarios, including the long-lived $a \to \gamma\gamma$ decays, which are probed here for the first time. The signal region examines the reconstructed invariant mass of all photon candidates, $m^{\text{reco}}_{\text{inv}}$, which is expected to peak at the Higgs boson mass, while the background is estimated from the sidebands. The final discriminant is the ALP mass $m^{\text{reco}}_a$, where a narrow resonance is sought.

\paragraph{$H \to aa \to \gamma\gamma jj$}\mbox{}\\
This is the first search for a Higgs boson decaying into a pair of new (pseudo)scalar bosons where one of the new bosons decays into a pair of photons and the other decays into a pair of gluons~\cite{HIGG-2017-09}. It is complementary to the $H \to aa \to 4\gamma $ analysis, which is more sensitive when the new sector has enhanced photon couplings. The search envisages a new sector where the ratio of photon and gluon couplings to the $a$-boson is similar to the ratio of couplings to the SM Higgs boson. It is also applicable to the mixed $H \to aa' \to \gamma\gamma jj$ scenario. To enhance the sensitivity, the VBF Higgs boson production mode is selected by requiring two light-quark jets with a large opening angle and a large invariant mass, in addition to the jets which are decay products of the $a$-boson. The mass range explored for the $a$-boson is 20--60~\GeV and the search uses 36.7~\ifb of data. The two pairs of signal candidates (a diphoton and a dijet) form a Higgs boson candidate if their combined reconstructed invariant mass is in the range $100 \leq m_{\gamma\gamma jj} \leq 150~\GeV$. The final discriminant is the diphoton mass $m_{\gamma\gamma}$, which has excellent  resolution, and is the proxy for the $a$-boson mass $m_a$.

\paragraph{$\Phi \to SS \to \text{LLP} $}\mbox{}\\
This is a dedicated search for pair-produced neutral long-lived scalar particles, $SS$, where the progenitor is a Higgs boson or more generally an additional scalar boson~\cite{EXOT-2019-23}. The decay $\Phi \to SS \to \ffbar\ffbar$ has fermions in the final state. The mass range for the scalar boson progenitor $\Phi$ is 60--1000~\GeV and that for the LLP intermediate-state scalar bosons $S$ is  5--475~\GeV. The search uses 139~\ifb of data. In the hypothesized physics process where the scalar has the largest mass $m_S$ (475~\GeV), decays into top quarks dominate ($\mathcal{B} > 99$\%). Conversely, for the hypothesis with the lightest $m_S$ (5~\GeV), decays into charm quarks dominate ($\mathcal{B}  \approx 75$\%), followed by decays into $\tau$-leptons ($\mathcal{B}  \approx 25$\%). For the other $m_S$ hypotheses, the branching fractions are approximately constant and typically 85\%:8\%:5\% for $\bbbar$, $c\bar{c}$, and $\tau^+\tau^-$, respectively. The SM fermions from the LLP decay result in displaced jets, where the proper lifetime in distance units is $c\tau \, \in \, [20~\text{mm}, 10~\text{m}]$. The event selection in this analysis therefore requires two non-standard displaced jets. There are two selections, one optimized for $m_{\Phi} \leq 200~\GeV$ and the other for $m_{\Phi} \geq 200~\GeV$. The dominant background is SM multijet production. This search employs a new per-jet method to discriminate between signal-like displaced jets and the non-displaced jets or other background sources, with the help of a deep neural network using an adversarial training scheme (the first such deployment in ATLAS). Per-event boosted decision trees use the per-jet neural network scores and other event-level variables to select signal events. Finally, a data-driven ABCD method is applied to estimate the background. The statistical signal-hypothesis test is performed simultaneously with the data-driven background estimation in all regions of the ABCD plane, so that the amount of signal dynamically affects the background prediction. Further LLP searches are reviewed in a companion report in this journal~\cite{EXOT-2023-14}.

\paragraph{$H \to Za \to\ellell + \text{jet}$}\mbox{}\\
This search investigates Higgs boson decay into a $Z$ boson and a light bosonic resonance ($m_a < 4$~\GeV) or a charmonium state which further decay into the two-lepton\,+\,jet final state, using 139~\ifb of data~\cite{HDBS-2018-37}. The branching fractions $\mathcal{B}(H \to Za)$ and $\mathcal{B}(H \to aa)$ can be independent, and therefore searches for $H \to aa$ do not constrain $\mathcal{B}(H \to Za)$. By targeting the $H \to Za$, $a\to \text{hadrons}$ decay channel, this search accesses new, previously unexplored regions of the parameter space. Higgs boson candidates are reconstructed from the lepton-pair\,+\,jet system, which requires $m_{\ell\ell\text{jet}}$ to be compatible with the mass of the SM Higgs boson. With the goal of enhancing the signal, substantial progress was made in the use of jet-substructure techniques in boosted final states. Various jet-substructure variables are combined, using machine-learning techniques, to improve the reconstruction of a light, boosted, hadronic final state. The background is dominated by $Z + \text{jet}$ events and this is estimated with a modified data-driven ABCD method. The hadronic final states of the mesons in $H \to Z \eta_c$ and $H \to Z J/\Psi$ decays are included, but they have SM  branching fractions of $1 .4 \times 10^{-5}$ and $2.2 \times 10^{-6}$, respectively, and are therefore negligible. The final discriminant is the  $m_{\ell\ell\text{jet}}$ variable.

\paragraph{$H \to Za \to\ellell + \gamma\gamma$}\mbox{}\\
This search probes Higgs boson decay into a $Z$ boson and a light pseudoscalar particle ($0.1 < m_a <33~\GeV$) which further decay into a two-lepton\,+\,two-photon final state, using 139~\ifb of data~\cite{HDBS-2019-09}. Light pseudoscalars that couple to Higgs bosons appear in a wide range of BSM scenarios. As this is a relatively new search, it probes an unexplored parameter space for models with ALPs and extended scalar sectors. The photons are treated either as a single cluster ($m_a < 2$~\GeV) or as two resolved clusters. The dominant background is from SM $Z$-boson production in association with photons or jets. A SM Higgs boson compatibility requirement is placed on $m_{Z\gamma}$ or $m_{Z\gamma\gamma}$ for the merged or resolved categories, respectively. The statistical analysis relies on fitting the $m_{\gamma\gamma}$ and $\Delta R_{Z\gamma}$ distributions simultaneously. The latter variable is the angular isolation of the lepton--photon system.

\paragraph{$ZH$, $H \to \chi_1\chi_2$}\mbox{}\\
In this search, the Higgs boson is envisaged to decay into the two lightest neutralinos, $H \to \tilde{\chi}_1^0\tilde{\chi}_2^0$, in a NMSSM scenario~\cite{HDBS-2018-07}. The search also requires $\tilde{\chi}_2^0 \to a \tilde{\chi}_1^0$ with $a \to \bbbar$ where $a$ is the additional pseudoscalar in the NMSSM, assumed to be lighter than the SM Higgs boson. The analysis therefore focuses on the Peccei--Quinn symmetry limit of the NMSSM. The Higgs boson is produced in association with a $Z$ boson, and the search uses 139~\ifb of data. The final-state signature consists of the two oppositely charged leptons from the $Z$-boson decay (for the trigger), two or more jets with at least one from a $b$-quark, and \met from the two $\tilde{\chi}_1^0$ neutralinos. The search is performed for $m_a$ values between 20 and 65~\GeV, and for a few sets of fixed values of $m_{\tilde{\chi}_2^0}$ and $m_{\tilde{\chi}_1^0}$. The main SM backgrounds in this search are $Z$ bosons produced with heavy-flavour jets, and $\ttbar$ events. The final discriminant is the dijet invariant mass.

\subsubsection{Rare, exclusive Higgs boson decays}
\label{sec:ex-higgs:rare}
This section collects the searches for rare, exclusive decays of the Higgs boson. First described is the search for $H\to ee /e\mu$ and the search for the lepton-flavour-violating decays $H\to e\tau/\mu\tau$. Then follow searches for decays into a meson and a photon, $H \to \omega/K^* + \gamma$, $H \to (J/\Psi, \Psi(2S), \Upsilon(nS)) + \gamma$ and $H \to D^* + \gamma$, which probe some quark Yukawa couplings and two flavour-violating decays.

\paragraph{$H\to ee/e\mu$}\mbox{}\\
This is the first ATLAS search for Higgs boson decay into an electron--positron pair, $H\to ee$, which should have a SM rate too small to be observed with the current dataset. In the same search, Higgs boson decay into an electron--muon pair, $H\to e\mu$, is also considered, which could indicate BSM physics with lepton flavour violation (LFV)~\cite{HIGG-2018-58}. The search closely follows that for the SM Higgs boson decay $H\to  \mumu$~\cite{HIGG-2016-10} and uses 139~\ifb of data. The background in the $ee$ search is dominated by Drell--Yan events, top-quark pair ($\ttbar$) events and diboson ($ZZ$, $WZ$ and $WW$) events. In the $e\mu$ search, a much smaller yield of SM  background events is expected. The final discriminants are  $m_{ee}$ and $m_{e\mu}$.

\paragraph{$H\to e\tau/\mu\tau$}\mbox{}\\
This is a direct search for LFV in Higgs boson decays, $H\to  e\tau$ and $H\to  \mu\tau$, using 138~\ifb of data~\cite{HIGG-2019-11}. It may be possible for Higgs boson decays to exhibit flavour-changing dynamics, leading to the discovery of LFV at the LHC. Both the leptonic $(\tau \to \ell\nu_\ell\nu_{\tau})$ and hadronic $(\tau \to \text{hadrons } \nu_{\tau})$ decay channels of the $\tau$-lepton are studied, as shown in Figure~\ref{fig:H->etau_diagrams}. The background estimation techniques include the use of templates from simulations with input from data-driven methods, and a method based on exploiting the symmetry between prompt electrons and prompt muons in the SM backgrounds. A multivariate analysis technique is deployed to further improve the separation of signal and background.

\begin{figure}[tb!]
\begin{center}

\subfloat[]{
\includegraphics[width=0.4\textwidth,valign=c]{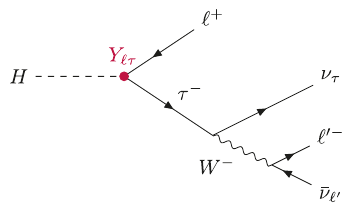}
}
\subfloat[]{
\includegraphics[width=0.4\textwidth,valign=c]{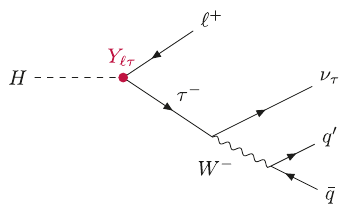}
}

\caption{Illustrative Feynman diagrams for LFV Higgs boson decays for the (a) $\tau\to\ell\nu_\ell\nu_{\tau}$ and (b) $\tau\to \text{hadrons}+\nu_{\tau}$ final states.
The off-diagonal Yukawa coupling term is indicated by the $Y_{\ell\tau}$ symbol.}
\label{fig:H->etau_diagrams}
\end{center}
\end{figure}

\paragraph{$H \to \omega/K^* + \gamma$}\mbox{}\\
Searches for the rare exclusive decays  $H/Z \to \omega\gamma$ and $H \to K^*\gamma$ can probe flavour-conserving and flavour-violating Higgs boson couplings to light quarks~\cite{HDBS-2019-33}. The search  used  89.5~\ifb and 134~\ifb of data, respectively, for the two decays. This search is interesting because it looks for possible BSM Higgs boson couplings to light first- and second-generation quarks. Most of the previous investigations target the heavier second- and third-generation fermions, with stronger couplings to the Higgs boson. The $\omega$ and $K^*$ mesons are reconstructed via their dominant decays into $\pi^+\pi^-\pi^0$ and $K^{\pm}\pi^{\pm}$ final states, respectively. The dominant background arises from inclusive $\gamma$\,+\,jet or multijet processes. This background model is derived using a fully data-driven template approach and validated in a number of control regions. The final discriminants are the reconstructed meson masses.

\paragraph{$H \to (J/\Psi, \Psi(2S), \Upsilon(nS)) + \gamma $}\mbox{}\\
This analysis searches for exclusive Higgs boson decays into a vector quarkonium state and a photon. The final state is $\mu^+\mu^-\gamma$ and the search uses 139~\ifb of data~\cite{HDBS-2018-53}. The Higgs boson decays  $H \to (J/\Psi, \Psi(2S), \Upsilon(nS)) + \gamma $ explore the charmonium and bottomonium sectors. Branching fractions are very low in the SM:
$\mathcal{B} (H \to J/\Psi\,\gamma) \approx 10^{-6}$ and
$\mathcal{B} (H \to \Upsilon(nS)\,\gamma) \approx  10^{-9}{-}10^{-8}$.
Deviations of the quark Yukawa couplings from SM expectations can lead to significant enhancements in the branching fractions of these radiative decays in BSM theories. The cross-section for $Z$-boson production at the LHC is approximately 1000 times larger than that for Higgs boson production. Examining similar radiative decays of the $Z$ boson could therefore provide additional sensitivity. The muons are well identified in ATLAS. Drell--Yan production of dimuons with significant final-state radiation is the main exclusive background. The dominant background, however, is mostly from inclusive $\gamma$\,+\,jet and multijet processes, and is estimated similarly to the $H \to \omega/K^* + \gamma$ search discussed above. The discriminating variable is  $m_{\mu^+\mu^-\gamma}$.

\paragraph{$H \to D^* + \gamma$}\mbox{}\\
This analysis searches for exclusive Higgs boson decays into a $D^*$ meson and photon final state to probe flavour-violating Higgs boson couplings to light quarks. The final state is $K^-\pi^+\gamma$ and the search uses 136.3~\ifb of data~\cite{HDBS-2018-52}. The signature includes a high-energy photon and a meson. A di-track signal is used to reconstruct the $D^0$ meson from the $D^* \to D^0\pi^0 / D^0\gamma, D^0  \to  K^-\pi^+ $ decay chain, leading to a three-body mass to reconstruct the initial Higgs boson. The $\pi^0$ or photon in the $D^*$ decay are soft and are not explicitly reconstructed. An additional feature is the displaced $D^0$ decay vertex, which provides a particularly distinct signature to use in rejecting multijet events, the dominant contribution to the background. The final discriminant is $m_{K\pi\gamma}$.


\section{Results}
\label{sec:results}

In this section the most important results of the ATLAS searches for additional Higgs bosons or exotic Higgs boson decays are reported. This section makes no attempt to be complete, but rather focusses on interesting excesses as well as model-independent and model-dependent constraints.


\subsection{Neutral heavy Higgs bosons}
\label{sec:results:neutralhiggs}

\subsubsection{Heavy Higgs bosons decaying into fermions}
\label{sec:results:neutralhiggs:fermions}

The \textbf{$A/H\to\tau^+\tau^-$} analysis~\cite{HDBS-2018-46} examines one of the most sensitive channels for type-II models. The transverse mass ($m_\text{T}$) distribution in the $b$-tag category of the \tauhad{}\tauhad channel is shown in Figure~\ref{fig:results:neutral:tautau:mT}. Since the $m_\text{T}$ variable does not fully reconstruct the resonance mass (due to the presence of neutrinos), the signal peaks below the hypothesized mass of the heavy Higgs boson. A slight excess is observed for $m_A$ of 400~\GeV, contributed by the $b$-tag category of the \tauhad{}\tauhad channel and the $b$-veto category of the \taulep{}\tauhad channel. The local significances are 2.2$\sigma$ for ggF production and 2.7$\sigma$ for $b$-associated production. The global significance takes into account the look-elsewhere-effect~\cite{Cowan:2010js} and is 1.9$\sigma$. The exclusions in the $m_A$--$\tan\beta$ plane of the $M_h^{125}$ scenario of the MSSM are presented in Figure~\ref{fig:results:neutral:tautau:mh125} at 95\% confidence level (CL). To obtain these results, the statistical procedure used is the CL$_\text{s}$ modified frequentist method~\cite{Read:2002hq}, as is the case for all other results in this report. The mass degeneracy of the $A$ and $H$ bosons is valid in almost the entire phase space, but not at low values of $m_A$ and $\tan\beta$. The excess is clearly visible in the MSSM exclusions, resulting in weaker constraints around 400~\GeV. Results in the hMSSM are displayed in Figure~\ref{fig:results:summary:hmssm}.

\begin{figure}[tb!]
\centering
\subfloat[]{
\label{fig:results:neutral:tautau:mT}
\includegraphics[width=0.48\textwidth,valign=c]{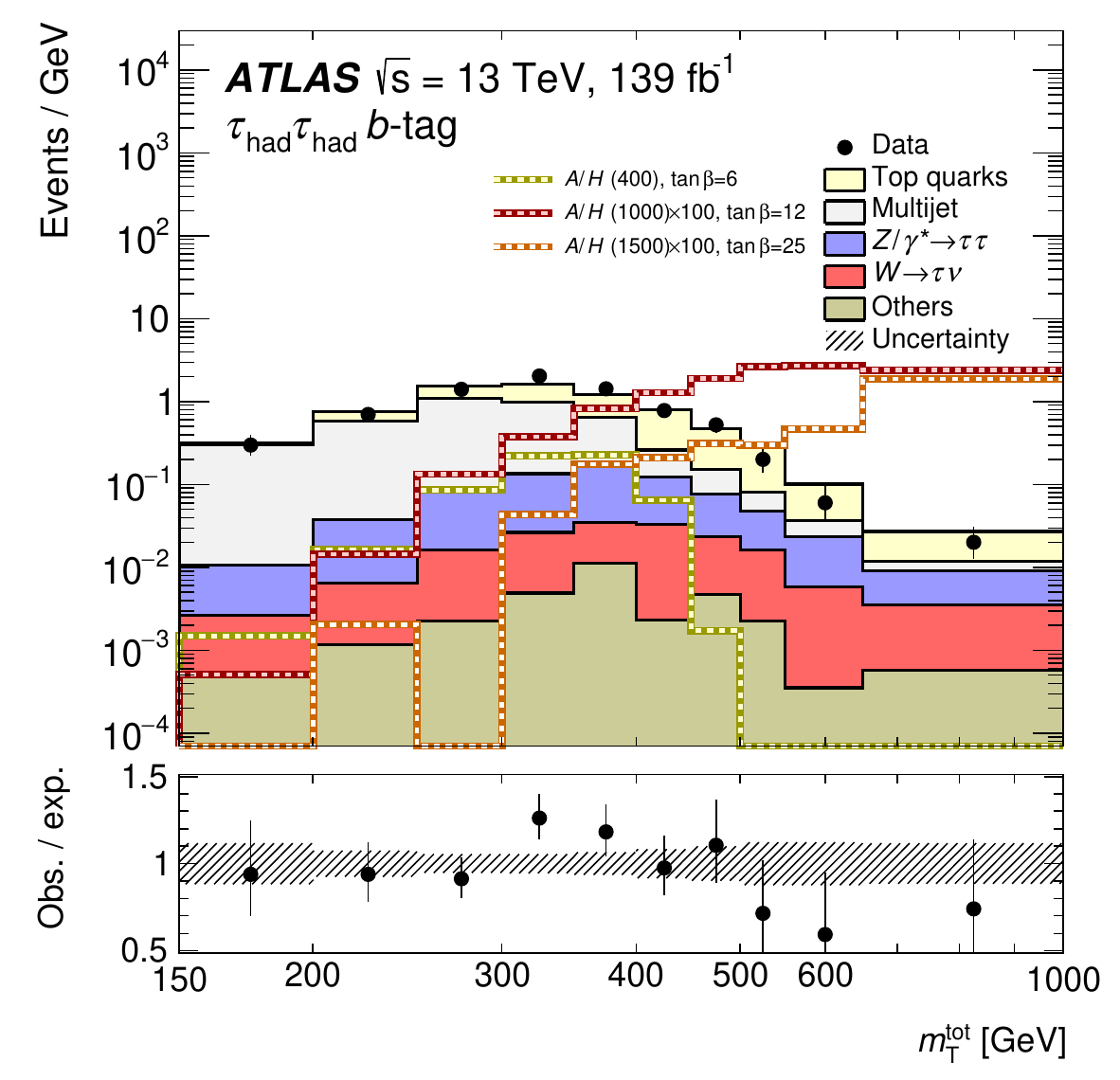}
}
\subfloat[]{
\label{fig:results:neutral:tautau:mh125}
\includegraphics[width=0.5\textwidth,valign=c]{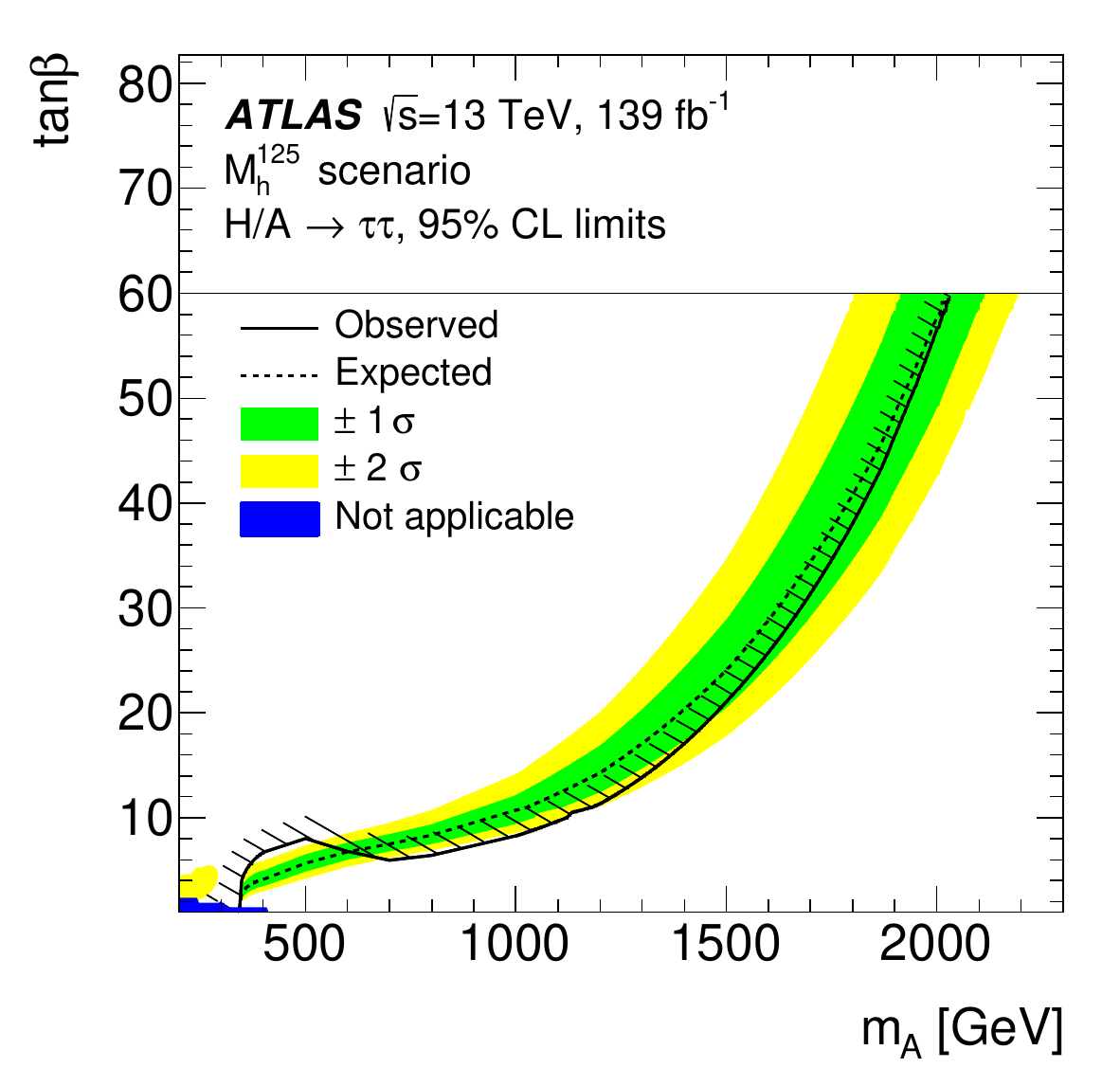}
}
\caption{\label{fig:results:neutral:tautau}$A/H\to\tau^+\tau^-$: (a) The transverse mass distribution in the $b$-tag category of the \tauhad{}\tauhad channel. The predictions and uncertainties for the background processes are obtained from a combined fit to all categories, assuming the background-only hypothesis. Expectations from signal processes are superimposed. Overflows are included in the last bin of the distribution. (b) The 95\% CL exclusions in the $M_h^{125}$ scenario of the MSSM in the $m_A$--$\tan\beta$ plane. The region above the solid black line is excluded. The small blue-shaded corner at low $m_A$ and $\tan\beta$ values is the region where the mass difference between the $A$ the $H$ bosons is larger than 50\% of the experimental mass resolution. Figures are taken from Ref.~\cite{HDBS-2018-46}.}
\end{figure}

The search for \textbf{$A/H\to\mu^+\mu^-$}~\cite{HIGG-2017-10} was carried out with the 2015+2016 data of \RunTwo. The dimuon mass in the $b$-tag category and the limits on $b$-associated production are displayed in Figure~\ref{fig:results:neutral:mumu}. A slight excess is observed at 480~\GeV which comes almost entirely from the $b$-tag category. The local significance of that excess is 2.3$\sigma$ for $b$-associated production. The significance decreases to 0.6$\sigma$ when the look-elsewhere-effect is included. In MSSM scenarios, the expected signal rate in the $\mu^+\mu^-$ channel is much smaller than that in the $\tau\tau$ channel. The sensitivity of this $A/H\to\mu^+\mu^-$ search is not sufficient to constrain any of the MSSM scenarios.

\begin{figure}[tb!]
\centering
\subfloat[]{
\includegraphics[width=0.48\textwidth,valign=c]{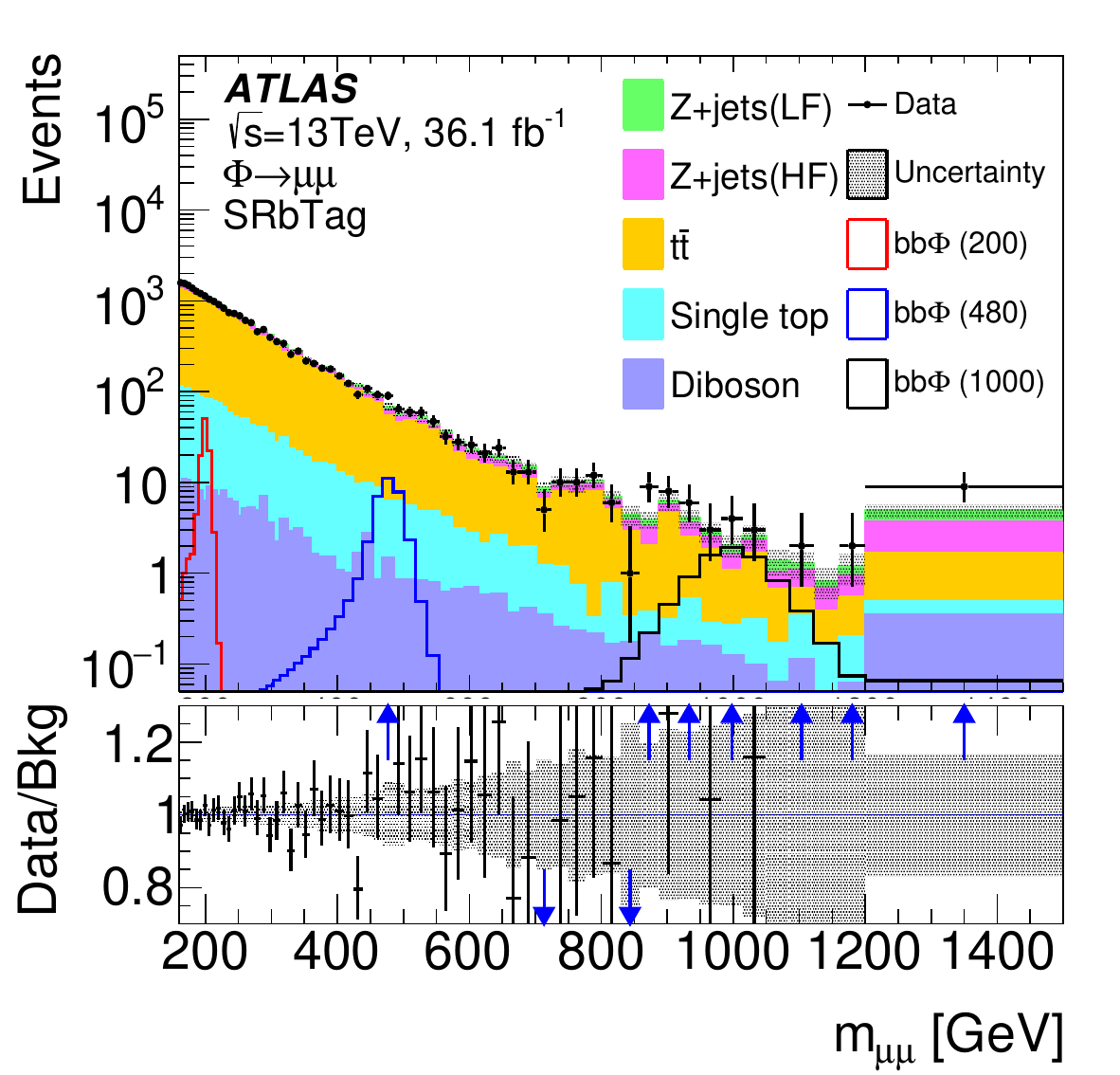}
}
\subfloat[]{
\includegraphics[width=0.5\textwidth,valign=c]{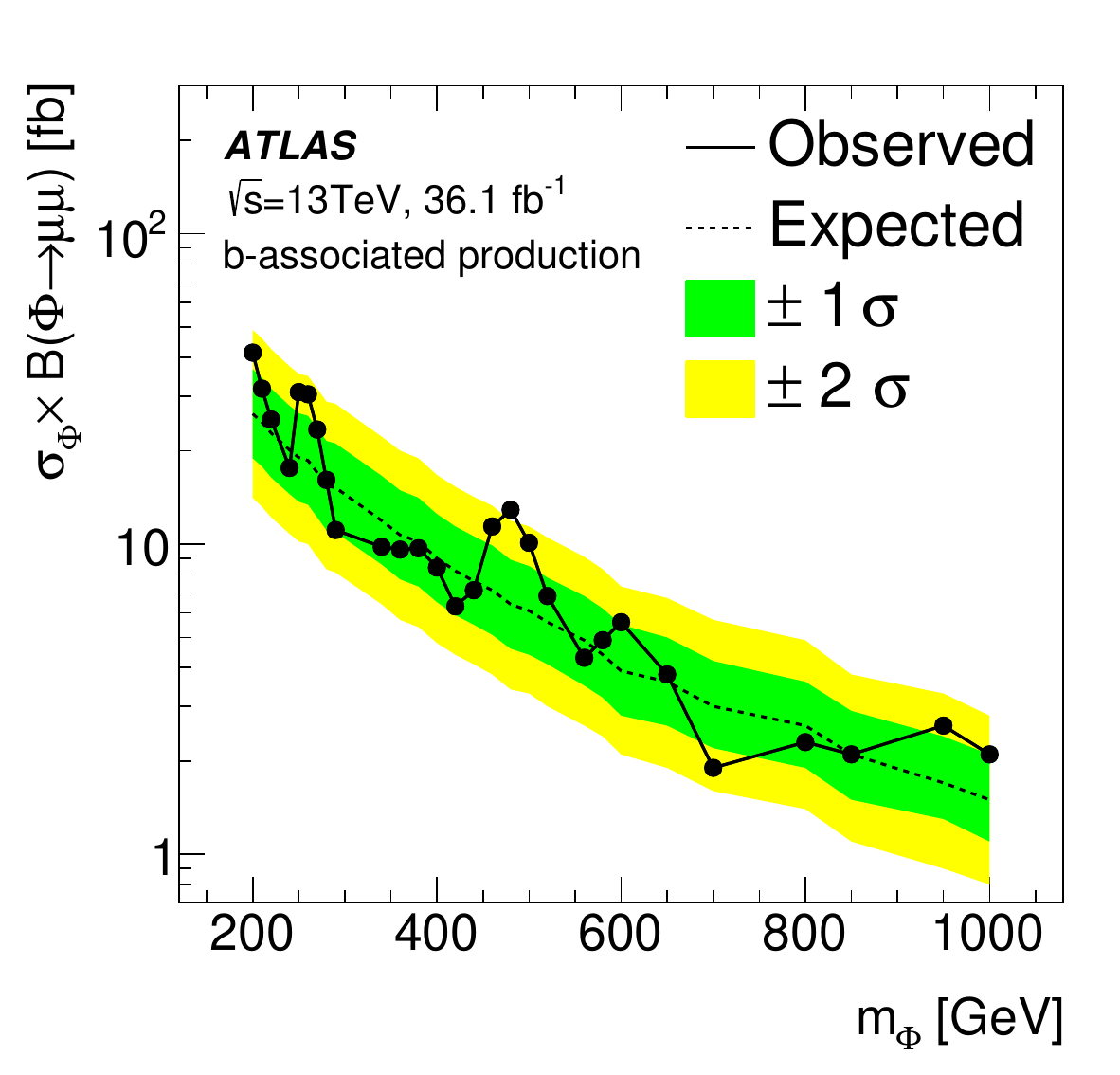}
}
\caption{\label{fig:results:neutral:mumu}$A/H\to\mu^+\mu^-$: (a) The dimuon mass in the $b$-tag category of the search for $A/H\to\mu^+\mu^-$, with three signal hypotheses overlaid. (b) The limits on the cross-section times branching fraction, assuming $b$-associated production. A slight excess at 480~\GeV is visible. Figures are taken from Ref.~\cite{HIGG-2017-10}.}
\end{figure}

The search for \textbf{$A/H\to b\bar{b}$}~\cite{HIGG-2016-32} yields sensitivity to type-II and flipped 2HDMs due to their enhanced coupling of the heavy Higgs bosons to $b$-quarks, which is exploited twice in this channel: in the $b$-associated production and in the decay. The discriminant $m'_{b\bar{b}}$ in the 5-jet category is presented in Figure~\ref{fig:results:neutral:bb:mass} for a heavy Higgs boson mass hypothesis of 600~\GeV. This distribution is $m_A$-dependent because a transformation is performed for each $A$ mass hypothesis to decrease the correlation between the $b\bar{b}$ mass and the \pt of the $b$-jets. This transformation increases the sensitivity of the analysis. The exclusions in the flipped model are displayed in Figure~\ref{fig:results:neutral:bb:limit} for a heavy Higgs boson mass of 450~\GeV. A specific mass was chosen to allow a graphical representation of the exclusion in an otherwise 3-dimensional parameter space. Results in the hMSSM are displayed in Figure~\ref{fig:results:summary:hmssm}.

\begin{figure}[tb!]
\centering
\subfloat[]{
\label{fig:results:neutral:bb:mass}
\includegraphics[width=0.53\textwidth,valign=c]{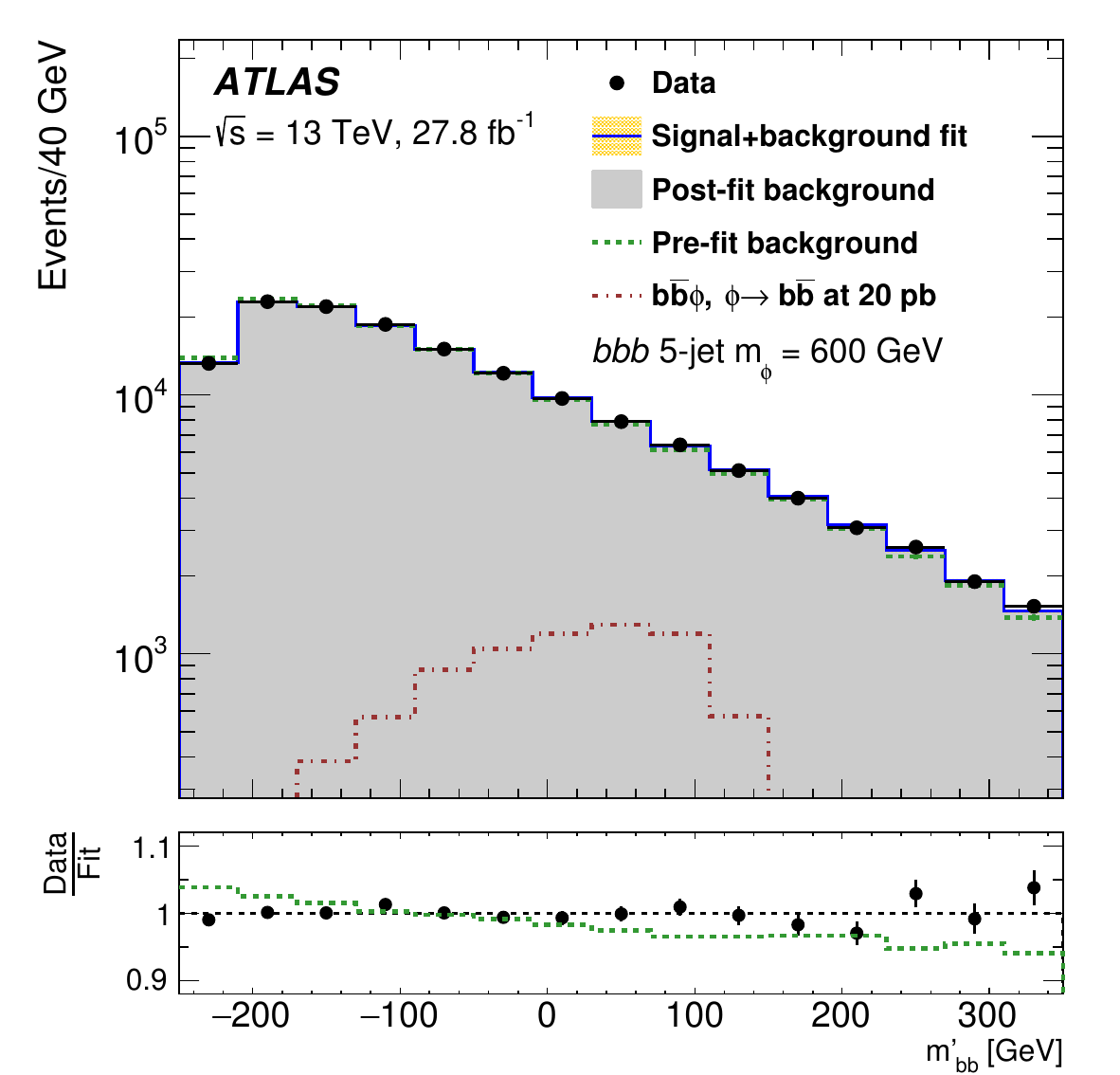}
}
\subfloat[]{
\label{fig:results:neutral:bb:limit}
\includegraphics[width=0.45\textwidth,valign=c]{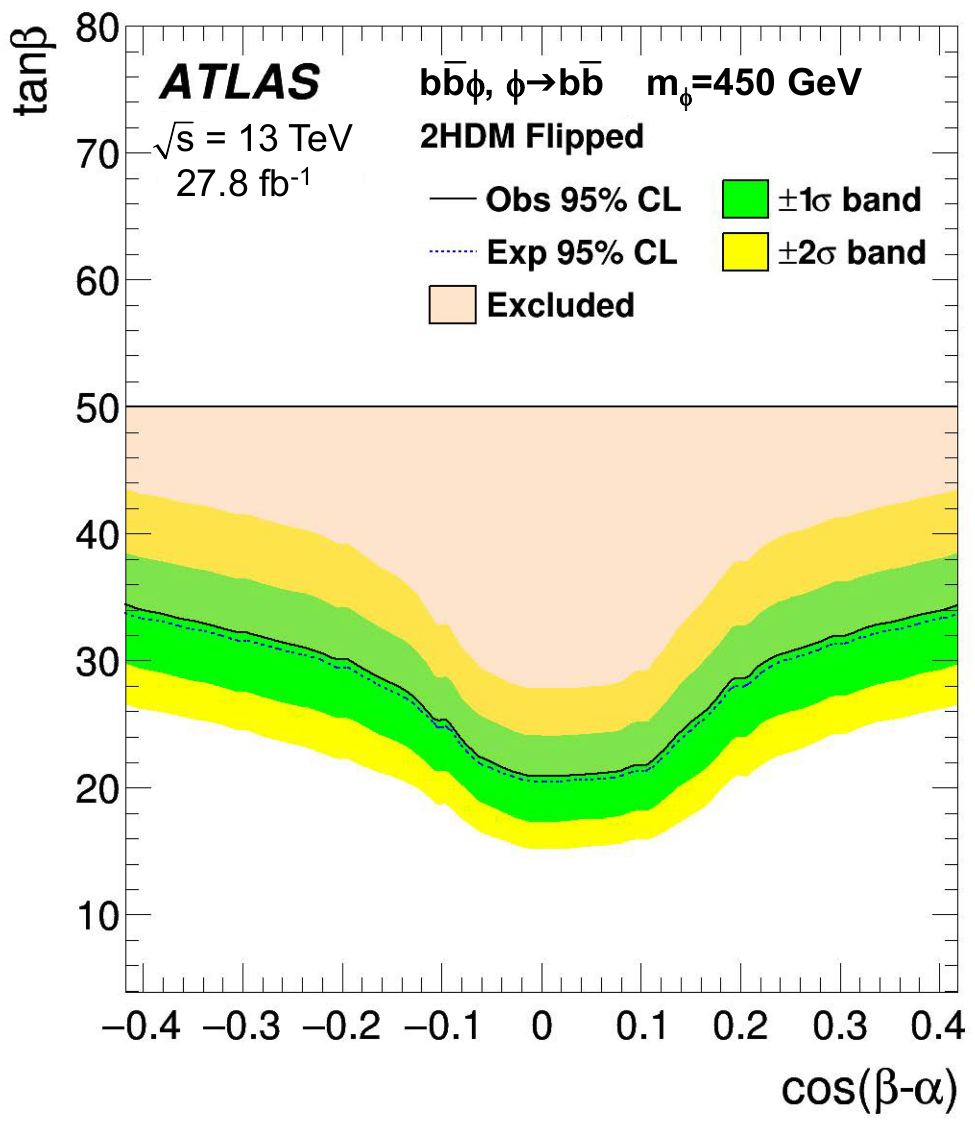}
}
\caption{\label{fig:results:neutral:bb}$A/H\to b\bar{b}$: (a) Post-fit distribution of $m'_{b\bar{b}}$ for the 600~\GeV mass hypothesis in the 5-jet category. The pre-fit background shape and its ratio to the post-fit shape are also shown. The signal shape (red dashed line) is overlaid for illustration. (b) The 95\% CL exclusion in the flipped 2HDM for a fixed $A$ mass hypothesis of 450~\GeV. The high $\tan\beta$ region above the line is excluded. Figures are taken from Ref.~\cite{HIGG-2016-32}.}
\end{figure}

The results of the search for \textbf{$A/H\to t\bar{t}$}~\cite{EXOT-2020-25} are in good agreement with the SM in all categories. The reconstructed $t\bar{t}$ mass in the resolved one-lepton category with two $b$-jets for a specific slice of the variable $|\cos\theta^{\ast}|$ is displayed in Figure~\ref{fig:results:neutral:ttbar:mass} after the fit to all categories. Here $\theta^{\ast}$ denotes the angle between the momentum of the leptonically decaying top-quark in the resonance centre-of-mass frame and the momentum of the reconstructed $t\bar{t}$ system in the laboratory frame. The peak--dip structure of the signal--background interference is visible for two signal hypotheses displayed in the lower ratio panel. The largest deviation from the background-only prediction has a local significance of $2.3\sigma$ and is obtained for $m_{A} = 800~\GeV$ and a generic signal width of 10\%. The 95\% CL limits in a type-II 2HDM in the alignment limit are presented in Figure~\ref{fig:results:neutral:ttbar:limit}. In this model, but not all those reported on, the $A$ and $H$ bosons are mass degenerate. The cross-section times branching fraction for the signal process rises for lower values of $\tan\beta$. Values of $\tan\beta$ smaller than 3.5 are excluded for a heavy Higgs boson mass of 400~\GeV, and mass values up to 1240~\GeV are excluded for the lowest tested $\tan\beta$ value of 0.4.

\begin{figure}[tb!]
\centering
\subfloat[]{
\label{fig:results:neutral:ttbar:mass}
\includegraphics[width=0.45\textwidth,valign=c]{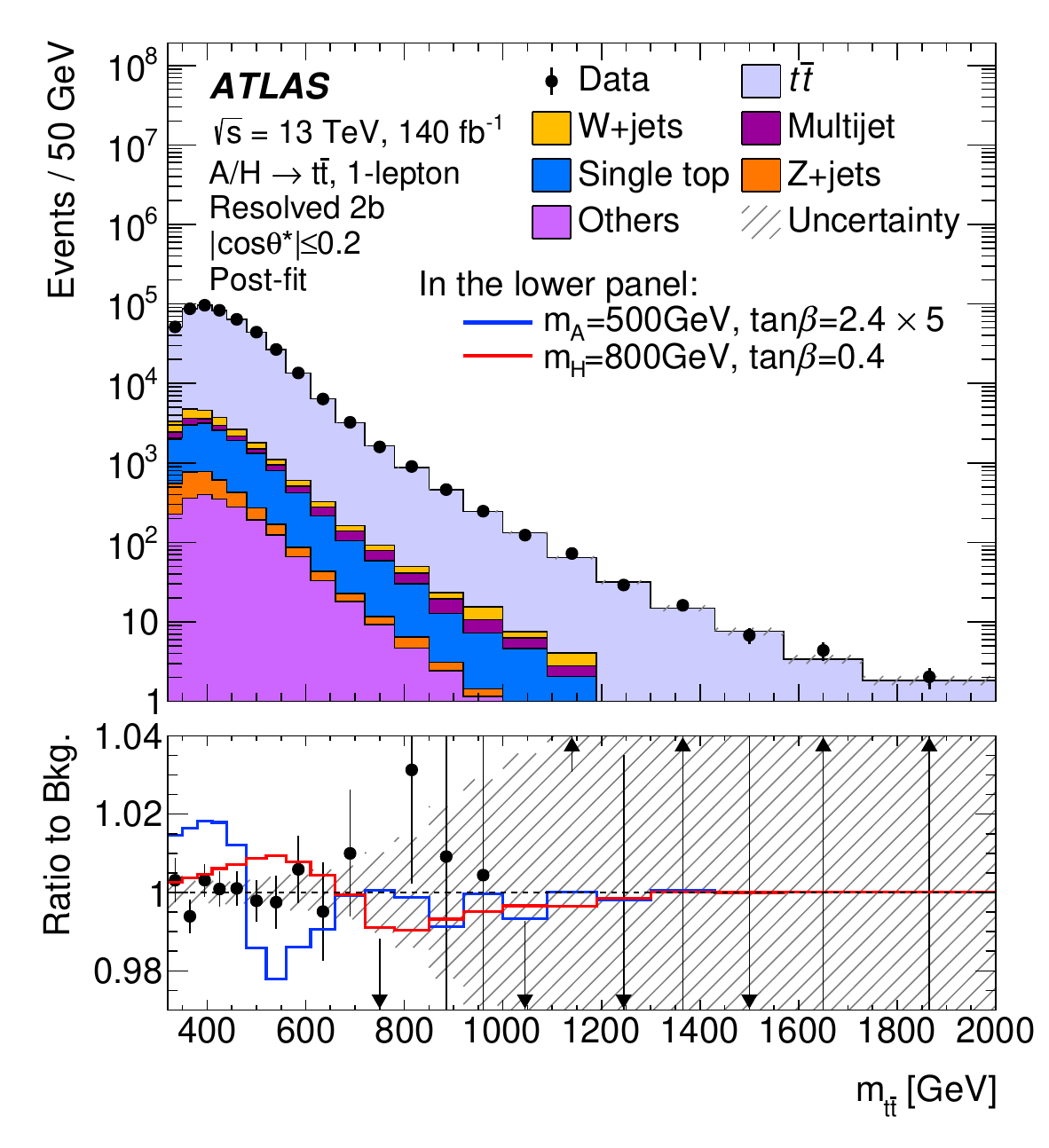}
}
\subfloat[]{
\label{fig:results:neutral:ttbar:limit}
\includegraphics[width=0.54\textwidth,valign=c]{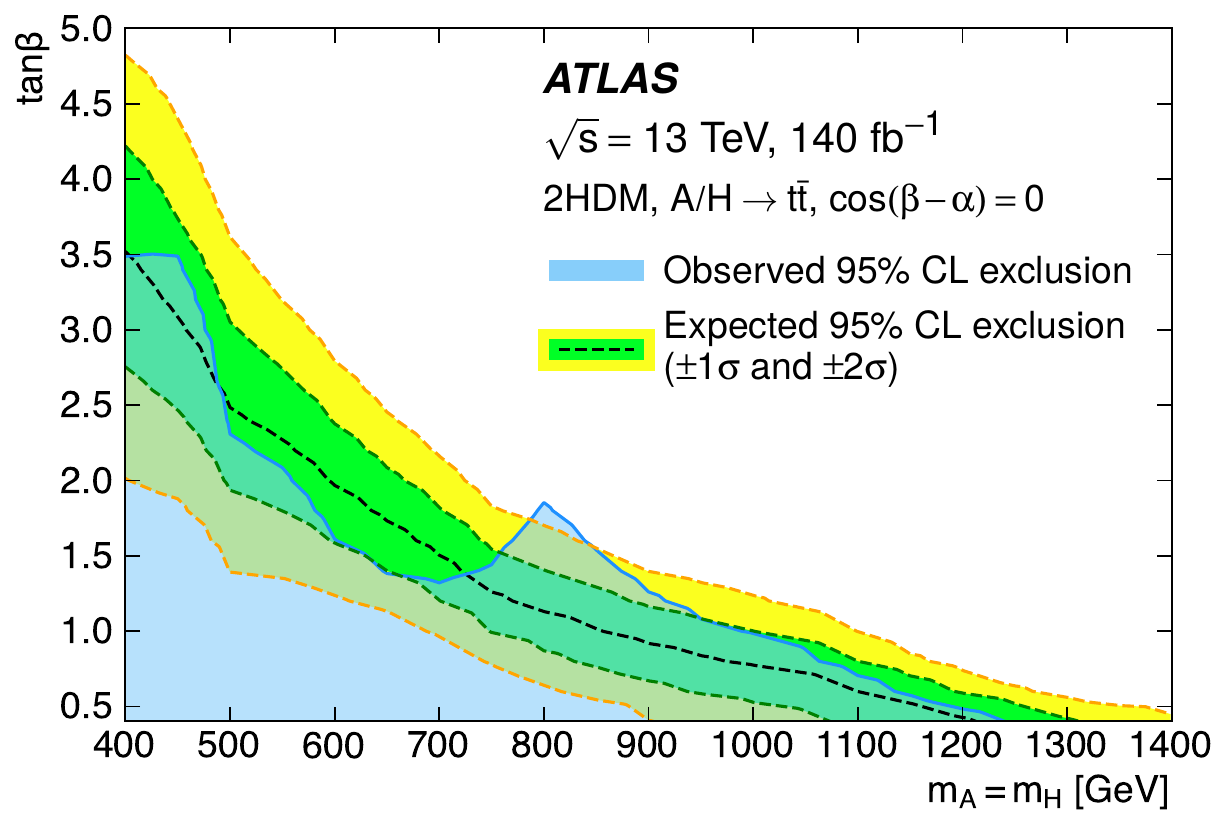}
}
\caption{\label{fig:results:neutral:ttbar}$A/H\to t\bar{t}$: (a) Post-fit distribution of $m_{t\bar{t}}$ in the resolved one-lepton category with two $b$-jets. The signal shape with the interference effect is visible in the lower panel for two signal hypotheses (red and blue lines). (b) The 95\% CL exclusion in the aligned 2HDM in the $m_A$--$\tan\beta$ plane. The low $\tan\beta$ region below the line is excluded. Figures are taken from Ref.~\cite{EXOT-2020-25}.}
\end{figure}

The \textbf{$ttH/A$, $H/A\to t\bar{t}$} search~\cite{EXOT-2019-26} suffers from a small cross-section compared to the ggF production model discussed briefly in Section~\ref{sec:neutralhiggs:fermions}, but there is no interference between the signal and the background in this case, which simplifies the search. However, the four-top topology leads to a high jet multiplicity that calls for multivariate analysis techniques. The distribution of the BDT discriminant for a heavy Higgs boson mass of 1000~\GeV is shown in Figure~\ref{fig:results:neutral:4tops:BDT}. The limits on the production cross-section times branching fraction are shown in Figure~\ref{fig:results:neutral:4tops:limits}, assuming that $H$ and $A$ are mass-degenerate and both Higgs bosons contribute to the production. The cross-section predicted in the type-II 2HDM is also shown, and the analysis constrains this model for very low values of $\tan\beta$, where the Yukawa coupling to top quarks is strongly enhanced. This analysis is less sensitive than ggF $A/H\to t\bar{t}$ at low mass, but is comparable for high Higgs boson masses. Results in the hMSSM are displayed in Figure~\ref{fig:results:summary:hmssm}.

\begin{figure}[tb!]
\centering
\subfloat[]{
\label{fig:results:neutral:4tops:BDT}
\includegraphics[width=0.43\textwidth,valign=c]{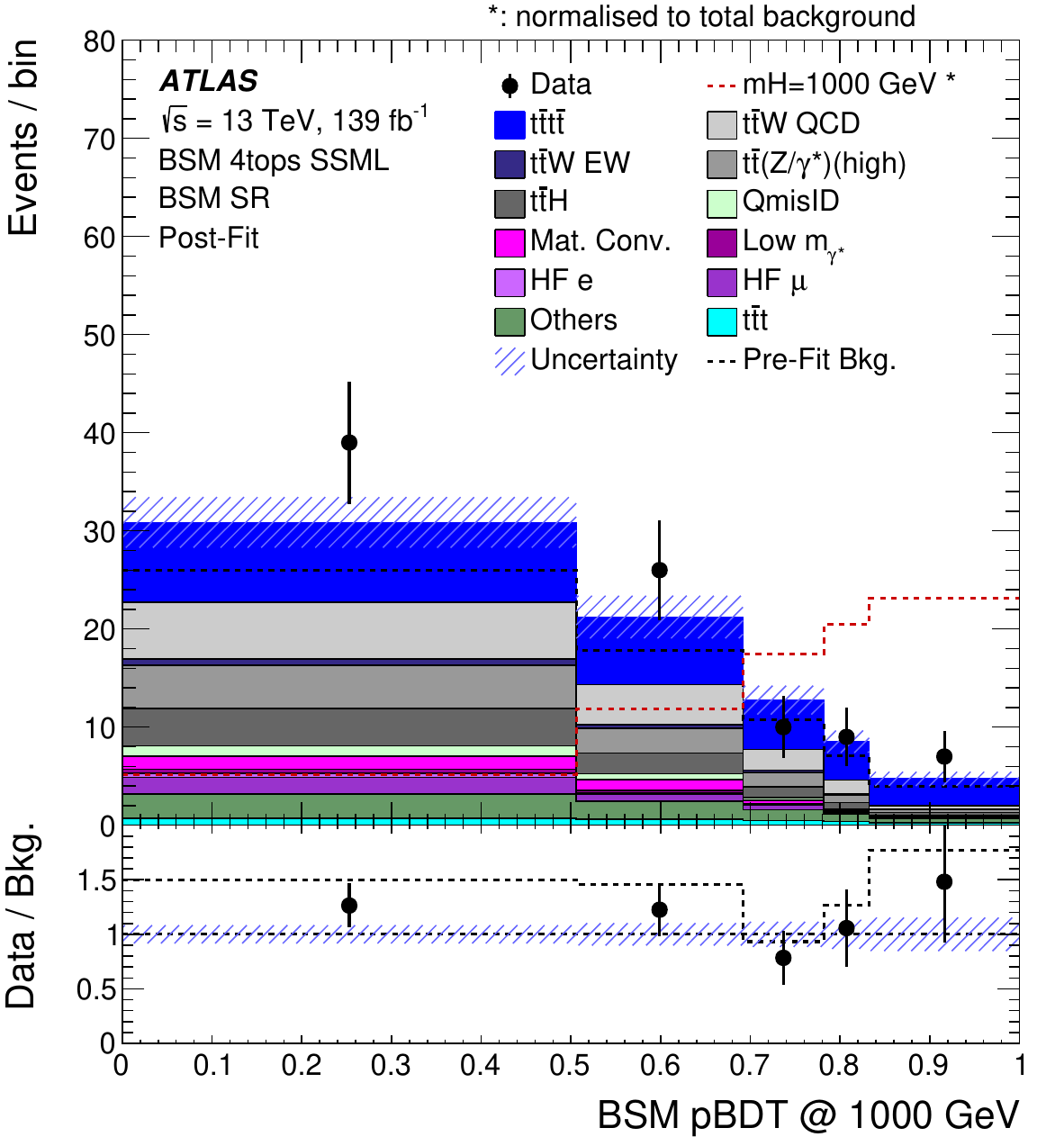}
}
\subfloat[]{
\label{fig:results:neutral:4tops:limits}
\includegraphics[width=0.55\textwidth,valign=c]{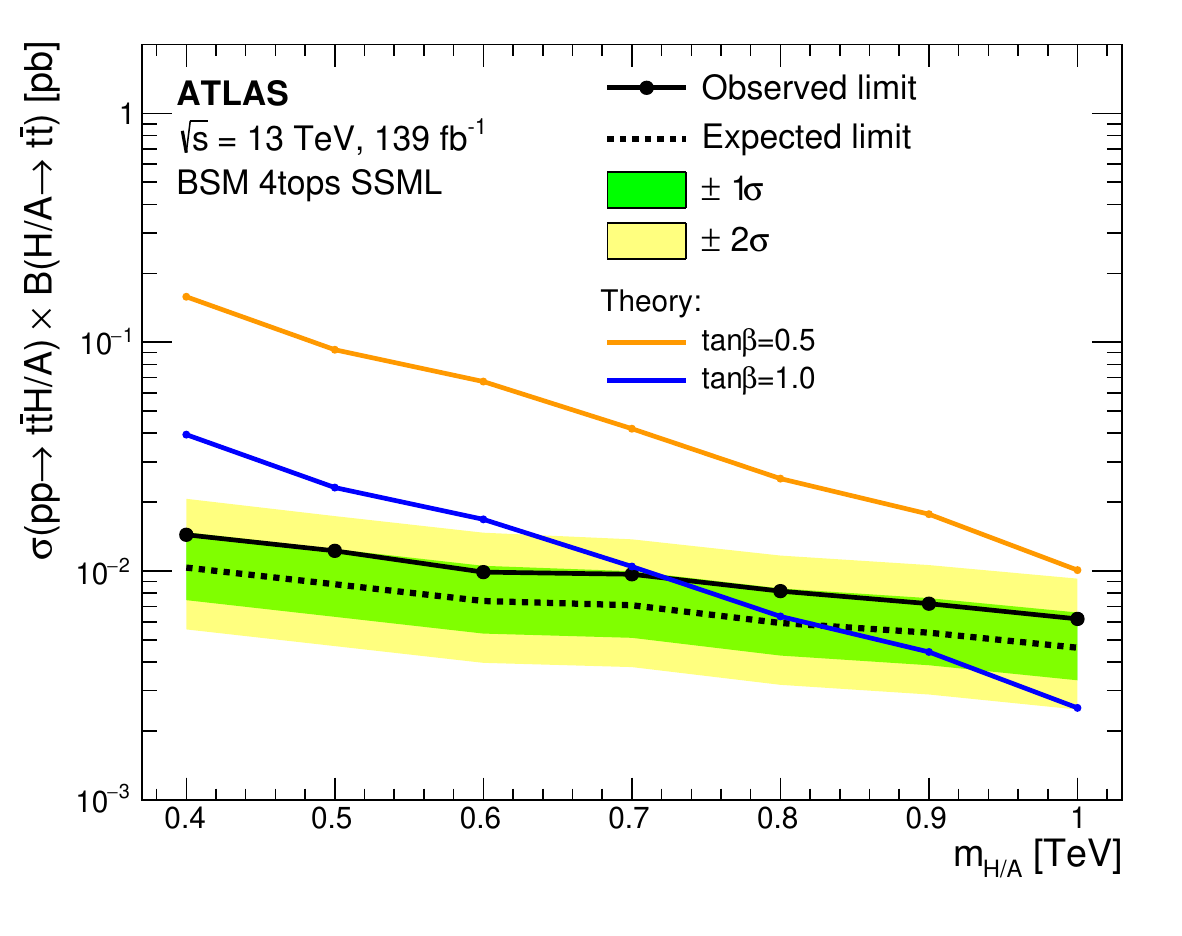}
}
\caption{\label{fig:results:neutral:4tops}$(t\bar{t})H/A, H/A\to t\bar{t}$: (a) The output score of the parameterized BDT for a heavy Higgs boson mass of 1000~\GeV. The signal accumulates at large values of the score, while the background peaks towards lower values. (b) The 95\% CL limits on the cross-section times branching fraction, with two cross-section predictions overlaid for the type-II 2HDM. Figures are taken from Ref.~\cite{EXOT-2019-26}.}
\end{figure}

The search for \textbf{$H\to t\bar{t}/t\bar{q}$}~\cite{HDBS-2020-03} that introduces the BSM couplings $\rho_{tt}$, $\rho_{tu}$ and $\rho_{tc}$ yields an excess with a local significance of 2.8$\sigma$ for coupling values of $\rho_{tt}=0.6$, $\rho_{tu}=1.1$ and $\rho_{tc}=0$. This excess is largest for a heavy Higgs boson mass hypothesis of 900~\GeV but the dependence on the mass is very weak, and the excess appears essentially for any hypothesized mass. The event yields in all 17 signal regions and the limits on the cross-section for a heavy Higgs boson are displayed in Figure~\ref{fig:results:neutral:g2HDM}. Assuming the coupling values for the largest excess, the hypothesis of a heavy Higgs boson in the g2HDM is nonetheless excluded at 95\% CL for the mass range of 200--1500~\GeV.

\begin{figure}[tb!]
\centering
\subfloat[]{
\label{fig:results:neutral:g2HDM:yields}
\includegraphics[width=0.48\textwidth,valign=c]{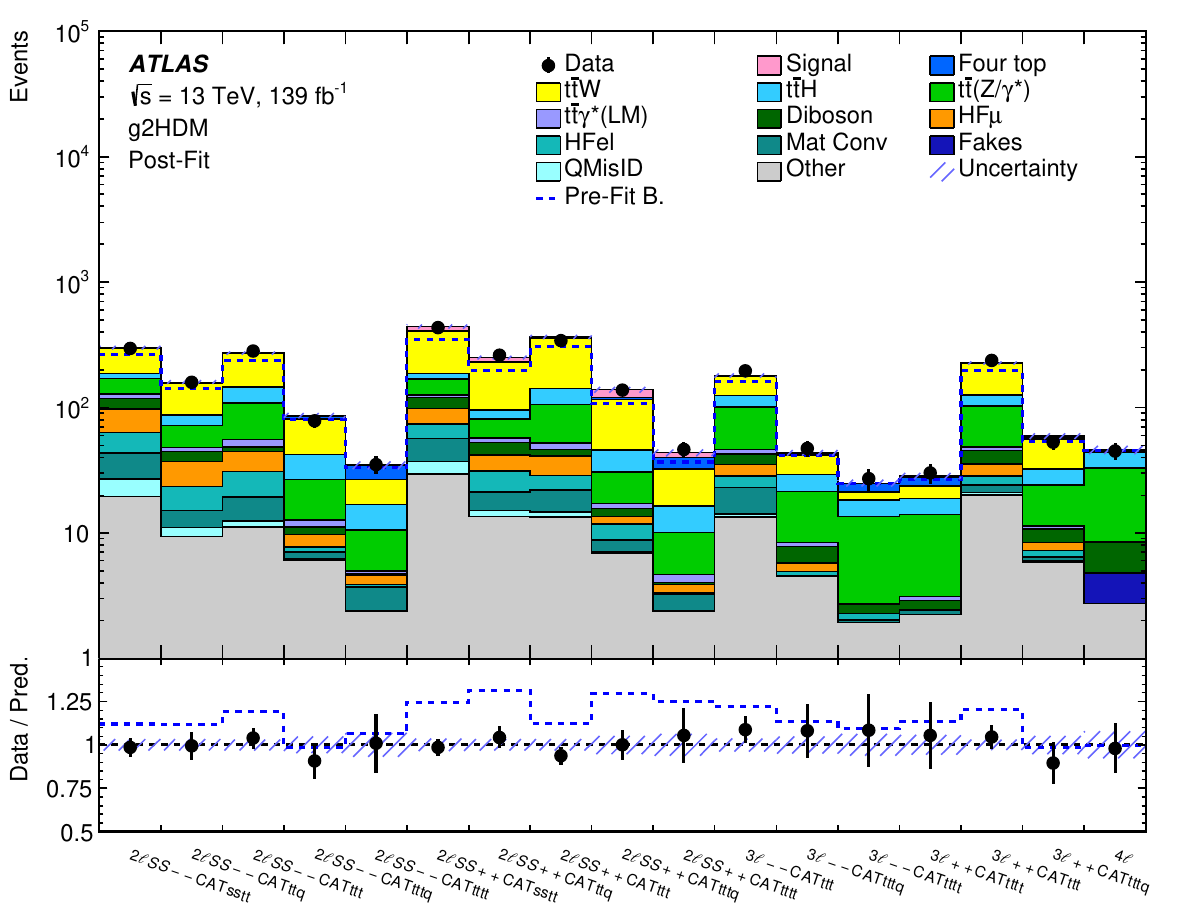}
}
\subfloat[]{
\label{fig:results:neutral:g2HDM:limits}
\includegraphics[width=0.5\textwidth,valign=c]{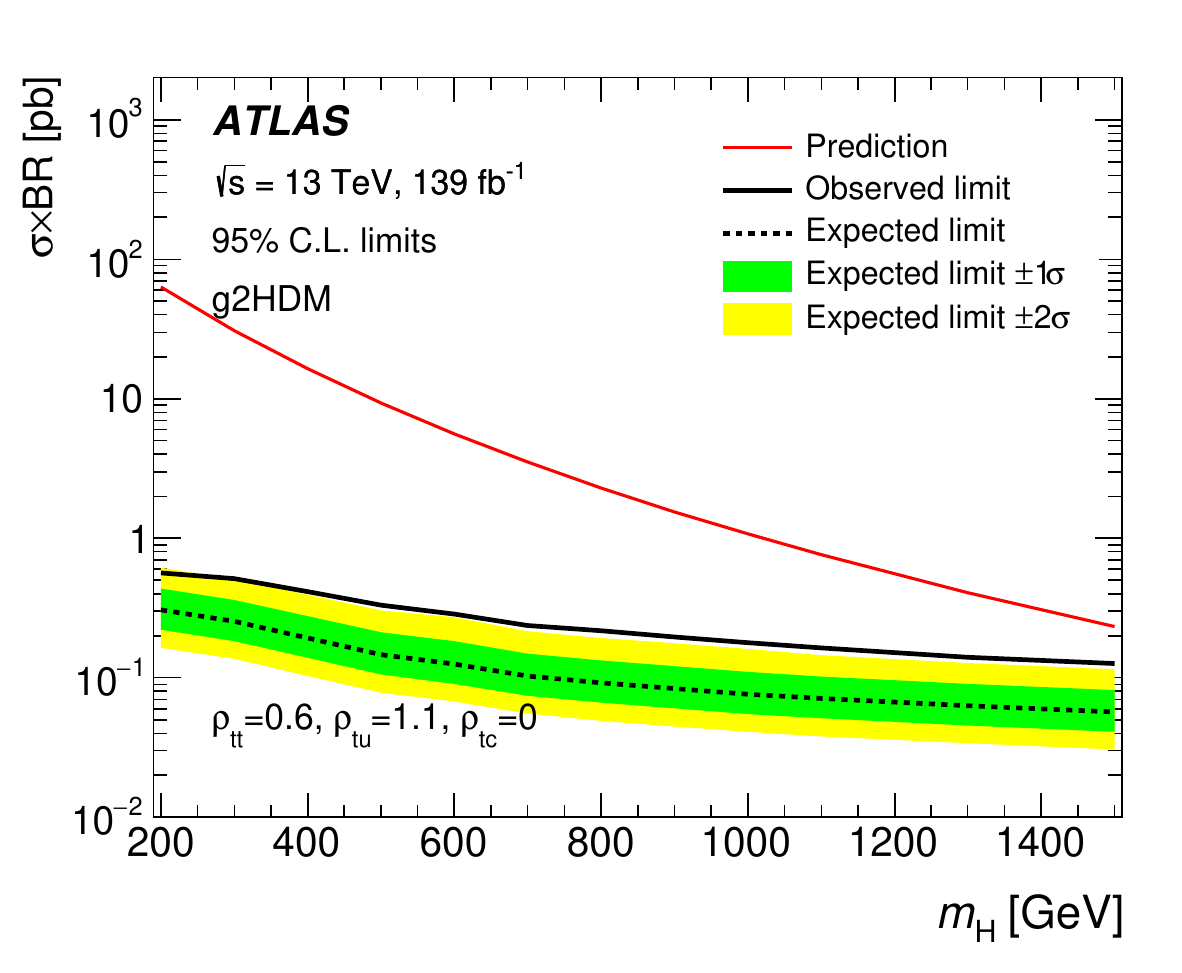}
}
\caption{\label{fig:results:neutral:g2HDM}$H\to t\bar{t}/t\bar{q}$: (a) Comparison between data and the background prediction for the event yields in the 17 signal regions after the fit under the signal-plus-background hypothesis. The signal displayed assumes $m_H=900$~\GeV and coupling values $\rho_{tt}=0.6$, $\rho_{tu}=1.1$ and $\rho_{tc}=0$. (b) The 95\% CL limits on the cross-section times branching fraction for a heavy Higgs boson for the same coupling values (corresponding to the largest excess). The red line indicates the predicted signal cross-section in the g2HDM, meaning this signal hypotheses is excluded for the mass range considered. Figures are taken from Ref.~\cite{HDBS-2020-03}.}
\end{figure}

\FloatBarrier

\subsubsection{Heavy Higgs bosons decaying into bosons}
\label{sec:results:neutralhiggs:bosons}

The search for \textbf{$H\to W^+W^-$}~\cite{ATLAS-CONF-2022-066} explored ggF and VBF production and a large range of masses up to 4000~\GeV but yielded no excess. Limits on the cross-section were obtained assuming either a generic narrow-width scalar or the hypothesis of a $H_5$ scalar in the GM model. The transverse mass in the VBF category with two jets and the limits on VBF production are presented in Figure~\ref{fig:results:neutral:WW}. For the GM model, cross-section times branching fraction values above 0.35~pb at $m_H=250$~\GeV and above 0.024~pb at 1~\TeV are excluded at 95\% CL.

\begin{figure}[tb!]
\centering
\subfloat[]{
\label{fig:results:neutral:WW:mT}
\includegraphics[width=0.47\textwidth,valign=c]{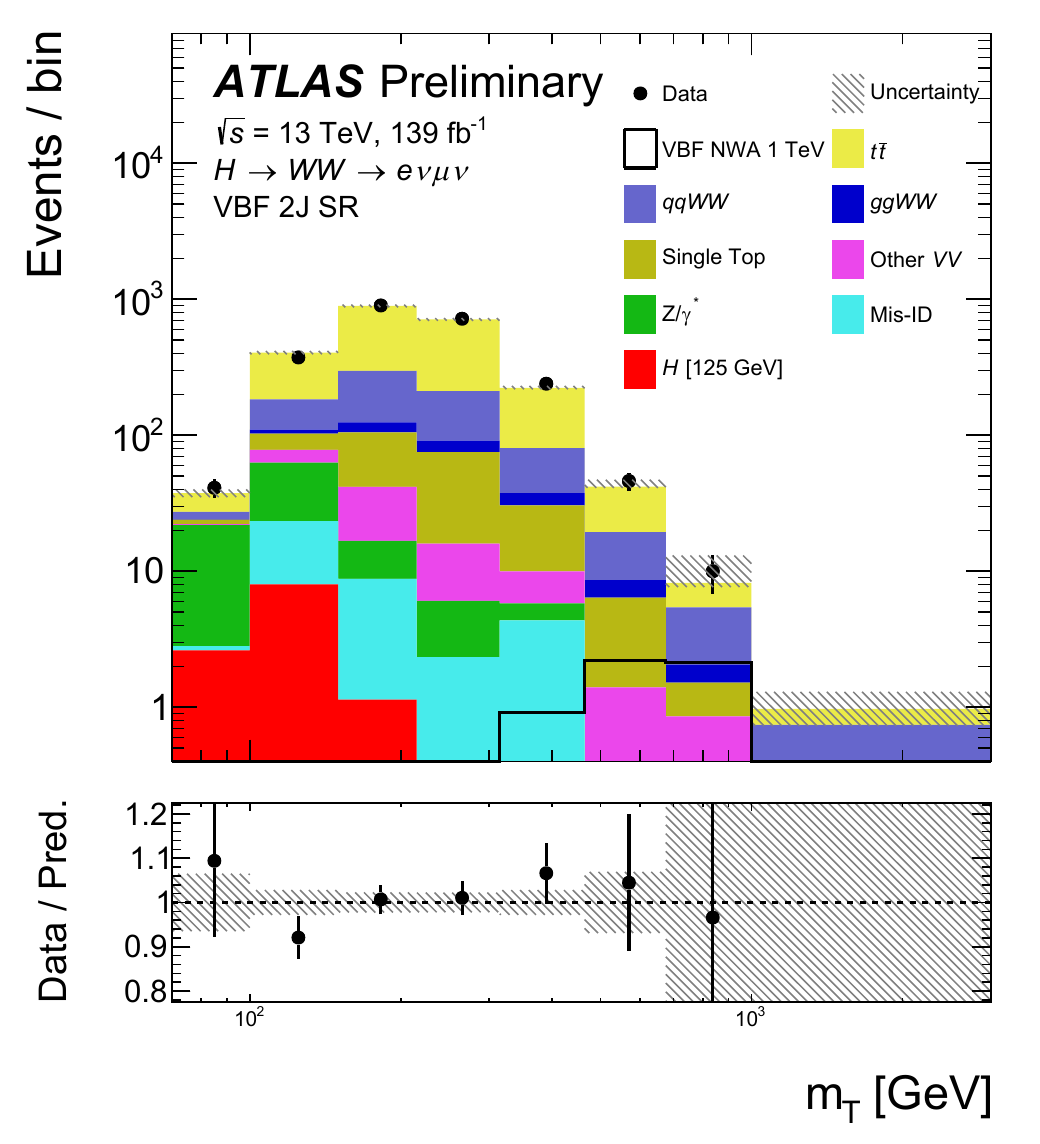}
}
\subfloat[]{
\label{fig:results:neutral:WW:limits}
\includegraphics[width=0.51\textwidth,valign=c]{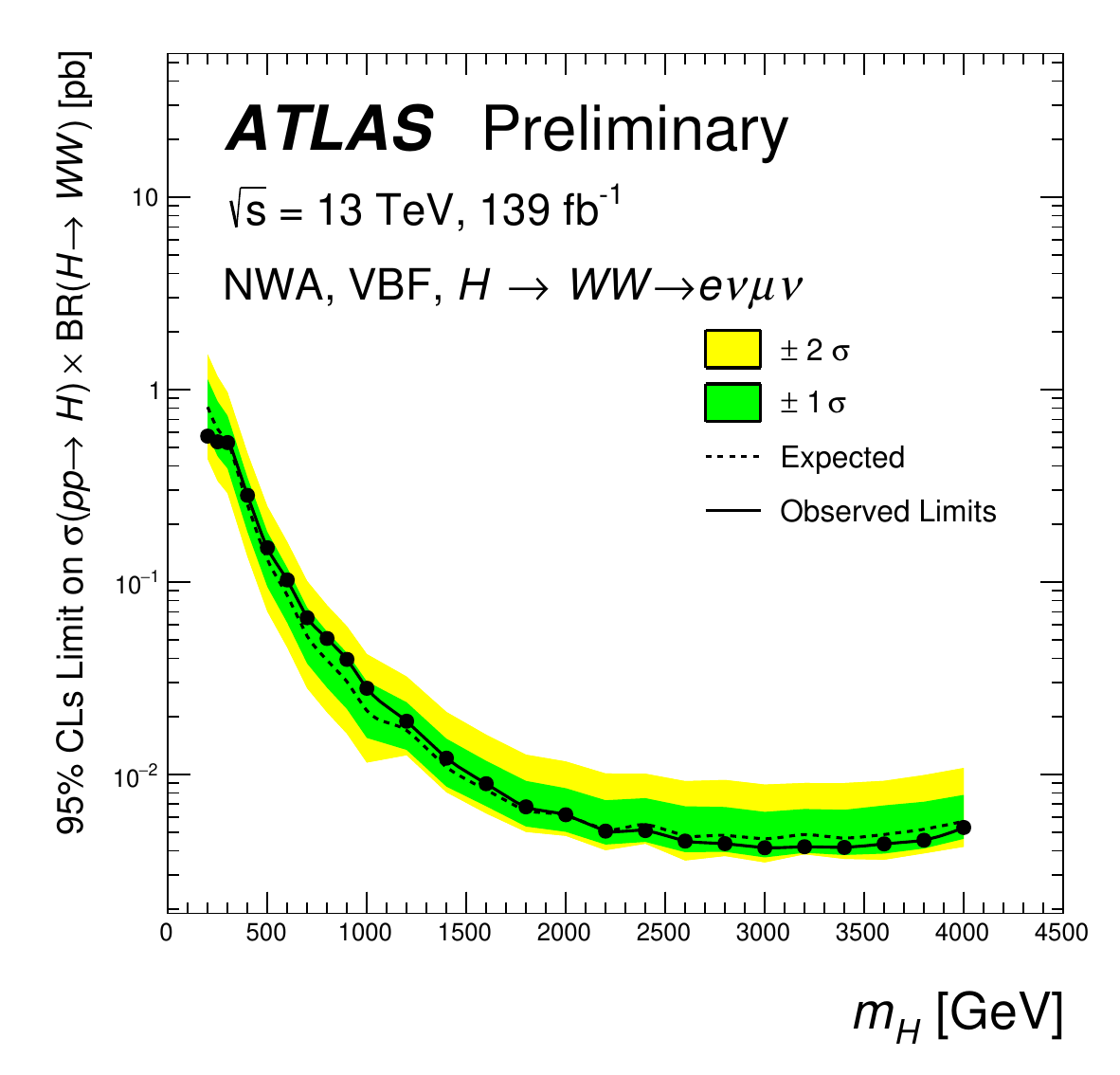}
}
\caption{\label{fig:results:neutral:WW}$H\to W^+W^-$: (a) The transverse mass in the VBF category with two jets, comparing data and background after the fit to all categories was performed assuming a signal mass of 1~\TeV. Very good agreement between data and SM backgrounds is observed. (b) The 95\% CL limits on the cross-section times branching fraction for VBF production of a scalar in the narrow-width approximation (NWA). Figures are taken from Ref.~\cite{ATLAS-CONF-2022-066}.}
\end{figure}

The search for \textbf{$H\to ZZ$}~\cite{HIGG-2018-09} was conducted in the $4\ell$ and $2\ell 2\nu$ final states from ggF and VBF production. For ggF, the maximum deviation from the background is observed around 240~\GeV with a local (global) significance of 2.0$\sigma$ (0.5$\sigma$), contributed by all four of the ggF-enriched categories in the $4\ell$ channel. For VBF, a slight excess at 620~\GeV is seen in the $4\ell$ channel with a local (global) significance of 2.4$\sigma$ (0.9$\sigma$). Narrow and wider signals were probed; the excesses become smaller when the fit is performed assuming a wider signal. The reconstructed $4\ell$ mass in the VBF-enriched category is displayed in Figure~\ref{fig:results:neutral:ZZ:4lmass}; the $2\mu 2\nu$ mass in the ggF-category is presented in Figure~\ref{fig:results:neutral:ZZ:2l2numass}. The $2\ell 2\nu$ channel is more sensitive than the $4\ell$ channel at high values of $m_H$. The exclusions obtained in a type-I 2HDM are shown in Figure~\ref{fig:results:neutral:ZZ:typeI}, to which both ggF and VBF contribute according to the model predictions. This search is also able to constrain type-II models. For $\cos\!\left(\beta-\alpha\right)=0$, the heavy Higgs boson has couplings like those in the SM (the alignment limit), which is the thin area that is not excluded. Results in the hMSSM are displayed in Figure~\ref{fig:results:summary:hmssm}.

\begin{figure}[tb!]
\centering
\subfloat[]{
\label{fig:results:neutral:ZZ:4lmass}
\includegraphics[width=0.33\textwidth,valign=c]{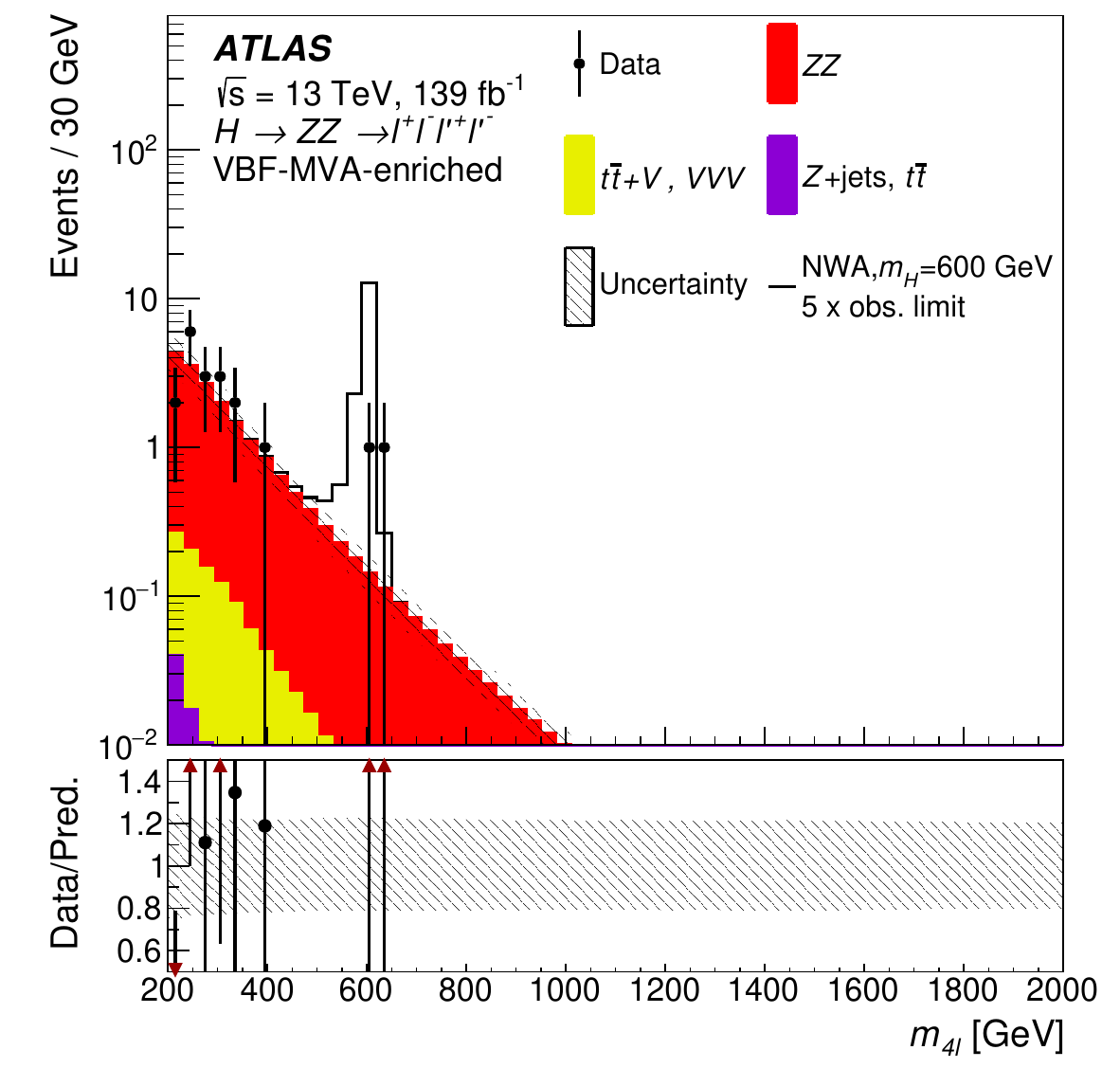}
}
\subfloat[]{
\label{fig:results:neutral:ZZ:2l2numass}
\includegraphics[width=0.33\textwidth,valign=c]{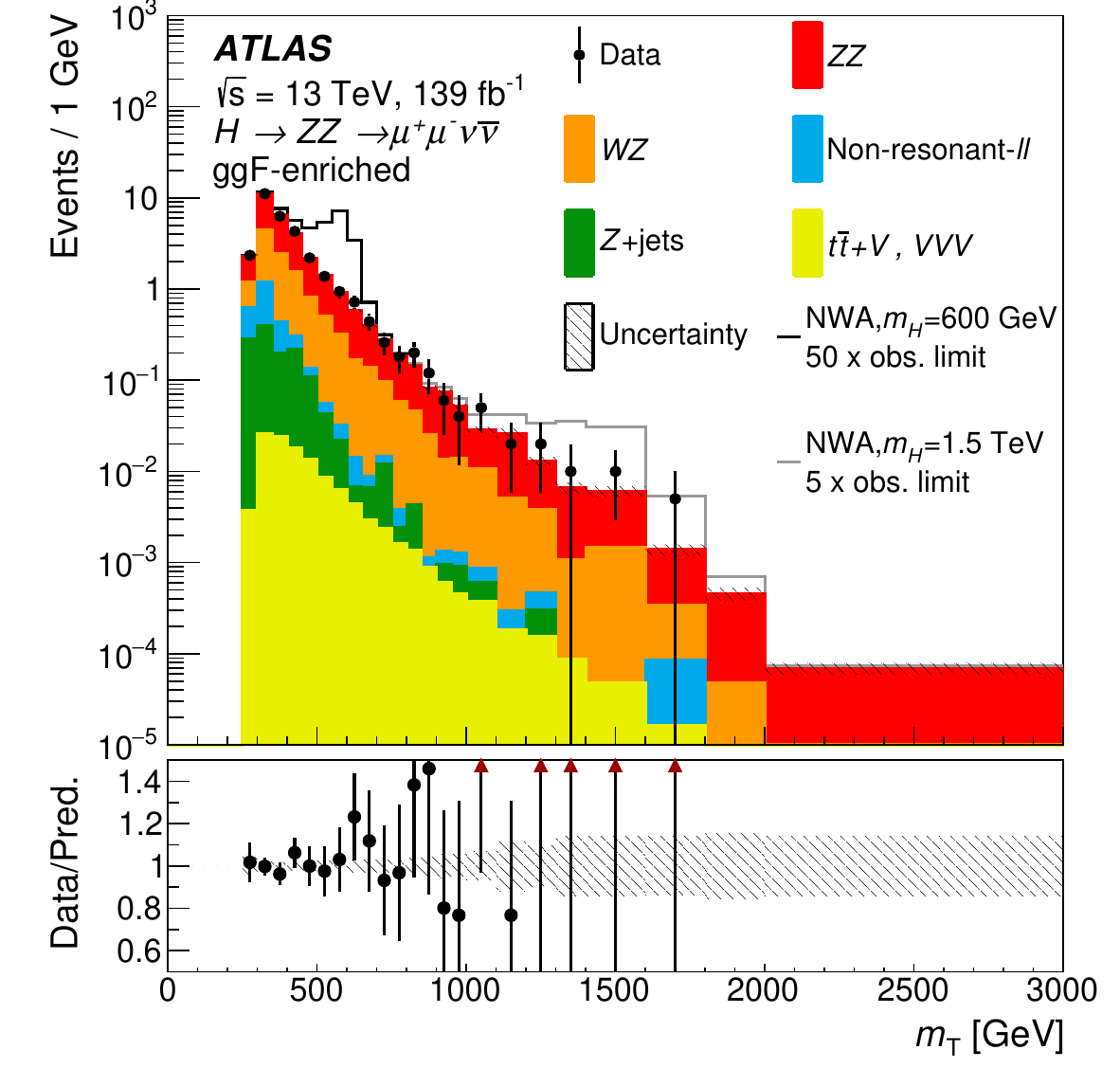}
}
\subfloat[]{
\label{fig:results:neutral:ZZ:typeI}
\includegraphics[width=0.33\textwidth,valign=c]{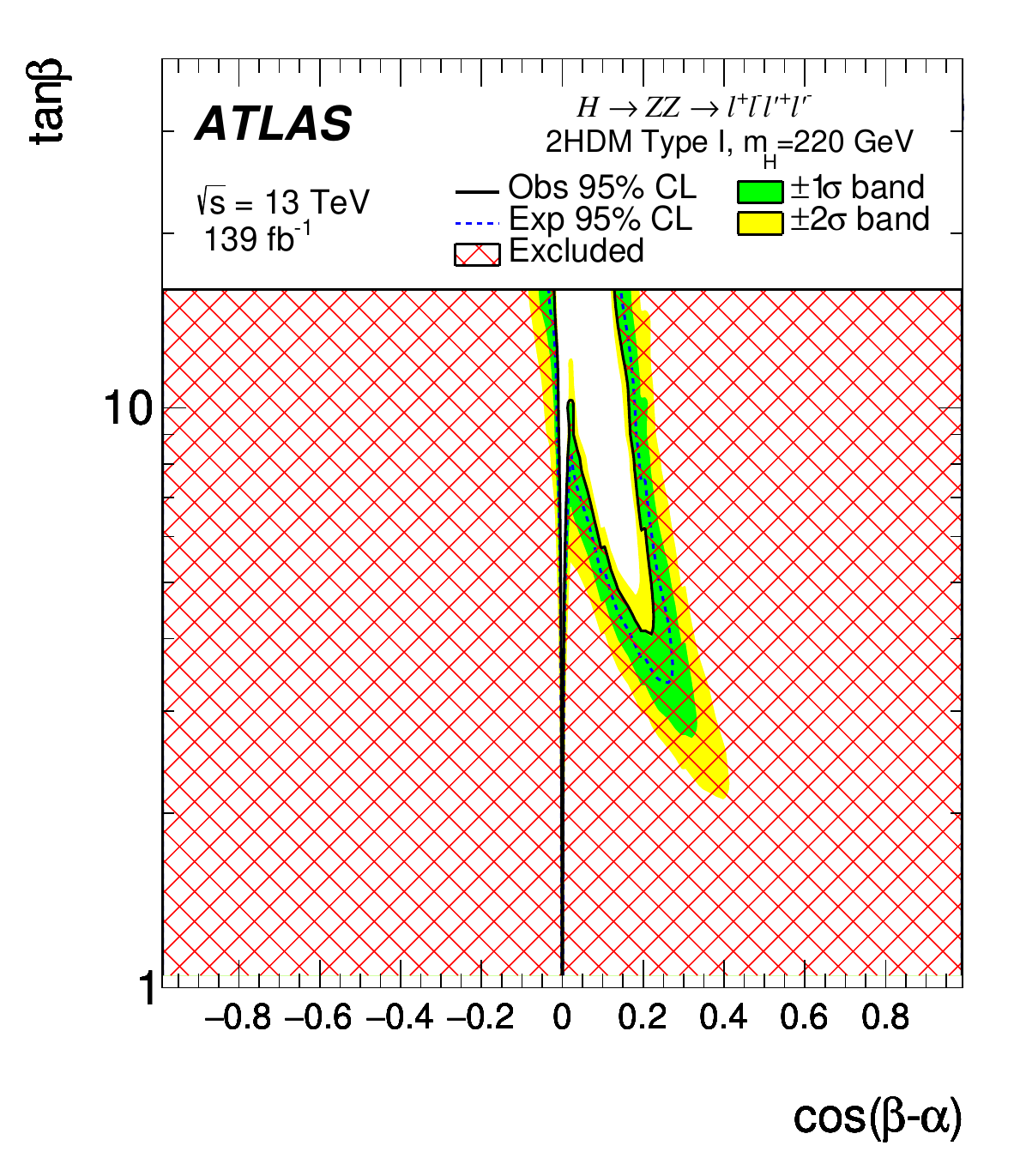}
}
\caption{\label{fig:results:neutral:ZZ}$H\to ZZ$: (a) The mass reconstructed from the four leptons in the VBF-enriched category. (b) The transverse mass in the $2\mu 2\nu$ channel in the ggF category. In both mass plots, various signal hypotheses are overlaid for illustration. (c) The 95\% CL exclusions in a type-I 2HDM for a heavy scalar mass of 220~\GeV. Figures are taken from Ref.~\cite{HIGG-2018-09}.}
\end{figure}

The high mass search for \textbf{$H\to\gamma\gamma$}~\cite{HIGG-2018-27} yielded a moderate narrow excess at a mass of 684~\GeV with a local significance of 3.3$\sigma$ that becomes 1.3$\sigma$ globally. The diphoton mass resolution is better than the width predicted by many models, so the fit not only tests a huge number of mass values, but also different widths. The significance of the excess decreases when the fit is performed assuming a model with a wider signal. Similarly, the limits on the production cross-section decrease for wider signal hypotheses. The intermediate mass search~\cite{HIGG-2023-12} was performed for generic scalars with either narrow or larger width hypotheses (\emph{model-independent analysis}), or assuming the width of a light SM-like Higgs boson (\emph{model-dependent analysis}). The model-dependent analysis is more sensitive to a light Higgs boson because it has additional BDT-based categories: it has a total of nine categories, instead of only three in the model-independent case. The largest excess for the model-independent narrow-width analysis is at 71.8~\GeV with a local significance of 2.2$\sigma$, whereas the largest excess for the model-dependent analysis is at 95.4~\GeV with a local significance of 1.7$\sigma$. Figure~\ref{fig:results:neutral:gamgam} presents the diphoton masses for the high mass search, the $\gamma\gamma$ masses for the category with two unconverted photons in the intermediate mass search, and the $p$-value for the model-dependent intermediate mass search.

\begin{figure}[tb!]
\centering
\subfloat[]{
\label{fig:results:neutral:gamgam:highmass}
\includegraphics[width=0.33\textwidth,valign=c]{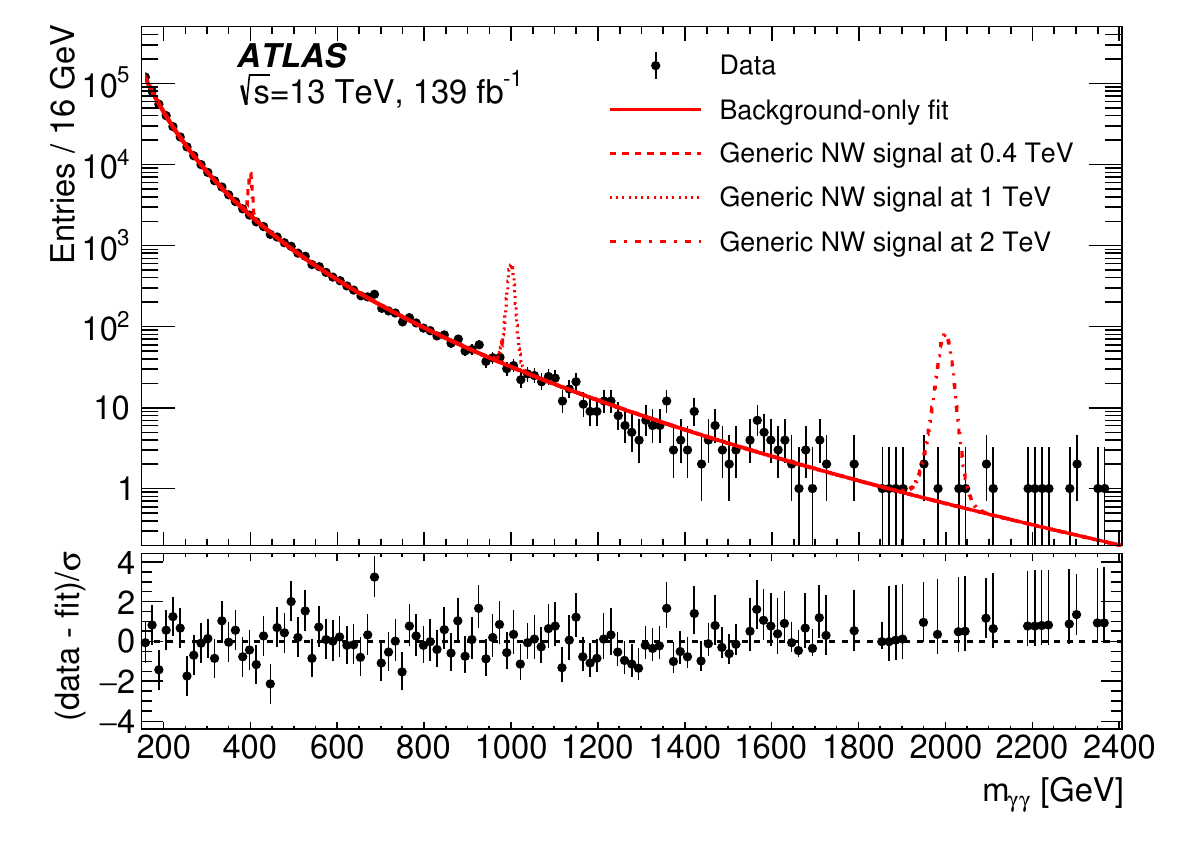}
}
\subfloat[]{
\label{fig:results:neutral:gamgam:lowmass:mass}
\includegraphics[width=0.33\textwidth,valign=c]{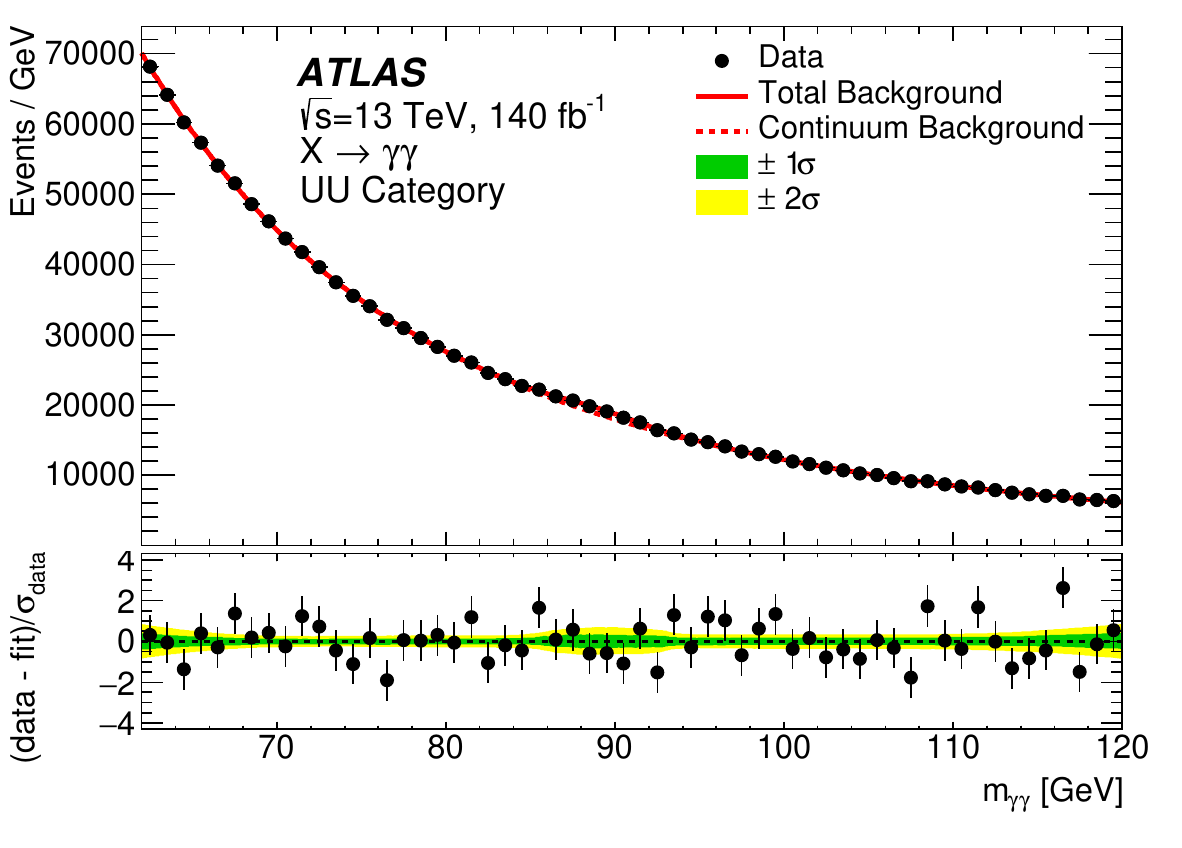}
}
\subfloat[]{
\label{fig:results:neutral:gamgam:lowmass:p}
\includegraphics[width=0.33\textwidth,valign=c]{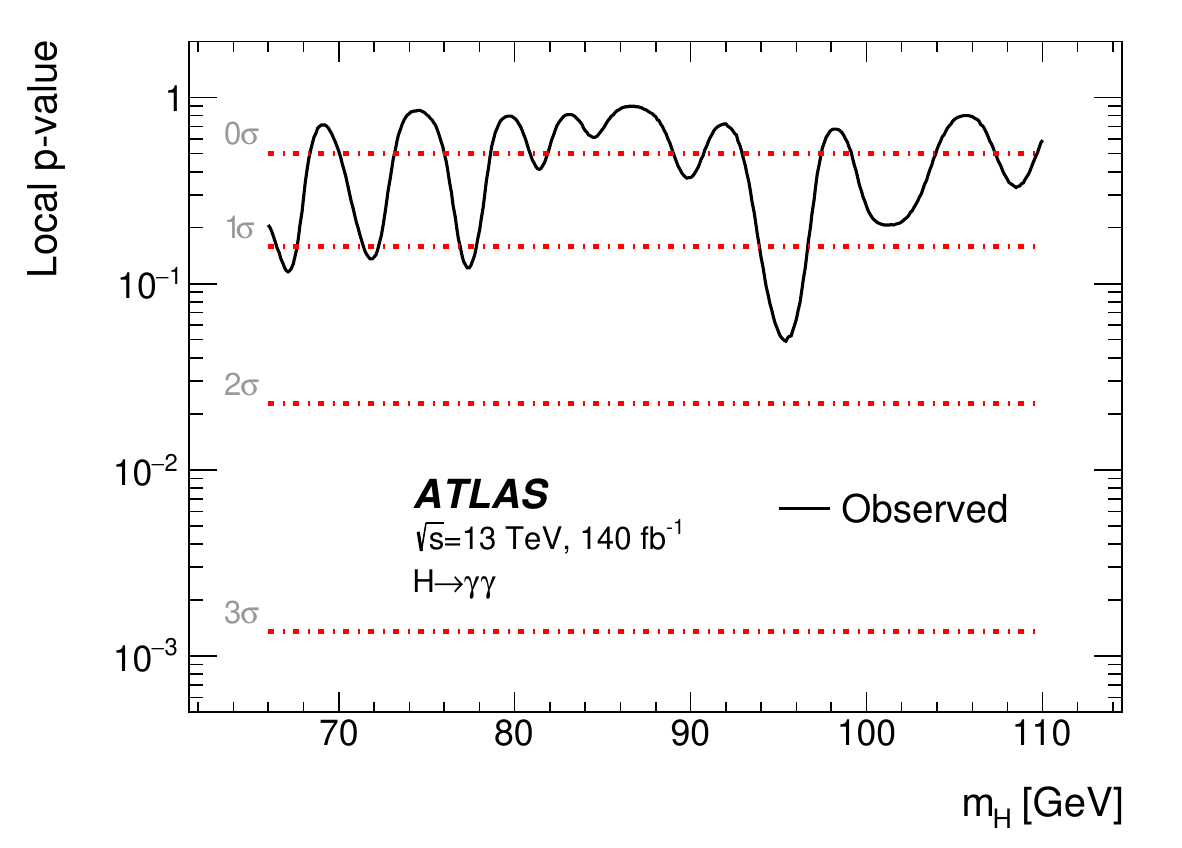}
}
\caption{\label{fig:results:neutral:gamgam}$H\to \gamma\gamma$: (a) The observed diphoton mass and a background-only fit for the high mass search. Various signal hypotheses are overlaid for illustration. (b) The $\gamma\gamma$ mass for the intermediate mass search in the category where both photons are unconverted. (c) The $p$-value for the model-dependent analysis in the intermediate mass search. Figures are taken from Ref.~\cite{HIGG-2018-27} or~\cite{HIGG-2023-12} for the high mass or intermediate mass search, respectively.}
\end{figure}

The search for a generic scalar $X$ with \textbf{$X\to Z\gamma$} in the leptonic $Z$ decay mode~\cite{HIGG-2018-44} significantly improves on previous results in the same channel and extends the mass range to 3400~\GeV. The largest excess is observed at 420~\GeV with a local significance of 2.3$\sigma$, where the $e^+e^-\gamma$ and $\mu^+\mu^-\gamma$ channels contribute with 2.1$\sigma$ and 1.1$\sigma$, respectively. For the hadronic $Z$ decay mode~\cite{HDBS-2019-10}, a local 2.5$\sigma$ excess is found at 3640~\GeV. The limits on the production cross-section for $Z\gamma$ are shown in Figure~\ref{fig:results:neutral:Zgam} for both decays. The hadronic channel is more sensitive for mass hypotheses above 2.1~\TeV. In both analyses, only narrow signals are investigated.

\begin{figure}[tb!]
\centering
\subfloat[]{
\label{fig:results:neutral:Zgam:limits_lep}
\includegraphics[width=0.49\textwidth,valign=c]{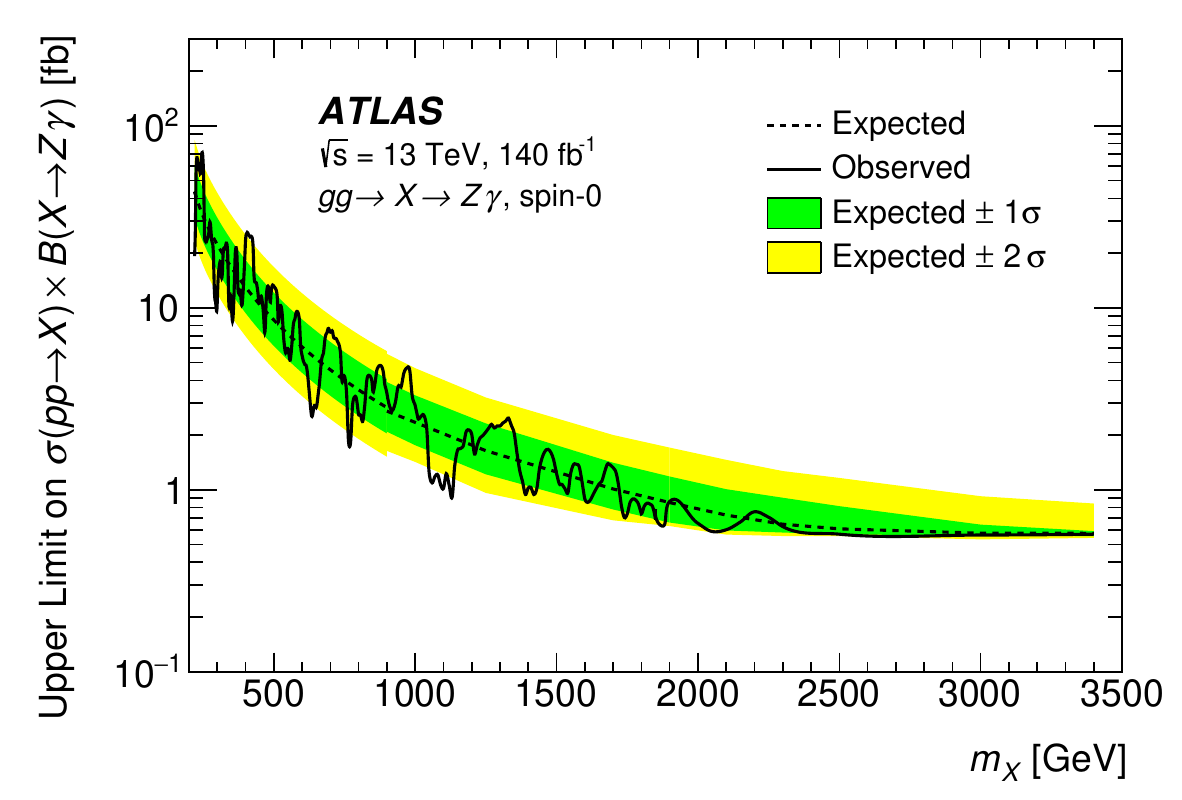}
}
\subfloat[]{
\label{fig:results:neutral:Zgam:limits_had}
\includegraphics[width=0.49\textwidth,valign=c]{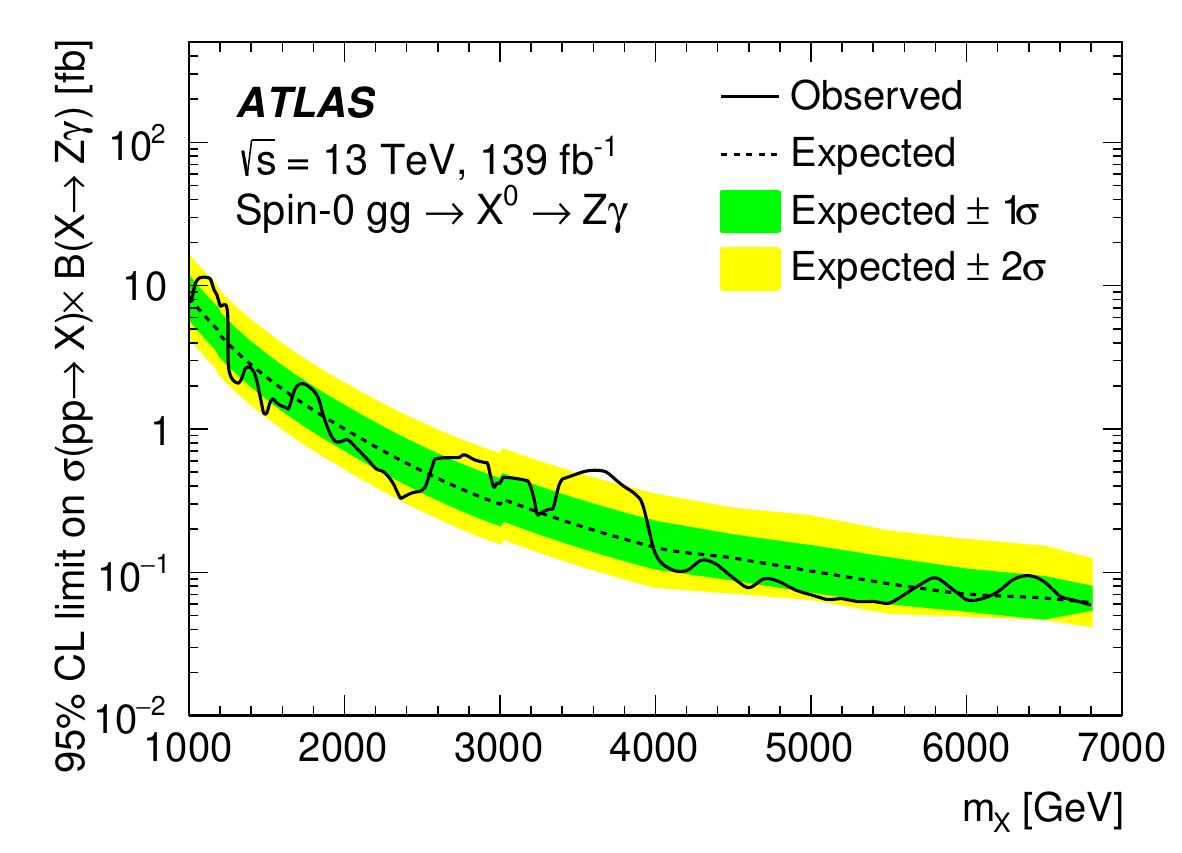}
}
\caption{\label{fig:results:neutral:Zgam}$X\to Z\gamma$: The 95\% CL limit on the cross-section times branching fraction as a function of the mass of the narrow-width scalar $X$ in the case where the $Z$ boson decays (a) into either electrons or muons, or (b) via the hadronic decay mode. The limits in (b) have a small discontinuity at 3~\TeV because the category exploiting $b$-tagging is dropped for higher masses. Figures are taken from Refs.~\cite{HIGG-2018-44} and~\cite{HDBS-2019-10}.}
\end{figure}

\subsubsection{Higgs-to-Higgs decays}
\label{sec:results:neutralhiggs:higgstohiggs}

The search for \textbf{$A\to Zh_{125}$} with $h_{125}\to b\bar{b}$~\cite{HDBS-2020-19} was performed for ggF and $b$-associated production and spans a large mass range involving resolved or boosted final states. The largest deviation from the SM expectations is found at an $A$-boson mass of 500~\GeV, originating mostly from the resolved 2-$b$-tag category of the 2-lepton channel. Assuming ggF production, this excess corresponds to a local (global) significance of 2.1$\sigma$ (1.1$\sigma$). For the signal hypothesis of $b$-associated production the local significance is 1.6$\sigma$ at the same resonance mass value. In Figure~\ref{fig:results:neutral:AZh125:mass} the reconstructed $A$ mass is displayed for one of the boosted categories with two leptons. Figure~\ref{fig:results:neutral:AZh125:limitsggF} shows the 95\% CL limit on the production cross-section for ggF, and Figure~\ref{fig:results:neutral:AZh125:typeII} shows the exclusions in a type-II 2HDM for $m_A=700$~\GeV. Results in the hMSSM are displayed in Figure~\ref{fig:results:summary:hmssm}.

\begin{figure}[tb!]
\centering
\subfloat[]{
\label{fig:results:neutral:AZh125:mass}
\includegraphics[width=0.33\textwidth,valign=c]{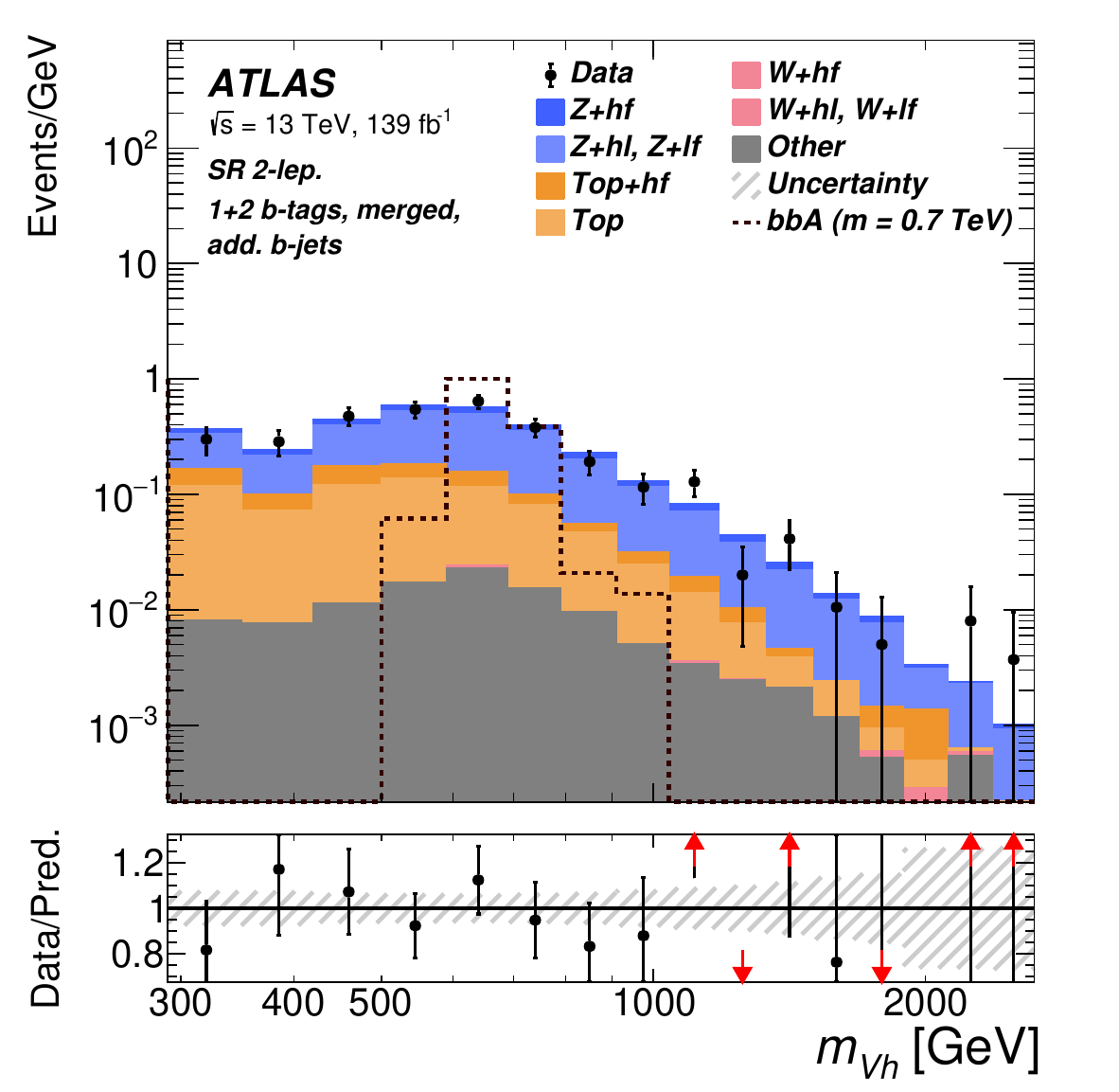}
}
\subfloat[]{
\label{fig:results:neutral:AZh125:limitsggF}
\includegraphics[width=0.36\textwidth,valign=c]{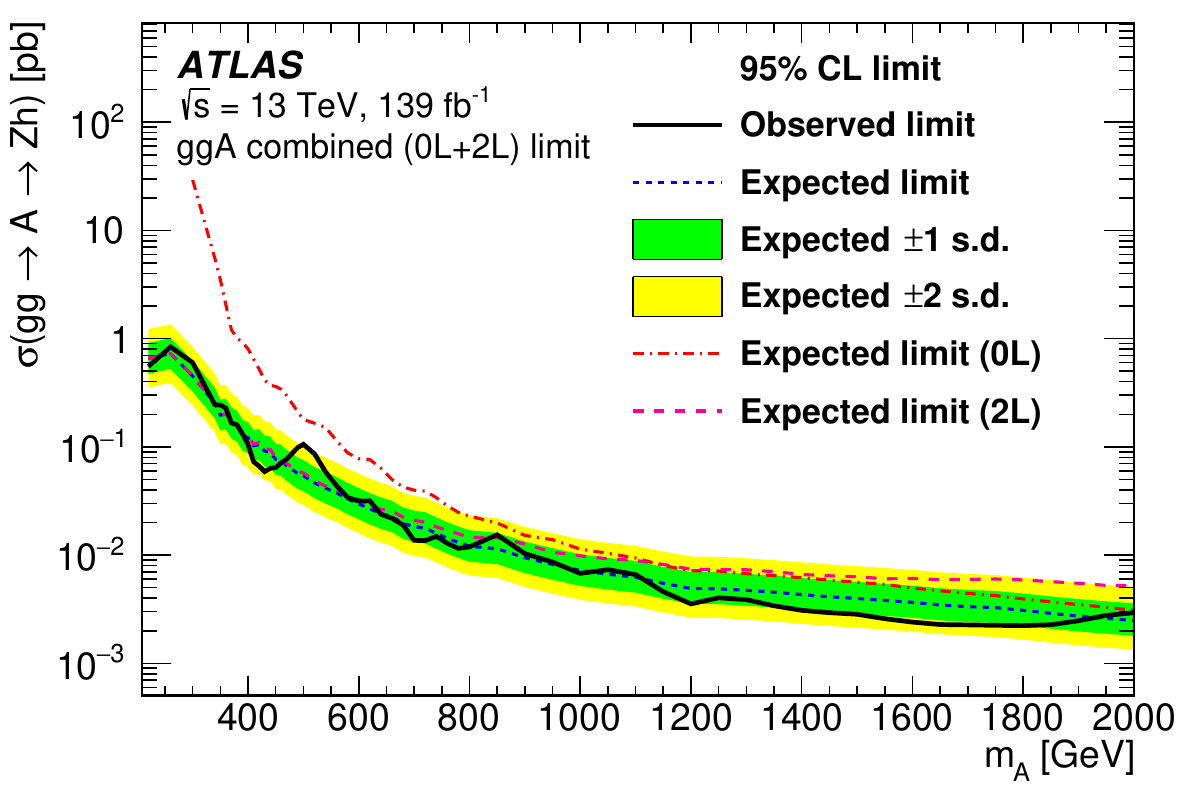}
}
\subfloat[]{
\label{fig:results:neutral:AZh125:typeII}
\includegraphics[width=0.3\textwidth,valign=c]{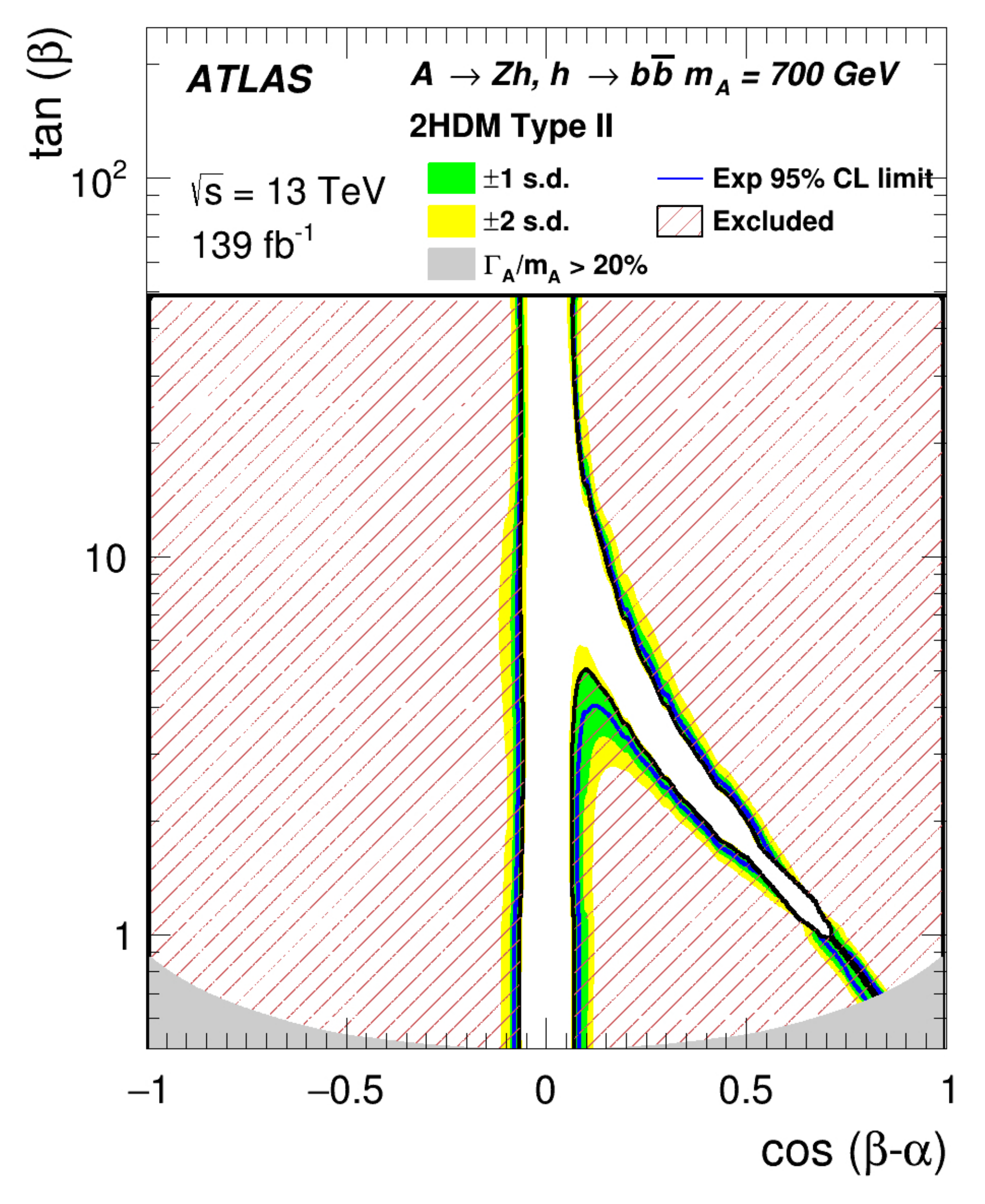}
}
\caption{\label{fig:results:neutral:AZh125}$A\to Zh_{125}$: (a) The reconstructed $A$ mass in the category with two leptons and at least three $b$-tags. A signal model at $m_A=700$~\GeV is overlaid as a dashed line. (b)~The 95\% CL limits on the cross-section times branching fraction for ggF production; the expected limits from the categories with no or two leptons is also shown. (c)~The exclusions in the type-II 2HDM for $m_A$=700~\GeV, combining ggF and $b$-associated production according to the model predictions. Figures are taken from Ref.~\cite{HDBS-2020-19}.}
\end{figure}

The search for \textbf{$A\to ZH$}~\cite{HDBS-2018-13} was performed in the \textbf{$2\ell 2b$} and \textbf{$2\ell W^+W^-\to 2\ell4q$} final states. The channels explore different aspects of the 2HDM. The $2\ell 2b$ channel is strong at the weak decoupling limit~\cite{PhysRevD.67.075019}, where the $H$ decay into vector bosons is suppressed and the decay into fermions favoured. In contrast, the $2\ell W^+W^-$ channel is interesting in the region close to, but not exactly at, the weak decoupling limit. In the $2\ell 2b$ channel, the largest excess for ggF production is at $\left(m_A,m_H\right)=\left(610,290\right)$~\GeV with a local (global) significance of 3.1$\sigma$ (1.3$\sigma$). For $b$-associated production, the most significant excess is at $\left(m_A,m_H\right)=\left(440,220\right)$~\GeV with a local (global) significance of 3.1$\sigma$ (1.3$\sigma$). In the case of the $2\ell W^+W^-$ channel (which only considers ggF), the largest excess is at $\left(m_A,m_H\right)=\left(440,310\right)$~\GeV with a local (global) significance of 2.9$\sigma$ (0.8$\sigma$). Both channels are able to constrain the 2HDM. The final discriminant for each channel as well as the exclusions in the type-I 2HDM are displayed in Figure~\ref{fig:results:neutral:AZH}.

\begin{figure}[tb!]
\centering
\subfloat[]{
\label{fig:results:neutral:AZH:massllbb}
\includegraphics[width=0.31\textwidth,valign=c]{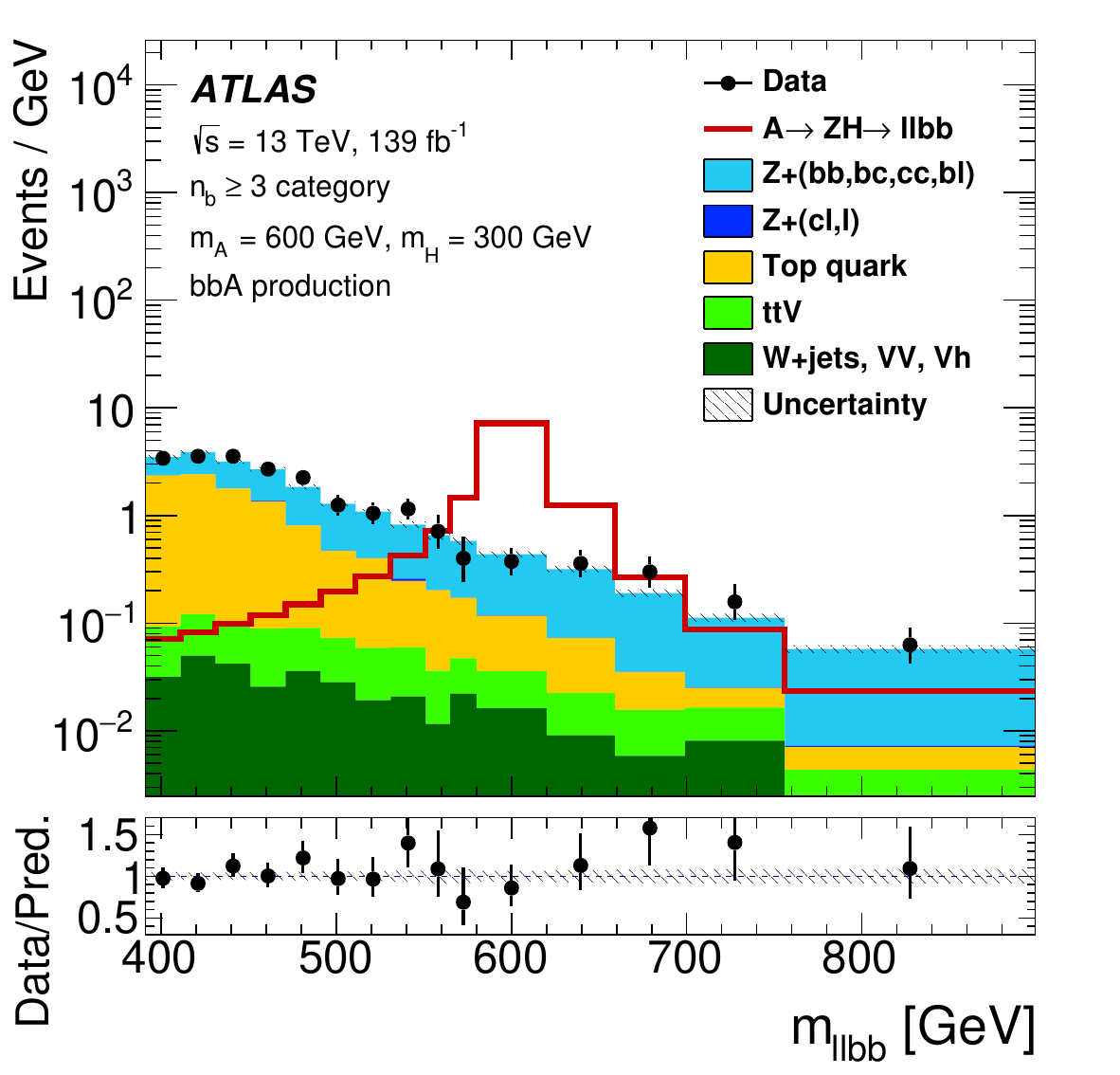}
}
\subfloat[]{
\label{fig:results:neutral:AZH:mass2l4q}
\includegraphics[width=0.31\textwidth,valign=c]{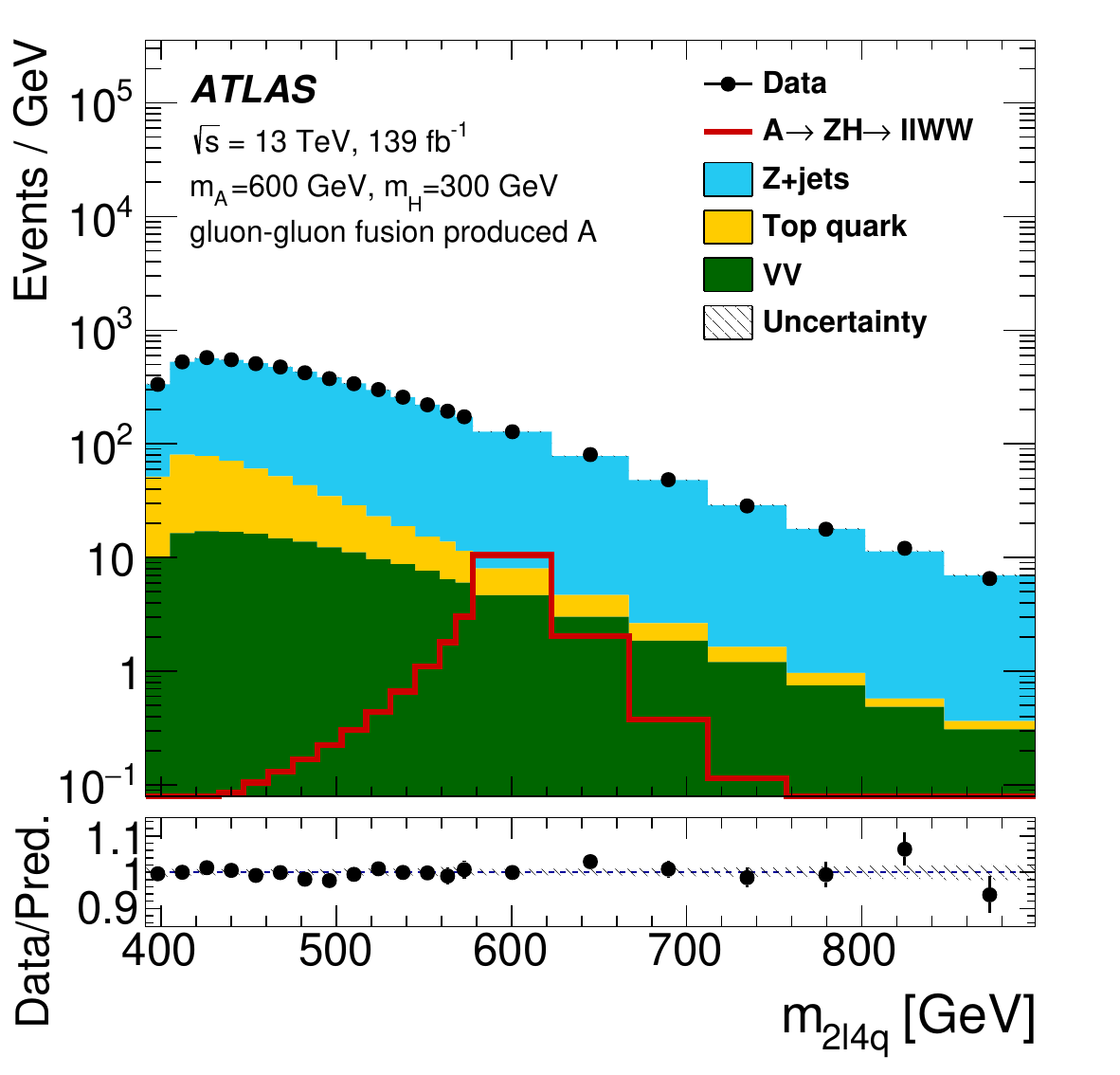}
}
\subfloat[]{
\label{fig:results:neutral:AZH:typeI}
\includegraphics[width=0.37\textwidth,valign=c]{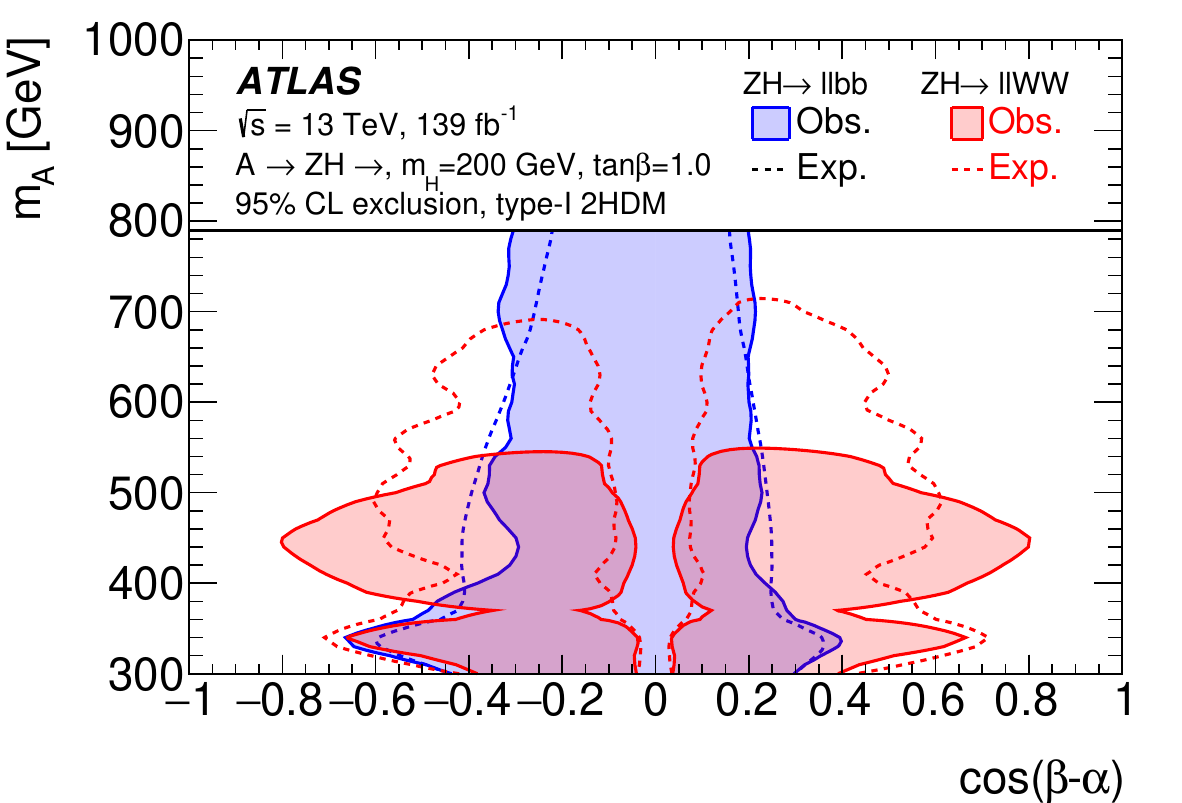}
}
\caption{\label{fig:results:neutral:AZH}$A\to ZH$: The reconstructed $A$ mass in (a) the $2\ell 2b$ channel for the category with at least three $b$-jets and $b$-associated production, and (b) the $2\ell W^+W^-$ channel. (c) An overlay of the 95\% CL exclusions in the type-I 2HDM for both channels, assuming $m_H=200$~\GeV and $\tan\beta=1$. Figures are taken from Ref.~\cite{HDBS-2018-13}.}
\end{figure}

Other channels that were investigated in the context of an \textbf{$A\to ZH$} signature are \textbf{$\ell^+\ell^- t\bar{t}$} and \textbf{$\nu\bar{\nu} b\bar{b}$}~\cite{HDBS-2021-02}. Also, these channels are complementary: the Higgs boson decay to $t\bar{t}$ is strong at low $\tan\beta$ and favoured in type-I models, while the $\nu\bar{\nu} b\bar{b}$ channel is strong at high $\tan\beta$ in type-II or flipped models. This complementarity is visible in the 2HDM exclusions displayed in Figure~\ref{fig:results:neutral:AZH_lltt_nnbb}, where $\cos\!\left(\alpha-\beta\right)=0$ is assumed. The largest excess, with a local (global) significance of 2.9$\sigma$ (2.4$\sigma$), was observed for $\ell^+\ell^- t\bar{t}$ at $\left(m_A,m_H\right)=\left(650,450\right)$~\GeV.

\begin{figure}[tb!]
\centering
\subfloat[]{
\label{fig:results:neutral:AZH_lltt_nnbb:typeI}
\includegraphics[width=0.48\textwidth,valign=c]{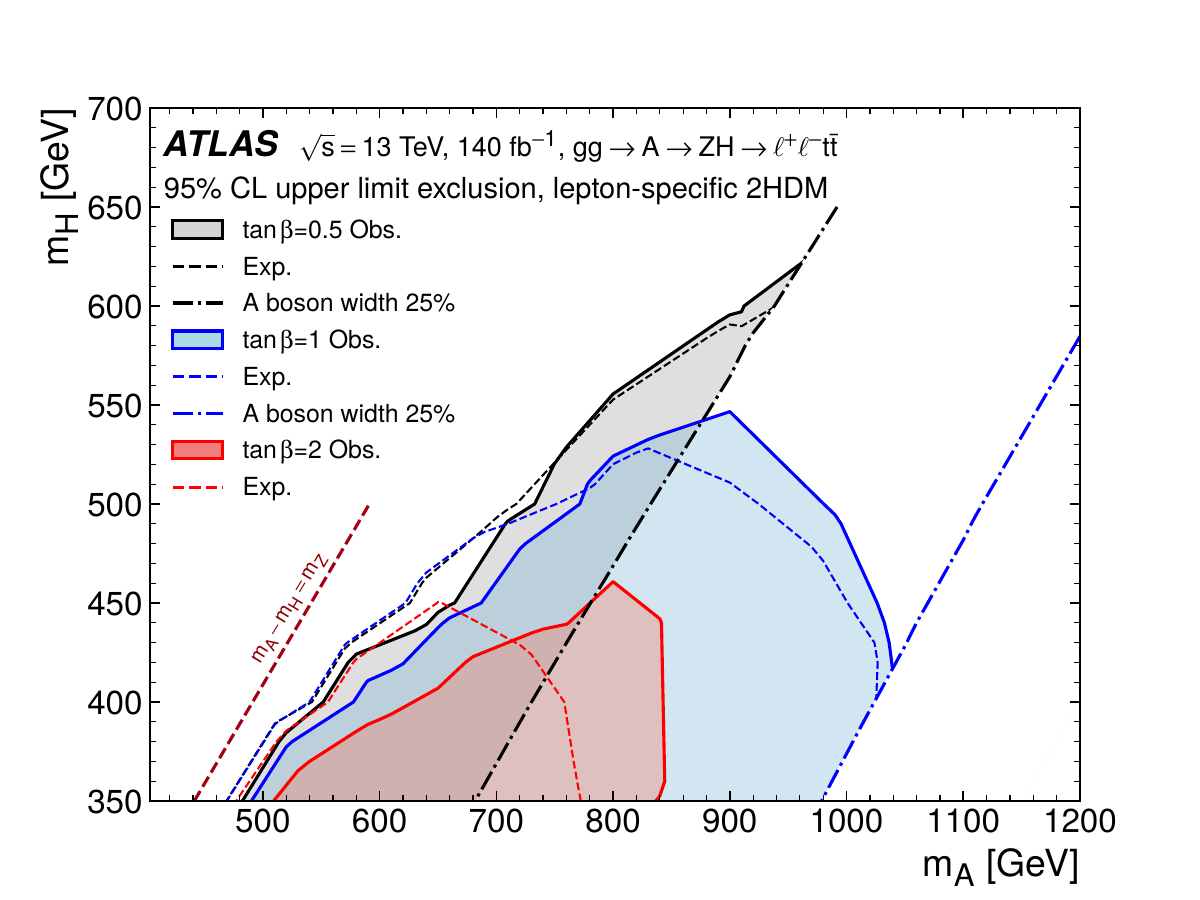}
}
\subfloat[]{
\label{fig:results:neutral:AZH_lltt_nnbb:flipped}
\includegraphics[width=0.48\textwidth,valign=c]{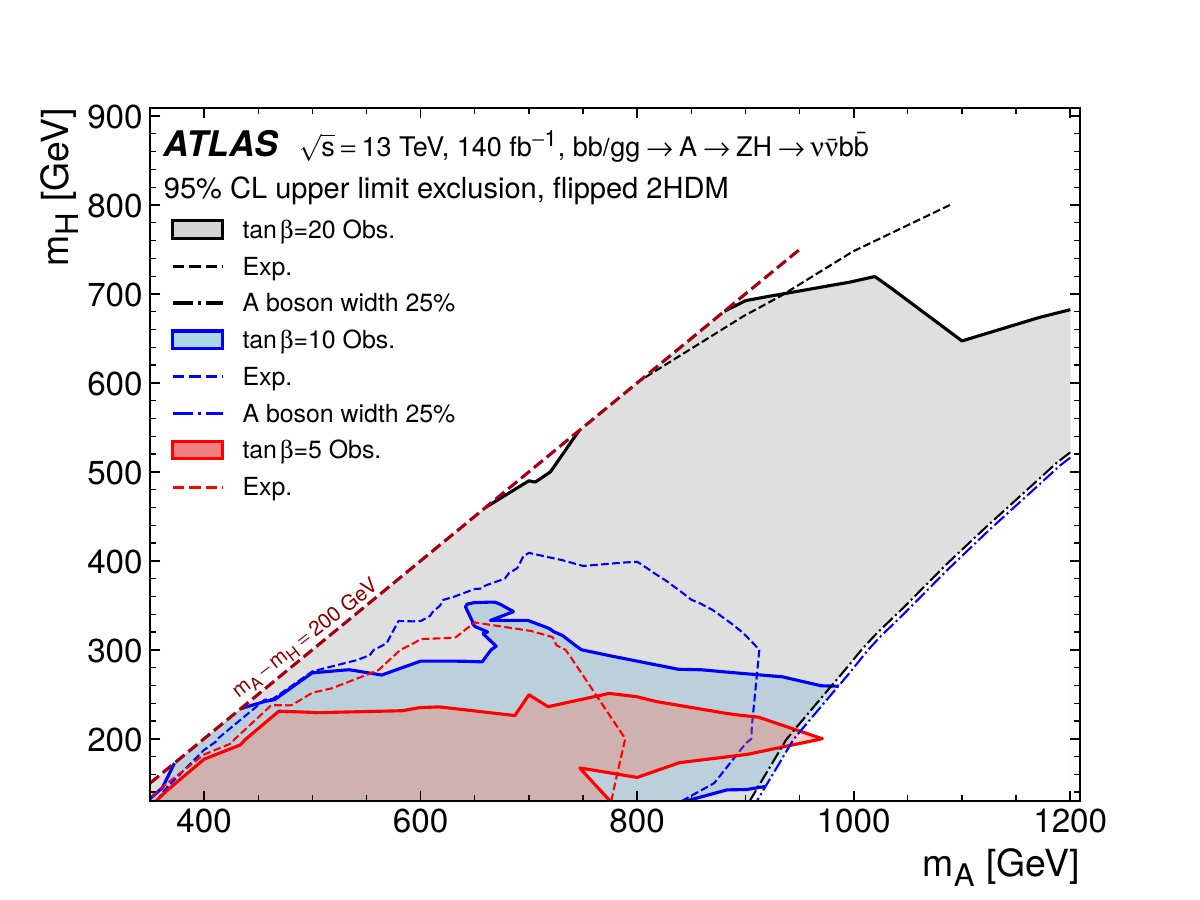}
}
\caption{\label{fig:results:neutral:AZH_lltt_nnbb}$A\to ZH$: 95\% CL exclusions in the 2HDM for $A\to ZH$ presented in the 2D mass plane of $A$ vs $H$. Here, figure (a) considers the type-I model for the $\ell\ell t\bar{t}$ channel, which is more sensitive towards lower values of $\tan\beta$, and figure (b) displays the flipped model for the $\nu\bar{\nu} b\bar{b}$ channel, which is more sensitive towards higher $\tan\beta$ values. Figures are taken from Ref.~\cite{HDBS-2021-02}.}
\end{figure}

Some results of the search for \textbf{$VH$ with $H\to hh\to\bbbar\bbbar$}~\cite{HDBS-2019-31} are displayed in Figure~\ref{fig:results:dihiggs:Vbbbb}. In this search the $h$ is identified with the 125~\GeV Higgs boson. The heavy Higgs boson $H\to hh$ is either produced in association with a $V$ boson or comes from the decay of heavier pseudoscalar $A$ boson. The data are in good agreement with the estimated SM background contributions, except for a few notable excesses. The most significant deviation is observed in the $A \to ZH \to Zhh$ channel for a large-width $A$ boson at $(m_A, m_H) = (420, 320)$~\GeV, where the local (global) significance is $3.8\sigma$ ($2.8\sigma$). Here, \enquote{large-width} means that the width of the $A$ is 20\% of its mass value. Upper limits on the $Vhh$ production cross-sections are obtained as a function of $m_H$ in the range 260--1000~\GeV for $WH$ and $ZH$ separately, and in the $(m_A, m_H)$ plane for $A \to ZH$, covering the $m_A$ range 360--800~\GeV and $m_H$ range 260--400~\GeV as shown in Figure~\ref{fig:results:dihiggs:Vbbbb}. The constraints on $A \to ZH$ production are also interpreted in the $(\cos(\beta - \alpha), m_A)$ parameter space of type-I and lepton-specific 2HDMs.

\begin{figure}[tb!]
\begin{center}

\subfloat[]{
\includegraphics[width=0.49\textwidth,valign=c]{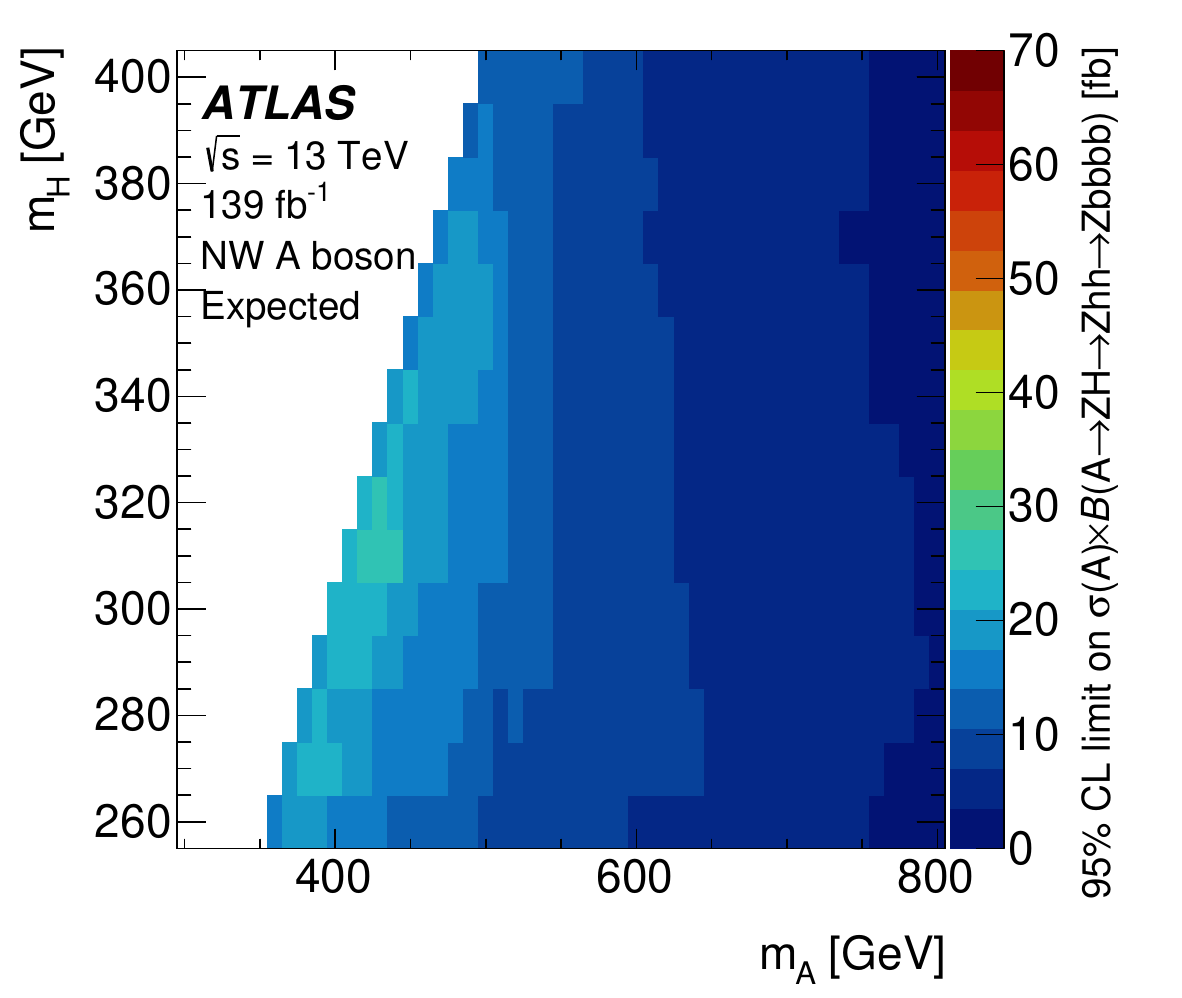}
}
\subfloat[]{
\includegraphics[width=0.49\textwidth,valign=c]{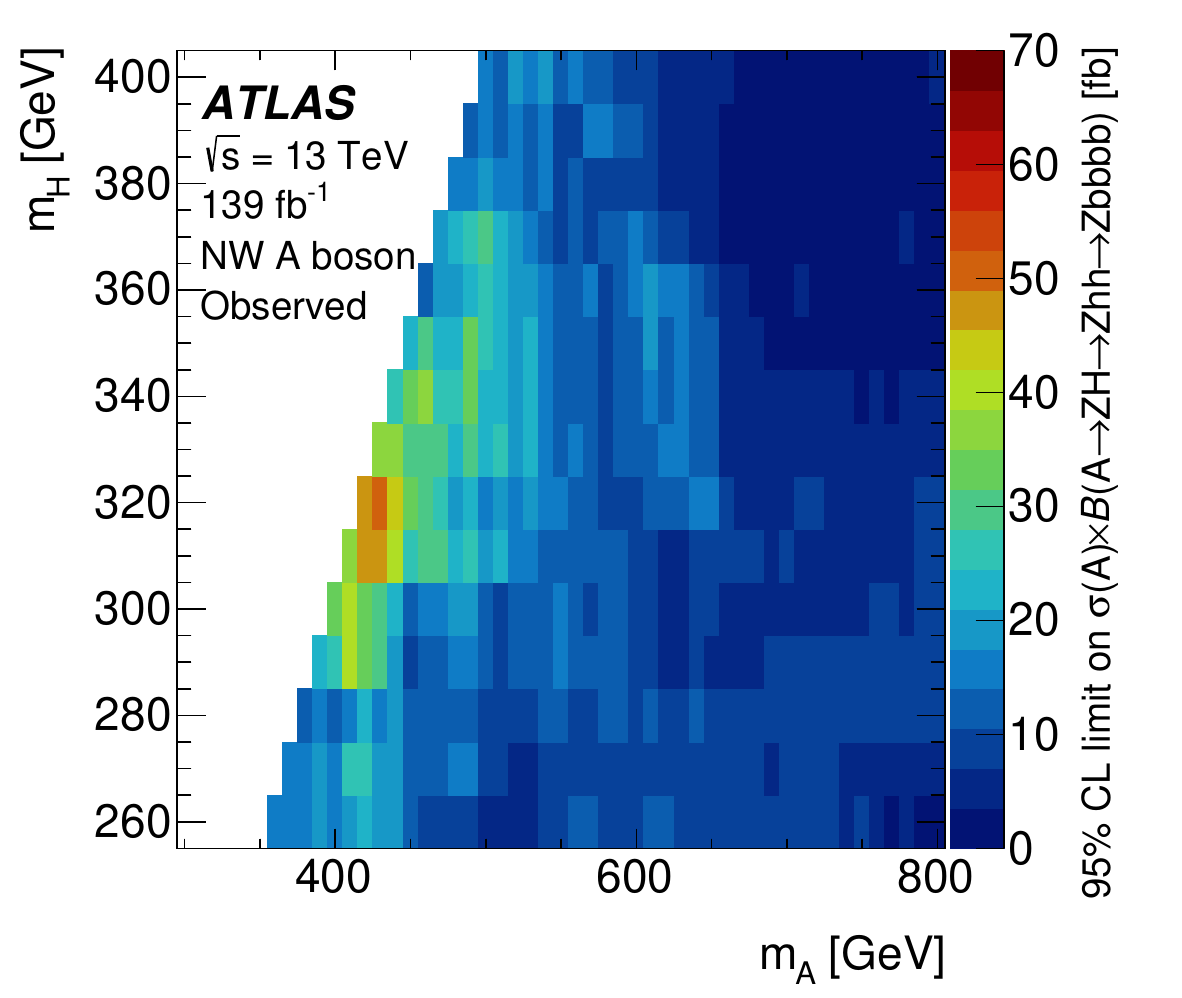}
}\\
\subfloat[]{
\includegraphics[width=0.49\textwidth,valign=c]{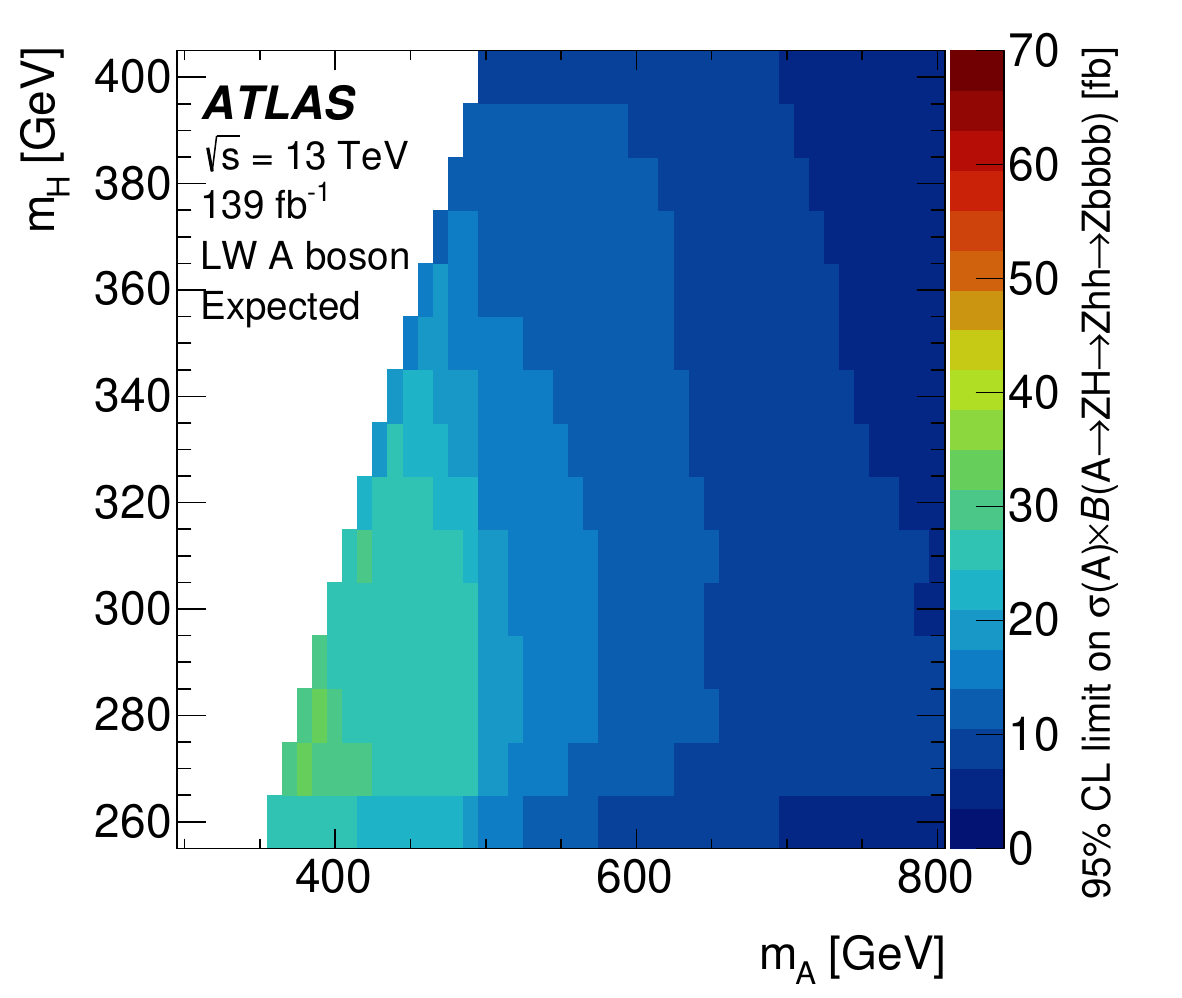}
}
\subfloat[]{
\includegraphics[width=0.49\textwidth,valign=c]{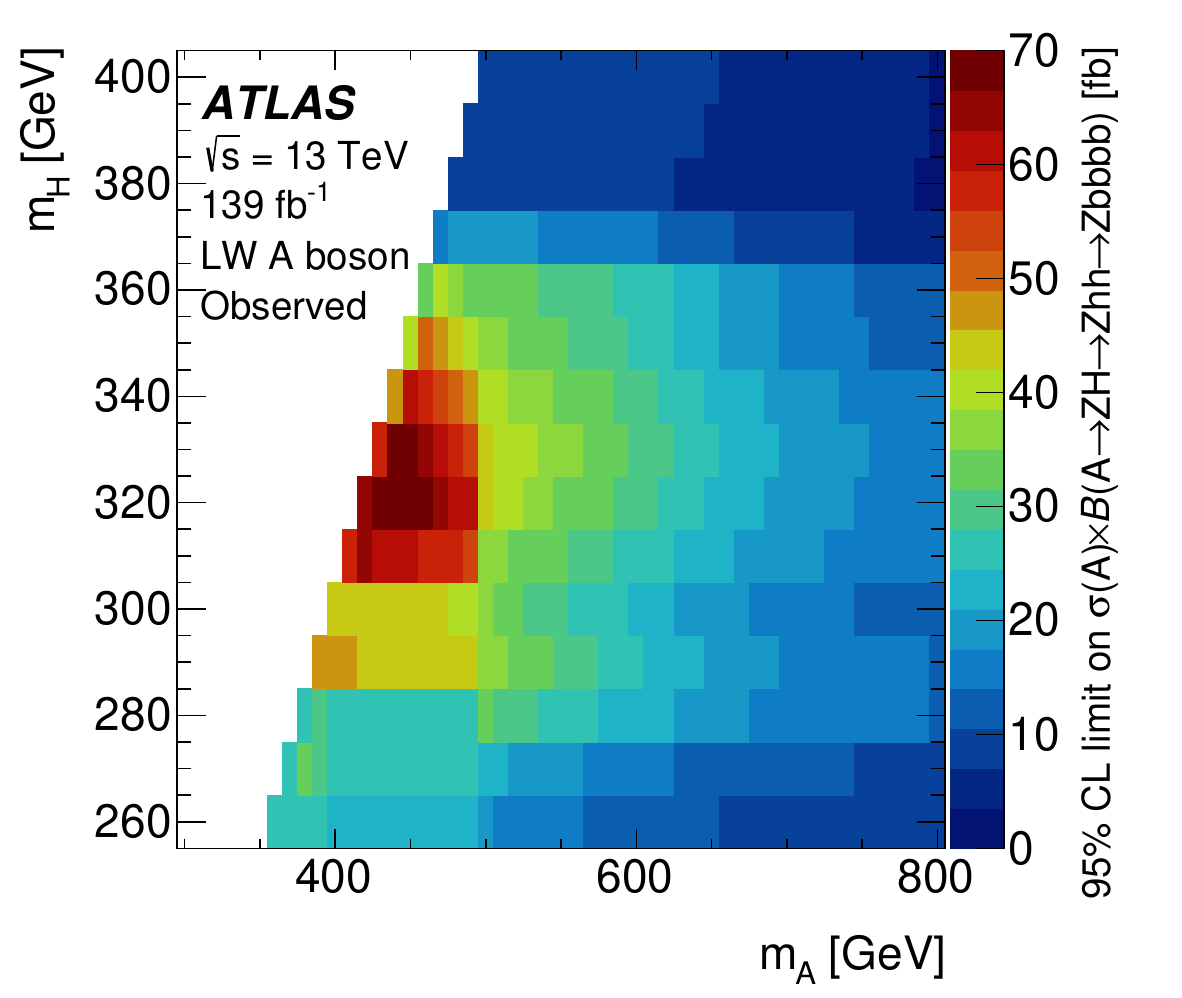}
}

\caption{$VH\to Vhh \to V\bbbar\bbbar$: Upper bounds at 95\% CL on $\sigma(A) \times \mathcal{B}(A \to ZH \to Zhh \to Z\bbbar\bbbar)$ in the $(m_A, m_H)$ plane for (a, b) a narrow-width (NW) $A$ boson and (c, d) a large-width (LW) $A$ boson. The expected upper limits are shown in (a) and (c), while the observed limits are shown in (b) and (d). The $A$ boson has a total decay width that is negligible compared to the experimental mass resolution in the NW case and is 20\% of its mass in the LW case. Figures are taken from Ref.~\cite{HDBS-2019-31}.}
\label{fig:results:dihiggs:Vbbbb}
\end{center}
\end{figure}

\FloatBarrier


\subsection{Charged Higgs bosons}
\label{sec:results:chargedhiggs}

\subsubsection{Charged Higgs bosons decaying into fermions}
\label{sec:results:chargedhiggs:fermions}

The search for \textbf{$H^+\to\tau^+\nu$}~\cite{HIGG-2016-11} was carried out over a large mass range of 90--2000~\GeV, seamlessly covering light and heavy $H^+$, including the intermediate mass range where $m_{H^+}\sim m_t$, which had not been explored previously by ATLAS. Its sensitivity and ability to consistently constrain $H^+$ production over this large mass range make this channel very powerful. Some MSSM scenarios that predict a light $H^+$ with $90 \leq m_{H^+}\leq 160$~\GeV are excluded by this search. Figure~\ref{fig:results:charged:taunu:BDT} displays the output score of a BDT trained to discriminate between SM backgrounds and $H^+$ signals in the intermediate mass range. Figure~\ref{fig:results:charged:taunu:limits} shows the limits on the cross-section times branching fraction of the $H^+$ as a function of its hypothesized mass. Constraints on the hMSSM are presented in Figure~\ref{fig:results:summary:hmssm}.

\begin{figure}[tb!]
\centering
\subfloat[]{
\label{fig:results:charged:taunu:BDT}
\includegraphics[width=0.48\textwidth,valign=c]{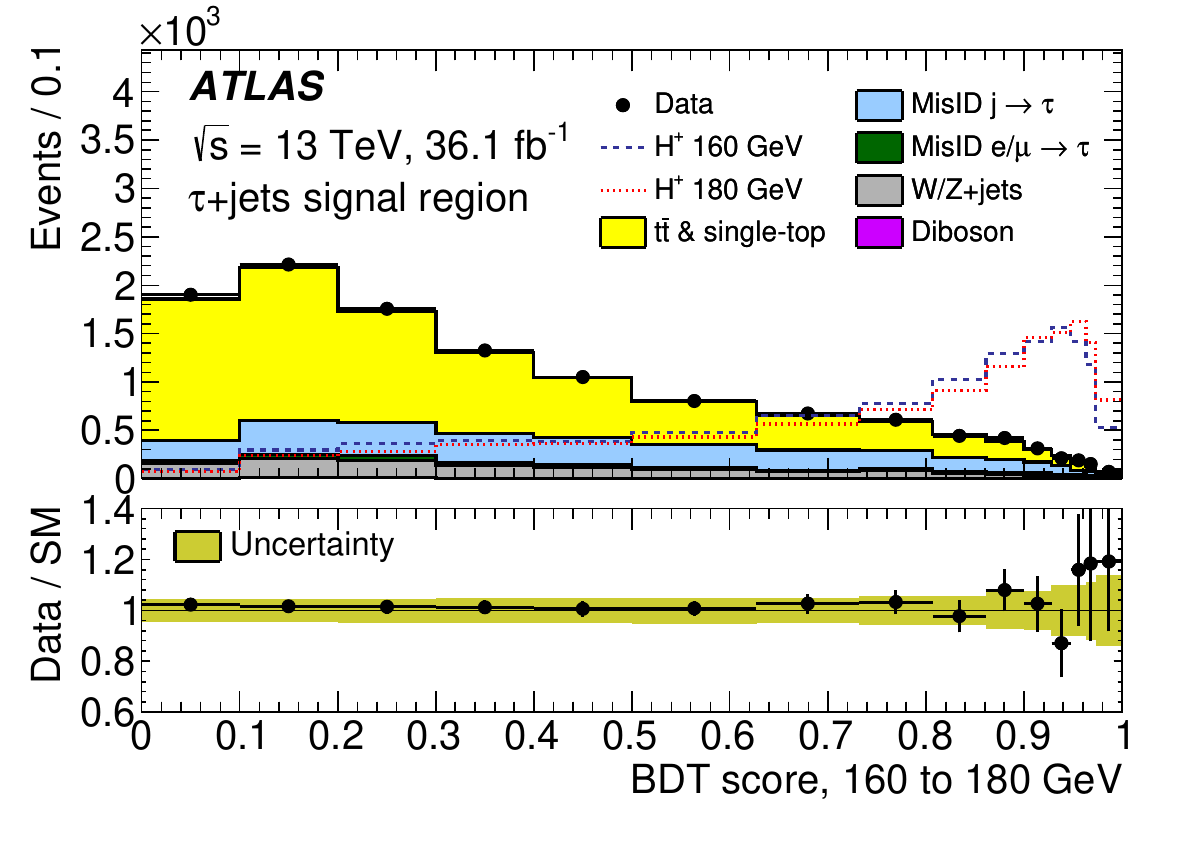}
}
\subfloat[]{
\label{fig:results:charged:taunu:limits}
\includegraphics[width=0.48\textwidth,valign=c]{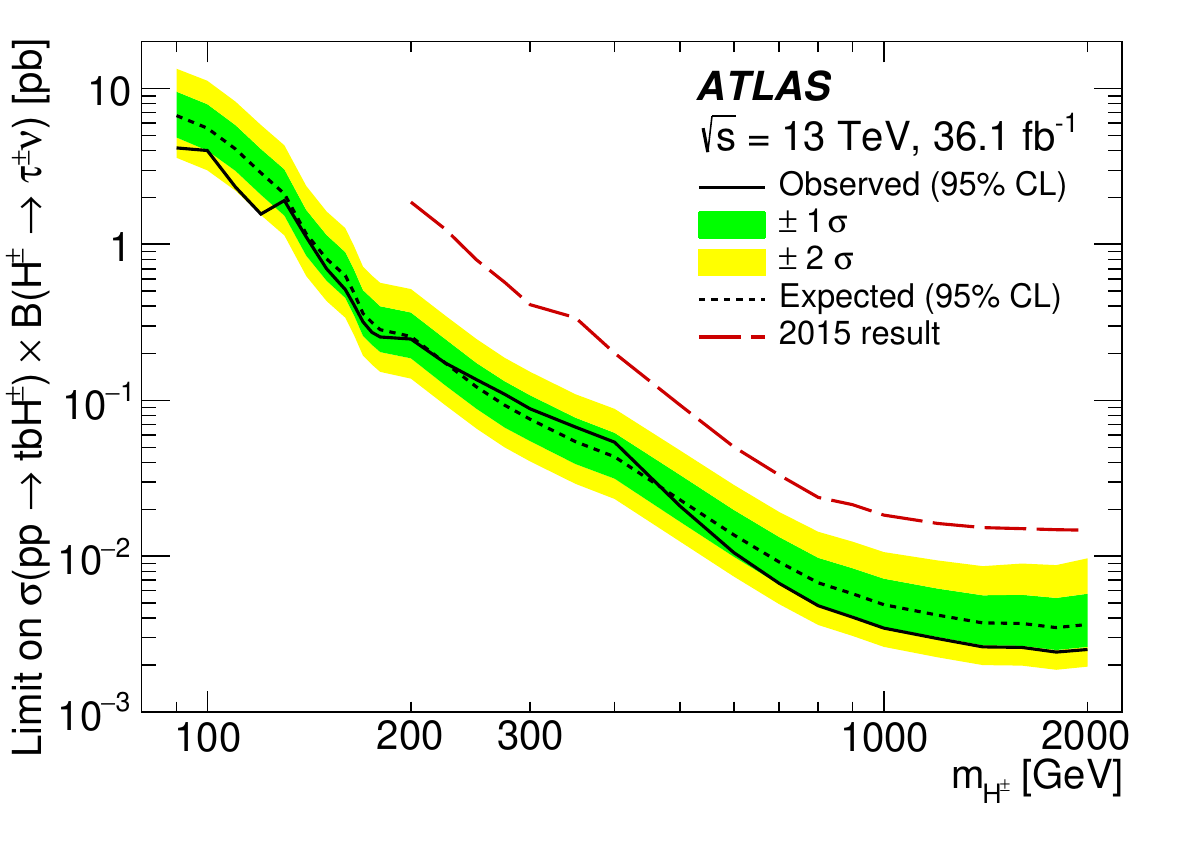}
}
\caption{\label{fig:results:charged:taunu}$H^+\to \tau^+\nu$: (a) BDT output score in the $\tau$+jets category for the intermediate mass range. The BDT discriminates between background (peaking at lower values) and signal (peaking at higher values). (b) The 95\% CL limits on the production cross-section times branching fraction as a function of $m_{H^+}$. These limits are a substantial improvement on the previous ones from data collected in 2015. For masses below 160~\GeV, the $H^+$ are produced in top-quark decay, and the $H^+$ cross-section is expressed as the $t\bar{t}$ cross-section times the branching fraction of $t\to H^+b$. Figures are taken from Ref.~\cite{HIGG-2016-11}.}
\end{figure}

In type-II models like the MSSM, the heavy charged Higgs boson decays mostly as \textbf{$H^+\to tb$}, and search results were published by ATLAS in Ref.~\cite{HDBS-2018-51}. Figure~\ref{fig:results:charged:tb:NN} shows the output score of the neural network in the category with at least six jets, of which at least four are $b$-tagged, after the fit to the data in all categories. Good agreement between the background predictions and the data is observed after the fit. The exclusions in the $M_h^{125}(\tilde \chi)$ scenario are shown in Figure~\ref{fig:results:charged:tb:lightchargino}, and results in the hMSSM are displayed in Figure~\ref{fig:results:summary:hmssm}. The $H^+\to tb$ channel involves couplings to up- and down-type fermions and thus has sensitivity to both low and high $\tan\beta$ values.

\begin{figure}[tb!]
\centering
\subfloat[]{
\label{fig:results:charged:tb:NN}
\includegraphics[width=0.46\textwidth,valign=c]{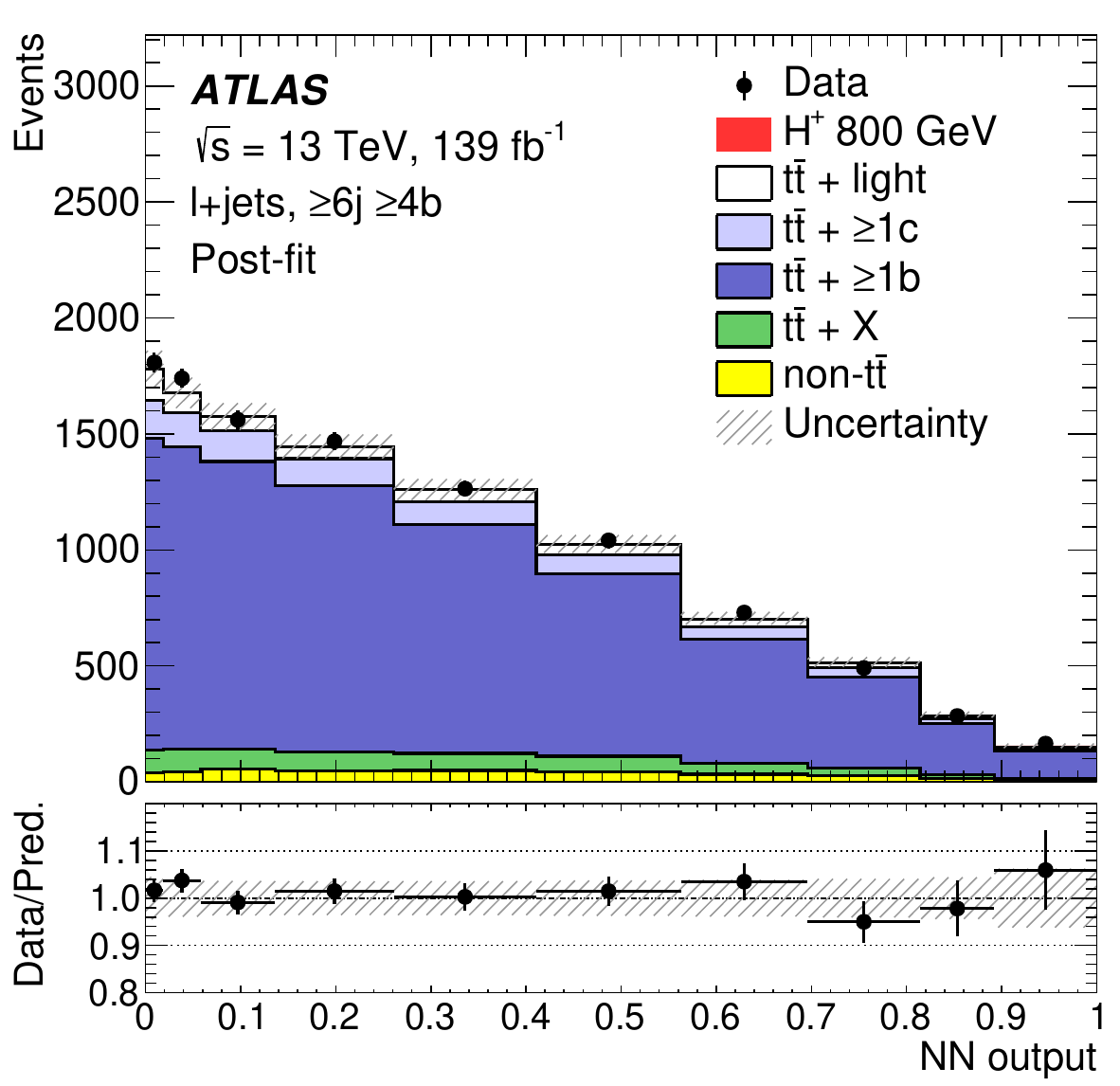}
}
\subfloat[]{
\label{fig:results:charged:tb:lightchargino}
\includegraphics[width=0.50\textwidth,valign=c]{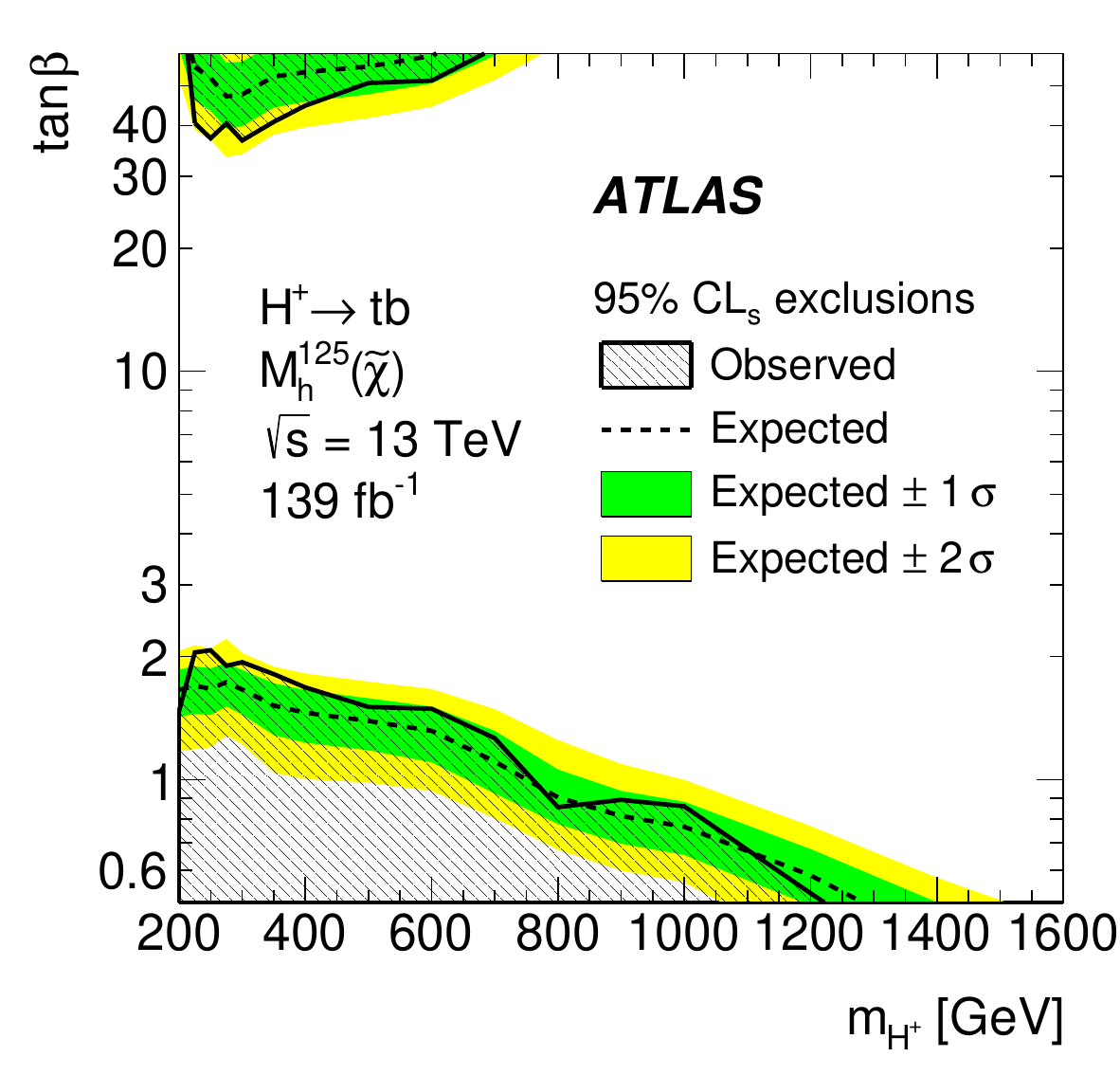}
}
\caption{\label{fig:results:charged:tb}$H^+\to tb$: (a) NN output score in the $\geq 6j\geq 4b$ category after the fit to the data, for a $H^+$ mass hypothesis of 800~\GeV. A signal would accumulate at high values of the NN score. (b) 95\% CL exclusions in the $M_h^{125}(\tilde \chi)$ scenario of the MSSM. Regions at both low and high $\tan\beta$ values are excluded. Figures are taken from Ref.~\cite{HDBS-2018-51}.}
\end{figure}

The light $H^+$ search, using \textbf{$H^+\to cb$}~\cite{HDBS-2019-24}, where the $H^+$ is produced via top-quark decays, also leads to a final state with many jets and $b$-jets. A NN discriminates between signal and background, which are displayed for the 4j3b category in Figure~\ref{fig:results:charged:cb:NN} after the fit to data. A moderate excess of signal events is observed in the vicinity of 130~\GeV, with a largest local significance of 3$\sigma$. This corresponds to a $t\to H^+b$ branching fraction of ($0.16 \pm 0.06)$\%, which assumes the branching fraction of $H^+\to cb$ is 100\%. The global significance was computed to be 2.5$\sigma$. The excess is consistent with the mass resolution of the hypothesized signal. The limits on the top-quark branching fraction are displayed in Figure~\ref{fig:results:charged:cb:limits}. The branching fractions predicted in the 3HDM for various parameter values are overlaid, showing that the search is able to constrain this model.

\begin{figure}[tb!]
\centering
\subfloat[]{
\label{fig:results:charged:cb:NN}
\includegraphics[width=0.4\textwidth,valign=b]{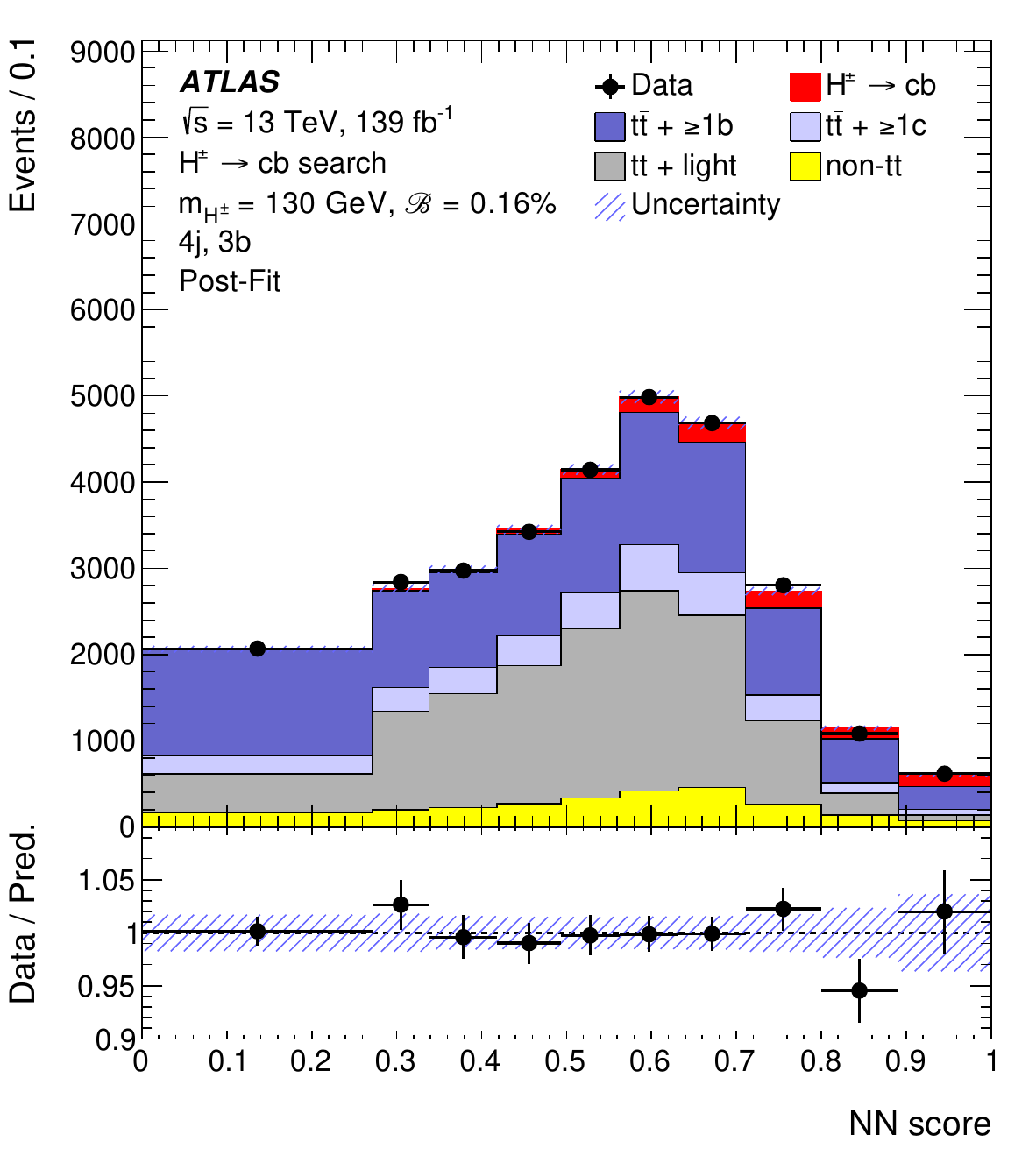}
}
\subfloat[]{
\label{fig:results:charged:cb:limits}
\includegraphics[width=0.56\textwidth,valign=b]{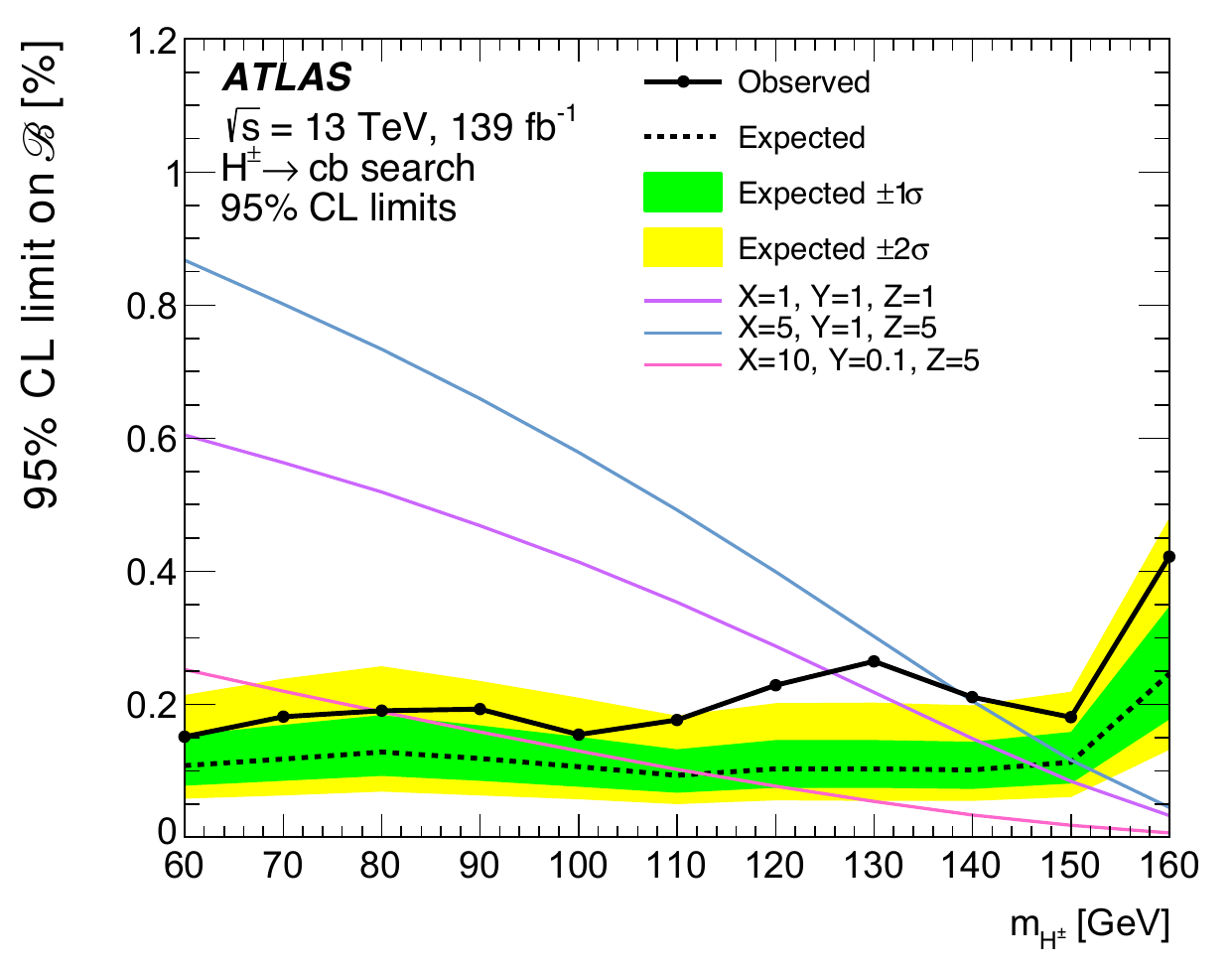}
}
\caption{\label{fig:results:charged:cb}$H^+\to cb$: (a) The output score of the NN in the $4j3b$ category after the fit to the data. The background is dominated by $t\bar{t}$+$b$-jets, and a slight excess of signal events (in red) is visible. (b) 95\% CL limits on the branching fraction of $t\to H^+b$ as a function of $m_{H^+}$, assuming the branching fraction of $H^+\to cb$ is 100\%. Theory predictions for the 3HDM are overlaid, showing that some of this model's phase space can be excluded. Figures are taken from Ref.~\cite{HDBS-2019-24}.}
\end{figure}

\FloatBarrier

\subsubsection{Charged Higgs bosons decaying to bosons}
\label{sec:results:chargedhiggs:bosons}

The search for \textbf{$H^+\to W^+a$}, with $a\to\mu^+\mu^-$, was explored for light $H^+$ hypotheses~\cite{HDBS-2020-12}. This search is more focused on $a$ than it is on $H^+$, although the decay of $H^+$ into a $W$ boson and a scalar is relevant in many models and deserves mentioning. The largest excess is found at $m_a=27$~\GeV, with a local significance of 2.4$\sigma$, and both the $e\mu\mu$ and $\mu\mu\mu$ categories contribute. This excess is independent of the $H^+$ mass hypothesis in the range that was investigated (120--160~\GeV). The dimuon mass in the $e\mu\mu$ category and the limits on the production cross-section times branching fraction are presented in Figure~\ref{fig:results:charged:Wa}.

\begin{figure}[tb!]
\centering
\subfloat[]{
\label{fig:results:charged:Wa:mass}
\includegraphics[width=0.46\textwidth,valign=c]{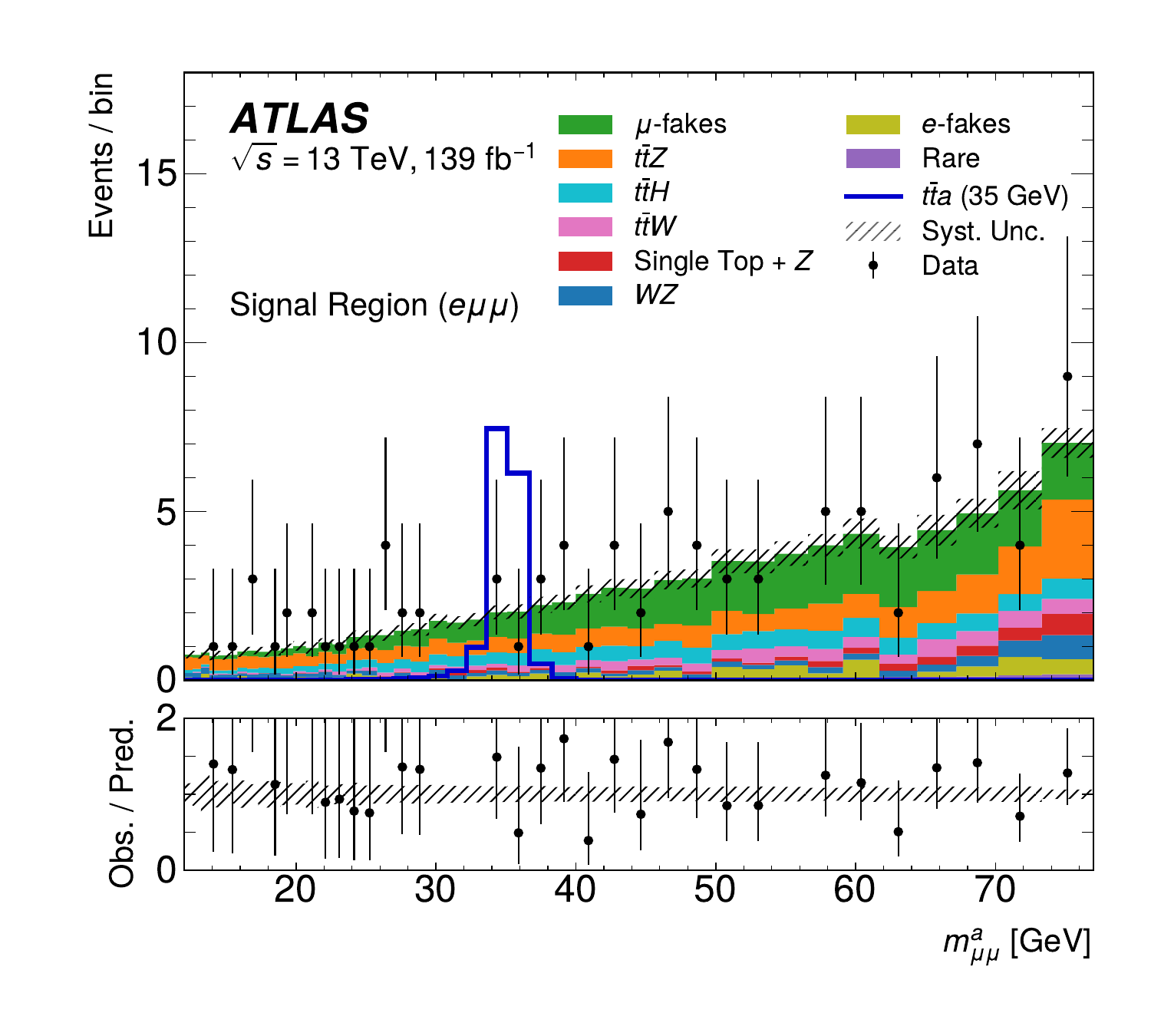}
}
\subfloat[]{
\label{fig:results:charged:Wa:limits}
\includegraphics[width=0.5\textwidth,valign=c]{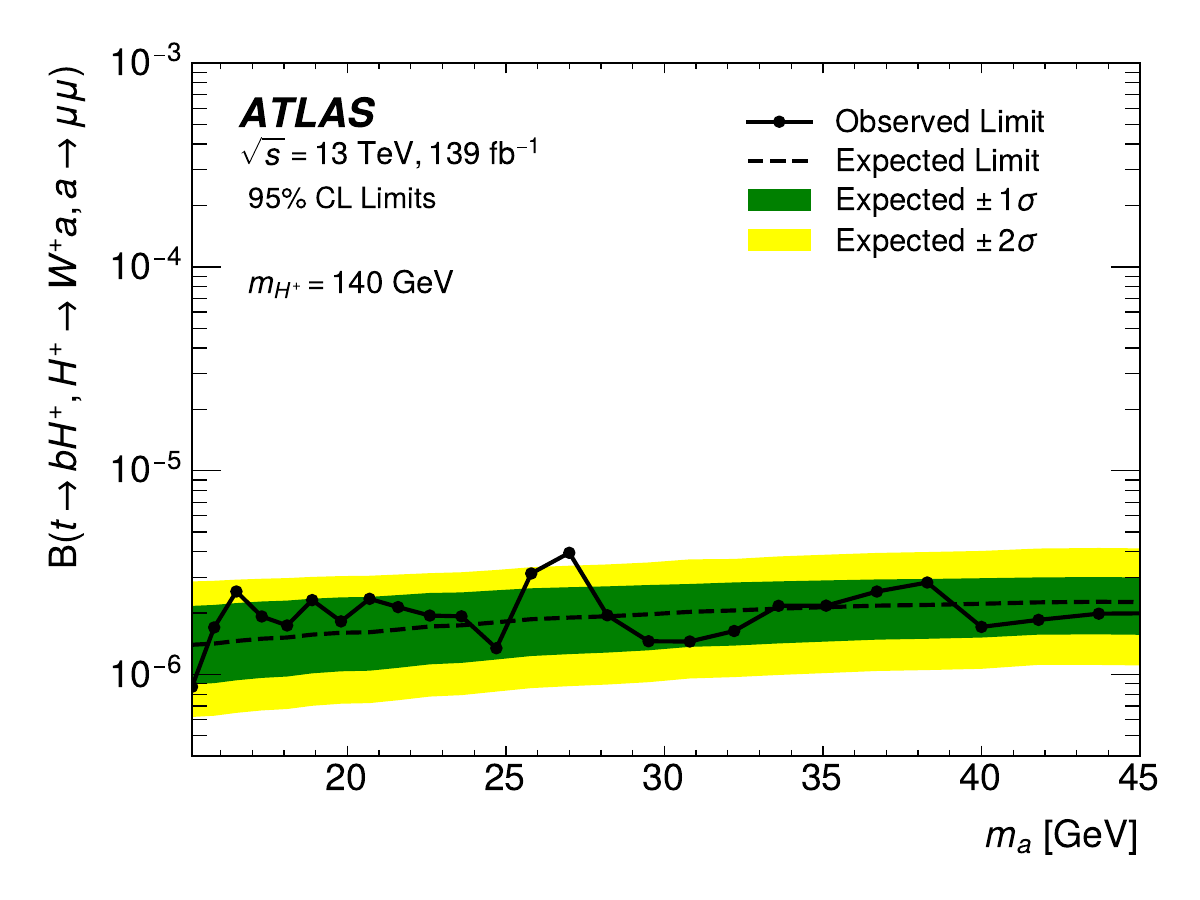}
}
\caption{\label{fig:results:charged:Wa}$H^+\to W^+a$: (a) The dimuon mass, which reconstructs the $a$-boson in the $e\mu\mu$ channel, with a signal hypothesis overlaid for illustration. (b) 95\% CL limits on the production cross-section times branching fraction as a function of the $a$ mass, displayed here for a fixed value of $m_{H^+}=140$~\GeV. Figures are taken from Ref.~\cite{HDBS-2020-12}.}
\end{figure}

The search for \textbf{$H^+\to W^+Z$} from VBF production was carried out for leptonic decays of the vector bosons~\cite{HDBS-2018-19}. The analysis found an excess at 375~\GeV with a local (global) significance of 2.8$\sigma$ (1.6$\sigma$). The reconstructed $H^+$ mass in the signal region is shown in Figure~\ref{fig:results:charged:WZ:mass}. The signal yield depends upon the $\sin\theta_H$ parameter of the GM model, and the analysis is able to constrain its value, as shown in Figure~\ref{fig:results:charged:WZ:limits}.

\begin{figure}[tb!]
\centering
\subfloat[]{
\label{fig:results:charged:WZ:mass}
\includegraphics[width=0.36\textwidth,valign=c]{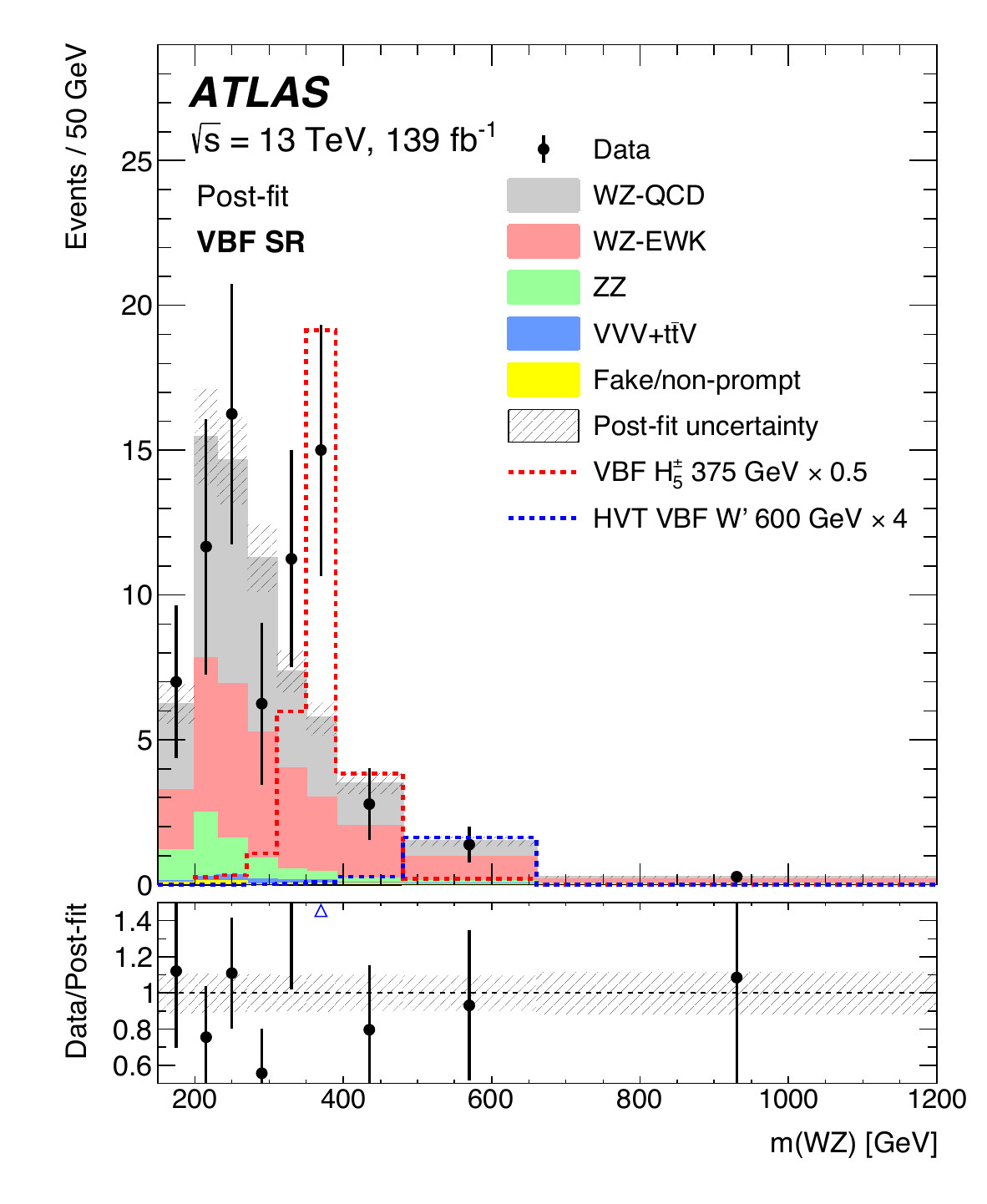}
}
\subfloat[]{
\label{fig:results:charged:WZ:limits}
\includegraphics[width=0.6\textwidth,valign=c]{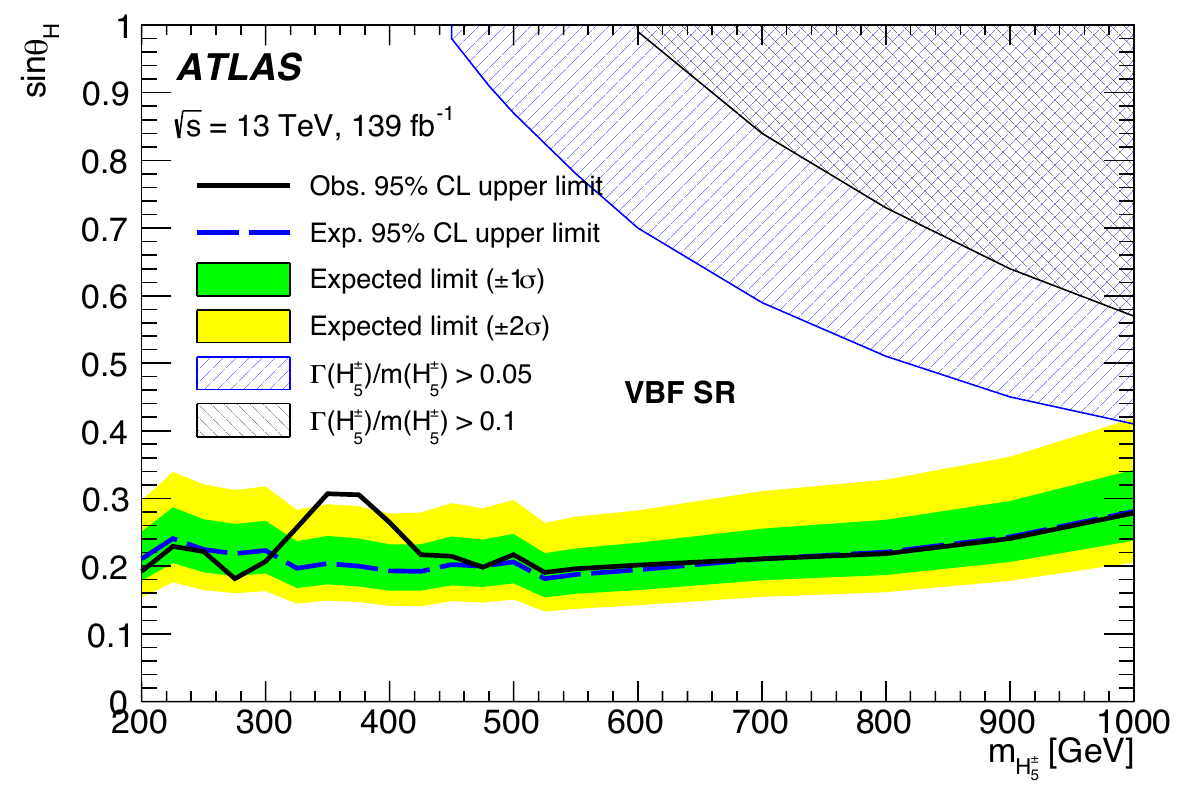}
}
\caption{\label{fig:results:charged:WZ}$H^+\to W^+Z$: (a) The reconstructed $H^+$ mass in the VBF-enriched signal region after the fit to data. An excess of data over background is visible, and a signal hypothesis assuming $\sin\theta_H=0.5$ is overlaid. (b) 95\% CL limits on the $\sin\theta_H$ parameter of the GM model as a function of the $H^+$ mass. The region above the line is excluded. In the blue-shaded area the width of the $H^+$ predicted by the model becomes large and the exclusion is not valid there. Figures are taken from Ref.~\cite{HDBS-2018-19}.}
\end{figure}

\FloatBarrier

\subsubsection{Doubly charged Higgs bosons}
\label{sec:results:chargedhiggs:doubly}

The search for \textbf{$H^{++}\to W^+W^+$} was conducted in multi-lepton final states~\cite{HDBS-2019-06}. The yields in all categories, after a selection optimized for various $H^{++}$ mass hypotheses, are displayed in Figure~\ref{fig:results:charged:Hplusplus:WW:yields} and show good agreement between data and SM expectations. The limits on the cross-section for $H^{++}H^{--}$ pair production are presented in Figure~\ref{fig:results:charged:Hplusplus:WW:limits}. At 95\% CL, $H^{++}$ bosons in the type-II seesaw model are excluded  up to 350~\GeV and 230~\GeV for the pair- and associated-production modes, respectively. The analysis of VBF-produced $H^{++}\to W^+W^+$ events~\cite{STDM-2018-32} yielded limits on their production and can constrain the value of $\sin\theta_H$ in the GM model (displayed in Figure~\ref{fig:results:charged:Hplusplus:WW:VBF_GM}). An excess was found at 450~\GeV with a local (global) significance of 3.2$\sigma$ (2.5$\sigma$).

\begin{figure}[tb!]
\centering
\subfloat[]{
\label{fig:results:charged:Hplusplus:WW:yields}
\includegraphics[width=0.54\textwidth,valign=c]{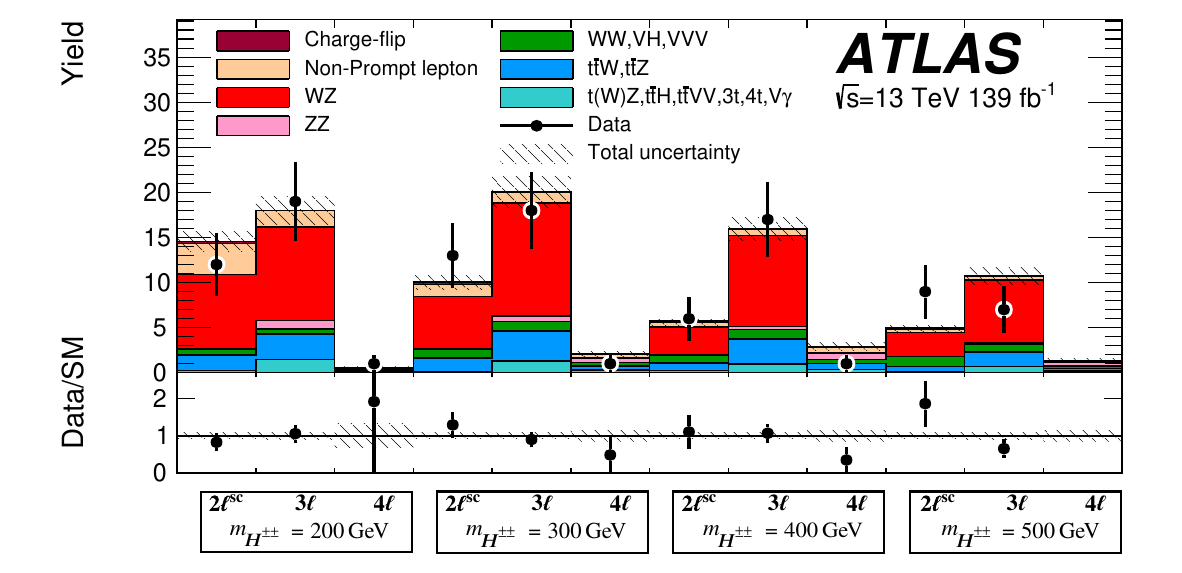}
}
\subfloat[]{
\label{fig:results:charged:Hplusplus:WW:limits}
\includegraphics[width=0.45\textwidth,valign=c]{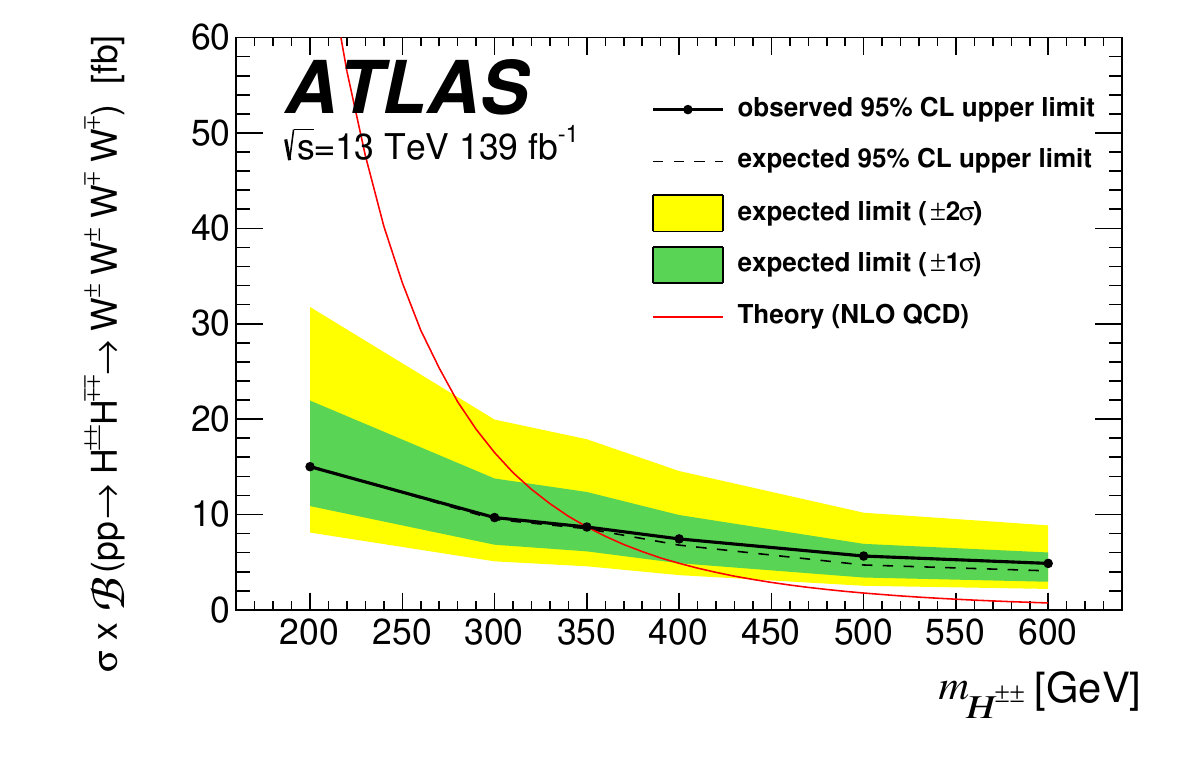}
}\\
\subfloat[]{
\label{fig:results:charged:Hplusplus:WW:VBF_GM}
\includegraphics[width=0.49\textwidth,valign=c]{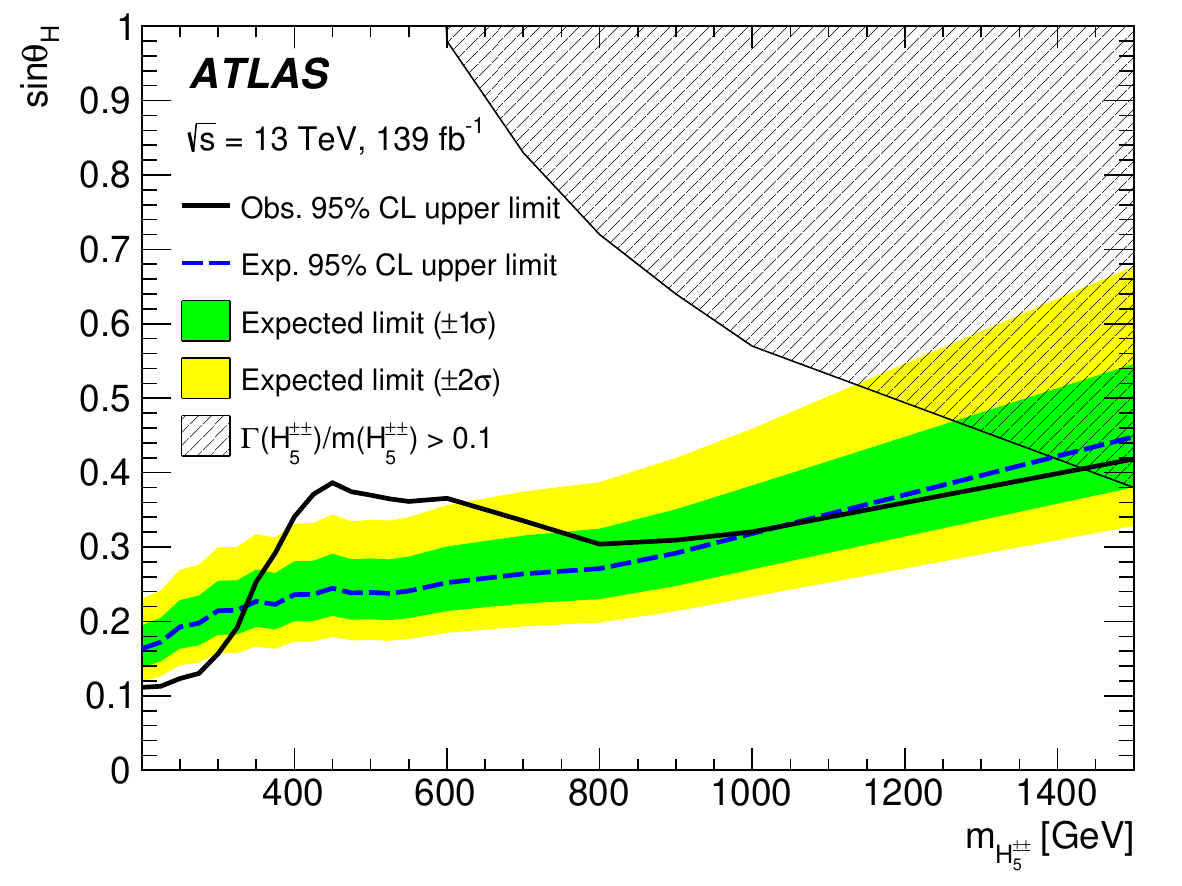}
}
\caption{\label{fig:results:Hplusplus:WW}$H^{++}\to W^+W^+$: (a) The event yields in the signal regions for $H^{++}$-mass-dependent selections for the analysis of pair production or associated production. (b) 95\% CL limits on the pair production of $H^{++}$ times branching fraction as a function of the hypothesized $H^{++}$ mass. The theory curve represents the predicted cross-section in the type-II seesaw model. (c) The 95\% CL exclusion limits on $\sin\theta_H$ in the analysis of VBF-produced $H^{++}\to W^+W^+$ in the GM model. The region where the predicted width becomes too large for the limits to be valid is indicated by the hatched area. Figures (a) and (b) are taken from Ref.~\cite{HDBS-2019-06}, and (c) is from Ref.~\cite{STDM-2018-32}.}
\end{figure}

The search for \textbf{$H^{++}\to\ell^+\ell^+$}~\cite{EXOT-2018-34} considered only the pair-production mode. The mass of the same-sign lepton pair in the electron channel and the limits on the cross-section are displayed in Figure~\ref{fig:results:Hplusplus:ll}. No data event is observed in the $4\ell$ category, which is consistent with the expectation. The observed lower limit on the mass of a $H^{++}$ is 1080~\GeV within the left-right symmetric type-II seesaw model. The Zee--Babu neutrino mass model is also constrained, the observed lower limit on the mass of the $H^{++}$ being 900~\GeV.

\begin{figure}[tb!]
\centering
\subfloat[]{
\label{fig:results:charged:Hplusplus:ll:mass}
\includegraphics[width=0.42\textwidth,valign=c]{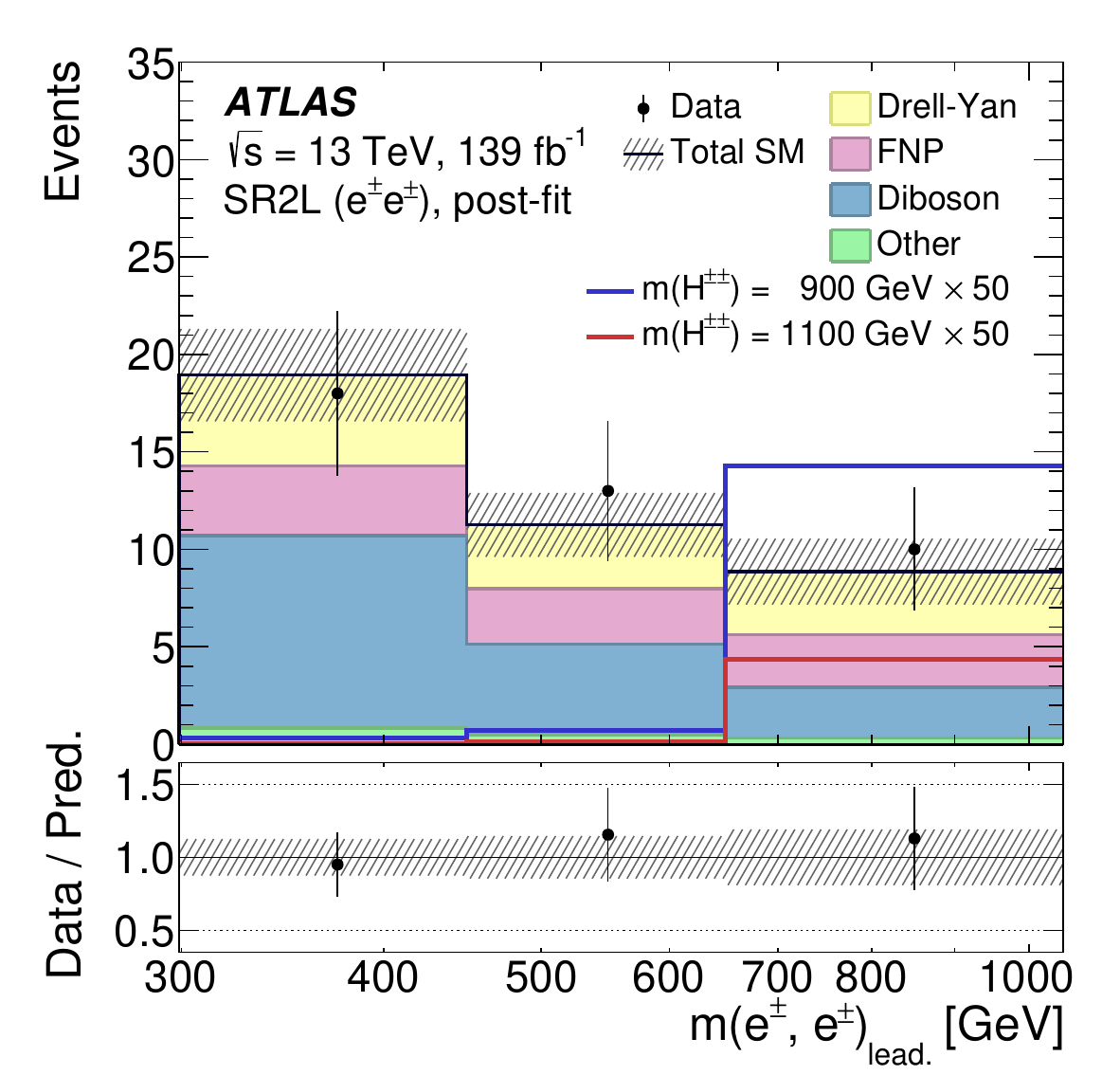}
}
\subfloat[]{
\label{fig:results:charged:Hplusplus:ll:limits}
\includegraphics[width=0.54\textwidth,valign=c]{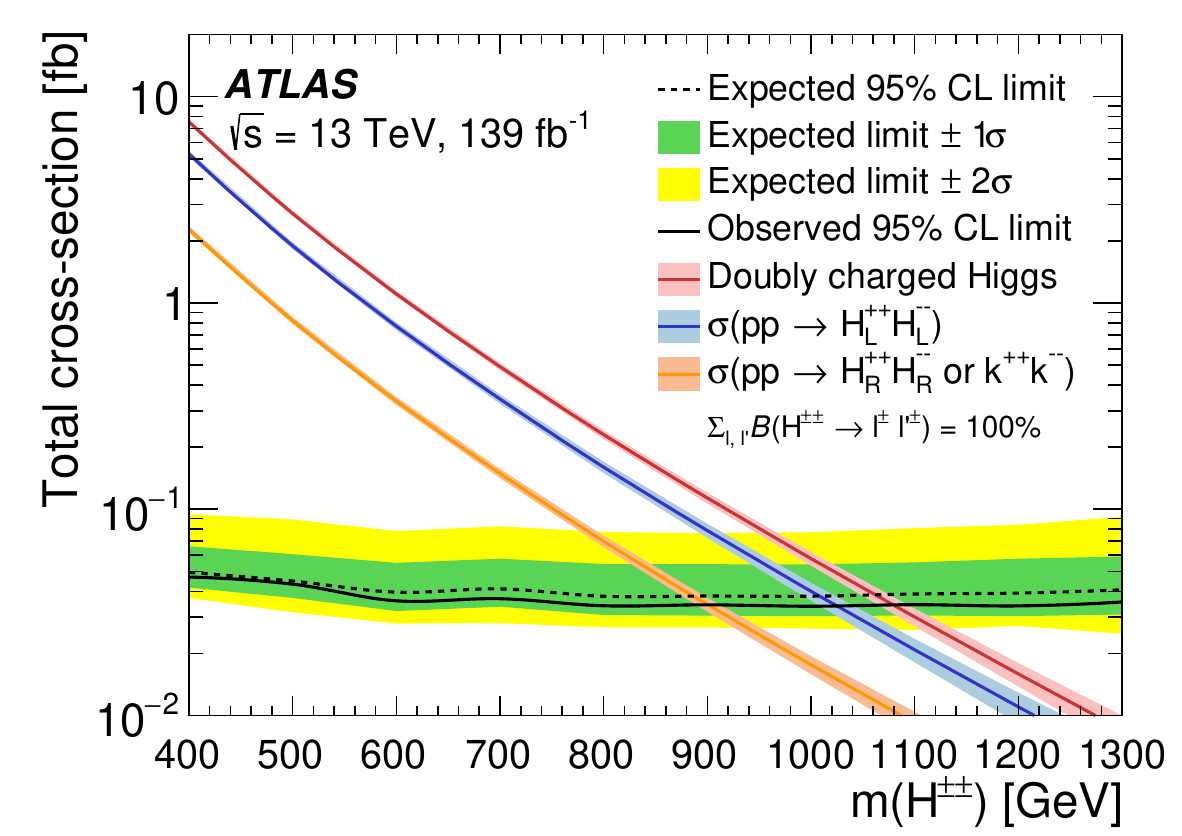}
}
\caption{\label{fig:results:Hplusplus:ll}$H^{++}\to \ell\ell$: (a) The invariant mass of the two same-sign leptons in the electron channel after a fit to the data under the background-only hypothesis. Two signal expectations are overlaid for illustration. (b) 95\% CL limits on the $H^{++}$ pair-production cross-section. The theory lines show the prediction for the left-handed $H^{++}_L$ (blue), the right-handed $H^{++}_R$ (orange), which is the same as predicted in the Zee--Babu model, and the sum of the two LRSM chiralities (red). Figures are taken from Ref.~\cite{EXOT-2018-34}.}
\end{figure}

\FloatBarrier


\subsection{Additional scalars decaying into Higgs boson pairs}
\label{sec:results:dihiggs}

\subsubsection{Resonant $HH$}
\label{sec:results:dihiggs:fermions}

The search  \textbf{ggF $X \to HH\to\bbbar\bbbar$} for a new boson revealed by resonant pair production of
Higgs bosons via ggF and with \bbbar{}\bbbar in the final state had results consistent with the
SM predictions~\cite{HDBS-2018-41}.
Upper limits are set on the cross-section for resonant Higgs boson pair production in a
benchmark model with a generic narrow spin-0 resonance,
as shown in Figure~\ref{fig:results:dihiggs:bbbb}.
The most significant excess is found for a signal mass of 1100~\GeV, where
the local (global) significance is 2.3$\sigma$ (0.4$\sigma$).
The results are therefore statistically consistent with the SM.
The expected upper limits on the cross-section improve on those
in the previous ATLAS search in this final state~\cite{EXOT-2016-31}
by approximately 20\% at low resonance masses and more than 80\% at high masses.
This search also covers resonance
masses in the range from 3 to 5~\TeV for the first time.
\begin{figure}[tb!]
\begin{center}{\includegraphics[width=0.47\textwidth]{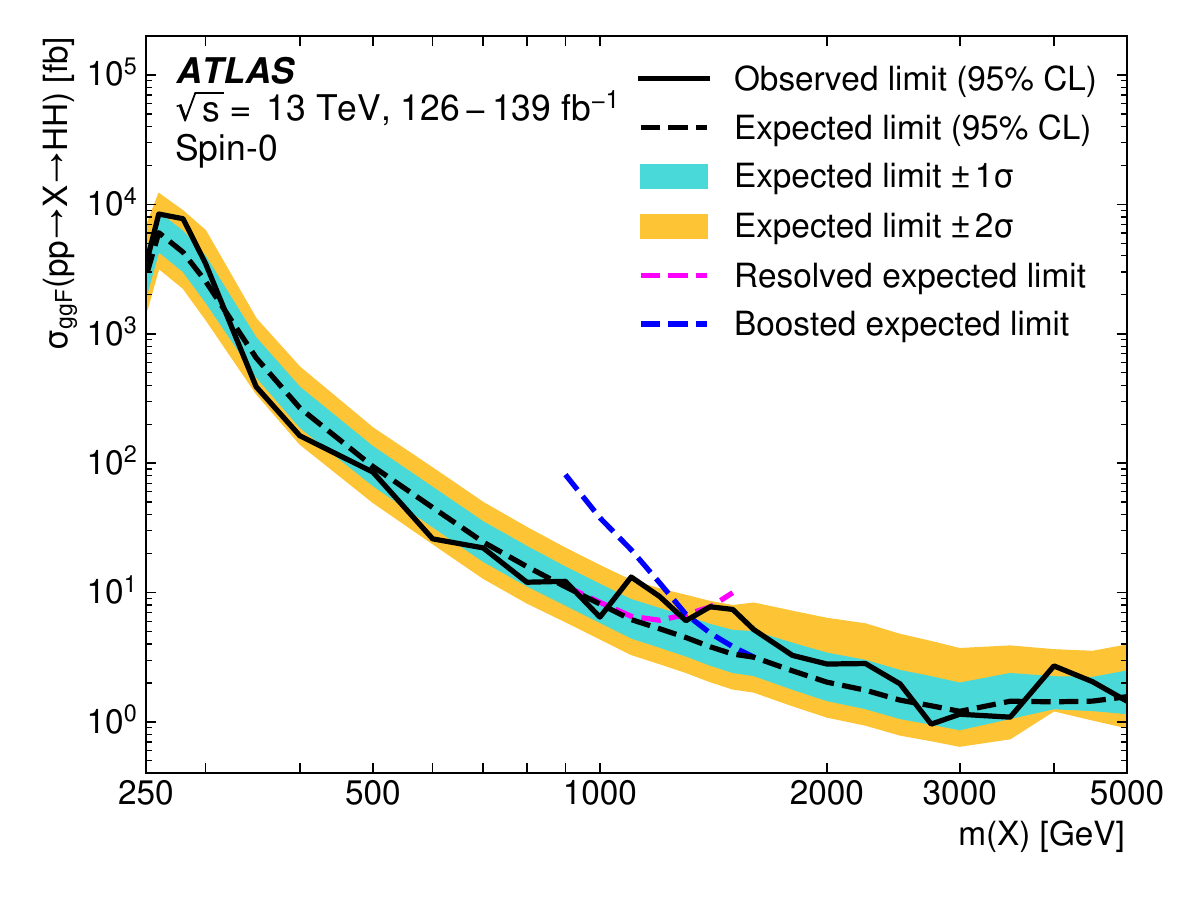}}%
\caption{ggF $X \to HH\to\bbbar\bbbar$:
Expected (dashed black lines) and observed (solid black lines) 95\%
CL upper limits on the cross-section for resonant
$HH\to 4b$ production in the
spin-0
signal model.
The $\pm 1\sigma$ and $\pm 2\sigma$ uncertainty ranges for the expected
limits (coloured bands) are shown.
Expected limits obtained by using the resolved and boosted channels separately
(dashed coloured lines) are shown.
Figures are taken from Ref.~\cite{HDBS-2018-41}.}
\label{fig:results:dihiggs:bbbb}
\end{center}
\end{figure}

The pair production of Higgs bosons via vector-boson fusion and with $b\bar{b}b\bar{b}$ in the final
state (\textbf{VBF $X \to HH\to\bbbar\bbbar$}) was used to search for a new boson in the mass range
of 260--1000~\GeV and revealed no significant excess relative to the SM expectation~\cite{HDBS-2018-18}.
The largest deviation from the background-only hypothesis is observed at 550~\GeV with a
local significance of $1.5\sigma$.
Upper limits on the production cross-section are set for narrow and broad
scalar resonances at 95\% CL as shown in Figure~\ref{fig:results:dihiggs:VBFbbbb}.
\begin{figure}[tb!]
\begin{center}
\subfloat[]{\includegraphics[width=0.47\textwidth]{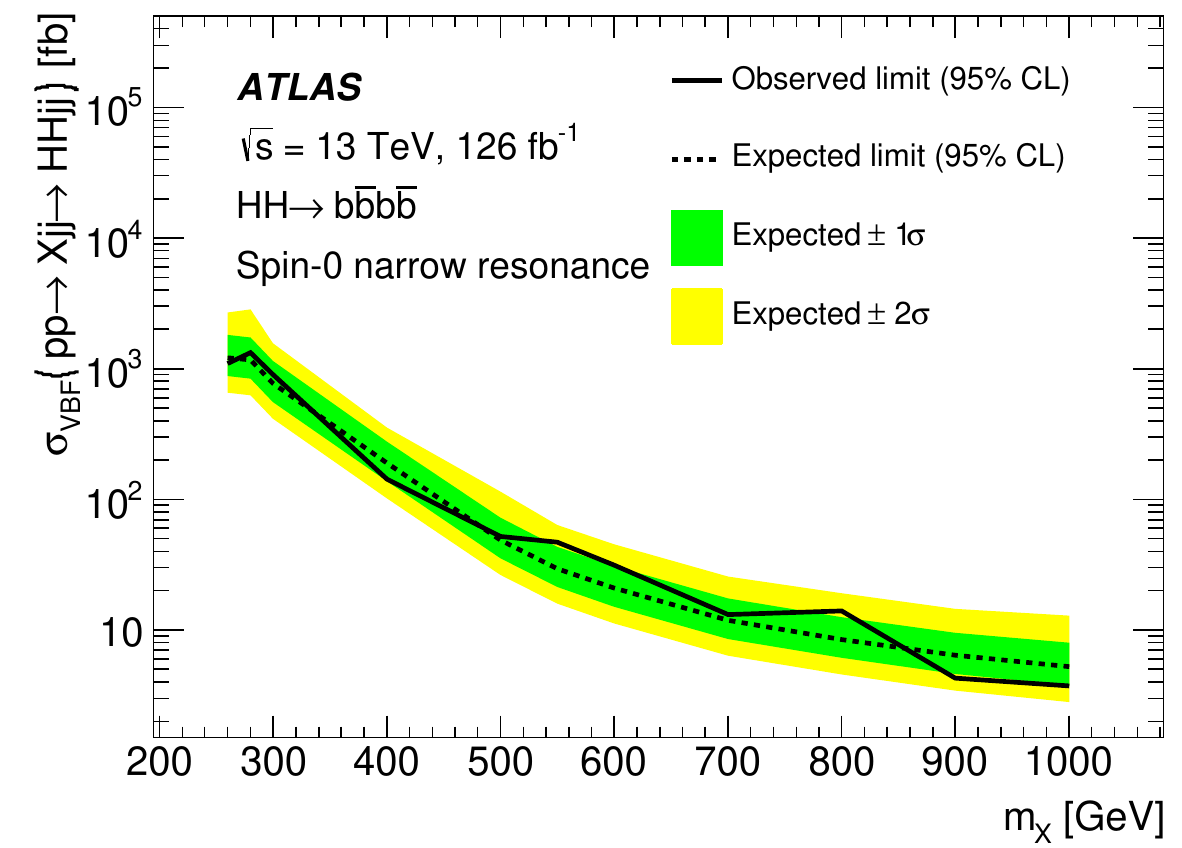}}%
\subfloat[]{\includegraphics[width=0.47\textwidth]{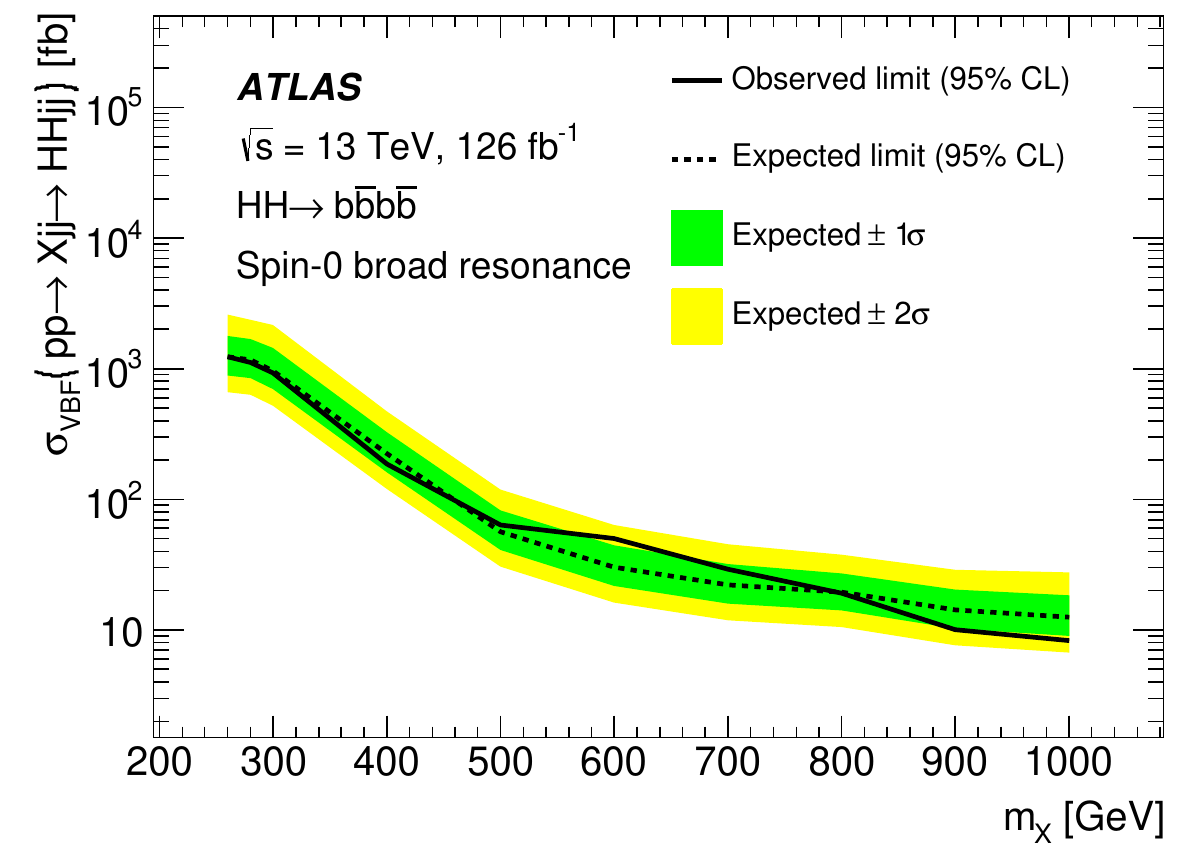}}%
\caption{VBF $X \to HH\to\bbbar\bbbar$:
Observed and expected 95\% CL upper limits on the cross-section for resonant
$HH\to 4b$ production via VBF as a function of the mass $m_X$.
The narrow-resonance hypothesis is shown in (a), and (b) shows the
broad-resonance hypothesis.
Figures are taken from Ref.~\cite{HDBS-2018-18}.}
\label{fig:results:dihiggs:VBFbbbb}
\end{center}
\end{figure}

The search targeting the decay of a narrow resonance \textbf{$X \to HH\to\bbbar\tautau$} in the range 251--1600~\GeV
found the data to be compatible with the background-only hypothesis~\cite{HDBS-2018-40}.
The largest deviation is at $m_X = 1$~\TeV, corresponding to a local (global)
significance of $3.1\sigma$ ($2.0\sigma$).
Observed (expected) upper limits are placed at 95\% CL on resonant Higgs boson production and exclude cross-sections
above 21--900~fb (12--840~fb),
depending on the mass of the resonance, as shown in Figure~\ref{fig:results:dihiggs:bbtautau}.
\begin{figure}[tb!]
\begin{center}{\includegraphics[width=0.6\textwidth]{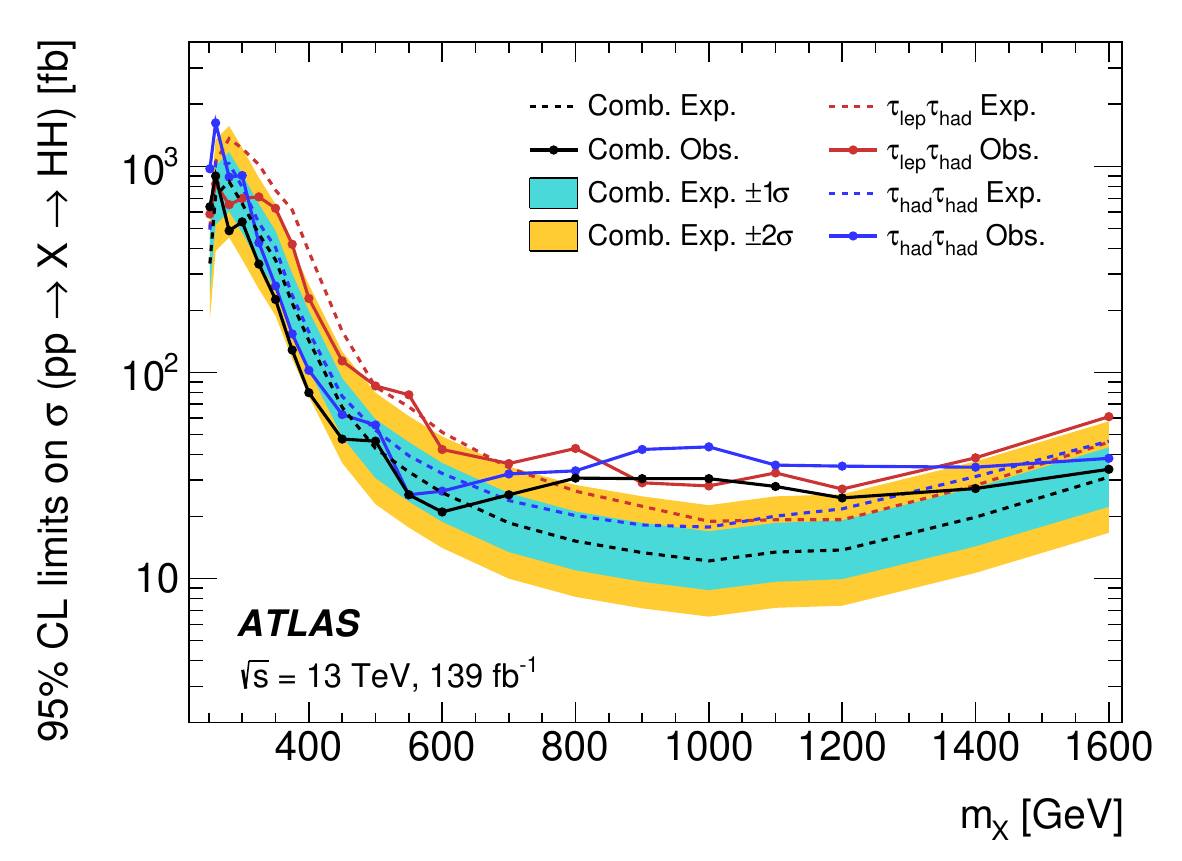}}%
\caption{$X \to HH\to\bbbar\tautau$:
Observed and expected limits at 95\% CL on the cross-section for resonant $HH$ production
as a function of the scalar resonance mass $m_X$.
The dashed lines show the expected limits, while the solid lines show the observed limits.
The blue and red lines are the limits for the \tauhad{}\tauhad channel
and \taulep{}\tauhad channel, respectively.
The black lines are the combined limits from the two channels.
Figures are taken from Ref.~\cite{HDBS-2018-40}.}
\label{fig:results:dihiggs:bbtautau}
\end{center}
\end{figure}

The  di-$\tau$ search \textbf{boosted $X\to HH\to\bbbar\tautau$} for a heavy, narrow, scalar resonance $X$
decaying into two boosted Higgs bosons in the high mass range $1 \leq m_X \leq 3$~\TeV did not find any
deviation from the SM predictions~\cite{HDBS-2019-22}.
Accordingly, 95\% CL upper limits are set.
Assuming SM branching fractions for the Higgs boson, the observed
(expected) upper limits on the production cross-section $\sigma(X \to HH)$ are 94--28~fb (74--32~fb) depending on the
resonance mass hypotheses, as shown in Figure~\ref{fig:results:dihiggs:bbtautau-boosted}.
This represents a first attempt with a novel di-$\tau$ tagger.
\begin{figure}[tb!]
\begin{center}{\includegraphics[width=0.6\textwidth]{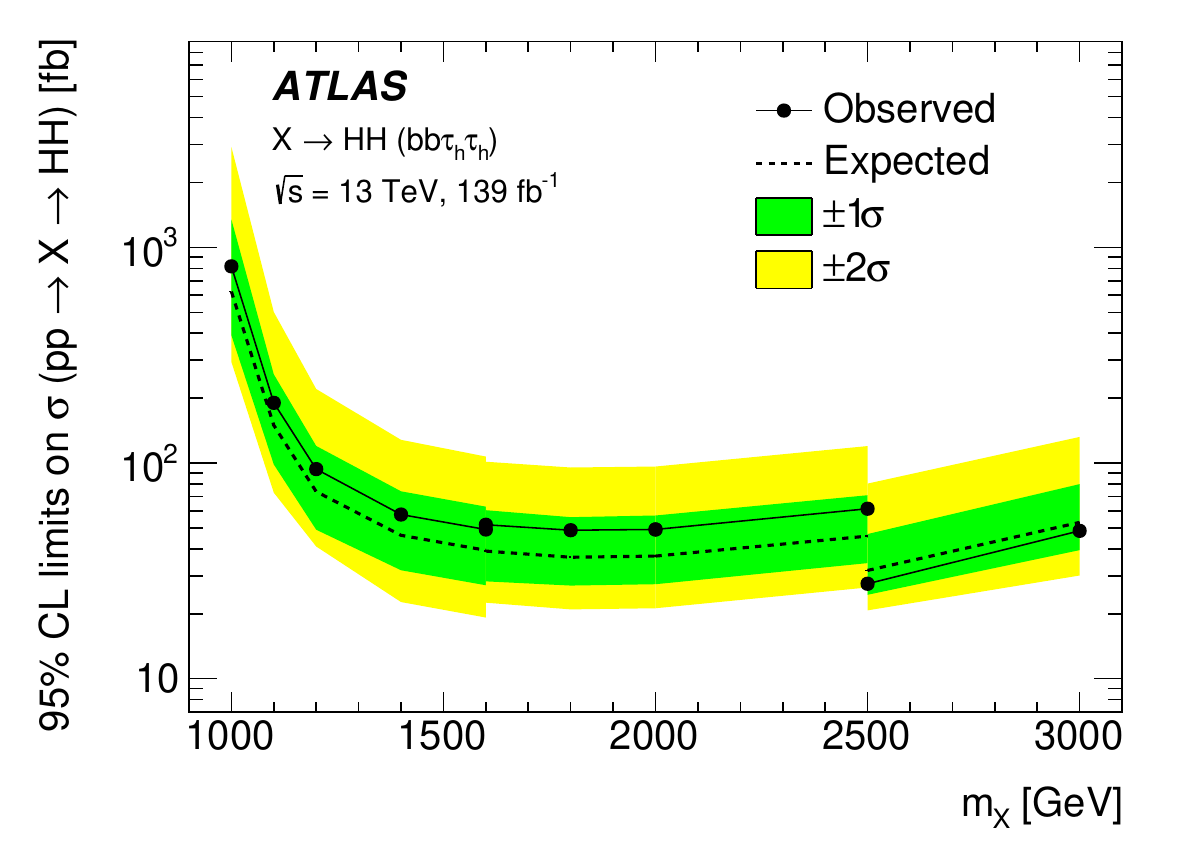}}%
\caption{Boosted $X \to HH\to\bbbar\tautau$:
Expected and observed 95\% CL upper limits on the production cross-section of a heavy,
narrow, scalar resonance decaying into a pair of Higgs bosons ($X \to HH)$.
The final state used in the search consists of a boosted $\bbbar$
pair and a boosted hadronically decaying $\tau^+\tau^-$ pair,
and the SM branching fractions of the Higgs boson are assumed.
Figures are taken from Ref.~\cite{HDBS-2019-22}.}
\label{fig:results:dihiggs:bbtautau-boosted}
\end{center}
\end{figure}

The search for \textbf{$X \to HH\to\bbbar\gamma\gamma$} Higgs boson pair production
(in both the ggF and VBF modes) by using the $\bbbar\gamma\gamma$ final
state did not observe any excess above the expected background~\cite{HDBS-2018-34}.
A 95\% CL upper limit on the cross-section for resonant production of a
scalar particle $X \to HH \to \bbbar\gamma\gamma$
is obtained for the narrow-width hypothesis as a function of $m_X$
as shown in Figure~\ref{fig:results:dihiggs:bbgg}.
The observed (expected) upper limits are in the
range 640--44~fb (391--46~fb) for $251 \leq m_X \leq 1000$~\GeV.
The expected limit on the resonant cross-section improves on the previous ATLAS search~\cite{HIGG-2016-15} by a
factor of two to three depending on the $m_X$ value.
Improvement by a factor of two arises from the increase in integrated luminosity,
while the additional improvement can be attributed to the use of multivariate techniques and
the more precise object reconstruction and calibration.
\begin{figure}[tb!]
\begin{center}{\includegraphics[width=0.6\textwidth]{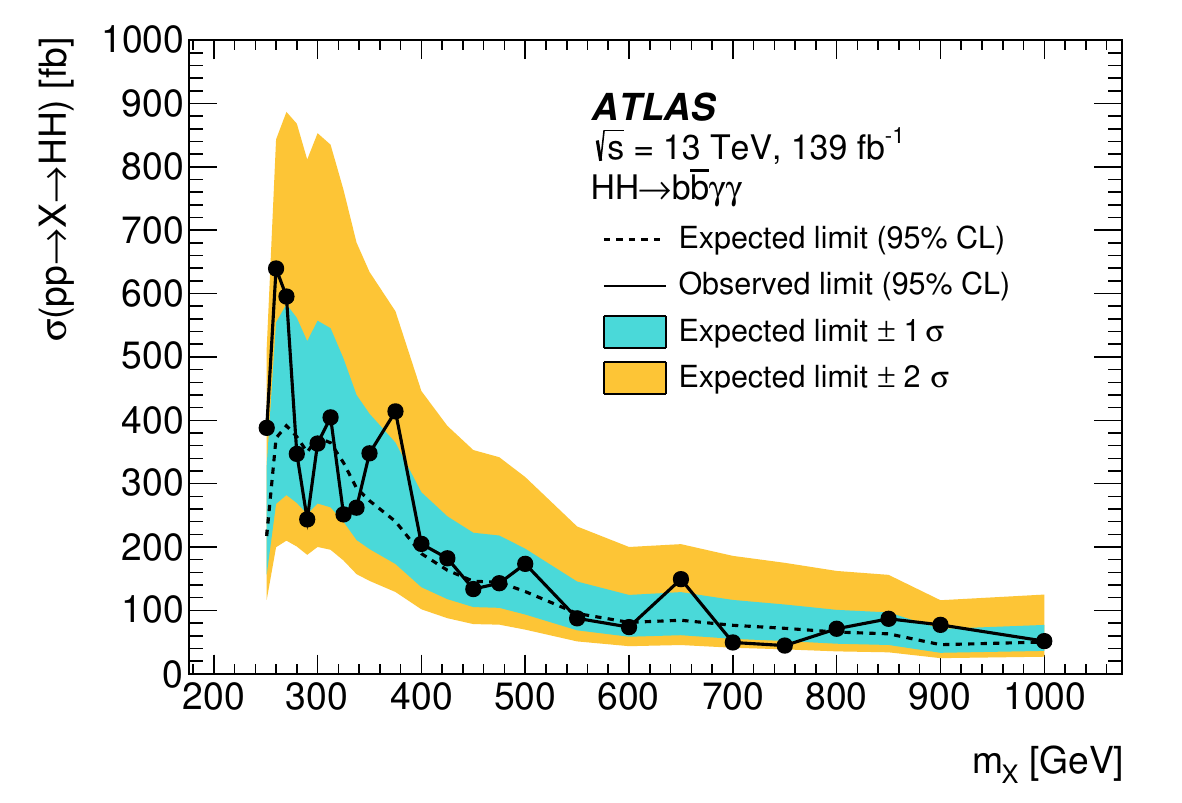}}%
\caption{$X \to HH\to\bbbar\gamma\gamma$:
Observed and expected limits at 95\% CL on the production cross-section of a narrow
scalar resonance $X$ as a function of its mass $m_X$.
The black solid line
represents the observed upper limits. The dashed line represents the expected upper limits.
Figures are taken from Ref.~\cite{HDBS-2018-34}.}
\label{fig:results:dihiggs:bbgg}
\end{center}
\end{figure}

The recent \textbf{$\text{Combination for } HH$} %
using three analyses with the final states $\bbbar\bbbar$,
$\bbbar\tautau$ and $\bbbar\gamma\gamma$  with the full \RunTwo dataset  did not find any
statistically significant excess beyond the SM predictions~\cite{HDBS-2023-17}.
The largest deviation is observed at 1.1~\TeV, corresponding to a local (global)
significance of 3.3$\sigma$ (2.1$\sigma$).
A 95\% CL upper limit is set on the resonant $X \to hh$ cross-section
for $251 \leq m _X \leq 5$~\TeV.
The observed (expected) upper limits are in the range 0.96--600~fb (1.2--390~fb).
This is an improvement by a factor of 2--5, depending on $m_X$,
relative to the previous ATLAS combined result~\cite{HDBS-2018-58}.
Some of the results are presented in Figure~\ref{fig:results:dihiggs:combo2}.
The results are also interpreted in the context of the type-I 2HDM and MSSM,
excluding parameter space that was hitherto allowed by the most sensitive
search results for these models.
\begin{figure}[tb!]
\begin{center}
\subfloat[]{\includegraphics[width=0.47\textwidth]{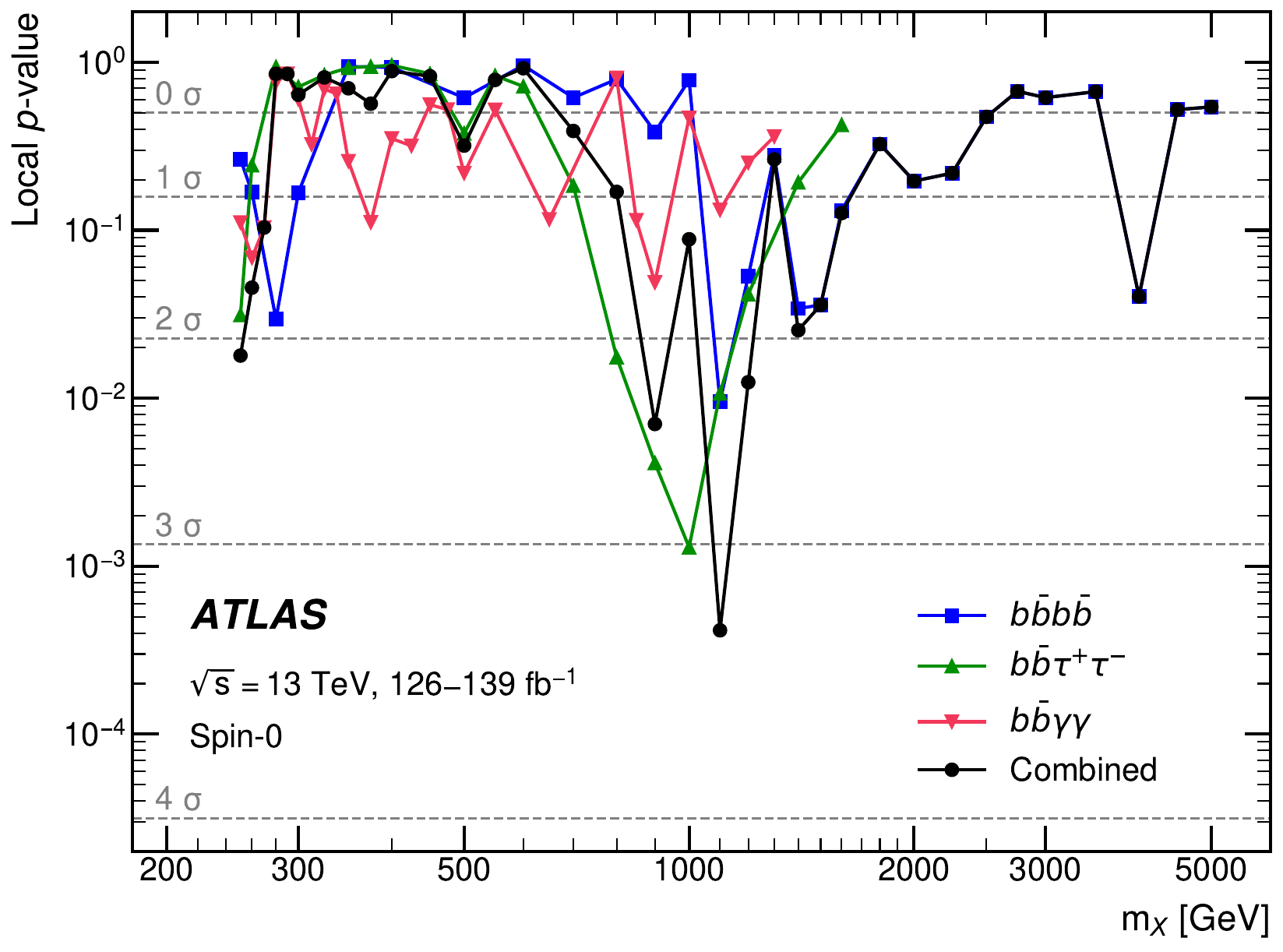}}%
\subfloat[]{\includegraphics[width=0.47\textwidth]{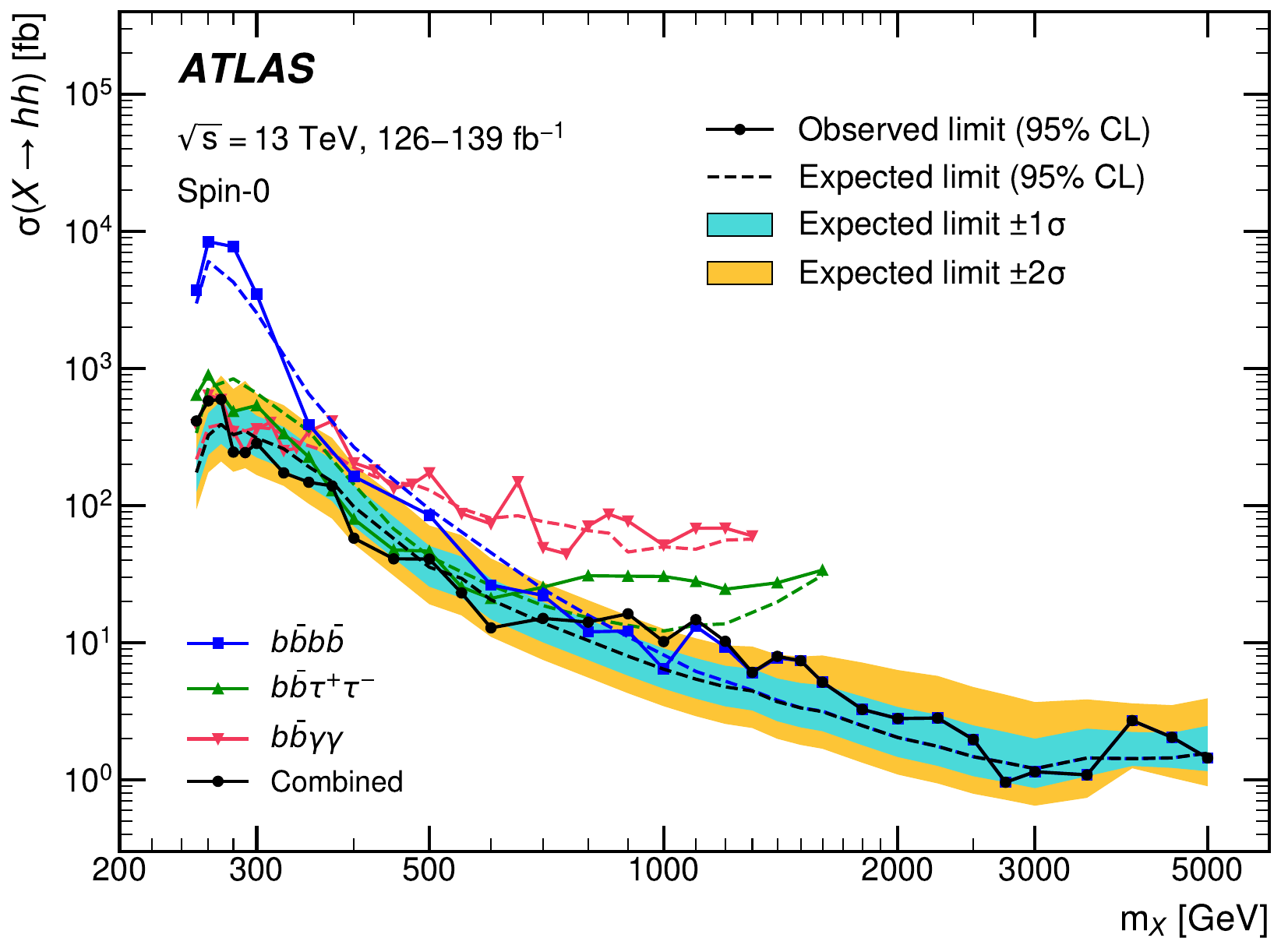}}%
\caption{$\text{Combination for } HH$:
(a) Local $p$-value and (b) observed and expected upper limits at the
95\% CL on the resonant Higgs boson
pair-production cross-section as a function of the resonance mass $m_X$.
Figures are taken from Ref.~\cite{HDBS-2023-17}.}
\label{fig:results:dihiggs:combo2}
\end{center}
\end{figure}

\subsubsection{Resonant $HH/SH/SS$ decaying to $W$ bosons}
\label{sec:results:dihiggs:bosons}

The search \textbf{$X \to HH/SS\to WW^* WW^*$} for a pair of neutral, scalar bosons each decaying into two $W$ bosons
did not observe any significant excess over the expected SM backgrounds~\cite{HIGG-2016-24}.
Upper limits are set on the production cross-section times branching fraction of a
heavy scalar $X$ that decays into two Higgs bosons for a mass range of $260  \leq m_X \leq 500$~\GeV and
the observed (expected) limits range from 9.3 (10)~pb to 2.8 (2.6)~pb
as shown in Figure~\ref{fig:results:dihiggs:HHWWWW}.
Upper limits are also set on the
production cross-section times branching fraction of a heavy scalar $X$ that decays into two heavy scalars $S$ for
mass ranges of $280  \leq m_X \leq 340$~\GeV and $135  \leq m_S \leq 165$~\GeV
and the observed (expected) limits range from 2.5 (2.5)~pb to 0.16 (0.17)~pb
as shown in Figure~\ref{fig:results:dihiggs:SSWWW}.
\begin{figure}[tb!]
\begin{center}{\includegraphics[width=0.60\textwidth]{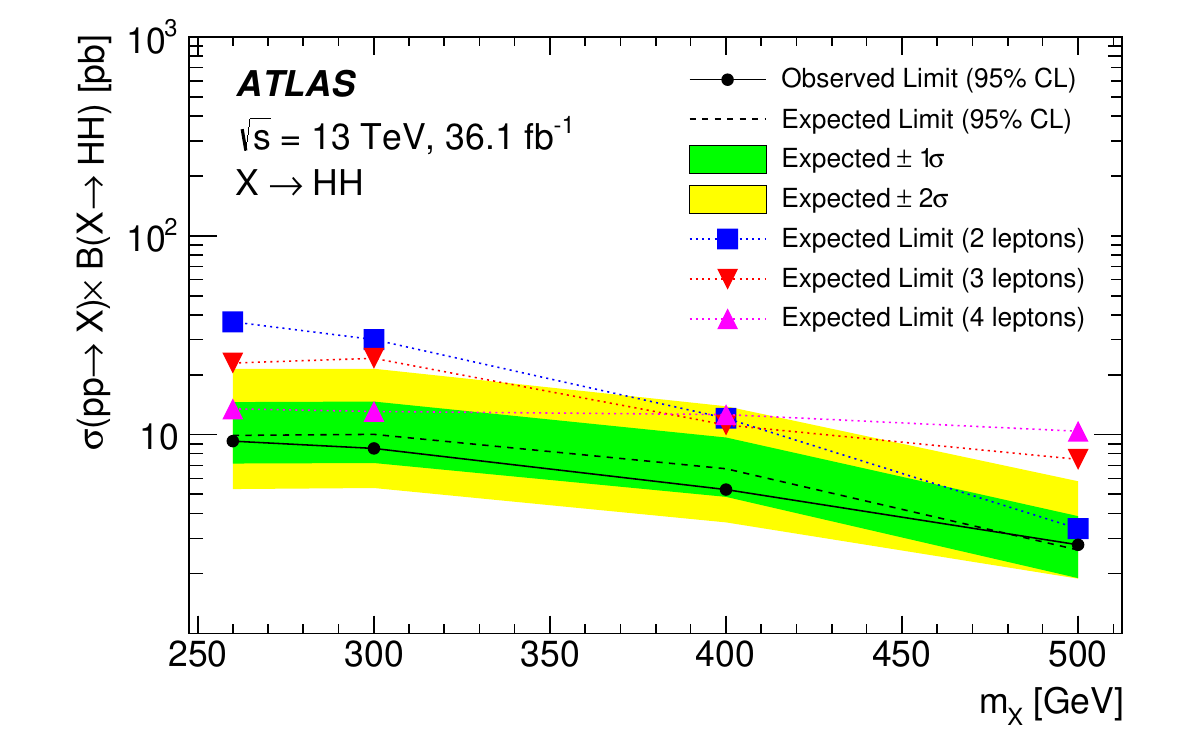}}%
\caption{$X \to HH\to WW^* WW^*$:
Expected and observed 95\% CL exclusion limits set on the cross-section times branching fraction
of resonant $HH$ production as a function of $m_X$. Limits are shown for each channel
individually as well as for the combination of the channels.
Figures are taken from Ref.~\cite{HIGG-2016-24}.}
\label{fig:results:dihiggs:HHWWWW}
\end{center}
\end{figure}
\begin{figure}[tb!]
\begin{center}

\subfloat[]{
\includegraphics[width=0.49\textwidth,valign=c]{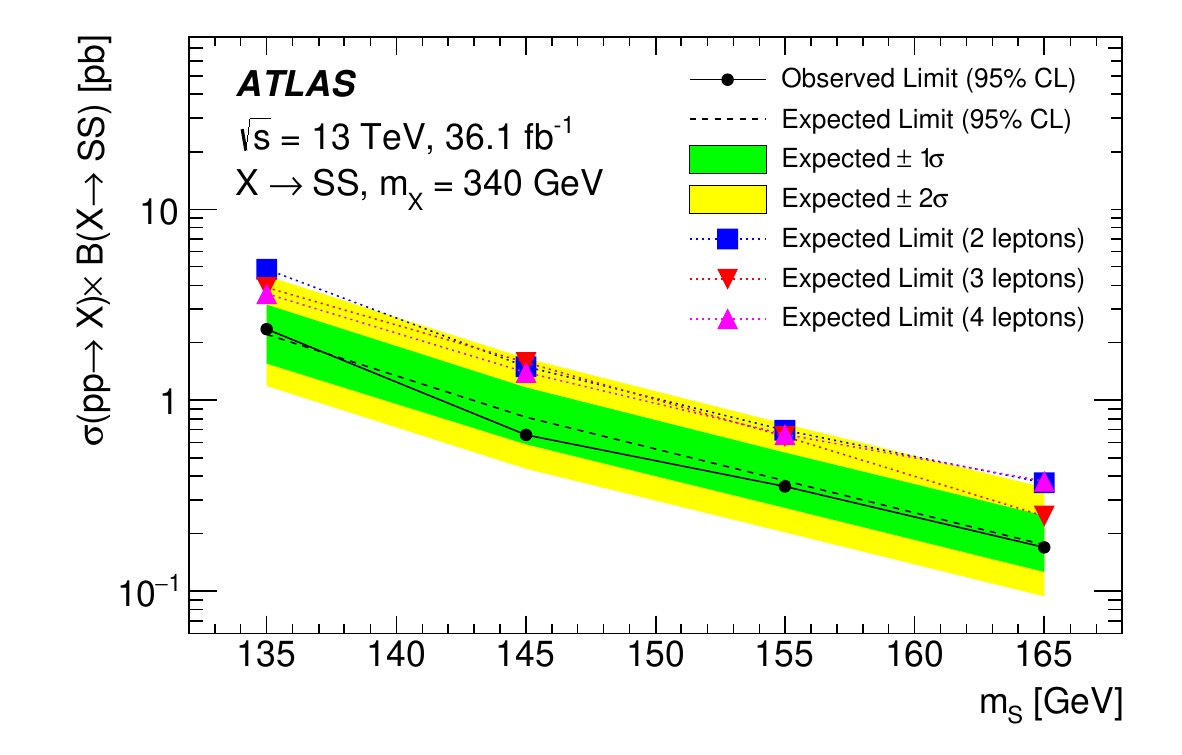}
}
\subfloat[]{
\includegraphics[width=0.49\textwidth,valign=c]{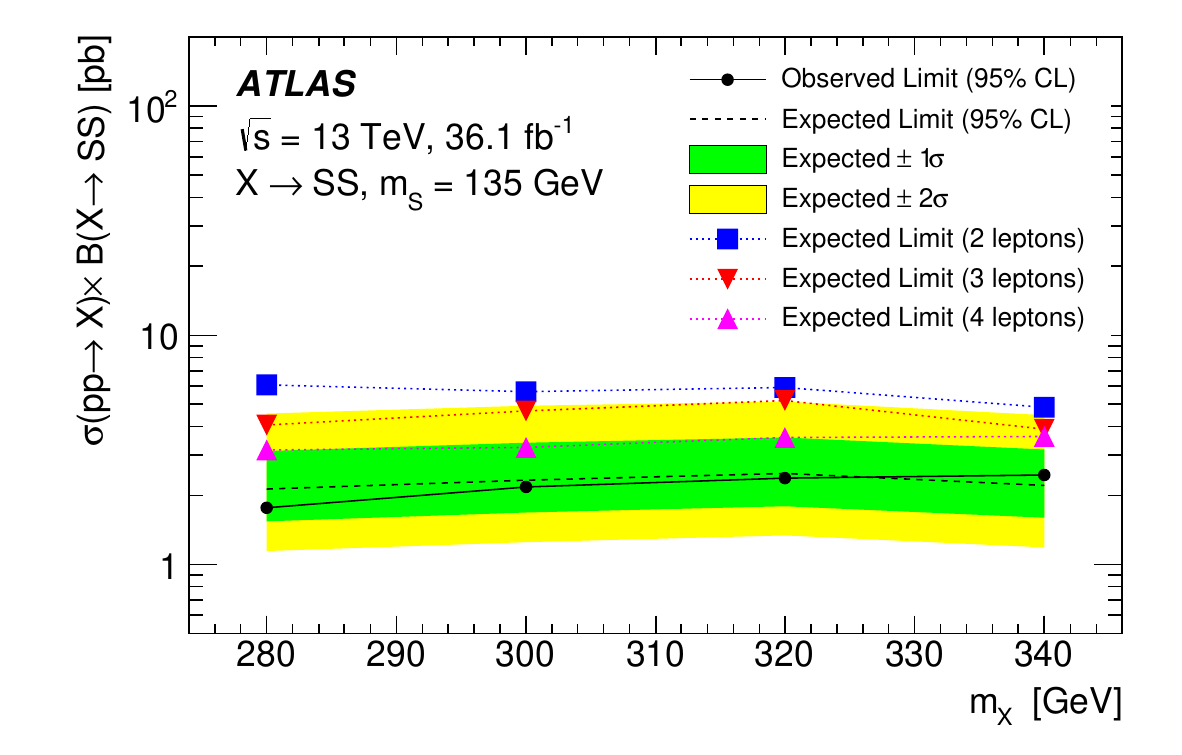}
}

\caption{$X \to SS\to WW^* WW^*$:
Expected and observed 95\% CL exclusion limits set on the cross-section times branching fraction of
resonant $X \to SS$ production as a function of (a) $m_S$ and (b) $m_X$.
Limits are shown for each channel individually as well as for the combination of the channels.
Figures are taken from Ref.~\cite{HIGG-2016-24}.}
\label{fig:results:dihiggs:SSWWW}
\end{center}
\end{figure}

The search for \textbf{$S \to HH\to\bbbar WW^*$}
with the  $\bbbar\ell\nu qq$ final state
did not observe any excess above the expected background~\cite{HIGG-2016-27}.
Limits at 95\% CL are set on the resonant production cross-section
$\sigma(pp \to S \to HH)$  as a function of the mass
of a scalar resonance
in the mass range 500 to 3000~\GeV,
as shown in Figure~\ref{fig:results:dihiggs:bbWW}.
The spin-0 scalar states were treated as narrow heavy neutral Higgs bosons.
The observed upper limits on the production cross-sections range from 5.6~pb for
$m_S = 500$~\GeV to 0.51~pb for $m_S = 3000$~\GeV in the case of a scalar hypothesis.
\begin{figure}[tb!]
\begin{center}{\includegraphics[width=0.5\textwidth]{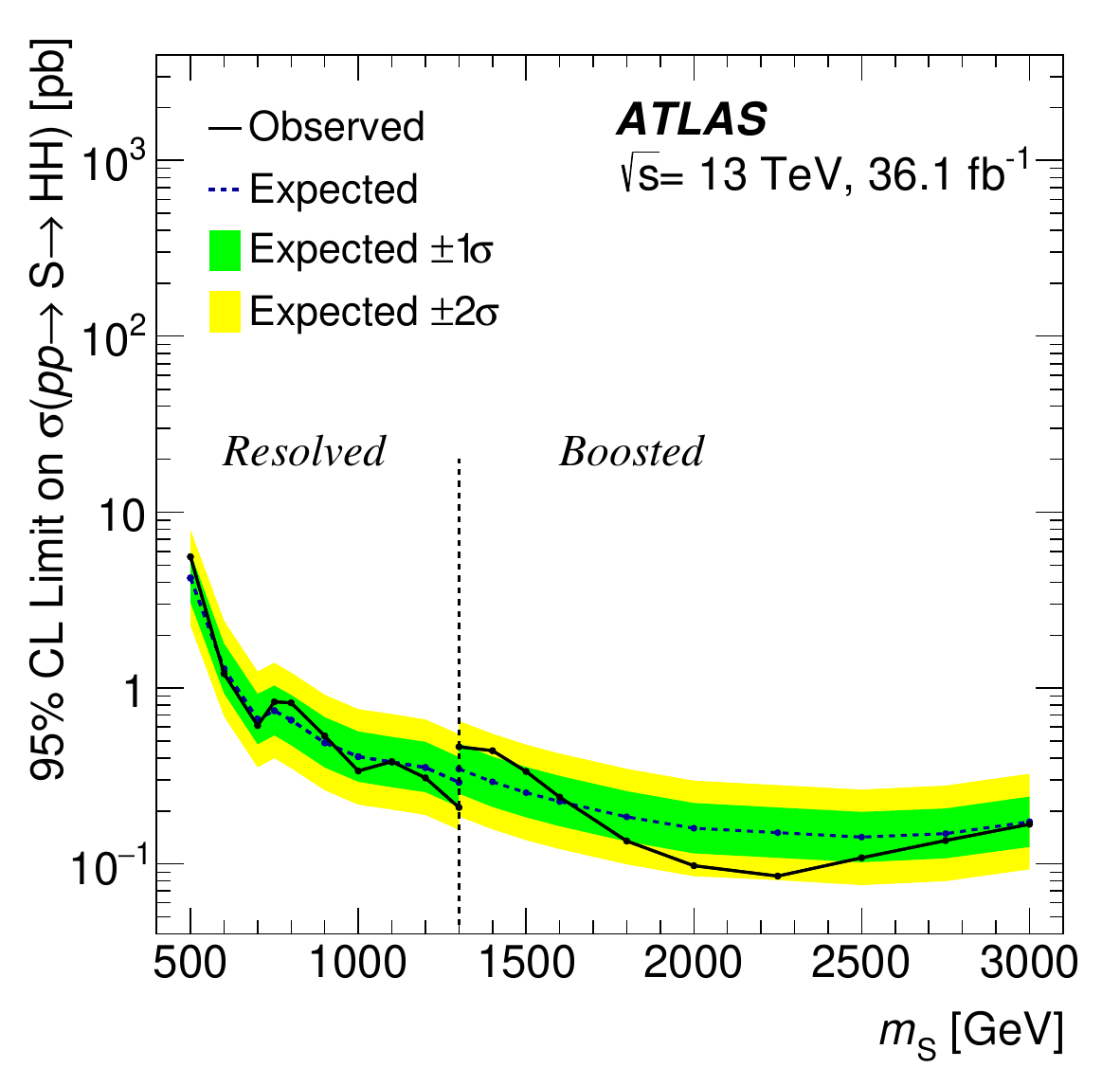}}%
\caption{$S \to HH\to\bbbar WW^*$: 95\% CL cross-section limits for resonant scalar production,
$\sigma(pp \to S \to HH)$.
The boosted case has a single large-radius jet representing
the $\bbbar$ pair.
The figure is taken from Ref.~\cite{HIGG-2016-27}.}
\label{fig:results:dihiggs:bbWW}
\end{center}
\end{figure}

The search for \textbf{$X \to HH\to\gamma\gamma WW^*$}
with the $\gamma\gamma\ell\nu jj$ final state did
not find any significant deviation from the SM prediction~\cite{HIGG-2016-20}.
The observed (expected) 95\% CL upper limit on the resonant
production cross-section times the branching fraction of
$X \to HH$ ranges between 40~pb and 6.1~pb (17.6~pb and 4.4~pb) for a hypothetical resonance
with a mass in the range of 260--500~\GeV, assuming SM branching fractions for
$H \to \gamma\gamma$ and $H \to WW^*$, as shown in Figure~\ref{fig:results:dihiggs:HH-ggWW}.
\begin{figure}[tb!]
\begin{center}

\subfloat[]{
\includegraphics[width=0.49\textwidth,valign=c]{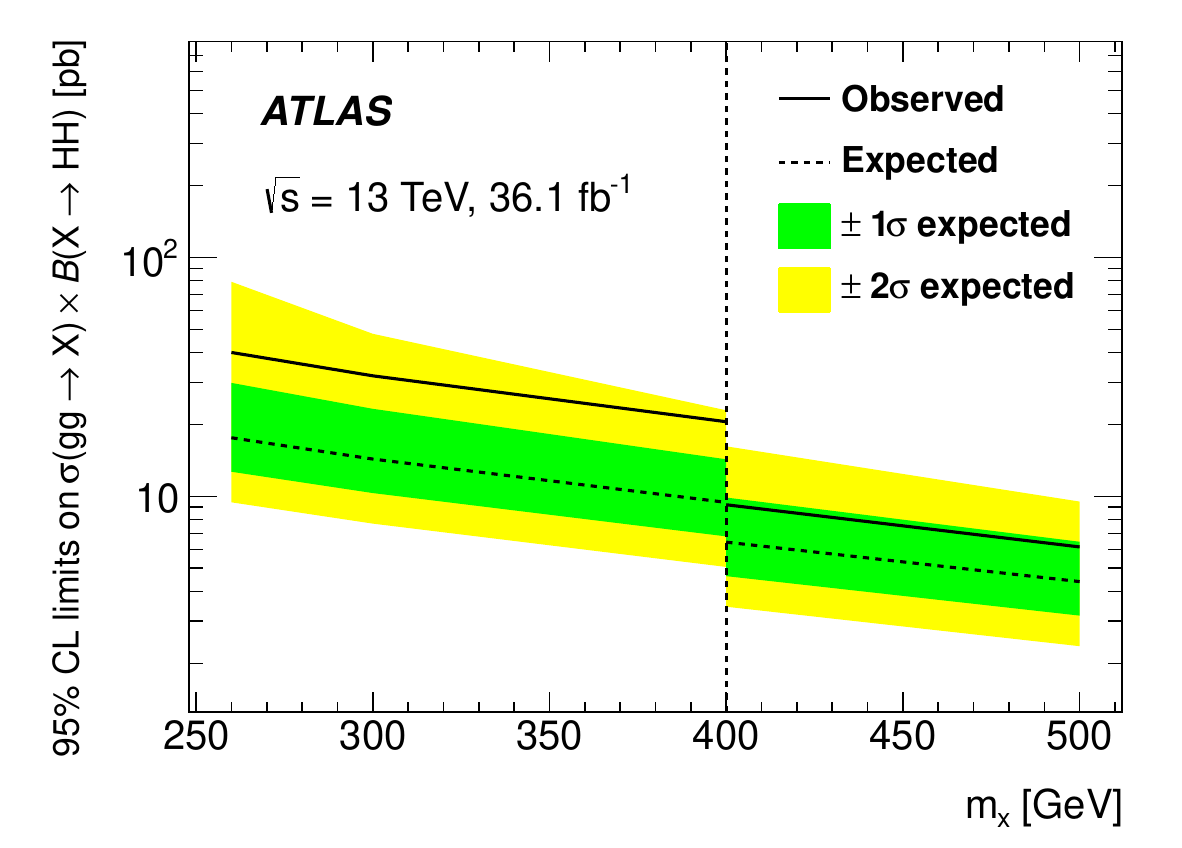}
}
\subfloat[]{
\includegraphics[width=0.49\textwidth,valign=c]{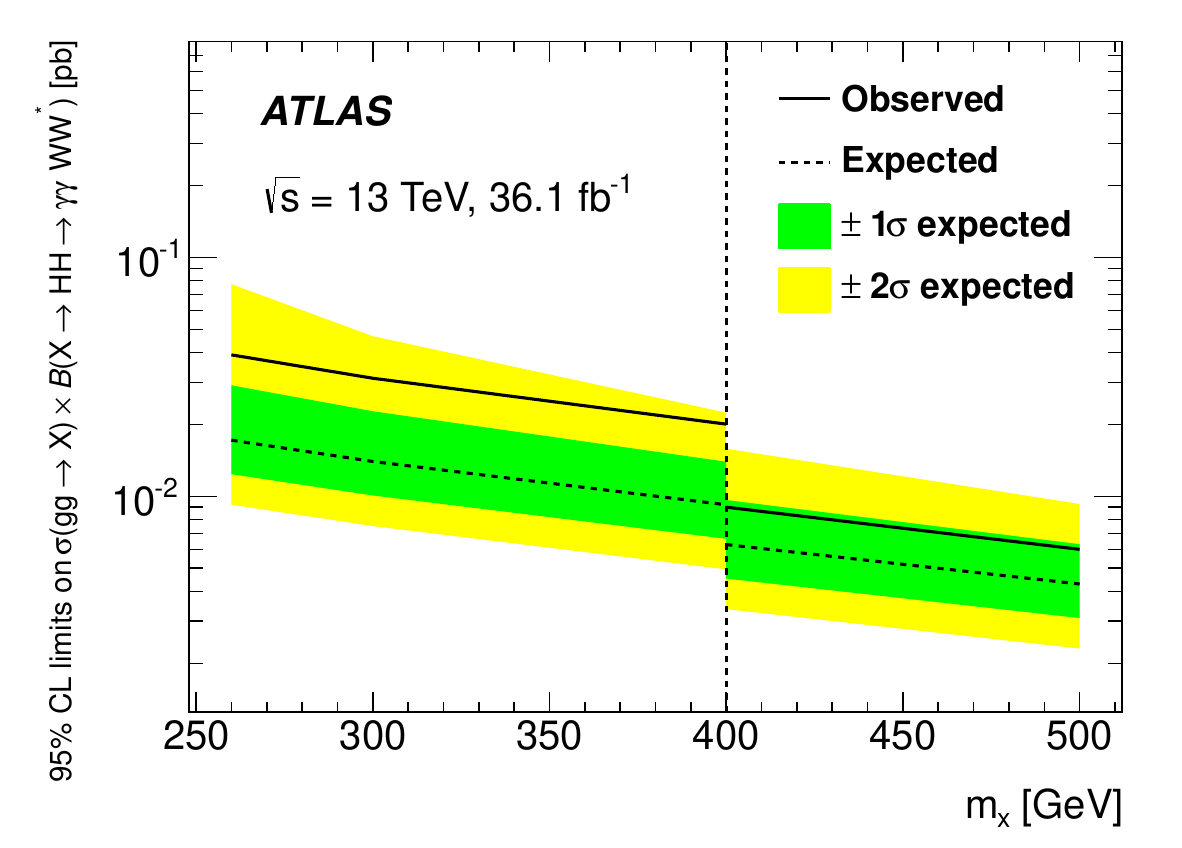}
}

\caption{$X \to HH\to\gamma\gamma WW^*$:
95\% CL expected (dashed line) and observed (solid line) limits on the resonant Higgs boson
pair production cross-section times the branching fraction of $X \to HH$ as a function of
$m_X$ (a) with and (b) without the assumption of SM branching fractions for
$H \to \gamma\gamma$ and $H \to WW^*$.
To the right, but not to the left, of the vertical dashed line at $m_X = 400$~\GeV, a
$p_\text{T}^{\gamma\gamma} > 100$~\GeV selection is applied in both plots.
Figures are taken from Ref.~\cite{HIGG-2016-20}.}
\label{fig:results:dihiggs:HH-ggWW}
\end{center}
\end{figure}

The search \textbf{$X\to SH\to VV\tautau$} ($V=W,Z$)
did not observe any excess beyond the expected SM background~\cite{HDBS-2022-44}.
The 95\% CL upper limits on the cross-section for $\sigma(pp \to X \to SH)$, assuming the
branching fractions for $S \to VV$ decay are the same as for SM Higgs boson decay, are between 72~fb and 542~fb.
Upper limits on the visible cross-sections $\sigma(pp \to X \to SH \to W^+W^-\tautau)$
and $\sigma(pp \to X \to SH \to ZZ\tautau)$ are set in
the ranges 3--26~fb and 5--33~fb, respectively, as shown
in Figure~\ref{fig:results:dihiggs:X-SH-TautauVV}.
The visible cross-section refers to events with \tauhad candidate jets that an
identification algorithm distinguishes from jets initiated by quarks or gluons.
\begin{figure}[tb!]
\begin{center}

\subfloat[]{
\includegraphics[width=0.7\textwidth,valign=c]{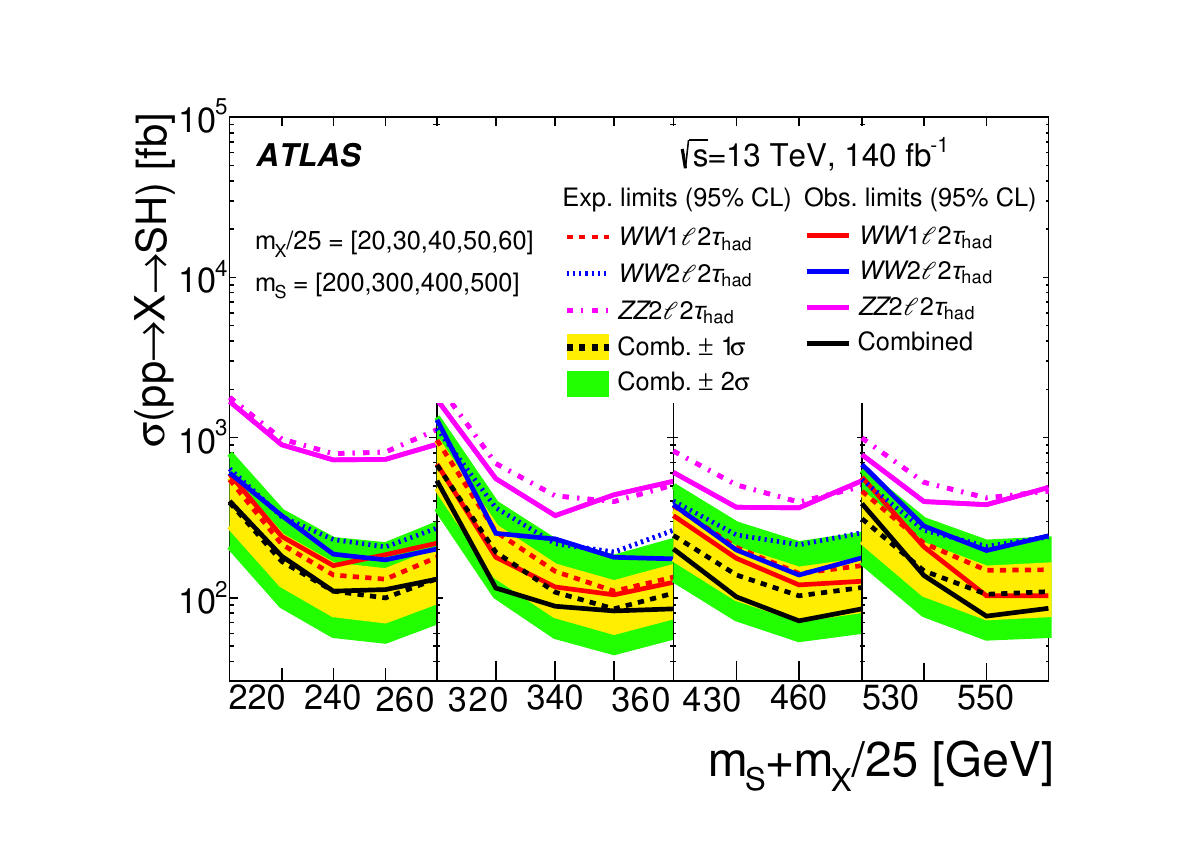}
}\\
\subfloat[]{
\includegraphics[width=0.49\textwidth,valign=c]{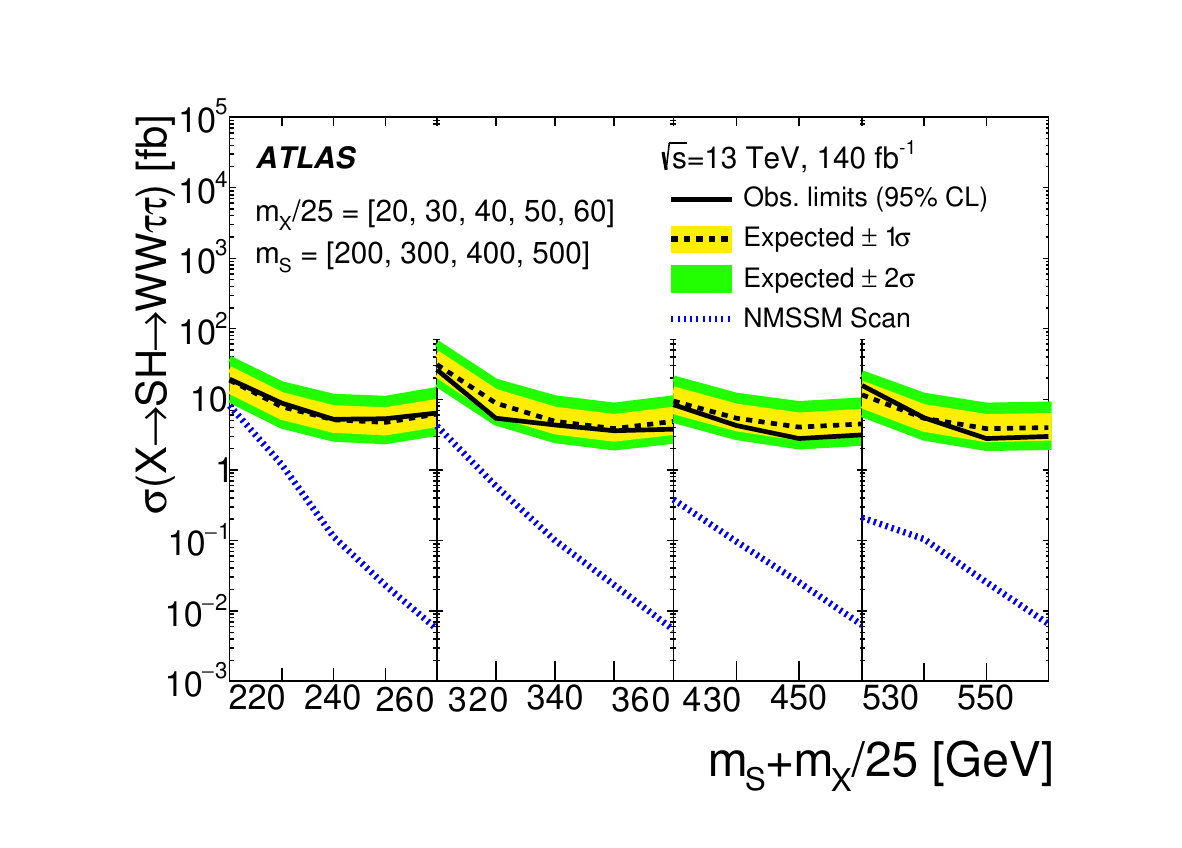}
}
\subfloat[]{
\includegraphics[width=0.49\textwidth,valign=c]{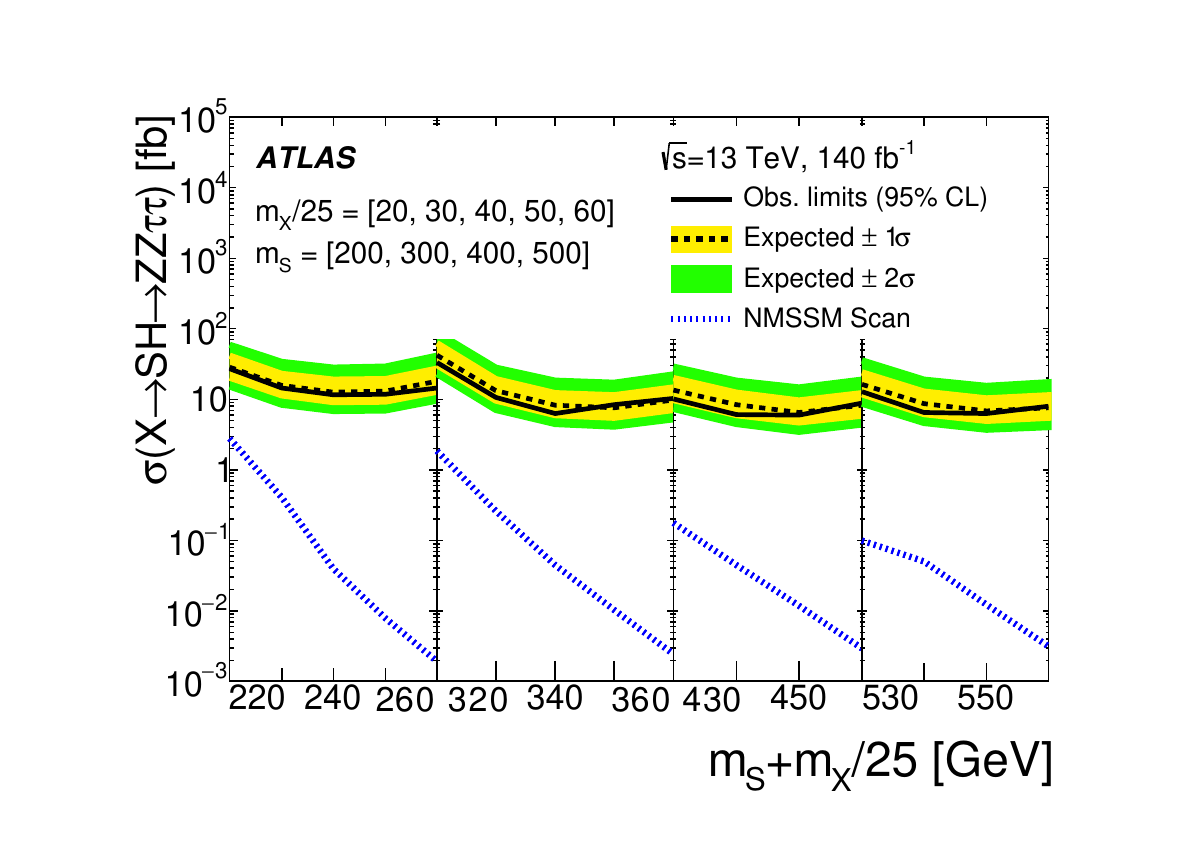}
}

\caption{$X\to SH\to VV\tautau$:
Observed and expected 95\% CL upper limits are shown for (a) $\sigma(pp \to X \to SH)$
obtained from three channels and their combination;
(b) $\sigma(pp \to X \to SH \to W^+W^-\tautau)$ obtained from the combination of
the $WW 1\ell 2\tauhad$ and $WW 2\ell 2\tauhad$ channels;
(c) $\sigma(pp \to X \to SH \to ZZ\tautau)$ obtained from the $ZZ 2\ell 2\tauhad$ channel,
as a function of combined $m_S$ and $m_X$ masses $(m_S + m_X/25)$ in \GeV.
The NMSSM scans of the allowed cross-sections
for $\sigma(pp \to X \to SH \to W^+W^-\tautau)$ and $\sigma(pp \to X \to SH \to ZZ\tautau)$
are also shown.
Figures are taken from Ref.~\cite{HDBS-2022-44}.}
\label{fig:results:dihiggs:X-SH-TautauVV}
\end{center}
\end{figure}

The search \textbf{$X\to SH\to  \bbbar\gamma\gamma$} did not observe any excess beyond the expected SM background~\cite{HDBS-2021-17}.
Accordingly, 95\% CL upper limits are set on
$\sigma(X\to SH\to  \bbbar\gamma\gamma)$.
However, some deviation is observed for a few points in the $(m_X, m_S)$ plane.
The largest of these is at $(m_X, m_S) = (575,200)$~\GeV with a
local (global) significance of 3.5$\sigma$ (2.0$\sigma$).
A point of particular interest is $(m_X, m_S) = (650,90)$~\GeV where the
CMS Collaboration reports an excess with a local (global)
significance of 3.8$\sigma$ (2.8$\sigma$)~\cite{CMS-HIG-21-011}.
A test signal injected at this point with the 0.35~fb cross-section
reported by the CMS experiment produced a local excess of 2.7$\sigma$.
This demonstrates consistency between the two analyses.
The observed and expected upper limits for $\sigma(X\to SH\to  \bbbar\gamma\gamma)$
are presented in Figure~\ref{fig:results:dihiggs:X-SH-bbgg}.
The observed (expected) limits range from 39 (25)~fb at $m_X = 170$~\GeV and
$m_S = 30$~\GeV to 0.09 (0.14)~fb at $m_X = 1000$~\GeV and $250 \leq m_S \leq 300$~\GeV.
The observed upper limit on $\sigma(X\to SH\to \bbbar\gamma\gamma)$ for the
point described above for the CMS result is 0.2~fb.
\begin{figure}[tb!]
\begin{center}
\subfloat[]{
\includegraphics[width=0.49\textwidth,valign=c]{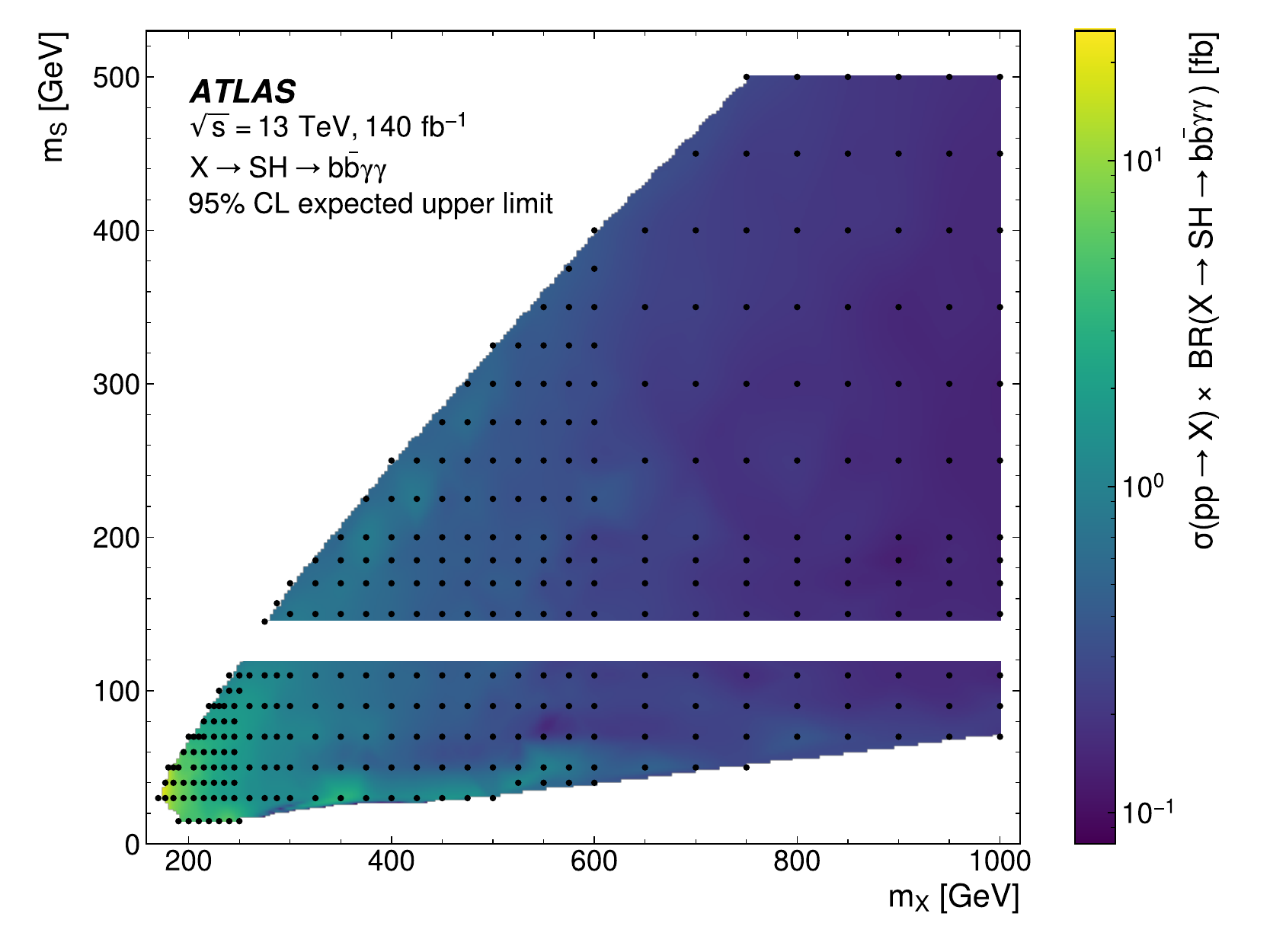}
}
\subfloat[]{
\includegraphics[width=0.49\textwidth,valign=c]{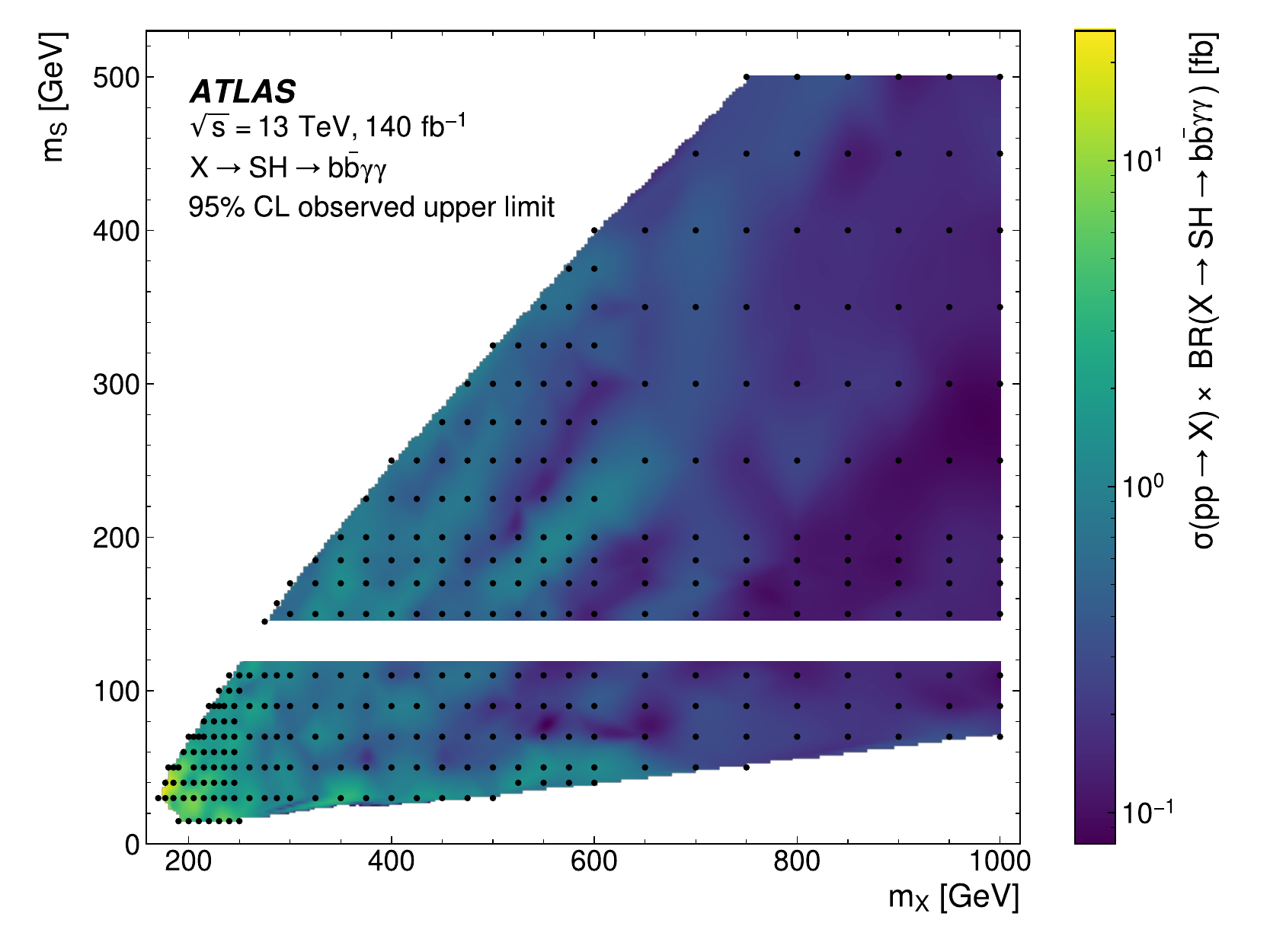}
}
\caption{$X\to SH\to \bbbar\gamma\gamma$:
Expected (a) and observed (b) upper limits on the signal cross-section times
branching fraction for the $X\to SH$ signal, in the $(m_X, m_S)$ plane.
The points show where the limits were evaluated.
The band at $m_S = 125$~\GeV is not shown because those points are
equivalent to those already probed in Ref.~\cite{HDBS-2018-34}.
The figures are taken from Ref.~\cite{HDBS-2021-17}.}
\label{fig:results:dihiggs:X-SH-bbgg}
\end{center}
\end{figure}

\FloatBarrier


\subsection{Summary of heavy Higgs boson searches}
\label{sec:results:summaryheavyhiggs}

The excluded regions in the $m_A$--$\tan\beta$ plane for the hMSSM are displayed in Figure~\ref{fig:results:summary:hmssm}~\cite{ATL-PHYS-PUB-2022-043}. The results for various channels from searches for neutral or charged Higgs bosons and resonant Higgs boson pair production are overlaid, but they are not statistically combined. This is a useful form of presentation for understanding the regions in which the various channels are sensitive in the same model.

The $A/H\to\tau\tau$ channel dominates the sensitivity over a large mass range. It is the strongest channel at high $\tan\beta$ values for $m_A$ above 400 GeV. The low $\tan\beta$ region is constrained by decays to top-quark pairs as well as decays to vector bosons or Higgs boson pairs. The search for $H^+\to tb$ has a unique sensitivity to both low and high values of $\tan\beta$. The region around $\tan\beta\sim 6$ is difficult to constrain since the coupling of the heavy Higgs bosons to SM particles has a minimum there, impacting both the production and decay rates. This region could be accessible via searches involving supersymmetric decay chains if the particle mass spectra are favourable~\cite{Gori_2019}.

The coupling analysis from the SM Higgs boson measurements also leads to constraints on BSM parameters. The almost vertical lines in pink in Figure~\ref{fig:results:summary:hmssm} are the limits on $m_A$ (with only a slight dependence on $\tan\beta$) after reparameterizing the coupling modifiers corresponding to the measured production and decay rates. The coupling modifiers are expressed as model-dependent cross-section scale factors and ratio scale factors that are calculated for discrete points in the $m_A$ vs $\tan\beta$ plane. A more complete set of plots obtained from the SM Higgs boson coupling analysis with the full \RunTwo dataset can be found in Ref.~\cite{HIGG-2022-17}.

The hMSSM summary plot displayed here is a specific benchmark; the sensitivities of the performed analyses and patterns of excluded areas will be different in other benchmarks.

\begin{figure}[tb!]
\centering
\includegraphics[width=0.8\textwidth]{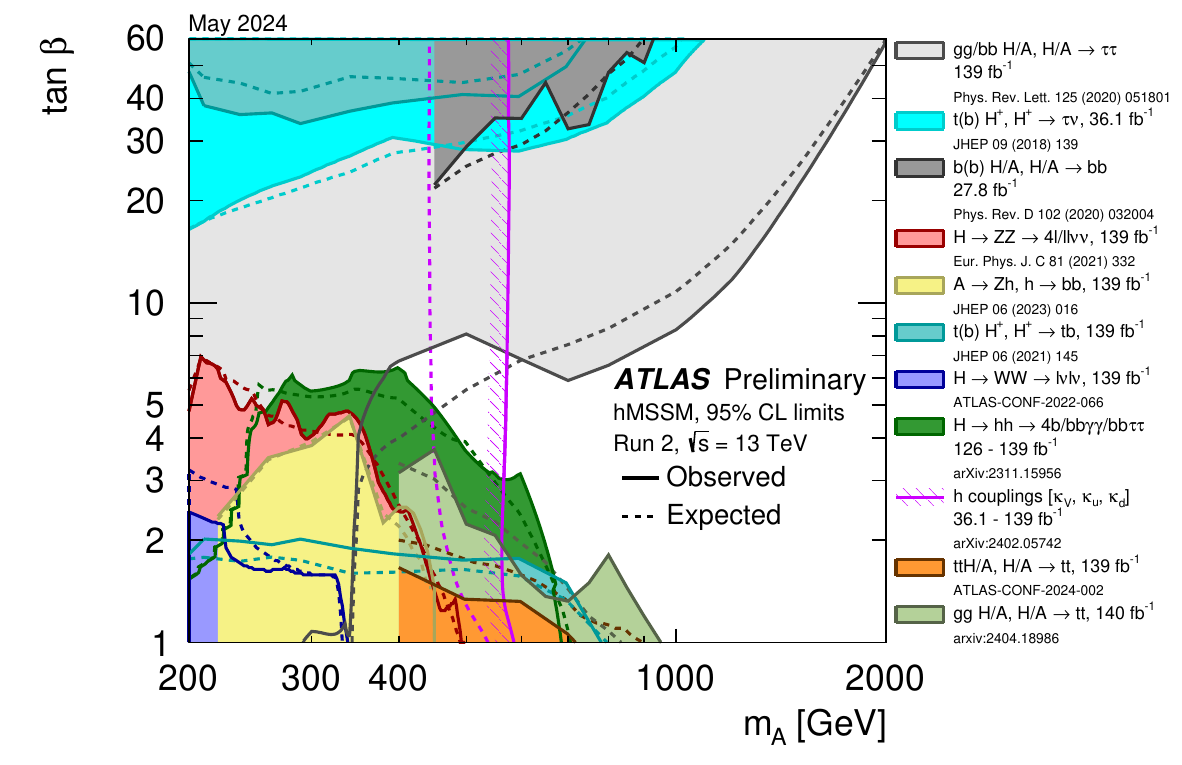}
\caption{The expected (dashed lines) and observed (filled areas) exclusions at 95\% CL for the hMSSM. The figure is taken from Ref.~\cite{ATL-PHYS-PUB-2022-043}.}
\label{fig:results:summary:hmssm}
\end{figure}

\FloatBarrier


\subsection{Additional scalars and exotic decays of the Higgs boson}
\label{sec:results:exoticdecays}
The Higgs BSM searches which involve exotic decays of the Higgs boson,
and possibly also additional scalars, are collected here.

\subsubsection{Exotic decays of the Higgs boson to invisible final states}
\label{sec:results:ex-higgs:inv}

The search for \textbf{$ZH$, $H\to$ invisible}~\cite{HIGG-2018-26} sets an
upper limit of 19\% on the branching fraction of the Higgs boson to invisible
particles at the 95\% CL
(assuming SM cross-sections for $ZH$ production).
The corresponding expected limit of 19\%
represents an improvement of about 45\% in comparison with a projection of the previous analysis scaled
to the present integrated luminosity.
Exclusion limits were also set for simplified dark-matter models
and 2HDM+$a$ models for a number of benchmark parameters, one of which is shown in
Figure~\ref{fig:results:ex-decays:fid_08372}.
\begin{figure}[tb!]
\begin{center}
\subfloat[]{\includegraphics[width=0.47\textwidth]{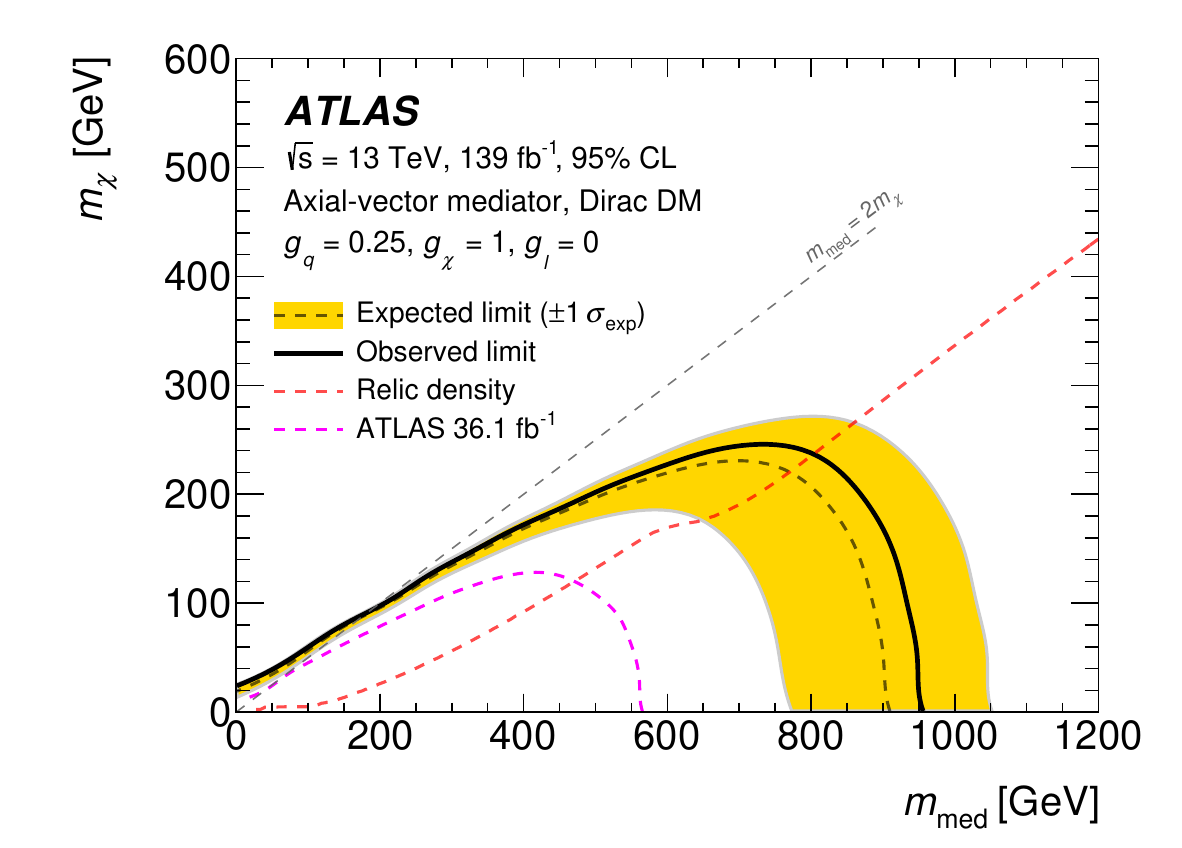}}%
\subfloat[]{\includegraphics[width=0.47\textwidth]{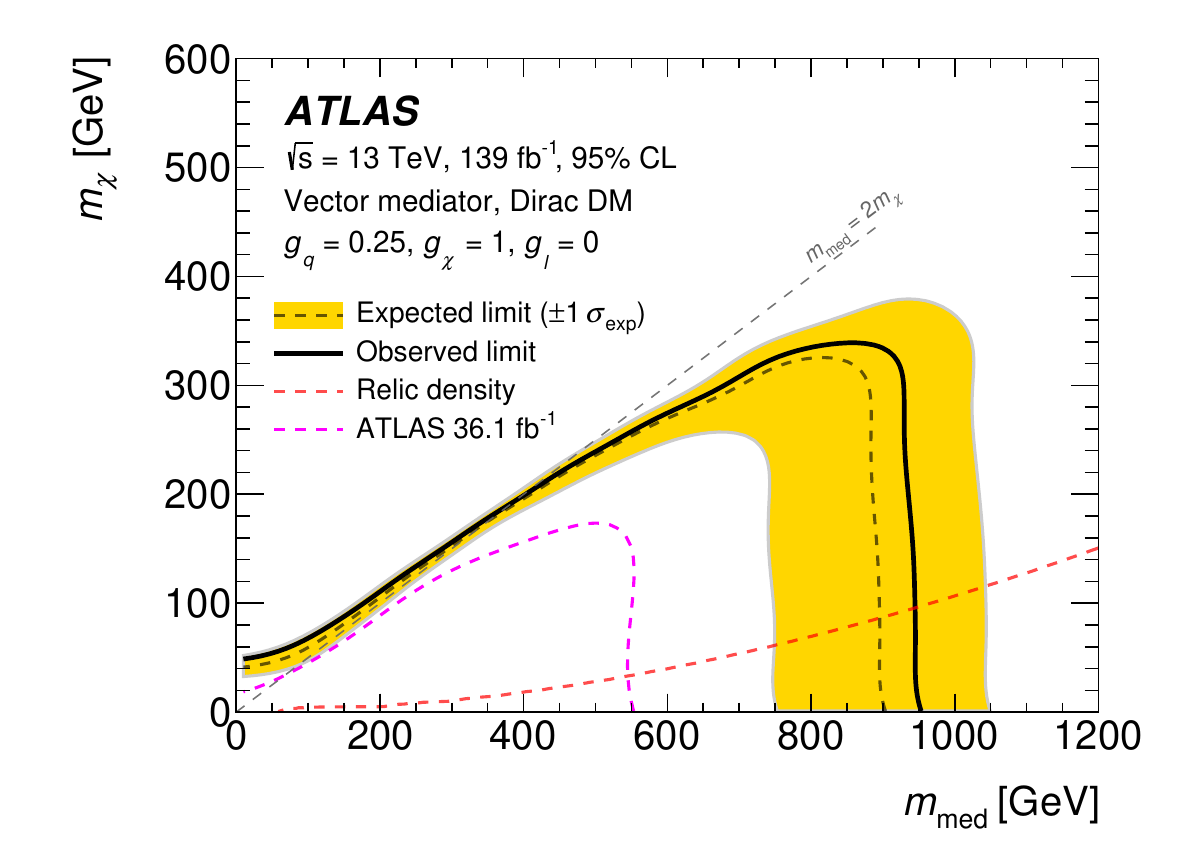}}%
\caption{$ZH$, $H\to$ invisible:
Exclusion limits for simplified DM models with $g_{\chi} = 1.0$, $g_{q} = 0.25$,
and $g_{\ell} = 0$~\cite{boveia2016recommendations,ALBERT2019100377}, when assuming
(a) an axial-vector mediator or (b) a vector mediator.
The region below the solid black line is excluded at the 95\% CL.
The dashed black line indicates the expected limit in the absence of signal, and the yellow band
the corresponding $\pm 1\sigma$ uncertainty band.
The dashed red line labelled  \enquote{Relic density} corresponds to
combinations of DM and mediator mass values that are consistent with a DM density of
$\Omega h^2 = 0.118$ and a standard thermal history, as computed in Ref.~\cite{ALBERT2019100377}.
Below the line, annihilation
processes described by the simplified model mostly predict too high a relic density while regions with
too low a relic density are mostly found for $m_{\text{med}}$ closer to the DM mass.
The dashed magenta line indicates the previous ATLAS result from a 36.1~\ifb
dataset~\cite{HIGG-2016-28}.
Figures are taken from Ref.~\cite{HIGG-2018-26}.}
\label{fig:results:ex-decays:fid_08372}%
\end{center}
\end{figure}

The search for \textbf{VBF $H\to$ invisible}
determined an observed (expected)
95\% CL upper limit of
$\mathcal{B}_{H\to \text{invisible}} < 0.145\,(0.103)$~\cite{EXOT-2020-11},
which is an improvement on the previous analysis~\cite{EXOT-2016-37}.
The result is interpreted using Higgs portal models to exclude regions in the parameter
space of $(\sigma_{\text{WIMP--nucleon}}, m_{\text{WIMP}})$ for various WIMP models~\cite{EXOT-2020-11}.
The obtained results are also interpreted as a search for invisible
decays of new scalar particles with masses of up to 2~\TeV, resulting in an upper limit of 1~pb on
$\sigma_{\text{VBF}} \times \mathcal{B}_{\text{invisible}}$
for a mediator mass of 50~\GeV,  decreasing to 0.1~pb for a mass of 2~\TeV, as shown in
Figure~\ref{fig:results:ex-decays:07953}.
\begin{figure}[tb!]
\begin{center}{\includegraphics[width=0.60\textwidth]{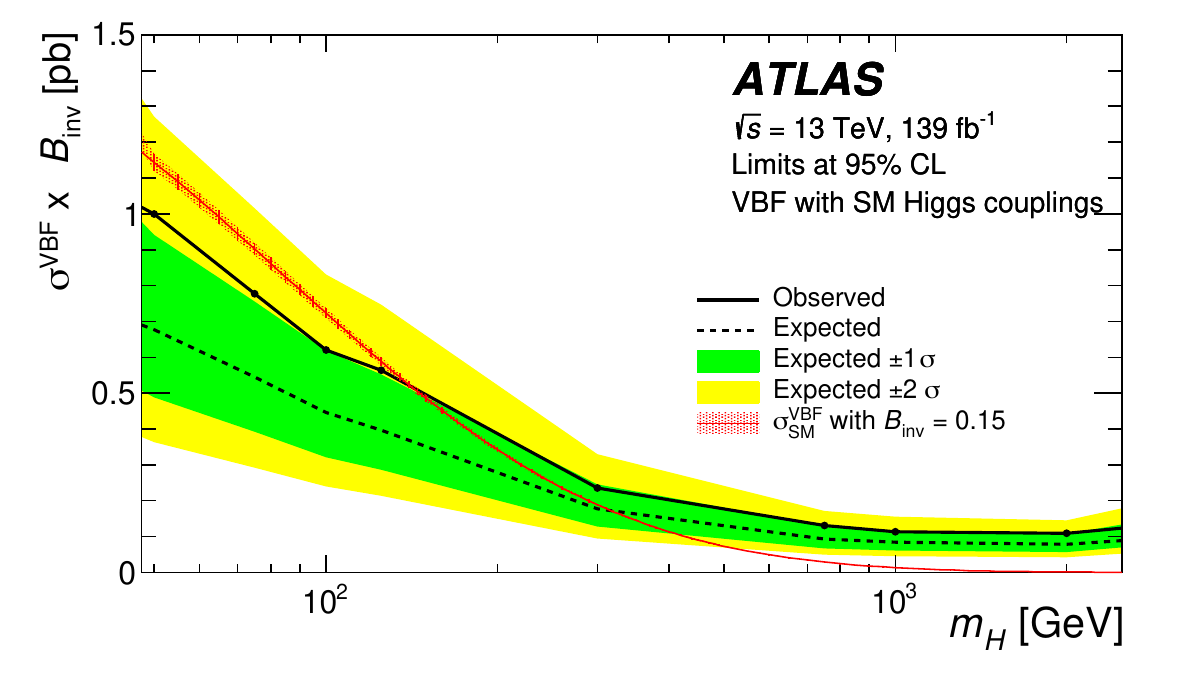}}%
\caption{VBF $H\to$ invisible:
Upper limit on the cross-section times branching fraction to invisible particles for a
scalar mediator
as a function of its mass.
For comparison, the VBF cross-section at NLO in QCD, i.e.\
without the electroweak corrections, for a particle with SM Higgs boson couplings,
multiplied by a $\mathcal{B}_{\text{invisible}}$ value of 15\%, is overlaid.
The figure is taken from Ref.~\cite{EXOT-2020-11}.}
\label{fig:results:ex-decays:07953}%
\end{center}
\end{figure}

Results of the \textbf{$H\to$ invisible combination}~\cite{HIGG-2021-05} are shown
in Figure~\ref{fig:results:ex-decays:10731}.
Figure~\ref{fig:results:decays:inv:limits} shows the observed and expected upper limits on
$\mathcal{B}_{H\to \text{invisible} }$ for the individual and combined searches.
The full combination, including the \RunOne result, gives the current most
sensitive observed (expected) ATLAS result: $\mathcal{B}_{H\to \text{invisible}}< 0.107\,(0.077)$ at the 95\% CL.
The most sensitive channels included in this combination are the
ones where the Higgs boson is produced via VBF or $ZH$, and are described in more detail above.
Figure~\ref{fig:results:decays:inv:wimp} shows the model-dependent Higgs portal interpretation where limits are set on the WIMP--nucleon
scattering cross-section, shown in a context of other related searches, highlighting the complementarity of
DM searches at the LHC and direct-detection experiments.
\begin{figure}[tb!]
\begin{center}
\subfloat[]{
\label{fig:results:decays:inv:limits}
\includegraphics[width=0.42\textwidth]{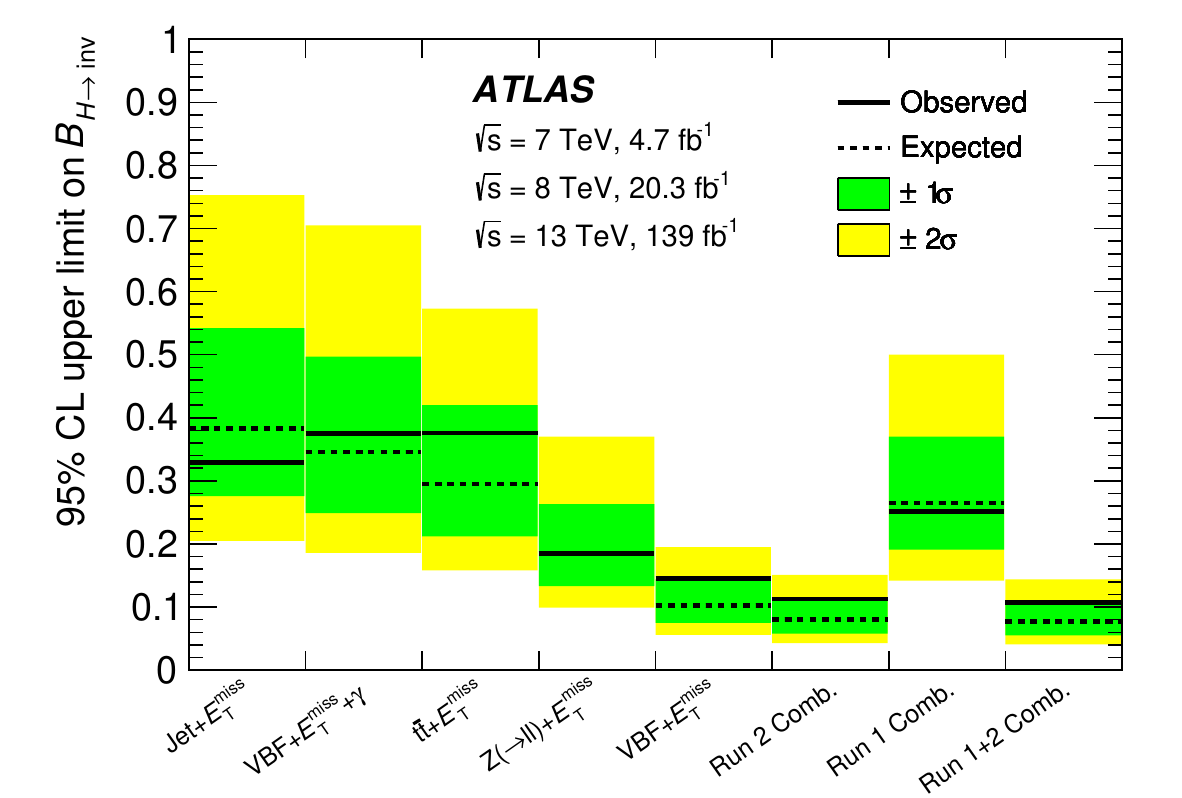}
}
\subfloat[]{
\label{fig:results:decays:inv:wimp}
\includegraphics[width=0.48\textwidth]{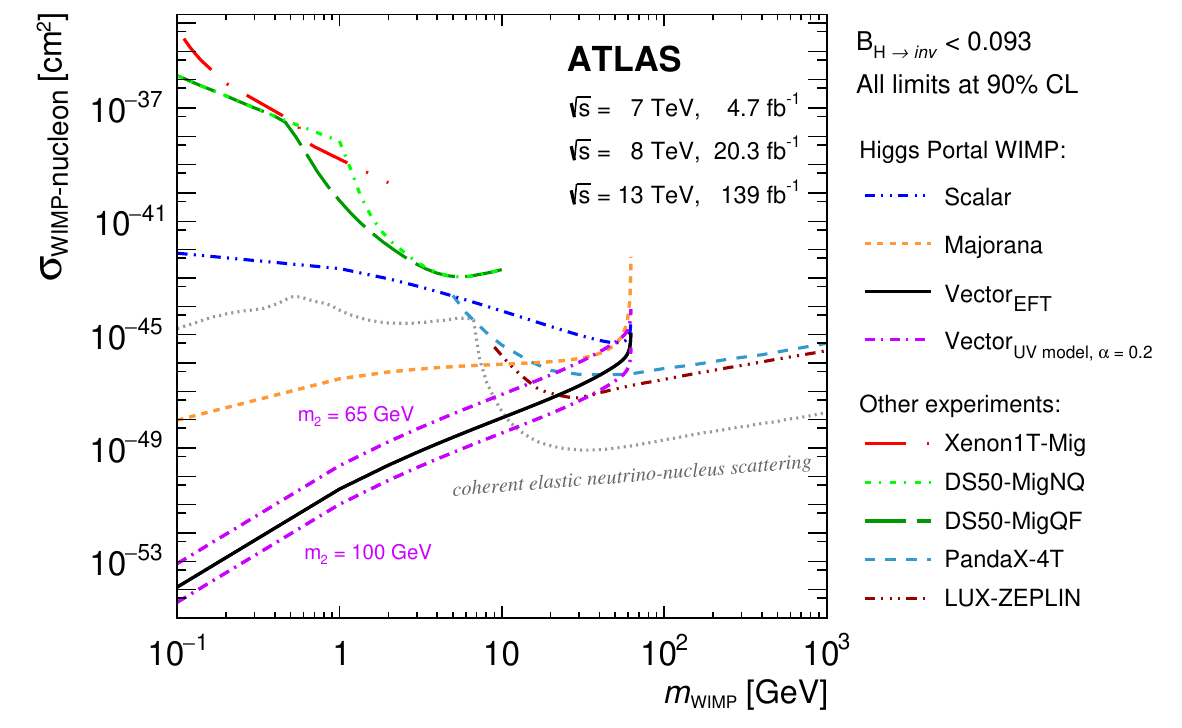}
}
\caption{$H\to$ invisible combination:
(a) The observed and expected upper limits on $\mathcal{B}_{H\to \text{invisible} }$ at 95\% CL
from the \RunTwo analyses targeting the production modes indicated on the $x$-axis
and their combination, the \RunOne
combination and the combined \RunOne and \RunTwo result; the $1\sigma$ and $2\sigma$ contours of the expected limit
distribution are also shown.
(b) Upper limit at the 90\% CL on the spin-independent WIMP--nucleon scattering cross-section as a
function of the WIMP mass for direct-detection experiments and the interpretation of the
$H\to \text{invisible}$ combination result in the context of Higgs portal models considering
scalar, Majorana and vector (WIMP) hypotheses.
For the vector case, results from UV-complete models are shown (pink curves) for
two representative values of the mass of the predicted dark Higgs particle ($m_2$)
and a mixing angle $\alpha=0.2$.
The uncertainties from the nuclear form factor are smaller than the line thickness.
Direct-detection results are taken from Refs.~\cite{PhysRevLett.121.081307,PhysRevLett.127.261802,
PhysRevLett.131.041002,PhysRevLett.123.241803}.
The neutrino floor for coherent elastic neutrino--nucleus scattering (solid grey line)
is taken from Refs.~\cite{PhysRevD.89.023524, PhysRevD.90.083510},
which assume that germanium is the target over the whole WIMP mass range.
The regions above the limit contours are excluded in the range shown in the plot.
Figures are taken from Ref.~\cite{HIGG-2021-05}.}
\label{fig:results:ex-decays:10731}%
\end{center}
\end{figure}

The search for \textbf{$H\to\gamma\gamma_d$} in associated production with a $Z$ boson did not reveal an excess~\cite{HDBS-2019-13}.
Exclusion limits are set on the branching fraction of SM Higgs boson decay into a photon and a dark
photon.
For a massless  $\gamma_d$, an observed (expected) 95\% CL upper limit of 2.28\%
was placed on $\mathcal{B}(H\to\gamma\gamma_d)$.
For a massive $\gamma_d$, observed (expected) upper limits are found to be within
the range 2.19\%--2.52\% (2.71\%--3.11\%) for masses from 1~\GeV to 40~\GeV.
This result is shown in Figure~\ref{fig:results:ex-higgs:gammad}.
\begin{figure}[tb!]
\begin{center}{\includegraphics[width=0.6\textwidth]{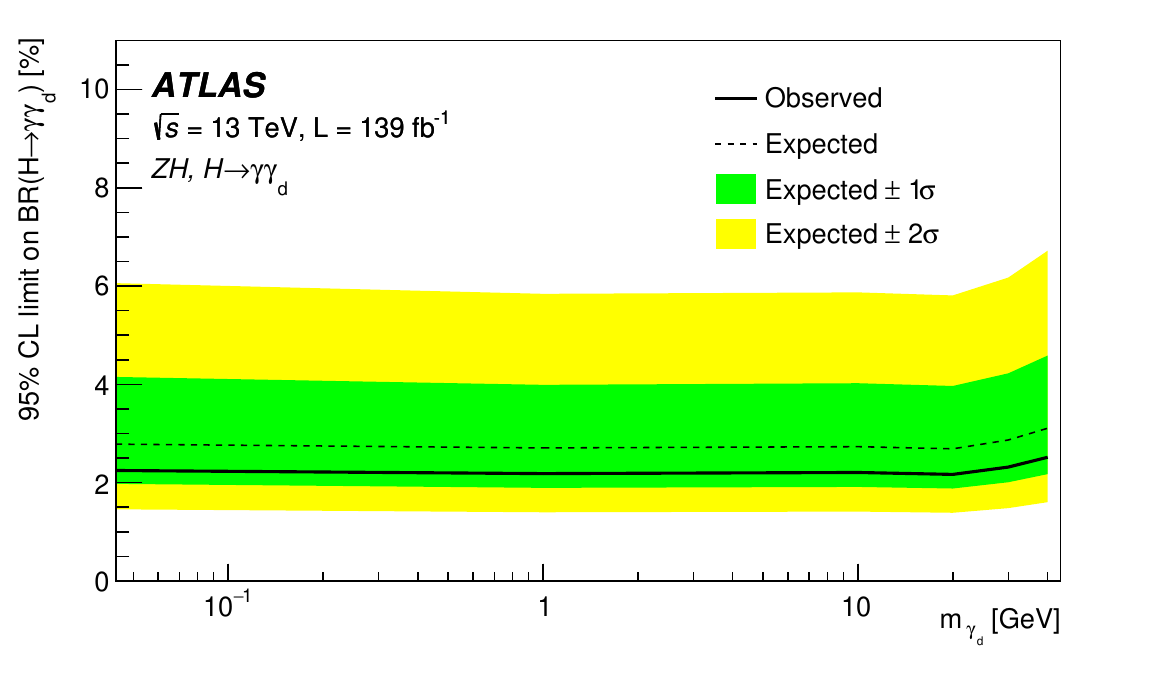}}%
\caption{$H\to\gamma\gamma_d$:
Observed and expected exclusion limits on $\mathcal{B}(H\to\gamma\gamma_d)$ at 95\% CL
as function of the $\gamma_d$ mass.
The figure is taken from Ref.~\cite{HDBS-2019-13}.}
\label{fig:results:ex-higgs:gammad}
\end{center}
\end{figure}

\subsubsection{Exotic decays of the Higgs boson or a heavy scalar to (pseudo)scalars or vector bosons}
\label{sec:results:ex-higgs:aa-XX}

The \textbf{$H\to XX/ZX \to 4\ell$} analysis~\cite{HDBS-2018-55} has three separate channels.
Some results are shown in Figure~\ref{fig:results:ex-higgs:fid_ZdZd}  for the   $H\to aa/ss$ and $H\to Z_dZ_d$
channels, where the target is the scalar/pseudoscalar of the 2HDM+S and the dark vector boson
of the HAHM, respectively.
Figure~\ref{fig:results:decays:4l:mass} displays the 95\% CL upper limits on the
fiducial cross-section in the full search range for the four-muon final state.
It is model-independent and would  be applicable to any model where the SM Higgs boson decays into four leptons via
two intermediate bosons that are narrow and on-shell, and decay promptly.
Figure~\ref{fig:results:decays:4l:limit} uses the HAHM for the model-dependent acceptances to display the 95\% CL
upper limit on the cross-section times branching fraction of the $H\rightarrow Z_dZ_d \rightarrow 4\ell$ process.
A slight excess with a local significance of 2.5$\sigma$ is found for a $Z_d$ mass hypothesis of 28~\GeV.
\begin{figure}[tb!]
\begin{center}
\subfloat[]{
\label{fig:results:decays:4l:mass}
\includegraphics[width=0.43\textwidth]{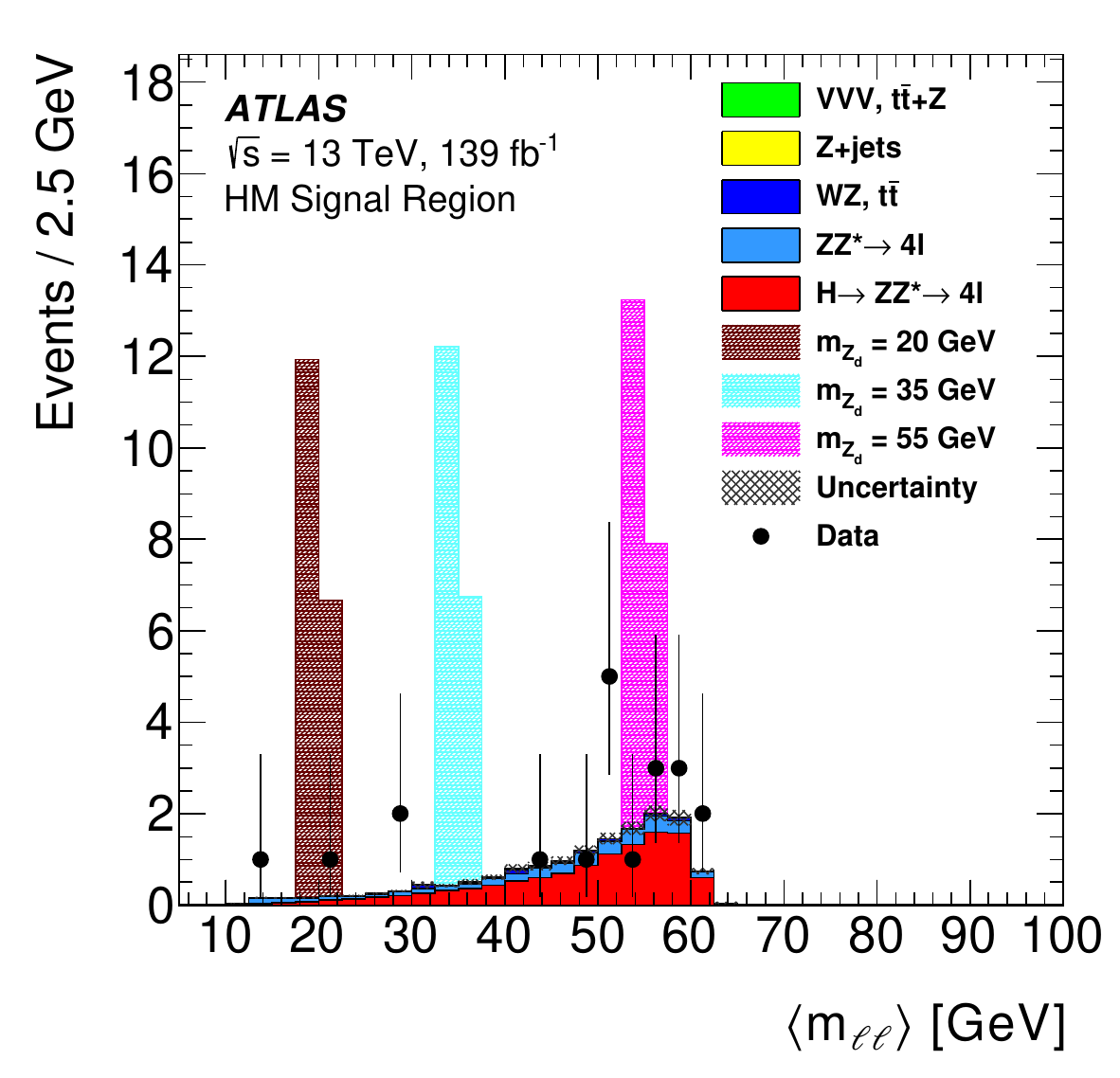}
}
\subfloat[]{
\label{fig:results:decays:4l:limit}
\includegraphics[width=0.56\textwidth]{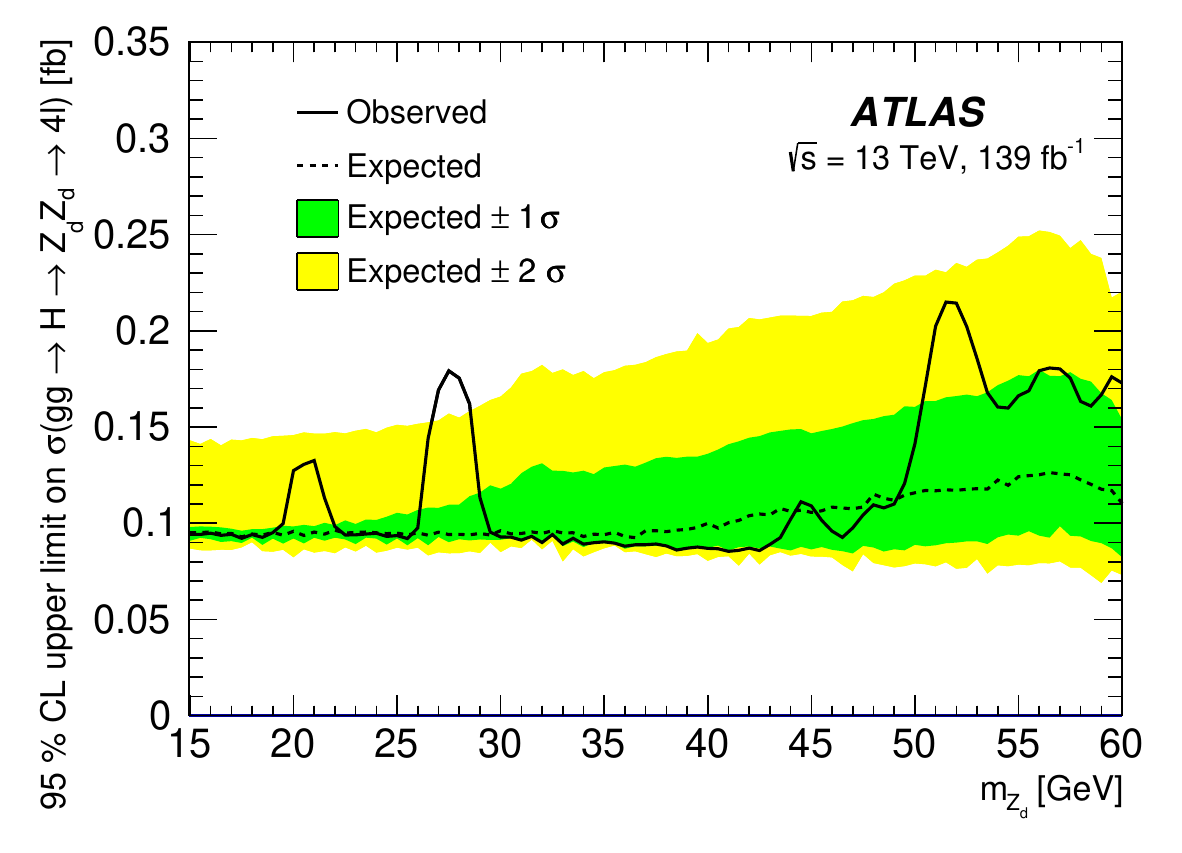}
}
\caption{$H\to XX \to 4\ell$:
(a)~The distribution of $\langle m_{\ell\ell}\rangle$,
including the (pre-fit) background expectations
and some stacked signal expectations.
The signal histograms' expected yields are normalized to
$\sigma(pp \to H \to Z_dZ_d \to 4\ell) =
\frac{1}{10} \sigma_{\text{SM}}(pp \to H \to ZZ^* \to 4\ell) = 0.60$~fb
(ggF process only).
(b)~The 95\% CL upper limits
on the cross-section of the $H\to Z_dZ_d\to 4\ell$ process,
assuming SM Higgs boson production via the ggF
process, with all final states combined.
Figures are taken from Ref.~\cite{HDBS-2018-55}.}
\label{fig:results:ex-higgs:fid_ZdZd}%
\end{center}
\end{figure}
Some limits from the $ZX$ process are shown in Figure~\ref{fig:results:ex-higgs:zx-limit}.
Figure~\ref{fig:results:ex-higgs:zx-limit}(a) shows the 95\% CL upper limit on the fiducial region cross-section.
Figure~\ref{fig:results:ex-higgs:zx-limit}(b) uses the HAHM for the model-dependent acceptances to display the 95\% CL upper limit
on the cross-section times branching fraction of the $H\rightarrow ZZ_d \rightarrow 4\ell$ process.
A slight excess with a local significance of 2.0$\sigma$ is found for a $Z_d$ mass hypothesis of 38~\GeV.
\begin{figure}[tb!]
\begin{center}
\subfloat[]{\includegraphics[width=0.45\textwidth]{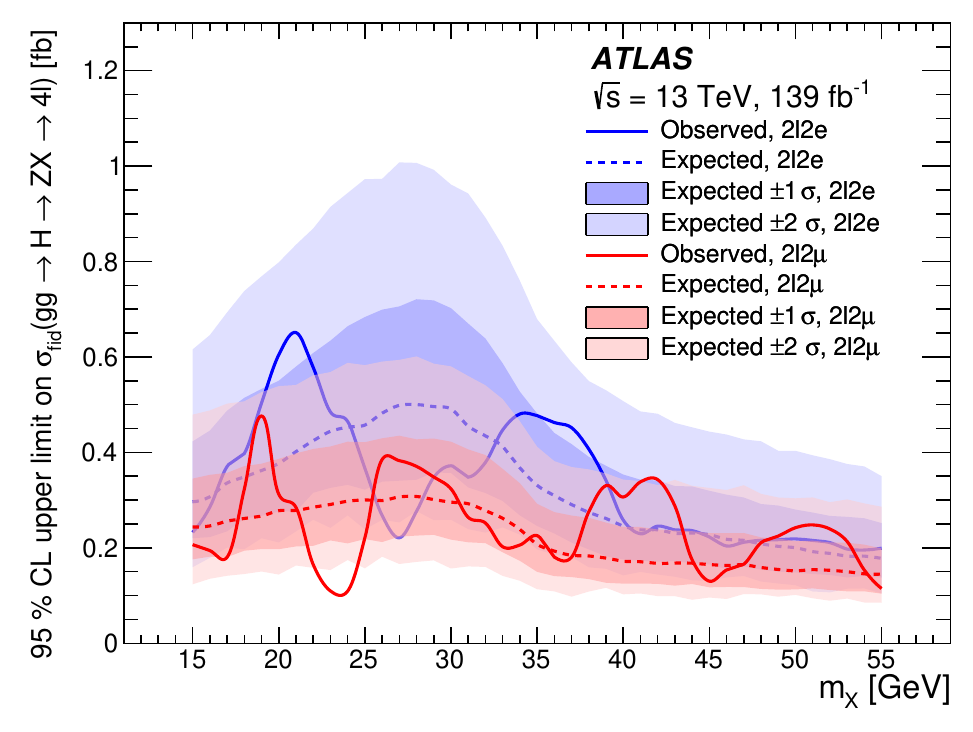}}%
\subfloat[]{\includegraphics[width=0.45\textwidth]{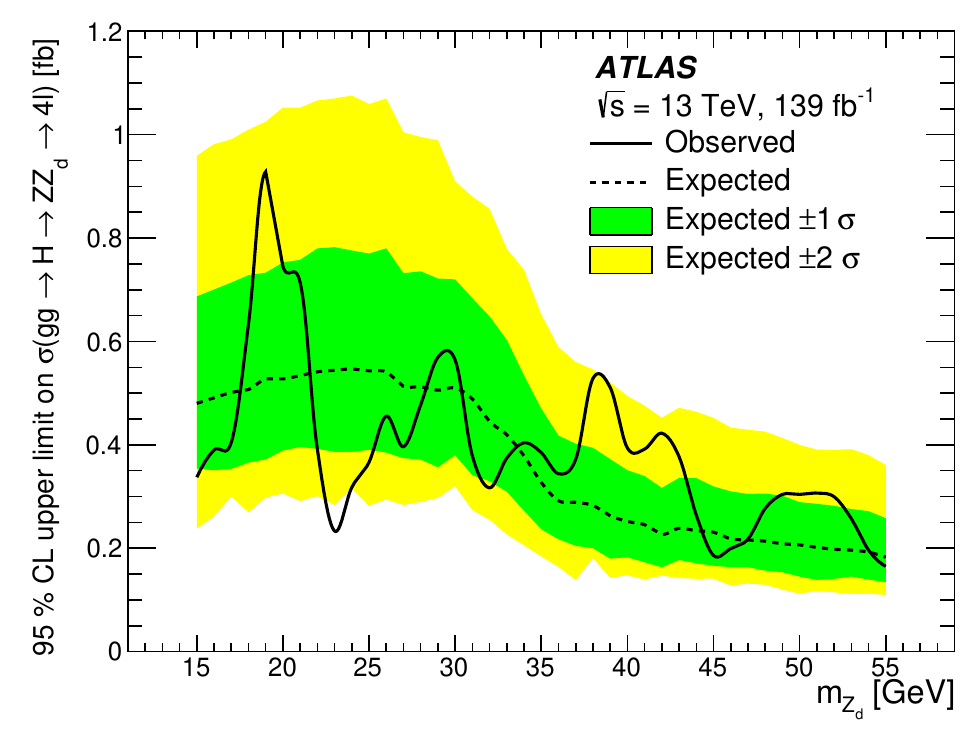}}%
\caption{$H\to ZX \to 4\ell$:
(a)~Per-channel 95\% CL upper limits on the fiducial cross-section
for the $H\rightarrow ZX \rightarrow 4\ell$
process.
(b)~Upper limits at 95\% CL
for the cross-section of the $H\to ZZ_d\to 4\ell$ processes,
assuming SM Higgs boson production via the ggF process.
All final states are combined.
Figures are taken from Ref.~\cite{HDBS-2018-55}.
}
\label{fig:results:ex-higgs:zx-limit}
\end{center}
\end{figure}
The same paper~\cite{HDBS-2018-55} also presents limits on total cross-sections and on the dark Higgs
boson mixing parameters.

The search \textbf{$H\to aa \to \bbbar\mumu$} for a light new pseudoscalar~\cite{HDBS-2021-03}
displays results with and
without the BDT classifiers for background rejection.
The purpose of the latter is to allow reinterpretation, and
also to accommodate the case where the $a$-boson might not be a pseudoscalar.
The results are shown in Figure~\ref{fig:results:ex-higgs:bbaa}.
While remaining statistically compatible with the SM, the largest excess of events above the
SM background is observed at a dimuon invariant mass of 52~\GeV
and corresponds to a local (global) significance of $3.3\sigma$ ($1.7\sigma$).
Otherwise, 95\% CL upper limits are placed
on the branching fraction of the Higgs boson to the $\bbbar\mumu$ final state,
$\mathcal{B}(H\to aa \to  \bbbar\mumu)$, and are in the range $(0.2{-}4.0) \times 10^{-4}$,
depending on the signal mass hypothesis.
Previous ATLAS results~\cite{HIGG-2016-03,CMS-HIG-18-011} are
improved on by a factor of $2{-}5$, for $m_a > 20$~\GeV,
while both results (with and
without the BDT) extend the search down to $m_a$ values of 16~\GeV.
\begin{figure}[tb!]
\begin{center}
\subfloat[]{\includegraphics[width=0.47\textwidth]{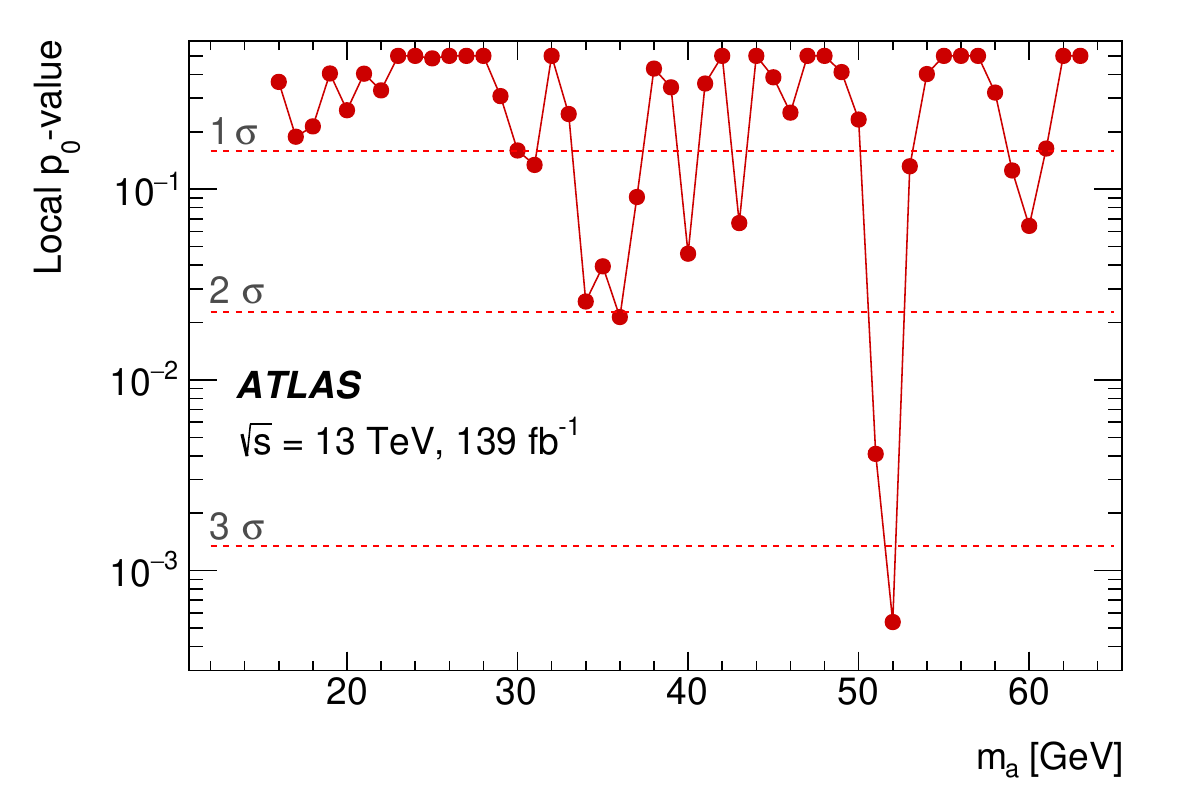}}%
\subfloat[]{\includegraphics[width=0.43\textwidth]{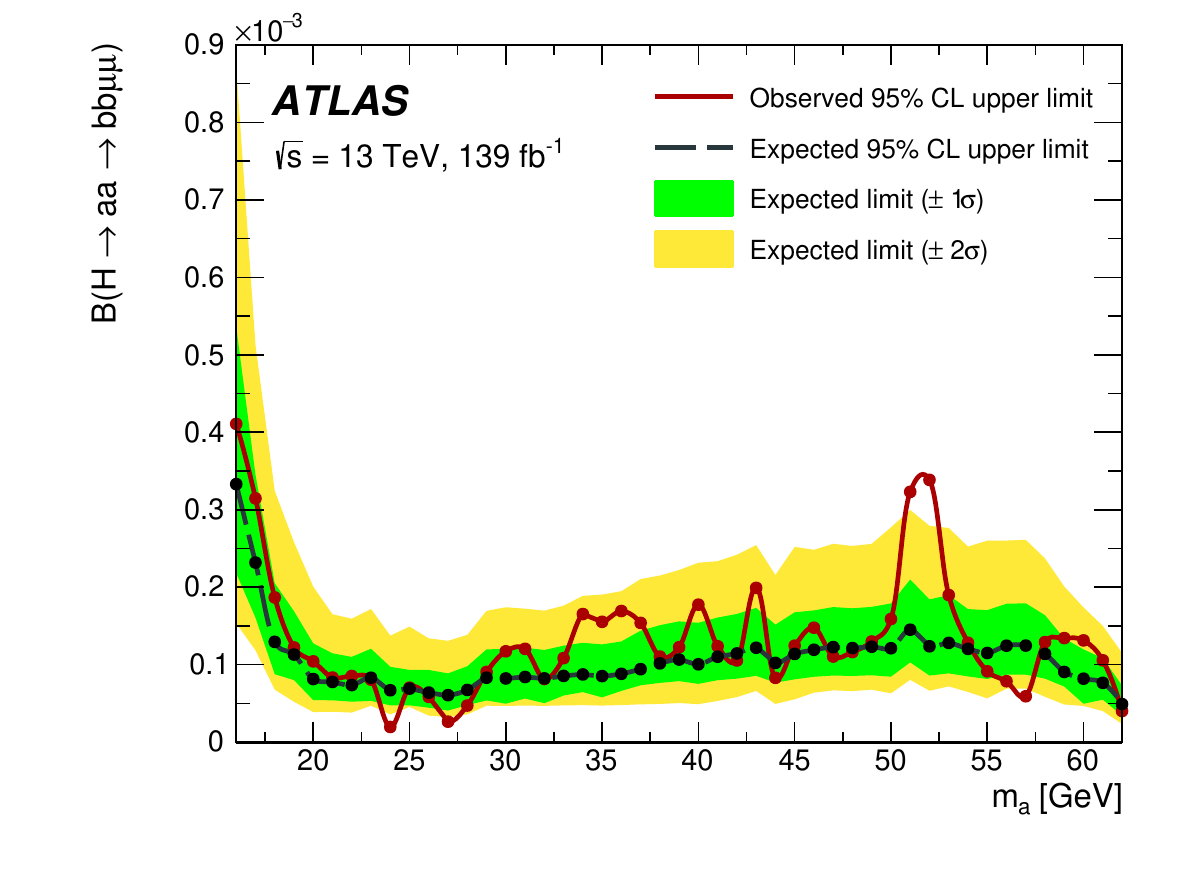}}%
\caption{$H\to aa \to \bbbar\mumu$:
(a)~The local $p_0$-values are also shown in standard deviations $\sigma$ and plotted as a
function of the signal
mass hypothesis. Between the points, the $p_0$-values are interpolated and
may not be fully representative of the actual sensitivity.
(b)~Upper limits on $\mathcal{B}(H\to aa \to  \bbbar\mumu)$ at 95\% CL,
including the BDT selection, as a
function of the signal mass hypothesis.
Black and red dots show masses for which the hypothesis testing was done.
Between these points, the limits are interpolated and may not be fully
representative of the actual sensitivity.
Figures are taken from Ref.~\cite{HDBS-2021-03}.
}
\label{fig:results:ex-higgs:bbaa}
\end{center}
\end{figure}

The search \textbf{$H\to aa \to (\bbbar)(\bbbar)$} for a new scalar boson via
Higgs boson decay~\cite{HDBS-2018-47}
yielded exclusion limits as shown in Figure~\ref{fig:results:ex-higgs:bbbb}.
No excess of data events consistent with $H\to aa \to (\bbbar)(\bbbar)$ is observed,
and 95\% CL upper limits on the production cross-section
$\sigma_{ZH}\mathcal{B}(H\to aa \to (\bbbar)(\bbbar))$ are obtained as a
function of the $a$-boson mass hypothesis.
This search explores a new extended low mass range of $15  \leq m_a \leq 30~\GeV$ by
using the di-$b$-quark tagger that is used in boosted topologies.
It improves the expected limit on $\sigma_{ZH}\mathcal{B}(H\to aa
\to (\bbbar)(\bbbar))$
for the mass hypothesis of $m_a = 20~\GeV$ by a factor of 2.5
relative to the previous search~\cite{HIGG-2017-05}.
\begin{figure}[tb!]
\begin{center}{\includegraphics[width=0.47\textwidth]{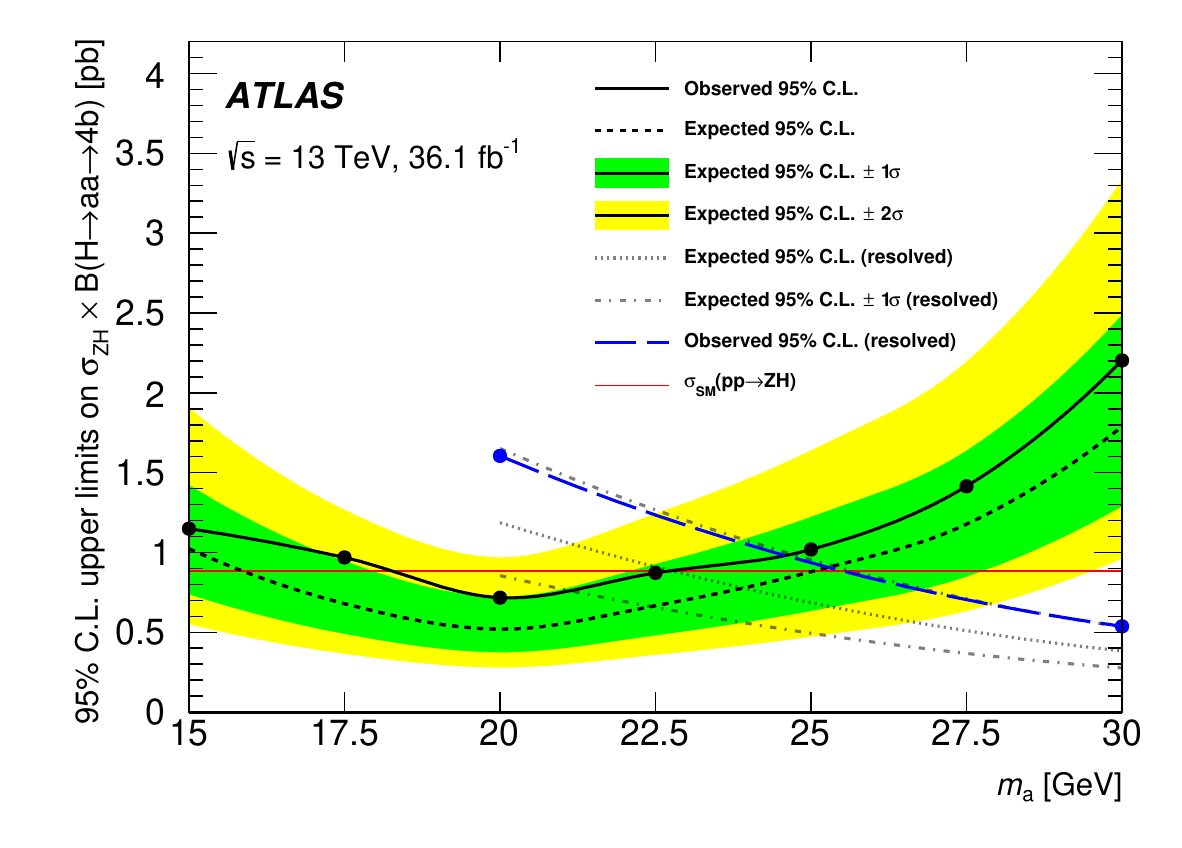}}%
\caption{$H\to aa \to (\bbbar)(\bbbar)$:
Upper limits on $\sigma_{ZH}\mathcal{B}(H\to aa \to (\bbbar)(\bbbar))$ at the 95\% CL
are shown. The current search explores a lower mass range by using a machine-learning-based
di-$b$-quark tagger designed for boosted topologies.
The previous higher-mass-range \enquote{resolved}
search~\cite{HIGG-2017-05} is also shown, along with the SM NNLO
cross-section of 0.88~pb for $pp \to ZH$.
The figure is taken from Ref.~\cite{HDBS-2018-47}.
}
\label{fig:results:ex-higgs:bbbb}
\end{center}
\end{figure}

This search \textbf{$H \to aa \to 4\gamma $}
did not find any significant excess over SM backgrounds~\cite{HDBS-2019-19}.
The largest deviation from the expected limit is $1.5\sigma$, which is observed in the
range of $10  \leq m_a \leq 25~\GeV$.
Upper limits at 95\% CL are set for $\mathcal{B}(H \to aa \to 4\gamma)$,
and range from $2\times 10^{-5}$ to $3\times 10^{-2}$, depending on $m_a$,
for the prompt axion-like particle search.
For the search for long-lived ALPs with significantly displaced decay vertices,
the 95\% CL upper limits range from
$2\times 10^{-5}$ to $6\times 10^{-5}$ for $10  \leq m_a \leq 62~\GeV$
and from $10^{-4}$ to $3 \times 10^{-2}$ for
$0.1  \leq m_a \leq 10~\GeV$.
The limits are summarized in the two-dimensional exclusion plot of
$C_{a\gamma\gamma}$ vs $m_a$ in Figure~\ref{fig:results:ex-higgs:aa4gam}.
These are the most stringent limits to date.
\begin{figure}[tb!]
\begin{center}{\includegraphics[width=0.6\textwidth]{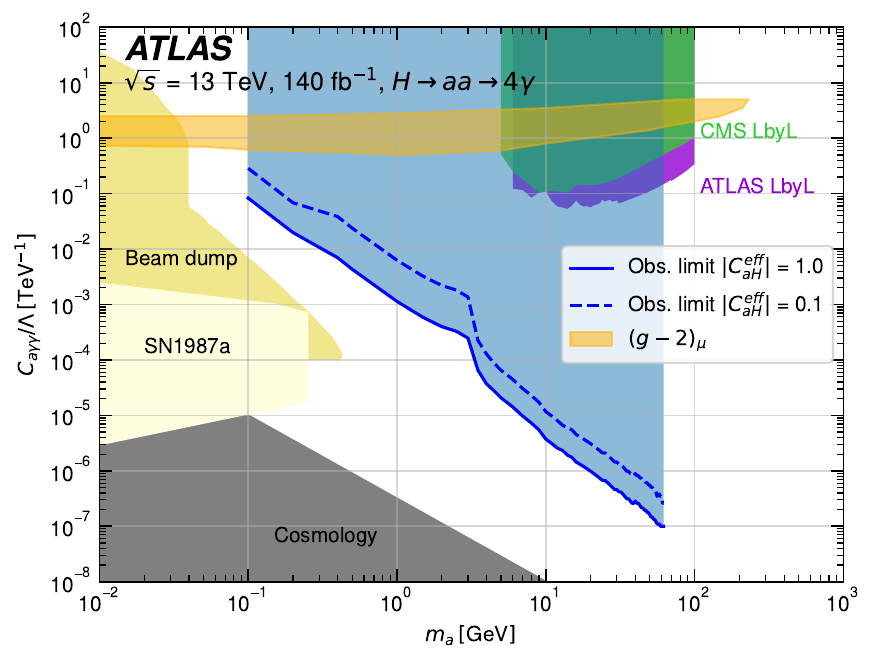}}%
\caption{$H \to aa \to 4\gamma $:
Limits on the ALP mass and coupling to photons at 95\% CL,
assuming $\mathcal{B}(a \to \gamma\gamma)$ = 1,
$\Lambda = 1$~\TeV with $|C^{\text{eff}}_{aH}| = 1$ (solid line)
and $|C^{\text{eff}}_{aH}| = 0.1$ (dashed line) as predicted in Ref.~\cite{Bauer2017}.
The shaded blue area represents the excluded region.
The nearly horizontal orange shaded area indicates the region favoured by an ALP
explanation for the $(g - 2)_\mu$ discrepancy~\cite{Bauer2017}.
Also shown are exclusion limits from the respective
ATLAS~\cite{HION-2019-08} and CMS~\cite{CMS-FSQ-16-012}
light-by-light (LbyL) scattering analyses, and beam dump experiments,
supernova SN1987a and cosmological observations adapted from
Ref.~\cite{JAECKEL2016482}.
The figure is taken from Ref.~\cite{HDBS-2019-19}.}
\label{fig:results:ex-higgs:aa4gam}
\end{center}
\end{figure}

The search \textbf{$H \to aa \to \gamma\gamma jj$}
found the data to be in agreement with the SM predictions~\cite{HIGG-2017-09}.
A  95\% CL upper limit is placed on the
production cross-section for $pp \to H$ times the branching fraction for
the decay $H \to aa \to \gamma\gamma jj$, normalized to the SM prediction, as shown in
Figure~\ref{fig:results:ex-higgs:ggjj}.
The upper limit ranges from 3.1~pb to 9.0~pb depending on $m_a$.
These results complement the previous upper limit on
$H \to aa \to \gamma\gamma\gamma\gamma$.
\begin{figure}[tb!]
\begin{center}{\includegraphics[width=0.6\textwidth]{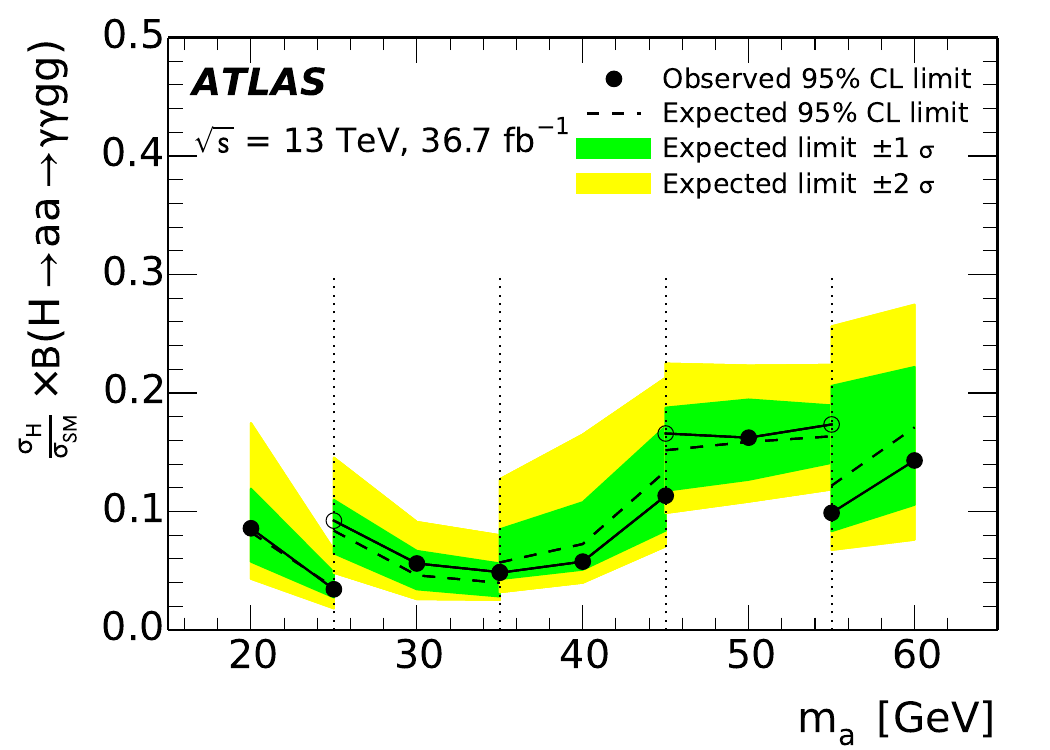}}%
\caption{$H \to aa \to \gamma\gamma jj$:
The observed (solid line) and expected (dashed line) 95\% CL
exclusion upper limits on the $pp \to H \to aa \to \gamma\gamma jj$
cross-section times branching fraction as a function of $m_a$,
normalized to the SM inclusive $pp \to H$ cross
section~\cite{deFlorian:2016spz}.
The vertical lines indicate the boundaries between the different
$m_{\gamma\gamma}$ analysis regimes.
At the boundaries, the $m_{\gamma\gamma}$ regime that yields the
better expected limit is used to provide the observed exclusion
limit (filled circles); the observed limit provided by the
regime that yields the poorer limit is also
indicated (empty circles).
The figure is taken from Ref.~\cite{HIGG-2017-09}.}
\label{fig:results:ex-higgs:ggjj}
\end{center}
\end{figure}

The search \textbf{$\Phi \to SS \to \text{LLP}$} for pair-produced neutral long-lived scalar
particles, $S$, did not find any significant excess of events in the signal region
relative to the data-driven background prediction~\cite{EXOT-2019-23}.
Upper limits at 95\% CL are set on the normalized cross-section times branching fraction
as a function of $c$ times the
long-lived particle mean proper lifetime $c\tau$.
These improve on the previous limits for mediator masses above or below 200~\GeV by a factor of
around 1.5--2 or 3--5, respectively.
An example where $(m_H,m_S)=(125,55)$ is shown in Figure~\ref{fig:results:ex-higgs:HSS}.
For models with a SM Higgs boson mediator, branching fractions to neutral scalars above 10\%
are excluded for $c\tau$ between approximately 20~mm and 10~m, depending on the model.
\begin{figure}[tb!]
\begin{center}{\includegraphics[width=0.7\textwidth]{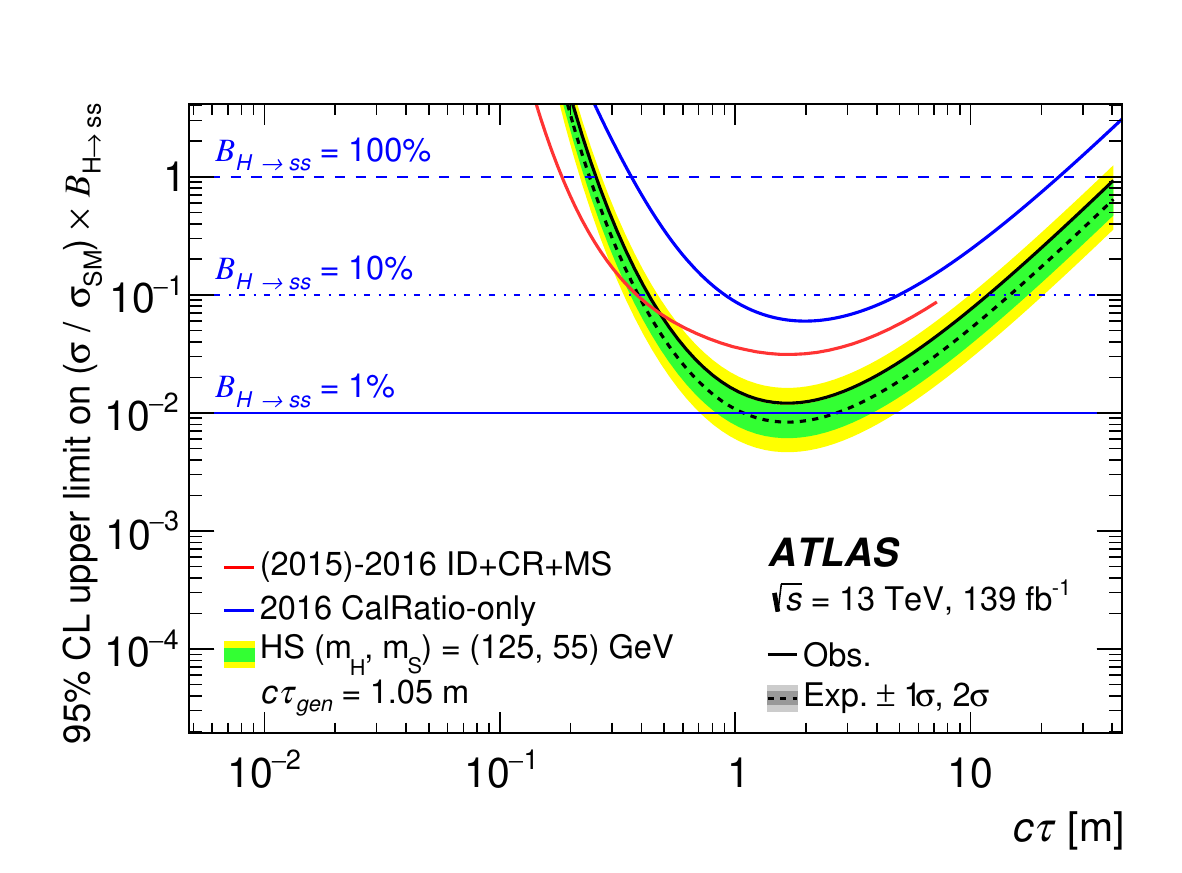}}%
\caption{ $\Phi \to SS \to \text{LLP}$:
95\% CL expected and observed limits on the branching fraction of SM Higgs bosons to
pairs of
neutral LLPs ($\mathcal{B}_{H \to SS}$), as well as a comparison with the results from
previous ATLAS searches~\cite{EXOT-2018-61, EXOT-2017-25}.
The figure is taken from Ref.~\cite{EXOT-2019-23}.}
\label{fig:results:ex-higgs:HSS}
\end{center}
\end{figure}
This search for the LLP, $S$, where the decay occurs in the ATLAS hadronic calorimeter is complemented by an additional
search aimed at longer lifetimes where the decay occurs in the muon
spectrometer~\cite{EXOT-2019-24}.
Other LLP searches from \RunTwo are reviewed in a companion report~\cite{EXOT-2023-14} in
this journal.
Overall, the lifetime range $c\tau$ has been covered from a scale of
mm to km for the decay of the Higgs boson $h_{125}$ into new scalars,
with a sensitivity comparable to $H\to\text{invisible}$.
The status of the \RunTwo LLP searches  is summarized in Figure~\ref{fig:results:ex-higgs:LLP_Summary}.
\begin{figure}[tb!]
\begin{center}{\includegraphics[width=0.9\textwidth]{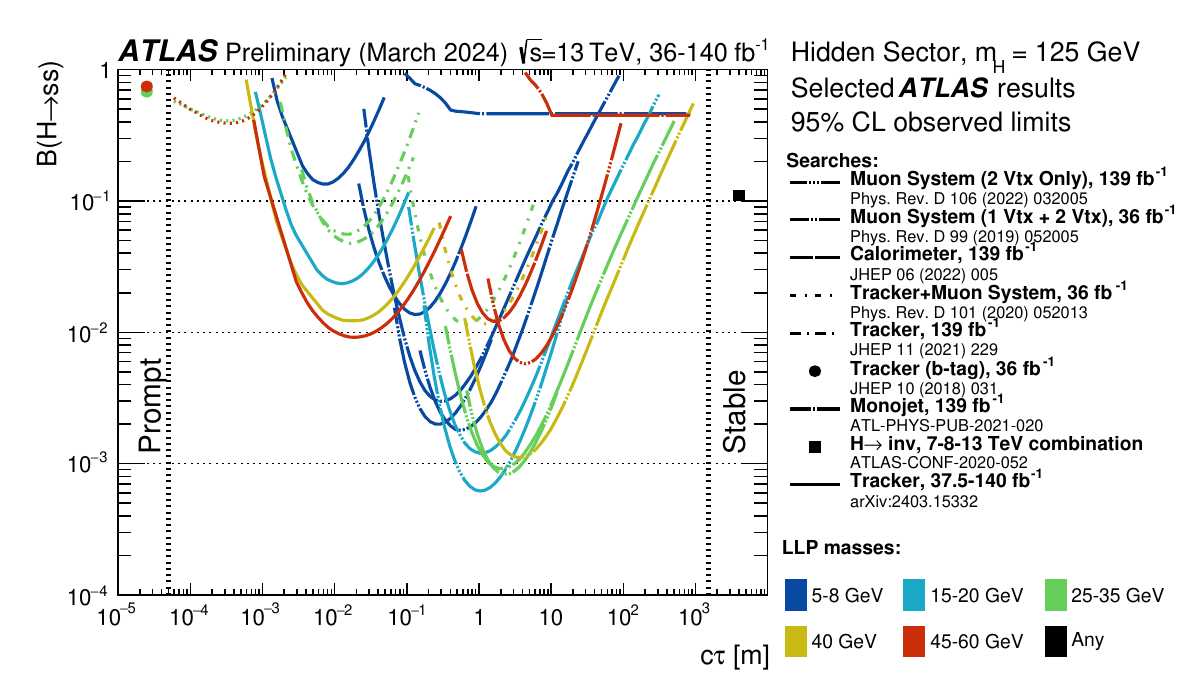}}%
\caption{ Regions in the Higgs branching fraction versus $c\tau$
plane excluded at 95\% CL, for a hidden-sector
model where a mediator Higgs boson of mass 125~\GeV
decays into a pair of long-lived neutral scalars ($s$).
The legend explains the coloured lines.
Also shown are exclusions for models where the neutral scalars are
prompt or detector-stable. The figure is taken from Ref.~\cite{ATL-PHYS-PUB-2024-003}.}
\label{fig:results:ex-higgs:LLP_Summary}
\end{center}
\end{figure}

The search \textbf{$H \to Za \to\ellell + \text{jet}$}  did not find an excess~\cite{HDBS-2018-37}.
Therefore, 95\% CL upper limits are set on $\sigma(pp\to H)\mathcal{B}(H \to Z(\eta_c\,/\,J/\Psi\,/\,a))$,
with observed values of 110~pb, 100~pb, and 17--340~pb for the $H \to Z \eta_c$, $H \to Z J/\Psi$,
and $H \to Za$ hypotheses, respectively.
The three-body mass distribution for $m_{\ell\ell\text{jet}}$ is shown in Figure~\ref{fig:results:ex-higgs:Za}
for data, the background prediction, and three signal hypotheses.
Assuming the SM prediction for inclusive Higgs boson production,
the limits on charmonium decay modes correspond to branching fraction limits in excess of 100\%.
This is the first direct limit on decays of the Higgs boson into light scalars
that decay into light quarks or gluons.
Because of the large value of $\mathcal{B}(a \to \text{hadrons})$ over the entire 2HDM(+S) parameter space,
these limits represent tight, direct constraints for low (high) $\tan\beta$
in the type-II and type-III (type-VI) 2HDM+S.
\begin{figure}[tb!]
\begin{center}{\includegraphics[width=0.6\textwidth]{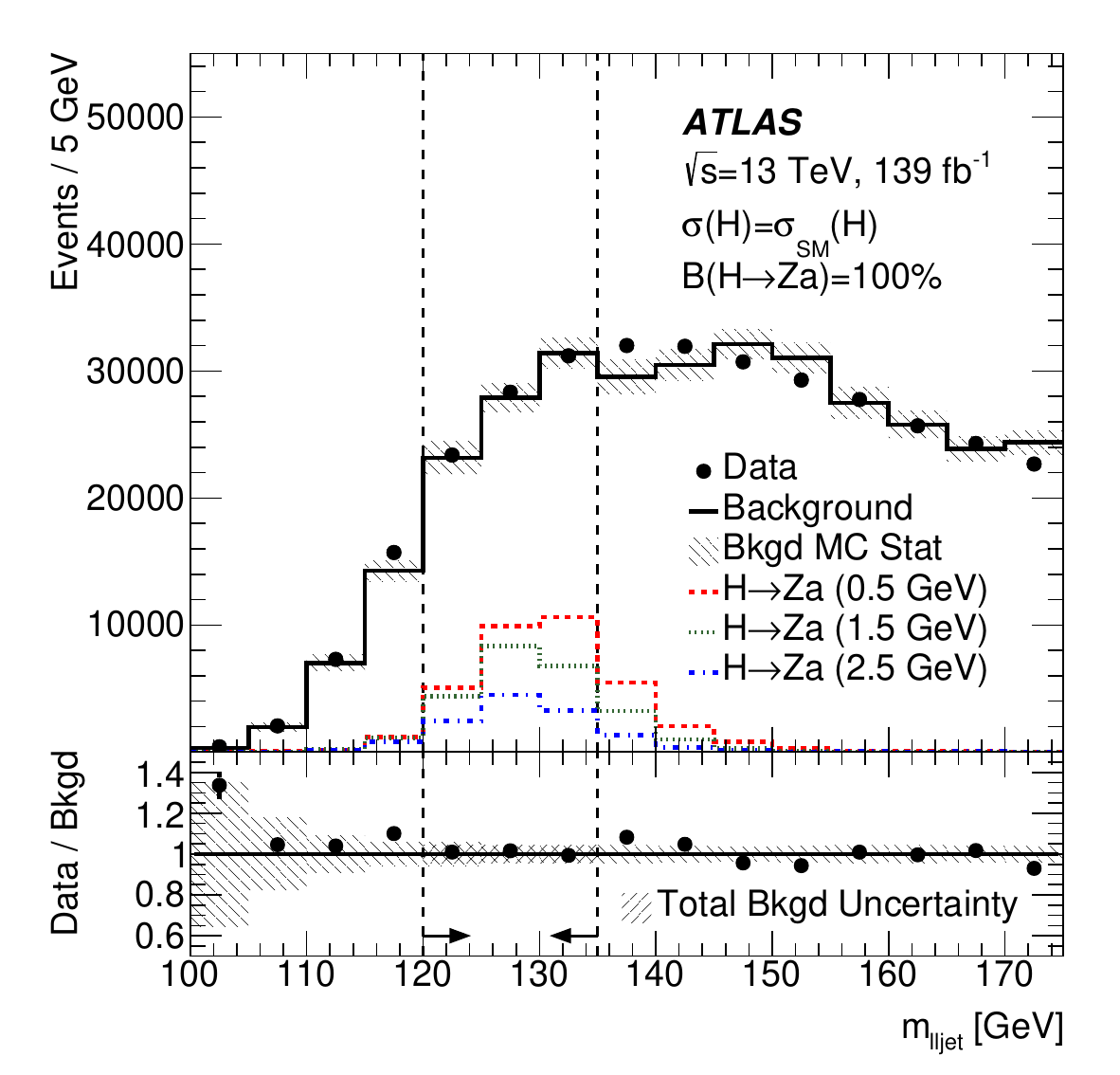}}%
\caption{$H \to Za \to\ellell + \text{jet}$:
Invariant mass of the lepton-pair\,+\,jet system for data, the predicted background, and three signal
hypotheses.
Events are required to pass the complete event selection,
including the multivariate jet selection requirement,
but not the requirement $120 \leq m_{\ell\ell\text{jet}} \leq 135$~\GeV for compatibility with the SM Higgs boson.
The background normalization is defined by the background estimate in the signal region,
and the signal normalizations assume the SM Higgs boson inclusive production cross-section
and $\mathcal{B}(H \to Za)$ = 100\%.
The error bars (hatched regions) represent the data (MC) sample statistical
uncertainty, in both the histograms and the ratio plots.
The region between the vertical dashed lines is the signal region.
The total background uncertainty in the signal region is also indicated.
Figures are taken from Ref.~\cite{HDBS-2018-37}.}
\label{fig:results:ex-higgs:Za}
\end{center}
\end{figure}

The \textbf{$H \to Za \to\ellell + \gamma\gamma$} analysis found no significant deviations from the
SM predictions~\cite{HDBS-2019-09}.
Upper limits are therefore set on the branching
fraction of the Higgs boson decay into $Za$ times the branching fraction
$a \to \gamma\gamma$, ranging from  0.08\% to 2\% depending on $m_a$,
as shown in Figure~\ref{fig:results:ex-higgs:Zagg}.
\begin{figure}[tb!]
\begin{center}
\subfloat[]{
\includegraphics[width=0.49\textwidth,valign=c]{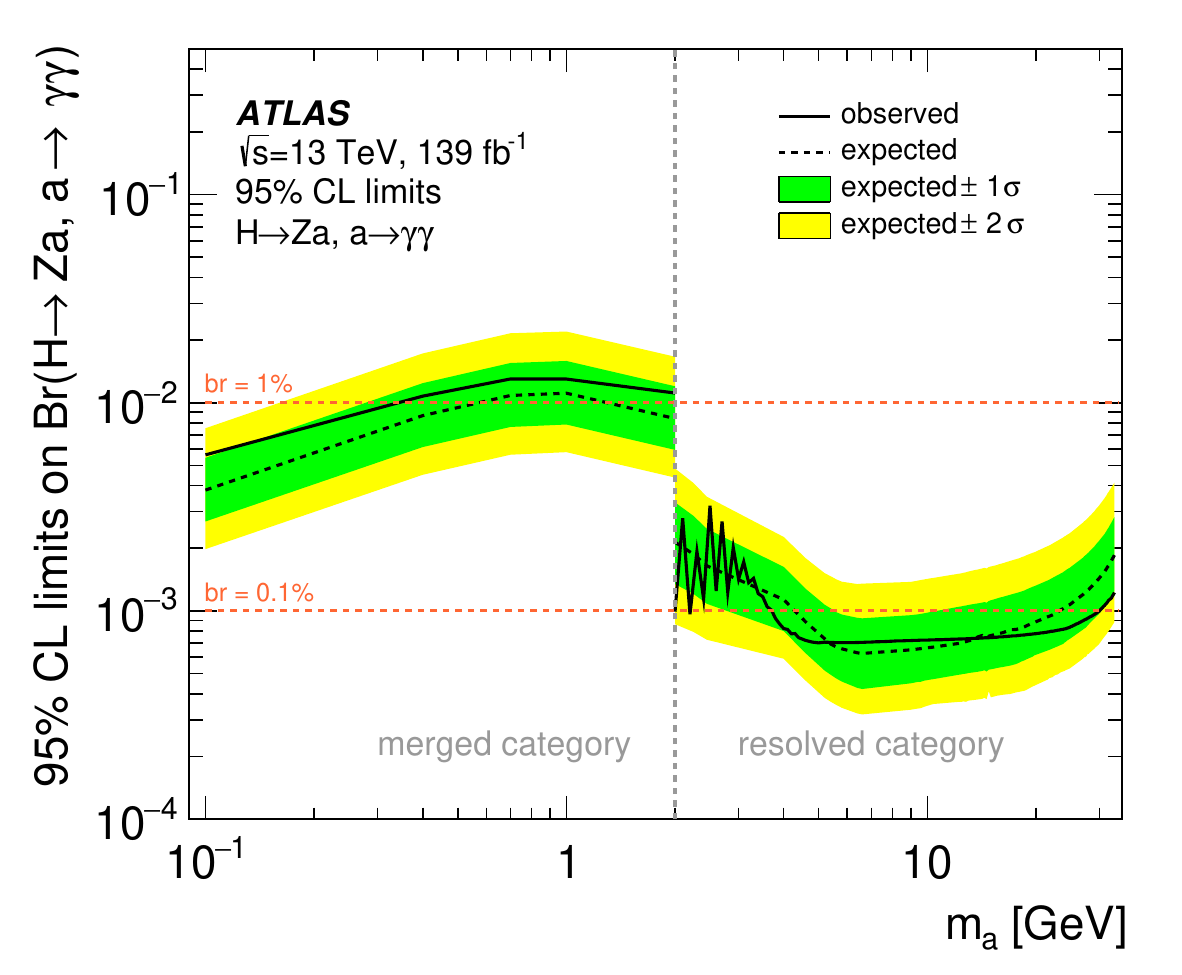}
}
\subfloat[]{
\includegraphics[width=0.49\textwidth,valign=c]{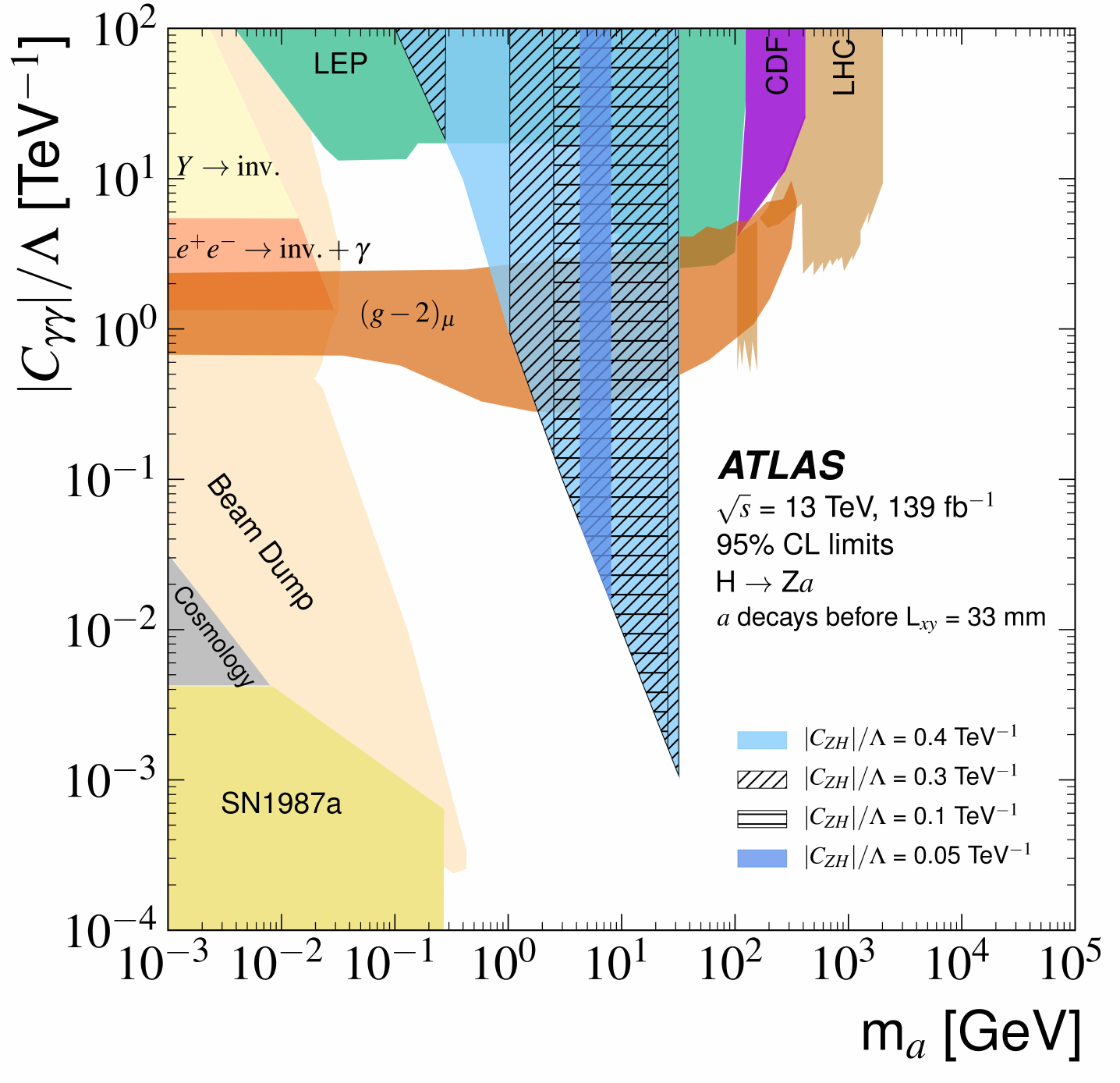}
}
\caption{$H \to Za \to\ellell + \gamma\gamma$:
(a) Expected and observed 95\% CL upper limits on the
branching fraction of the Higgs boson decay into $Za$
times the branching fraction $a \to \gamma\gamma$ as a function of the
$a$-boson mass in the merged ($m_a < 2$~\GeV)
and resolved
($m_a > 2$~\GeV) categories.
(b) ATLAS observed 95\% CL exclusion limit contours in terms
of the ALP's mass
and its effective coupling to photons,
$|C_{\gamma\gamma}|/\Lambda$, for different values of the Higgs
coupling to $Za$, $|C_{ZH}|/\Lambda$.
Limit contours from other direct experimental searches
are shown as well.
The collider bounds (LHC, LEP, CDF) are displayed at 95\% CL,
while the remaining bounds (SN1987a, Cosmology and Beam Dump)
are presented at 90\% CL. The red band shows
the preferred parameter space where the $(g-2)_\mu$ anomaly
can be explained at 95\% CL.
These contours are adapted from
Refs.~\cite{PhysRevLett.119.031802, JAECKEL2016482}.
Figures are taken from Ref.~\cite{HDBS-2019-09}.}
\label{fig:results:ex-higgs:Zagg}
\end{center}
\end{figure}

The \textbf{$H \to \chi_1\chi_2$} analysis searched for the exotic
Higgs boson decay into neutralinos (with $\tilde{\chi}_2^0 \to a\tilde{\chi}_1^0$   and  $a \to \bbbar$) after associated $ZH$ production~\cite{HDBS-2018-07}.
The signal is a $\bbbar$ resonance plus \met.
The search is novel in that it applied to the Peccei-Quinn
symmetry limit rather than the $R$-symmetry limit of the NMSSM.
The observations were consistent with SM.
Upper limits are therefore set on
the product of cross-section times branching fraction, using a three-dimensional
scan of the masses of the $\chi_1, \chi_2$ and $a$-boson, as shown in Figure~\ref{fig:results:ex-higgs:chi}.
These limits assume 100\%  branching fractions for the decays
$\tilde{\chi}_2^0 \to a\tilde{\chi}_1^0$   and  $a \to \bbbar$.
They represent the first $\tilde{\chi_1}\tilde{\chi_2}$
direct limits on this exotic Higgs boson decay obtained at the LHC.
\begin{figure}[tb!]
\begin{center}{\includegraphics[width=0.8\textwidth]{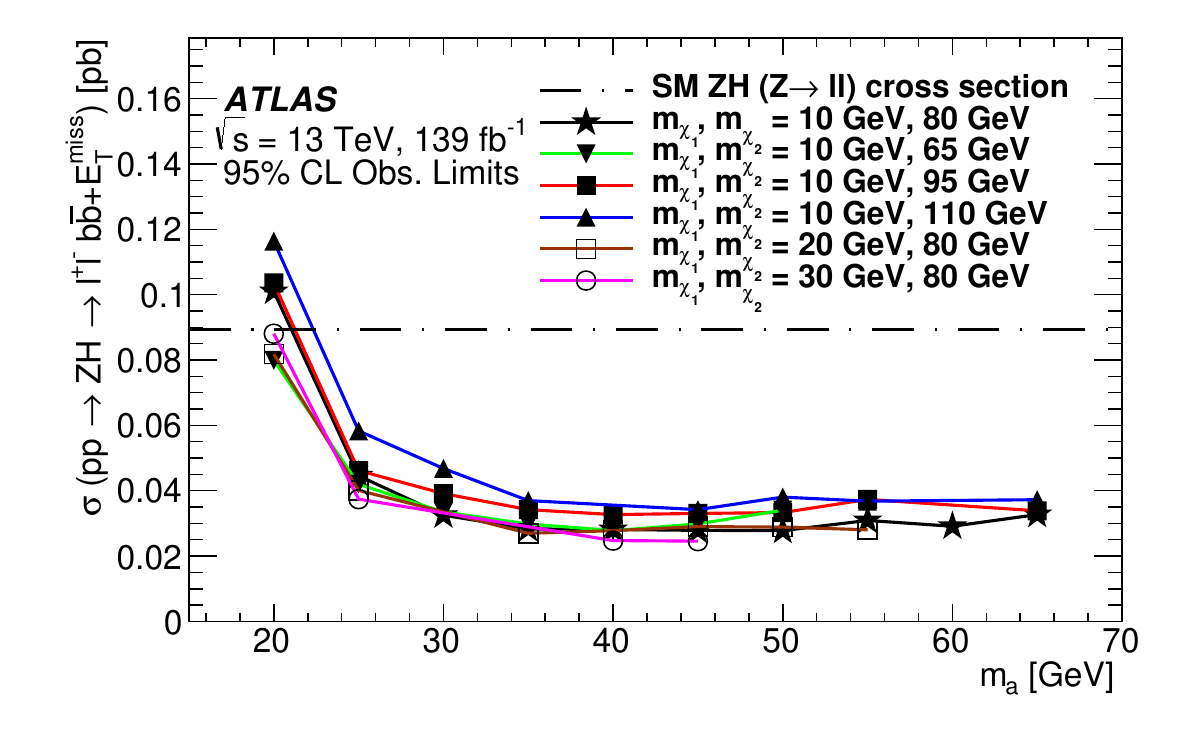}}%
\caption{$H \to \chi_1\chi_2$:
Upper limits at 95\% CL on the cross-section for $pp \to ZH$ times the branching fractions for
$Z \to \ell^+\ell^-$ (where $\ell = e,\mu,\tau$) and
$H \to \tilde{\chi}_1^0\tilde{\chi}_2^0 \to a\tilde{\chi}_1^0 \tilde{\chi}_1^0 \to \bbbar\tilde{\chi}_1^0 \tilde{\chi}_1^0$
as a function of $m_a$ for several values of $m_{\tilde{\chi}_1^0}$ and $m_{\tilde{\chi}_2^0}$
in the NMSSM scenario described in the text.
All branching fractions in the Higgs boson decay chain after the decay
$H \to \tilde{\chi}_1^0\tilde{\chi}_2^0$ are set to 100\%.
The different ranges in $m_a$ reflect differences in the allowed event kinematics.
The lines joining the $m_a$ points come from an assumed linear interpolation of the limits.
The SM value for the cross-section
$\sigma(pp \to ZH) \times \mathcal{B}(Z \to \ell^+\ell^-)$ is shown for reference.
The figure is taken from Ref.~\cite{HDBS-2018-07}.}
\label{fig:results:ex-higgs:chi}
\end{center}
\end{figure}

\subsubsection{Rare exclusive Higgs boson decays}
\label{sec:results:ex-higgs:rare}

The search for \textbf{$H\to ee/e\mu$},  which may reveal BSM enhancement of the $ee$ channel or
LFV in the $e\mu$ channel~\cite{HIGG-2018-58}, set exclusion limits on these processes.
Observed (expected) 95\% CL upper limits on the branching fractions,
$3.6 \times 10^{-4}~(3.5 \times 10^{-4})$
for $\mathcal{B}(H\to ee)$ and $6.2 \times 10^{-5}~(5.9 \times 10^{-5})$ for $\mathcal{B}(H\to e\mu)$,
are obtained for a Higgs boson with mass 125~\GeV.
These are the first such searches made by the ATLAS Collaboration and are
considerable improvements on previous measurements.

The direct searches \textbf{$H\to e\tau$} and \textbf{$H\to \mu\tau$} for LFV
in Higgs boson decays produced the
results detailed in Ref.~\cite{HIGG-2019-11}.
In particular, a small
excess was observed with respect to the SM background,
but below the threshold for evidence of a new
signal, when the two processes were treated independently.
When the two processes were treated
simultaneously, the excess was compatible with a branching fraction of zero within $2.1\sigma$.
Results of the fits are shown in Figure~\ref{fig:results:ex-higgs:H->etau}.
\begin{figure}[tb!]
\begin{center}
\subfloat[]{\includegraphics[width=0.47\textwidth]{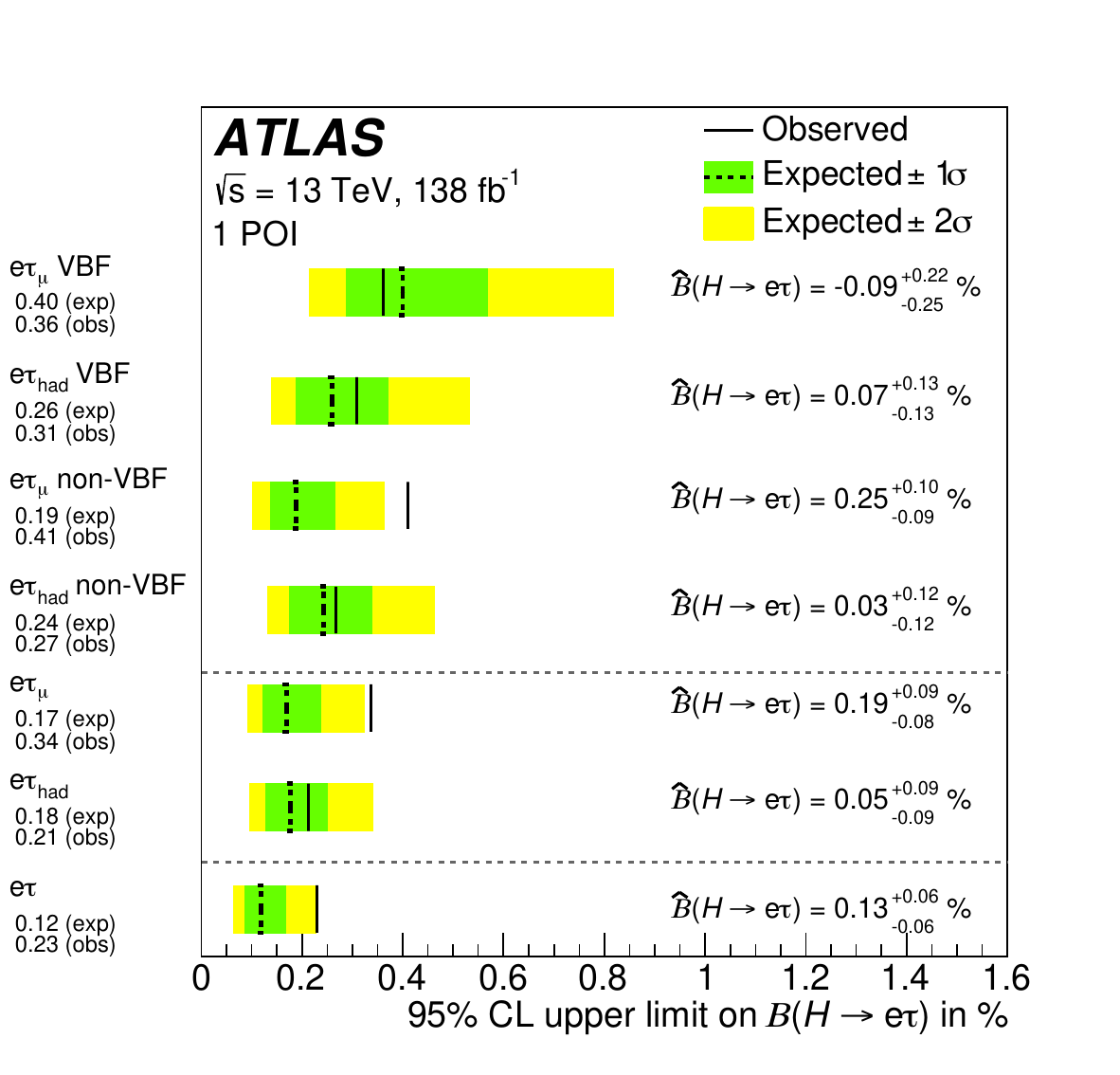}}%
\subfloat[]{\includegraphics[width=0.47\textwidth]{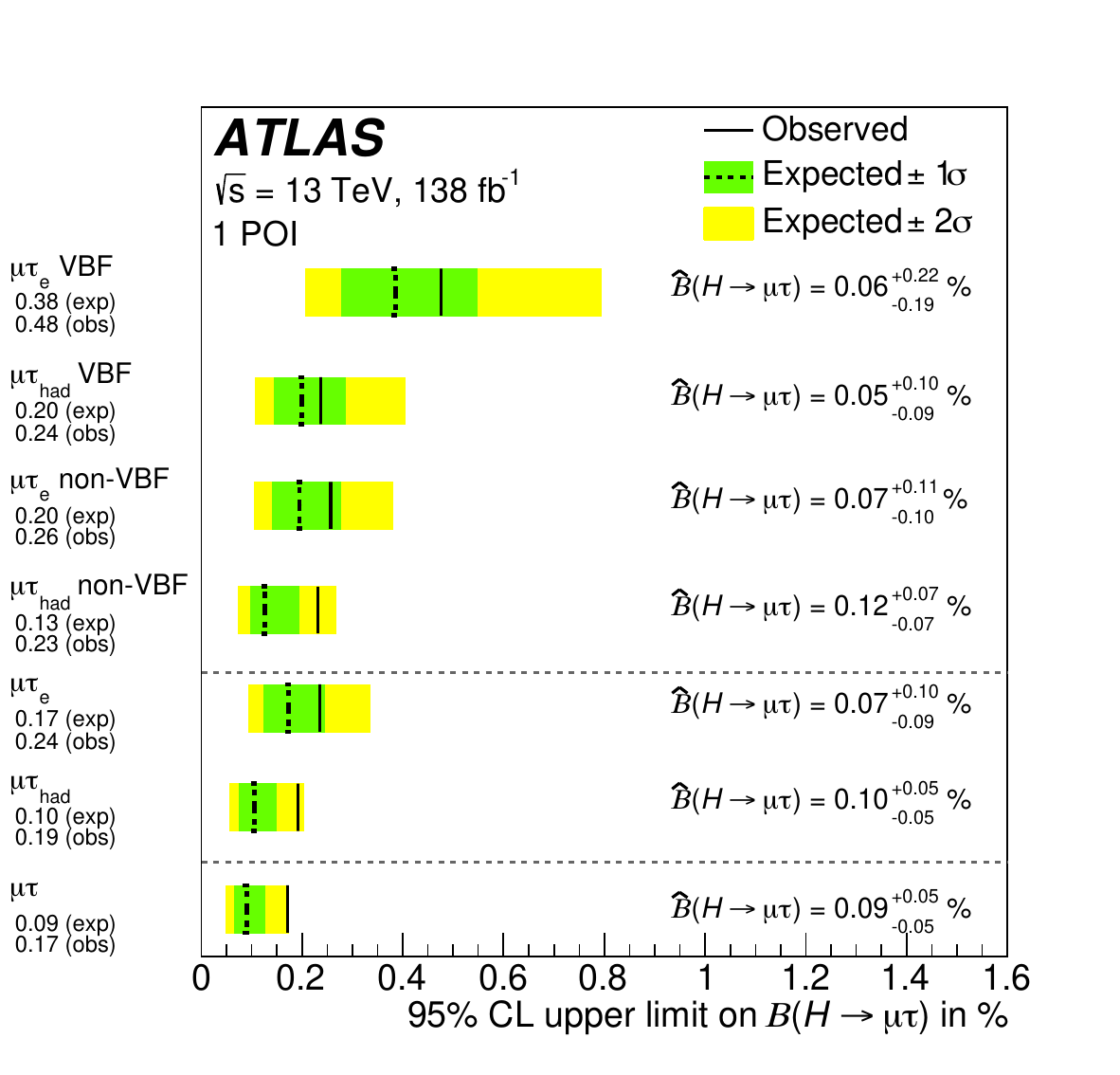}}%
\caption{$H\to e\tau$ and $H\to \mu\tau$:
Fit results of the independent searches, showing 95\% CL upper limits on the LFV branching
fractions of the Higgs boson, (a) $H\to  e\tau$ and (b) $H\to  \mu\tau$,
indicated by solid lines (observed results) or dashed lines (expected results).
Best-fit values of the branching fractions ($\mathcal{\hat{B}}$) are also provided, in \%.
The limits are computed while assuming that either
(a) $\mathcal{B}(H\to  \mu\tau)=0$ or (b) $\mathcal{B}(H\to  e\tau)=0$.
The channels and categories included in each
likelihood fit are shown on the $y$-axis, and the signal and control
regions from all other channels/categories are removed from the fit.
The results from stand-alone channel/category fits shown at the top are
compared with the results of the combined fit displayed in the last row.
Figures are taken from Ref.~\cite{HIGG-2019-11}.}
\label{fig:results:ex-higgs:H->etau}
\end{center}
\end{figure}

The searches  \textbf{$H/Z \to \omega\gamma$} and \textbf{$H \to K^*\gamma$} for these rare exclusive decays
study possible BSM Higgs boson couplings to
light quarks~\cite{HDBS-2019-33}.
They did not find any significant excess above the SM background.
Results for background-only fits performed in the signal region
are shown in Figure~\ref{fig:results:ex-higgs:omega}.
Exclusion limits were set on the branching fractions:
$\mathcal{B}(H \to \omega\gamma) < 1.5 \times 10^{-4}$ ($100\times$SM),
$\mathcal{B}(Z \to \omega\gamma) < 3.8 \times 10^{-7}$ ($17\times$SM),
and $\mathcal{B}(H \to K^*\gamma) < 8.9 \times 10^{-5}$ at 95\% CL.
The result for $Z \to \omega\gamma$ is a three-orders-of-magnitude
improvement over the limit previously set at DELPHI, while the $H \to \omega/K^* + \gamma$
results are new.
\begin{figure}[tb!]
\begin{center}
\subfloat[]{\includegraphics[width=0.47\textwidth]{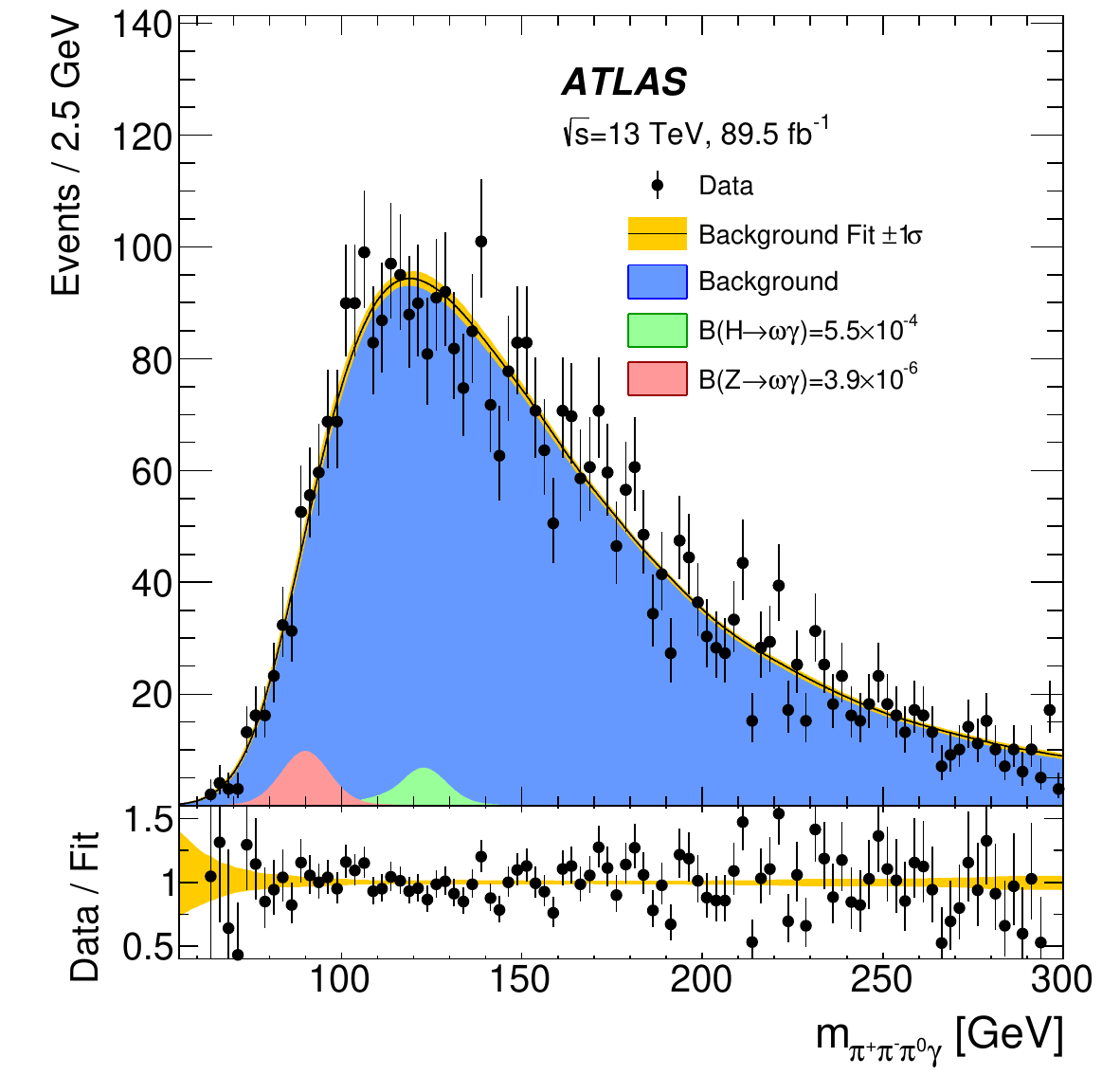}}%
\subfloat[]{\includegraphics[width=0.47\textwidth]{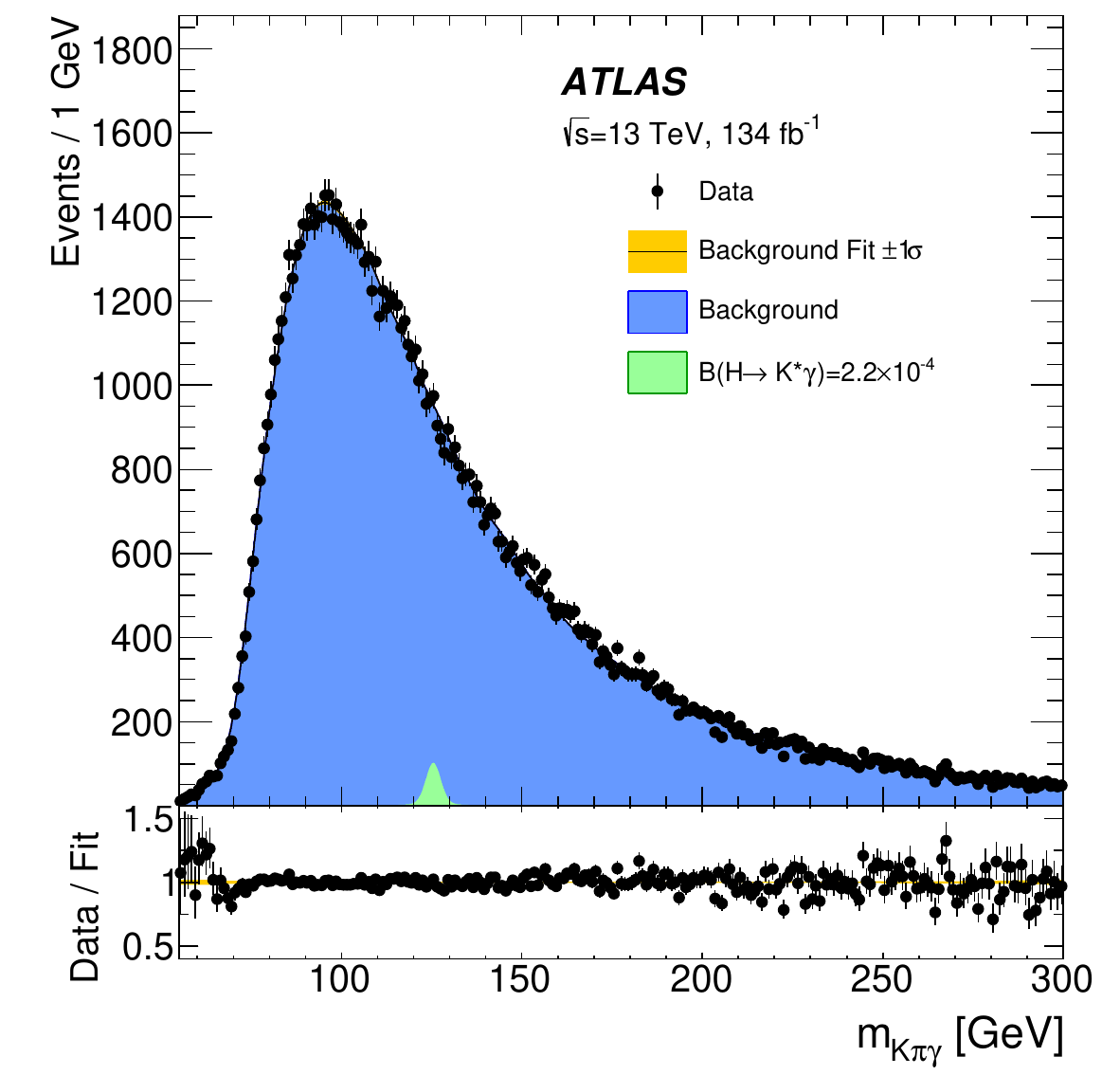}}%
\caption{$H/Z \to \omega\gamma$ and $H \to K^*\gamma$:
Background-only fits performed in the signal region for (a) $\omega\gamma$ and
(b) $K^*\gamma$ final states.
The branching fraction for each of the signals is set to the observed 95\% CL upper limit.
The yellow band represents the uncertainty in the fit arising from the
constrained background shape's systematic uncertainties.
Figures are taken from Ref.~\cite{HDBS-2019-33}.}
\label{fig:results:ex-higgs:omega}
\end{center}
\end{figure}

The search \textbf{$H \to (J/\Psi, \Psi(2S), \Upsilon(nS)) + \gamma $}
is for a Higgs boson decaying exclusively into a vector
quarkonium state and a photon, in the $\mu^+\mu^-\gamma$ final state~\cite{HDBS-2018-53}.
The data are compatible
with the background expectations.
The 95\% CL upper limits obtained for the $J/\Psi\,\gamma$ final state are
$\mathcal{B} (H \to J/\Psi\,\gamma) < 2.1 \times 10^{-4}$ and
$\mathcal{B} (Z \to J/\Psi\,\gamma) < 2.1 \times 10^{-6}$.
The corresponding upper limits for the
$\Psi(2S)\,\gamma$ final state are
$\mathcal{B} (H \to \Psi(2S)\,\gamma) < 10.9 \times 10^{-4}$ and
$\mathcal{B} (Z \to \Psi(2S)\,\gamma) < 2.3 \times 10^{-6}$.
The 95\% CL upper limits
$\mathcal{B} (H \to \Upsilon(nS)\,\gamma) < (2.6, 4.4, 3.5) \times 10^{-4}$ and
$\mathcal{B} (Z \to \Upsilon(nS)\,\gamma) < (1.0, 1.2, 2.3) \times 10^{-6}$
are set for the $\Upsilon(1S,2S,3S)\,\gamma$ final states.

The search \textbf{$H \to D^* + \gamma$} probed flavour-violating Higgs boson couplings
to light quarks~\cite{HDBS-2018-52}.
The results of the background-only fit are shown in Figure~\ref{fig:results:ex-higgs:D*},
where the signal distributions correspond to the extracted 95\% CL upper limit on the branching fraction.
The data are compatible with the expected background.
The observed 95\% CL upper limit is $\mathcal{B}(H \to D^* + \gamma) < 1.0 \times 10^{-3}$.
These results set the first limit on the decay $H \to D^* + \gamma$.
\begin{figure}[tb!]
\begin{center}{\includegraphics[width=0.47\textwidth]{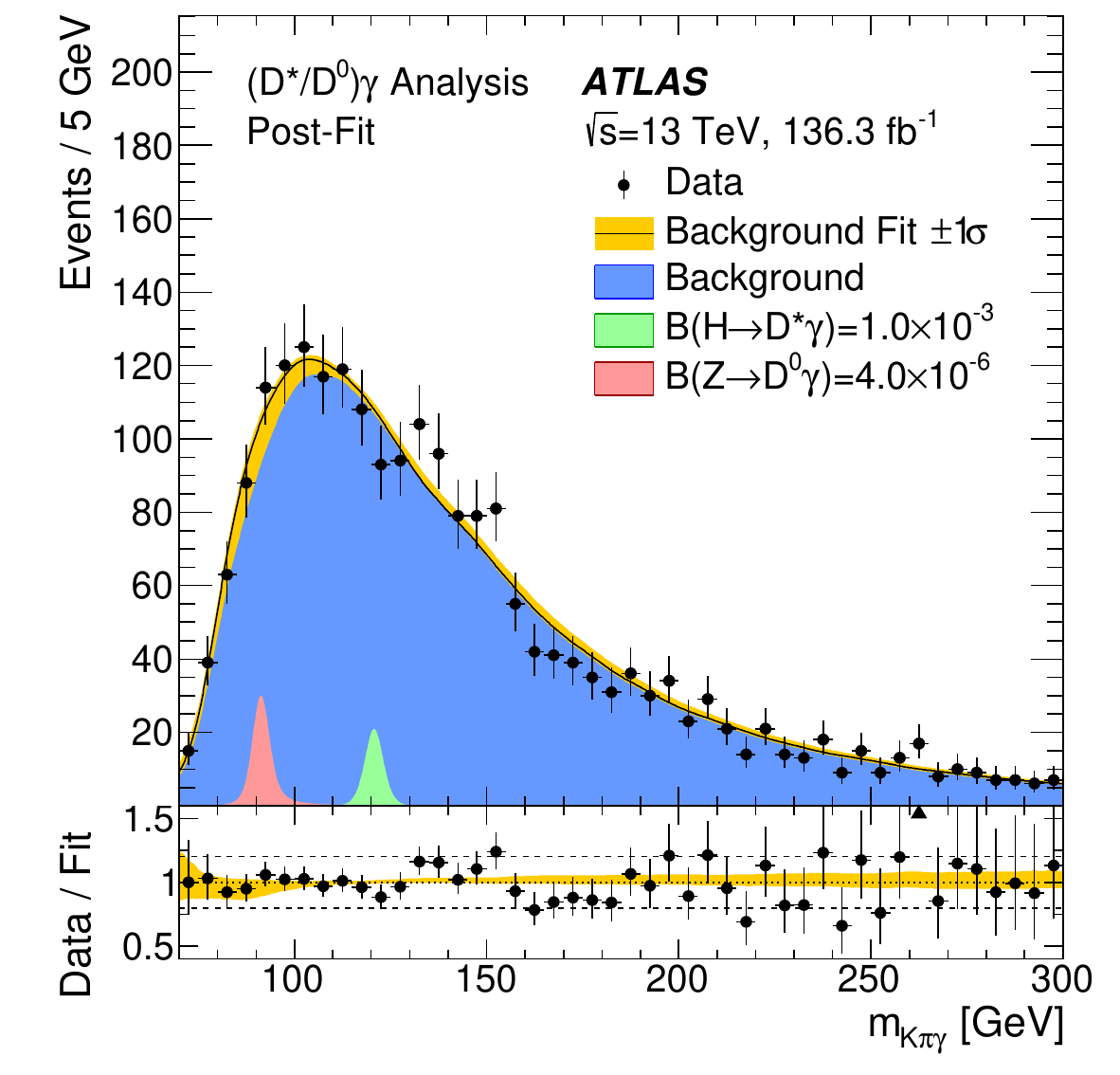}}%
\caption{$H \to D^* + \gamma$:
Comparison between data and the background prediction for the $m_{K\pi\gamma}$
distribution after the background-only fit
(\enquote{Post-Fit}) in the signal region for the $D^* + \gamma$ final state.
The unbinned background pdf is shown
with a yellow band that represents the uncertainty in the fit arising from the constrained background shape's
systematic uncertainties.
This uncertainty is largest in the region $m_{K\pi\gamma}< 100$~\GeV,
where the gradient of the distribution varies most.
The lower panel shows the ratio of the data to the background prediction.
The expected signal distributions are shown normalized to a branching fraction
corresponding to the observed 95\% CL upper limit. The results for the $Z$-boson decay are not discussed here.
The figure is taken from Ref.~\cite{HDBS-2018-52}.}
\label{fig:results:ex-higgs:D*}
\end{center}
\end{figure}

\FloatBarrier


\section{Discussion and outlook}
\label{sec:discussion}

To summarize this report, a list of the small excesses found in \RunTwo data relative to
the SM predictions is presented in Table~\ref{tab:discussions:summary}.
The aim is to be comprehensive and reflect the findings of each paper.
As yet, there is no statistically significant excess.
Despite the huge number of searches that ATLAS has conducted in \RunTwo data, several signatures remain uncovered and are topics for future investigations. Some of these signatures are listed and briefly discussed. Furthermore, limitations of some of the \RunTwo searches are discussed, and are often the sources of leading systematic uncertainties. Interesting analysis techniques that were used in \RunTwo are also reviewed, emphasizing which experimental challenges they are able to overcome. %



\begin{table}[h!]
\begin{center}
\caption{This table summarizes the small excesses recorded per channel, the mass hypotheses for which they occur, and the local and global significances. The integrated luminosity $L$ and the reference to each analysis are also given. In some cases, the global significance was not computed (indicated as n.a.).}
\label{tab:discussions:summary}
\resizebox{\textwidth}{!}{
\begin{tabular}{l|c|c|c|c|c|c}
\multirow{2}{*}{\textbf{Decay channel}} &	 \textbf{Production}	&	\multirow{2}{*}{\textbf{Mass [\GeV{}]}} &	\textbf{Significance} &	\textbf{Significance}	& \multirow{2}{*}{\textbf{$L$ [$\mathrm{fb}^{-1}$]}	}	&	\multirow{2}{*}{\textbf{Ref.}} \\
& \textbf{mode}					&																					& \textbf{local}				& \textbf{global}				& 																										&	\\
\hline
$H\to\tau\tau$ &											$b$-associated &		400 &											2.7$\sigma$	&		n.a. & 139 & 					\cite{HDBS-2018-46} \\
$H\to\tau\tau$ &											ggF &								400 &											2.2$\sigma$	&		n.a. & 139 & 					\cite{HDBS-2018-46} \\
$H\to\mu\mu$ &												$b$-associated &		480 &											2.3$\sigma$	&		0.6$\sigma$ & ~~36 & 	\cite{HIGG-2017-10} \\
$H\to t\bar{t}$ &											ggF &								800 &											2.3$\sigma$	&		n.a. &				140 & 	\cite{EXOT-2020-25} \\
$H\to t\bar{t}/t\bar{q}$ &						qq and qg &					900 &											2.8$\sigma$	&		n.a. &				139 & 	\cite{HDBS-2020-03} \\
$H\to ZZ\to 4\ell /2\ell 2\nu$ &			ggF &								240 &											2.0$\sigma$	&		0.5$\sigma$ & 139 & 	\cite{HIGG-2018-09} \\
$H\to ZZ\to 4\ell /2\ell 2\nu$ &			VBF &								620 &											2.4$\sigma$	&		0.9$\sigma$ & 139 & 	\cite{HIGG-2018-09} \\
$H\to \gamma\gamma$ &									ggF &								684 &											3.3$\sigma$	&		1.3$\sigma$ & 139 & 	\cite{HIGG-2018-27} \\
$H\to \gamma\gamma$ &									ggF &								~~~~~95.4 &								1.7$\sigma$	&		n.a. &				140 & 	\cite{HIGG-2023-12} \\
$H\to Z(\ell\ell)\gamma$ &						ggF &								420 &											2.3$\sigma$	&		n.a. &				140 & 	\cite{HIGG-2018-44} \\
$H\to Z(q\bar{q})\gamma$ &						ggF &								3640~~ &									2.5$\sigma$	&		n.a. &				139 & 	\cite{HDBS-2019-10} \\
$A\to Zh_{125}(b\bar{b})$ &						ggF &								500 &											2.1$\sigma$	&		1.1$\sigma$ & 139 & 	\cite{HDBS-2020-19} \\
$A\to Zh_{125}(b\bar{b})$ &						$b$-associated &		500 &											1.6$\sigma$	&		n.a. &				139 & 	\cite{HDBS-2020-19} \\
$A\to ZH\to \ell\ell b\bar{b}$ &			ggF	&								610 ($A$), 290 ($H$) &		3.1$\sigma$	&		1.3$\sigma$ & 139 & 	\cite{HDBS-2018-13} \\
$A\to ZH\to \ell\ell b\bar{b}$ &			$b$-associated &		440 ($A$), 220 ($H$) &		3.1$\sigma$	&		1.3$\sigma$ & 139 & 	\cite{HDBS-2018-13} \\
$A\to ZH\to \ell\ell WW$ &						ggF	&								440 ($A$), 310 ($H$) &		2.9$\sigma$	&		0.8$\sigma$ & 139 & 	\cite{HDBS-2018-13} \\
$A\to ZH\to \ell\ell t\bar{t}$ &			ggF	&								650 ($A$), 450 ($H$) &		2.9$\sigma$ &		2.4$\sigma$ & 140 & 	\cite{HDBS-2021-02} \\
$A\to ZH\to Zh_{125}(b\bar{b})h_{125}(b\bar{b})$ &	VH &	420 ($A$), 320 ($H$) &		3.8$\sigma$ &		2.8$\sigma$ & 139 & 	\cite{HDBS-2019-31} \\
$H^+\to cb$ &													$t\bar{t}$ decay &	130 &											3.0$\sigma$ &		2.5$\sigma$ & 139 & 	\cite{HDBS-2019-24} \\
$H^+\to Wa(\mu\mu )$ &								$t\bar{t}$ decay &	120--160 ($H^+$), 27 ($a$) &	2.4$\sigma$ &	n.a. &			139 & 	\cite{HDBS-2020-12} \\
$H^+\to WZ$ &													VBF &								375 &											2.8$\sigma$ &		1.6$\sigma$ & 139 & 	\cite{HDBS-2018-19} \\
$H^{++}\to WW$ &											VBF &								450 &											3.2$\sigma$ &		2.5$\sigma$ & 139 & 	\cite{STDM-2018-32} \\
$H\to h_{125}h_{125}\to 4b$ &					ggF &								1100~~ &									2.3$\sigma$ &		0.4$\sigma$ &	126--139 & 	\cite{HDBS-2018-41}~~ \\
$H\to h_{125}h_{125}\to 4b$ &					VBF &								550 &											1.5$\sigma$ &		n.a. &				126 & 	\cite{HDBS-2018-18}~~ \\
$H\to h_{125}h_{125}\to b\bar{b}\tau\tau$ &	ggF &					1000~~ &									3.1$\sigma$ &		2.0$\sigma$ & 139 & 	\cite{HDBS-2018-40}~~ \\
$H\to h_{125}h_{125} \, \text{combination}$ &	ggF &				1100~~ &									3.3$\sigma$ &		2.1$\sigma$ & 126--139 & 	\cite{HDBS-2023-17}~~ \\
$X\to Sh_{125}\to \bbbar\gamma\gamma$ &	ggF &							575 ($X$), 200 ($S$) &		3.5$\sigma$ &		2.0$\sigma$ &	140 & 	\cite{HDBS-2021-17}~~ \\
$h_{125}\to Z_dZ_d \to 4\ell$	&				ggF &								~~28 &										2.5$\sigma$ &		n.a. &				139 & 	\cite{HDBS-2018-55}~~ \\
$h_{125}\to ZZ_d \to 4\ell$ &					ggF &								~~39 &										2.0$\sigma$ &		n.a. &				139 & 	\cite{HDBS-2018-55}~~ \\
$h_{125}\to aa\to b\bar{b}\mu\mu$ &		ggF, VBF, VH &			~~52 &										3.3$\sigma$ &		1.7$\sigma$ & 139 & 	\cite{HDBS-2021-03}~~ \\
$h_{125}\to aa\to 4\gamma$ &					ggF &								10--25~~~~ &							1.5$\sigma$ &		n.a. &				140 & 	\cite{HDBS-2019-19}~~ \\
$h_{125}\to e\tau$ and $h_{125}\to \mu\tau$ &		ggF, VBF, VH &	125~~ &							2.1$\sigma$ &		n.a. &				138 & 	\cite{HIGG-2019-11}~~

\end{tabular}
}
\end{center}
\end{table}

\subsection{Summary of excesses}
\label{sec:discussions:summary}

Table~\ref{tab:discussions:summary} lists the small excesses seen in \RunTwo data relative to SM expectations, expressed as a local and/or global significance, $p_0$. The table contains an entry for each paper reviewed in this report when that information is available, but cases with a local significance below 1.5$\sigma$ are not mentioned. The common statistical procedure used is the CL$_\text{s}$ modified frequentist method~\cite{Read:2002hq}. An excess with $p_0 \ge 3\sigma$ is sufficiently significant to provide \enquote{evidence} of a signal. An excess with $p_0 \ge 5\sigma$ is by convention considered a \enquote{discovery}. None of these observations are significant enough to establish new physics. Awareness of these results is useful, as they can motivate future searches with the \RunThr dataset and help set priorities.


\subsection{Uncovered signatures}
\label{sec:discussion:uncoveredsignatures}

Several potential signal topologies for additional Higgs bosons or exotic decays of the 125~\GeV Higgs boson have not yet been explored (or only explored insufficiently) in ATLAS searches, despite being well motivated by phenomenological considerations. A non-exhaustive list of such signatures, separated into signatures for high or low mass searches, is discussed in the following.

\subsubsection{High mass searches}
\label{sec:discussion:uncoveredsignatures:highmass}

\begin{itemize}

\item $H^+\to Wh_{125}$: In the 2HDM, unless $\cos\left(\alpha-\beta\right)$ is very close to zero, the decay rate of the heavy charged Higgs boson with a mass of at least 200~\GeV into $Wh_{125}$ can become very large. The charged Higgs boson would be produced in association with $t$ and $b$, and the $h_{125}$ would decay, like the SM Higgs boson, mainly into $b\bar{b}$. The final state would be characterized by a large number of jets and $b$-jets, and the leptonic decay of the $W$ (either from the production or decay) could be used for triggering. Other decay modes of the $h_{125}$ are also worth exploring, as they might have a smaller signal rate, but a cleaner signature.\\
In the MSSM the $H^+\to Wh_{125}$ channel is less relevant, except for $H^+$ masses around 200~\GeV and for low values of $\tan\beta$~\cite{Moretti_2000}.

\item $H^+\to WH$: If there is a difference between the masses of the heavy CP-even $H$ and the $H^+$, then the decay $H^+\to WH$ may open up. Since this channel does not require the SM-like Higgs boson to couple to the BSM Higgs sector, this decay rate can be large even in the case of alignment ($\cos\left(\alpha-\beta\right)\approx 0$). If $H$ is heavier than twice the top-quark mass, then the decay into $t\bar{t}$ will be dominant, leading to a busy final state with many jets and $b$-jets. Other decay modes are also possible, depending on the specifics of the model.

\item $H\to WH^+$: This decay is essentially the reverse of the one discussed before. In this case, the heavy Higgs boson $H$ (or $A$) would decay into $WH^+$, if kinematically allowed. This decay would compete mostly with $H\to t\bar{t}$ if $H$ is heavy enough. The aligned 2HDM~\cite{Enomoto_2022}, or the Gildener--Weinberg 2HDM in the alignment limit~\cite{Eichten_2023}, strongly motivate such a search. The dominant $H^+$ decay mode is $H^+ \to tb$ for $H^+$ masses above 200~\GeV.

\item $H^+\to W\gamma$: In the case of a fermiophobic $H^+$, as predicted for example in the GM model for the five-plet, the decay of the charged Higgs boson to vector bosons becomes dominant. For large $H^+$ masses, above 180~\GeV, the decay into $WZ$ is explored in ATLAS searches, but for lighter $H^+$, where $WZ$ is kinematically not allowed, the decay into $W\gamma$ instead becomes important and therefore should be investigated too.

\item $H\to SS$: The heavy CP-even Higgs boson $H$ can also decay into two lighter scalars $S$, where $S$ is not the 125~\GeV Higgs boson. Relevant models for such a signature are those that predict two scalars with different masses and in addition a SM-like Higgs boson. In particular, in the case of alignment, with vanishing $H{-}h_{125}$ couplings, such decays can be more important than decays into two 125~\GeV bosons. ATLAS investigated this for the final state of $SS\to 4W$~\cite{HIGG-2016-24} with a partial \RunTwo dataset, but further decay modes of the $S$ should be explored.

\item $H\to\chi\chi$ and $H^+\to\chi\chi^+$: The decay of heavy neutral or charged Higgs bosons into SUSY particles can become relevant if the SUSY mass scale is low enough for light charginos and neutralinos to be predicted. In the MSSM for instance, the coupling of the heavy Higgs bosons to SUSY particles becomes large in the region around $\tan\beta\sim 6$ (also called the \emph{wedge} region), which is difficult to constrain otherwise.

\item $H^{++}$: Searches for doubly-charged Higgs bosons are also still undercovered. In particular, searches for light $H^{++}$ particles with a mass below 200~\GeV, with subsequent decay of the $H^{++}$ into same-sign $W$-boson pairs, have not yet been explored at the LHC. A promising phenomenolopical study can be found at Ref.~\cite{Ashanujjaman_2023} for a mass range of 84 -- 200~\GeV. Another possible search opportunity~\cite{Ashanujjaman_2023_2} is the decay of the $H^{++}$ to a $W$-boson and a charged Higgs boson $H^+$, with $H^+\to WH$, which is less constrained than the decays to $WW$ or $\ell\ell$.

\end{itemize}

\subsubsection{Low mass searches}
\label{sec:uncoveredsignatures:lowmass}

\begin{itemize}

\item Decays to axion-like particles involving higher-dimension operators: Reference~\cite{Biek_tter_2022} investigated rare multi-body decays of the 125~\GeV Higgs boson in the context of effective field theories, involving four- or five-point interactions of the Higgs boson with ALPs and other particles. The decays that were suggested are $h_{125}\to a\mu\mu$, $h_{125}\to aa\mu\mu$ and $h_{125}\to aaqq$ (with $a\to\gamma\gamma$ or $a\to\mu\mu$) for $a$-boson masses between 10~\GeV and 50~\GeV, and the results of this phenomenological study showed that good sensitivity could be achieved with the current dataset. Such signatures have not yet been explored experimentally.

\item (Semi-)invisible decays: Particles that traverse the detector without any interaction (for example dark matter, very long-lived particles, gravitinos, stable SUSY particles, or neutrinos) can only be indirectly inferred via the presence of missing transverse momentum. The Higgs boson can decay fully or partly into such invisible particles, e.g.\ via $h_{125}\to\chi_1\chi_2$ with $\chi_2\to a\chi_1$ ($\chi_1$ is the lightest neutralino), which was explored by ATLAS in a specific final state~\cite{EXOT-2018-57}. Another possible signature is $h_{125}\to Za\to 2\ell+\met$, studied in Ref.~\cite{Aguilar_Saavedra_2022}. These kinds of decays are challenging, in part because the decay cannot be fully reconstructed, but should be explored in more detail and also in other final states.

\item Exploring other production modes: So far, the experimental study of exotic Higgs boson decays and direct production of light scalars has focused on production via gluon--gluon fusion. The ggF mode of 125~\GeV Higgs boson production has the largest cross-section, but other production modes can help to distinguish potential signal from background through their distinctive signatures, such as the presence of forward jets in the VBF mode, or leptons (from Higgs boson production in association with top quarks or vector bosons) that can be exploited for triggering and event selection. One analysis that has explored this already is the search for $t\bar{t}a$ with $a\to\mu\mu$~\cite{HDBS-2020-12}. Another, still unexplored, possibility is the production of the unknown (pseudo)scalar $a$ in association with a $W$ boson and a photon, which was proposed in Ref.~\cite{Brivio_2017}. Such signatures require dedicated event selections to optimize the sensitivity.

\item Intermediate mass region $60~\GeV<m_a<125~\GeV$: The decay of the 125~\GeV Higgs boson into $aa$ opens up for $a$ masses below 60~\GeV, but off-shell effects may still contribute for heavier $a$-bosons~\cite{Gon_alves_2018}, which haven't been explored experimentally yet. The production of additional Higgs bosons (for example via ggF) in this mass range is also possible, but direct searches for additional Higgs bosons typically focus on the mass range above 125~\GeV because large SM background contributions (such as from decays of $W$ and $Z$ bosons, or \ttbar) complicate these lower mass searches.

\item Long-lived particles (LLPs): LLPs are exotic particles that do not decay promptly but travel into the detector before they eventually decay. LLPs can be produced in exotic Higgs boson decays for example. The lifetime is an unknown parameter, and if it is large enough the LLP may even exit the detector before it decays. If it decays before exiting, unusual detector signals may occur, such as calorimeter energy clusters without corresponding tracks (if the particle decays in the calorimeter), or tracks appearing in the muon spectrometer without corresponding signals elsewhere. In general, these searches are difficult and require dedicated techniques depending on the LLP type and lifetime. This is still a relatively new and promising area with many opportunities for future efforts that extend the existing searches for LLPs.

\end{itemize}


\subsection{Limitations of current searches, and interesting new techniques}
\label{sec:discussions:limitations}

In order to improve future analyses, it can be helpful to investigate the limiting factors in current searches. New analysis techniques are being developed to improve the searches for BSM Higgs bosons. Some of these limitations and promising new techniques -- without claiming completeness -- are discussed below.

\subsubsection{Data sample size and trigger}

The amount of data is the most important factor in a very large number of searches performed by ATLAS. The cross-sections for BSM Higgs boson production are expected to be very small compared to those for typical SM processes. For example, the total cross-section for $pp\to t\bar{t}$ at 13~\TeV is 834~pb~\cite{Czakon:2011xx} and the total SM Higgs boson production cross-section is 55~pb~\cite{deFlorian:2016spz}, whereas the cross-section for $H^+\to tb$ with a $H^+$ mass of 800~\GeV and $\tan\beta=1$ in the hMSSM is only 0.1~pb (which also marks the current sensitivity limit of the dedicated search~\cite{HDBS-2018-51}). Collecting huge amounts of data is therefore paramount in establishing the existence of such processes with sufficient statistical significance (in simplified terms a discovery is established when the measured signal is five times the size of the background uncertainty).

The total dataset from \RunTwo that is deemed suitable for physics analyses (by satisfying a list of quality criteria~\cite{DAPR-2018-01}) comprises 140~\ifb, although the exact integrated luminosity used in each analysis varies slightly due to the trigger choices or previous (less precise) luminosity determination methods. The dataset recorded in \RunThr by the end of 2023 amounted to 65~\ifb, and another 190~\ifb are expected to be recorded by the end of 2025.

While the amount of $pp$ collision data is determined by the operation of the LHC, the sample size available for a specific analysis depends strongly on the trigger. The trigger is a trade-off between signal efficiency and background rejection, but the main limitation is in the resources available for long-term data storage. The LHC bunch-crossing rate is 40~MHz, which is reduced to a recording rate of only 1~kHz by the trigger system, which accepts a variety of interesting events. This rate is distributed among approximately 1500 trigger chains in \RunTwo~\cite{TRIG-2019-04}. Specialized triggers may help to isolate signatures of BSM physics, e.g.\ by selecting multiple high-\pt objects. Some physics triggers also make increasing use of machine-learning (ML) algorithms, for example in online flavour tagging. Some \RunTwo analyses have used $b$-jet triggers, for example in the search for $H\to b\bar{b}$~\cite{HIGG-2016-32} or $h_{125}h_{125}\to 4b$~\cite{HDBS-2018-41}. This concept could be expanded to include other selection criteria, tailored to enhance the efficiency for rare processes.

A special form of analysis designed to overcome some of these limitations is called trigger-level analysis (TLA). The trigger uses a selective readout and then stores only a subset of the typical event information. The partial event recording allows a larger number of events to be stored, which means searches can be expanded into regions that would otherwise be rejected, especially regions of low \pt. However, special reconstruction algorithms and calibrations are required. The TLA technique was used, for example, in the ATLAS search for low-mass dijet resonances~\cite{EXOT-2016-20} and increased future usage is expected.

\subsubsection{Monte Carlo sample size as a leading systematic uncertainty}

MC simulations are a crucial ingredient in physics analyses looking for BSM signals. They are used to model the signal, and they are also needed to estimate backgrounds when data-driven techniques are not applicable (in particular, non-reducible backgrounds are difficult to select without a large contamination from signal events). For background modelling, huge MC event samples are needed. For analyses that rely on the spurious-signal technique~\cite{ATL-PHYS-PUB-2020-028}, e.g.\ $H\to\gamma\gamma$, $H\to Z\gamma$ and $(H\to )h_{125}h_{125}\to bb\gamma\gamma$, MC sample size is a pivotal parameter. The largest systematic uncertainty in these \RunTwo analyses is the spurious signal. The goal of this technique is to quantify how well an analytic function can model a background distribution shape when fitting a signal-plus-background model (where the background component is described by this function) to a background-only dataset (often a MC dataset). If the background-only dataset has large statistical fluctuations, the spurious signal will be dominated by these fluctuations rather than the actual modelling uncertainty. In \RunTwo, the ratio of the total number of simulated MC events to the number of recorded data events was approximately 1:1. Maintaining a similar ratio in the future is a huge challenge and requires increased usage of fast simulation, currently provided by \textsc{AtlFast3}~\cite{SIMU-2018-04}. Increasing the size of MC samples for spurious-signal studies to be much larger than the corresponding data samples requires even faster simulation techniques.

\subsubsection{Constraining systematic uncertainties using data}

Using data to decrease systematic uncertainties is a practice often used in searches for BSM Higgs bosons. Simulated MC data often does not model the real data perfectly, so correction factors need to be derived, or uncertainties are assigned to cover the differences, for example in terms of object reconstruction and selection efficiencies for signal and background processes, or background normalizations. Instead, real data may be used directly.

There are various ways to use the data for this purpose. Most commonly, the background can be extrapolated from a signal-depleted region to a signal-enhanced region. The signal-depleted control regions are typically designed to enhance the contribution of a specific background component, while still being sufficiently similar to the signal region. This can be achieved, for example, by reversing a selection cut or loosening the $b$-tagging criteria. Corrections are often needed because the control region does not match the signal region perfectly in some way. For example, object kinematics may need to be reweighted, or systematic uncertainties may need to be introduced to account for any differences between the true background and the extrapolated background. Typical approaches when using data to estimate the background are the fake-factor method and the matrix method~\cite{EGAM-2019-01}. ATLAS uses sophisticated methods to identify particle types and their charges, yet even a small probability of misidentifying objects can lead to large backgrounds that can be difficult to estimate from simulation. Therefore, data is used to measure this probability with better precision. This is used, for example, in the search for $A/H\to\tau^+\tau^-$~\cite{HDBS-2018-46}.

Another technique is the use of fits that are performed simultaneously on events in signal-enriched and signal-depleted categories. The nuisance parameters or scale factors connected to uncertainties in this background are correlated across categories. This means the fit will be able to use the information in the control region to constrain the nuisance parameters or normalization factors (that are connected to systematic uncertainties) in all regions, particularly the signal region. This technique is used in many analyses, but it has especially large benefits for searches that involve multiple top quarks and contributions from \ttbar+\,\qqbar (where $q$ is a heavy-flavour quark such as $b$ or $c$), for example in $H^+\to tb$~\cite{HDBS-2018-51} or $ttH\to t\bar{t}t\bar{t}$~\cite{EXOT-2019-26}.

\subsubsection{Machine learning to estimate or reject the background and improve the analysis sensitivity}

The usage of multivariate analysis (MVA) techniques to separate potential signal from background is well established in high-energy physics. These tools, such as boosted decision trees or different types of neural networks, are able to handle correlations between a large number of variables and are therefore far superior to optimizations done by hand on one variable at a time. Recently, parameterized MVAs were used successfully in searches for BSM Higgs bosons, for example in the search for $H\to h_{125}h_{125}\to bb\tau\tau$~\cite{HDBS-2018-40}. The MVA output scores are either used in defining cuts and/or categories, or used directly as final discriminants.

Neural networks can also be used for data-driven background estimation. In $H\to h_{125}h_{125}\to 4b$~\cite{HDBS-2018-41}, the background is dominated by multijet events, which are difficult to model with MC samples and lead to very large uncertainties. Therefore, the data is selected from a signal-depleted region and then extrapolated into the signal region. Differences affecting various kinematic variables in a correlated way cannot be avoided, and neural networks are used to reweight the data in the control region to closely resemble that in the signal region. Remaining non-closure uncertainties need to be evaluated carefully, and these are expected to become more important as the amount of data increases and statistical uncertainties decrease.

Anomaly detection techniques are expected to play a larger role in the future, as they enable searches for generic signals without specifying features of those signals, for example their mass or width. Larger phase-space regions can be covered in this way. Anomaly detection with weak supervision was used, for example, to search for dijet resonances~\cite{HDBS-2018-59}, and a fully unsupervised technique was applied in the generic search for $Y\to Xh_{125}$~\cite{HDBS-2019-23}.

Also interesting are procedures that make use of several different ML techniques in series. An example of this was applied in the ATLAS search for $h_{125}\to Za$ with $a\to\text{jet}$~\cite{HDBS-2018-37}, where a regression multilayer perceptron (MLP) was used to estimate the mass of the hypothesized $a$-boson and its output score fed into a MLP discriminant to improve the background suppression.

Another innovative approach uses ML in transformations of physics variables with a principal component analysis (PCA), as was done in the search for $H\to b\bar{b}$~\cite{HIGG-2016-32}. In that search the $b\bar{b}$ mass resolution is improved after a PCA transformation, which essentially corresponds to a reduction of correlations between the mass and other quantities. This then leads to stronger limits on the production of a heavy Higgs boson.

\subsubsection{Merged objects in boosted topologies}

If the decay products of new resonances are very light relative to the resonance mass, then they are boosted, and the decay products are emitted almost collinearly. For example, two jets from the decay of a boosted resonance can be so close that they are reconstructed as one jet with a larger radius. Other objects can merge similarly. These signatures are difficult to reconstruct and often require dedicated algorithms and calibrations. New taggers target these merged objects, exploiting low-level detector information such as that from overlapping energy deposits in the calorimeter. Special techniques for boosted particles were developed, for example, in the search for boosted $H\to h_{125}h_{125}\to bb\tau\tau$~\cite{HDBS-2019-22} and the search for $H\to Z\gamma$ (with leptonic decay of the $Z$ boson)~\cite{HIGG-2018-44}. The techniques vary depending on the mass of the resonance and therefore the \pt-regime; for example, although the approaches used to reconstruct heavy $H\to h_{125}h_{125}\to 4b$~\cite{HDBS-2018-41} or $h_{125}\to aa\to 4b$~\cite{HDBS-2018-47} are similar, the identification of $b$-jets depends on the \pt and requires specific optimizations.

\subsubsection{Improved flavour-tagging with advanced machine-learning techniques}

Flavour tagging, especially the correct identification of jets originating from $b$-quarks, is an essential technique in many searches involving BSM Higgs bosons. The enhanced coupling of heavy additional Higgs bosons to $b$-quarks means that event selections targeting such processes will include criteria for these objects. The correct identification of events with $b$-jets implies that backgrounds with jets originating from other quark types are suppressed, and this greatly enhances the sensitivity. Flavour-tagging algorithms make heavy use of machine learning, exploiting track properties, impact parameters or jet kinematics. During \RunTwo, the taggers used were based on deep neural networks~\cite{FTAG-2019-07}. Recent developments for \RunThr have led to huge advances in this area, notably making use of graph neural networks~\cite{ATL-PHYS-PUB-2022-027}. These new $b$-tagging algorithms have higher rejection rates for the same $b$-jet efficiency, particularly for very high \pt jets. Improved $b$-tagging for boosted jets will significantly increase the sensitivity of searches at higher resonance masses, which is currently limited.


\section{Conclusions}
\label{sec:discussion:conclusions}

The success of LHC \RunTwo is the result of tremendous progress in the accelerator, detector and computing technologies, theoretical advances, and the development of new analysis techniques. Compared to \RunOne, the \RunTwo dataset features much higher integrated luminosity and collision energy, as well as improved detector performance. Refined analysis algorithms include better object reconstruction, the introduction of boosted objects allowing extended search ranges, the capacity to identify long-lived particles, and machine-learning techniques, which are now widely used in several areas such as flavour tagging and signal purification. The \RunTwo dataset, together with all these advances, has therefore provided ATLAS with an outstanding opportunity to search for additional Higgs bosons and explore other BSM signatures.

A compendium of almost 50 searches for possible additional scalars and exotic Higgs boson decays is presented in this report. Stringent upper limits have been set on the production of heavy or light Higgs bosons and on exotic decays. Theoretical models have been constrained considerably, yet current searches are not sensitive enough to rule out these models completely. Several small excesses have been observed in these searches, but no statistically significant evidence has been uncovered for any of this much-anticipated new physics beyond the Standard Model.

Looking ahead, LHC \RunThr offers an even better opportunity to advance these searches. \RunThr data-taking started in 2022 and will continue until the end of 2025. Detailed analysis of the \RunThr dataset and exploitation of its full potential will continue long after data-taking has ended and while the ATLAS Collaboration prepares for the High Luminosity LHC era scheduled to begin in 2029. \RunThr brings increases not only in total integrated luminosity, but also in LHC beam energy, from 6.5~\TeV to 6.8~\TeV. Expected cross-sections, especially for very heavy BSM particles, are larger. \RunThr provides an independent dataset that may confirm or rule out some of the small excesses observed in \RunTwo, so very exciting times lie ahead.

Continued meticulous effort is needed to achieve the maximum possible coverage of the parameter space for new physics and to make new discoveries. Searches in currently unexplored channels or phase-space regions, including those for long-lived particles, cascade decays, and very light or very heavy states, should be performed, along with high-precision measurements of SM particles and the 125~\GeV Higgs boson. Advances in machine learning are extremely helpful in discriminating between background and well-hidden, small signals in dense environments. Refined particle identification and the tagging of signatures helps to reduce the background and constrain systematic uncertainties. Developments in software and computing are crucial for efficiently processing the large amounts of real or simulated data. Finally, creative new analysis ideas are likely to pave the way forward to the next generation of searches in \RunThr data and beyond.


\clearpage
\section*{Acknowledgements}

%
%

%
%

We thank CERN for the very successful operation of the LHC and its injectors, as well as the support staff at
CERN and at our institutions worldwide without whom ATLAS could not be operated efficiently.

The crucial computing support from all WLCG partners is acknowledged gratefully, in particular from CERN, the ATLAS Tier-1 facilities at TRIUMF/SFU (Canada), NDGF (Denmark, Norway, Sweden), CC-IN2P3 (France), KIT/GridKA (Germany), INFN-CNAF (Italy), NL-T1 (Netherlands), PIC (Spain), RAL (UK) and BNL (USA), the Tier-2 facilities worldwide and large non-WLCG resource providers. Major contributors of computing resources are listed in Ref.~\cite{ATL-SOFT-PUB-2023-001}.

We gratefully acknowledge the support of ANPCyT, Argentina; YerPhI, Armenia; ARC, Australia; BMWFW and FWF, Austria; ANAS, Azerbaijan; CNPq and FAPESP, Brazil; NSERC, NRC and CFI, Canada; CERN; ANID, Chile; CAS, MOST and NSFC, China; Minciencias, Colombia; MEYS CR, Czech Republic; DNRF and DNSRC, Denmark; IN2P3-CNRS and CEA-DRF/IRFU, France; SRNSFG, Georgia; BMBF, HGF and MPG, Germany; GSRI, Greece; RGC and Hong Kong SAR, China; ISF and Benoziyo Center, Israel; INFN, Italy; MEXT and JSPS, Japan; CNRST, Morocco; NWO, Netherlands; RCN, Norway; MNiSW, Poland; FCT, Portugal; MNE/IFA, Romania; MESTD, Serbia; MSSR, Slovakia; ARIS and MVZI, Slovenia; DSI/NRF, South Africa; MICIU/AEI, Spain; SRC and Wallenberg Foundation, Sweden; SERI, SNSF and Cantons of Bern and Geneva, Switzerland; NSTC, Taipei; TENMAK, T\"urkiye; STFC/UKRI, United Kingdom; DOE and NSF, United States of America.

Individual groups and members have received support from BCKDF, CANARIE, CRC and DRAC, Canada; PRIMUS 21/SCI/017, CERN-CZ and FORTE, Czech Republic; COST, ERC, ERDF, Horizon 2020, ICSC-NextGenerationEU and Marie Sk{\l}odowska-Curie Actions, European Union; Investissements d'Avenir Labex, Investissements d'Avenir Idex and ANR, France; DFG and AvH Foundation, Germany; Herakleitos, Thales and Aristeia programmes co-financed by EU-ESF and the Greek NSRF, Greece; BSF-NSF and MINERVA, Israel; Norwegian Financial Mechanism 2014-2021, Norway; NCN and NAWA, Poland; La Caixa Banking Foundation, CERCA Programme Generalitat de Catalunya and PROMETEO and GenT Programmes Generalitat Valenciana, Spain; G\"{o}ran Gustafssons Stiftelse, Sweden; The Royal Society and Leverhulme Trust, United Kingdom.

In addition, individual members wish to acknowledge support from CERN: European Organization for Nuclear Research (CERN PJAS); Chile: Agencia Nacional de Investigaci\'on y Desarrollo (FONDECYT 1190886, FONDECYT 1210400, FONDECYT 1230812, FONDECYT 1230987); China: Chinese Ministry of Science and Technology (MOST-2023YFA1605700), National Natural Science Foundation of China (NSFC - 12175119, NSFC 12275265, NSFC-12075060); Czech Republic: Czech Science Foundation (GACR - 24-11373S), Ministry of Education Youth and Sports (FORTE CZ.02.01.01/00/22\_008/0004632), PRIMUS Research Programme (PRIMUS/21/SCI/017); EU: H2020 European Research Council (ERC - 101002463); European Union: European Research Council (ERC - 948254, ERC 101089007), Horizon 2020 Framework Programme (MUCCA - CHIST-ERA-19-XAI-00), European Union, Future Artificial Intelligence Research (FAIR-NextGenerationEU PE00000013), Italian Center for High Performance Computing, Big Data and Quantum Computing (ICSC, NextGenerationEU); France: Agence Nationale de la Recherche (ANR-20-CE31-0013, ANR-21-CE31-0013, ANR-21-CE31-0022, ANR-22-EDIR-0002), Investissements d'Avenir Labex (ANR-11-LABX-0012); Germany: Baden-Württemberg Stiftung (BW Stiftung-Postdoc Eliteprogramme), Deutsche Forschungsgemeinschaft (DFG - 469666862, DFG - CR 312/5-2); Italy: Istituto Nazionale di Fisica Nucleare (ICSC, NextGenerationEU), Ministero dell'Università e della Ricerca (PRIN - 20223N7F8K - PNRR M4.C2.1.1); Japan: Japan Society for the Promotion of Science (JSPS KAKENHI JP21H05085, JSPS KAKENHI JP22H01227, JSPS KAKENHI JP22H04944, JSPS KAKENHI JP22KK0227); Netherlands: Netherlands Organisation for Scientific Research (NWO Veni 2020 - VI.Veni.202.179); Norway: Research Council of Norway (RCN-314472); Poland: Ministry of Science and Higher Education (IDUB AGH, POB8, D4 no 9722), Polish National Agency for Academic Exchange (PPN/PPO/2020/1/00002/U/00001), Polish National Science Centre (NCN 2021/42/E/ST2/00350, NCN OPUS nr 2022/47/B/ST2/03059, NCN UMO-2019/34/E/ST2/00393, UMO-2020/37/B/ST2/01043, UMO-2021/40/C/ST2/00187, UMO-2022/47/O/ST2/00148, UMO-2023/49/B/ST2/04085); Slovenia: Slovenian Research Agency (ARIS grant J1-3010); Spain: Generalitat Valenciana (Artemisa, FEDER, IDIFEDER/2018/048), Ministry of Science and Innovation (MCIN \& NextGenEU PCI2022-135018-2, MICIN \& FEDER PID2021-125273NB, RYC2019-028510-I, RYC2020-030254-I, RYC2021-031273-I, RYC2022-038164-I), PROMETEO and GenT Programmes Generalitat Valenciana (CIDEGENT/2019/023, CIDEGENT/2019/027); Sweden: Swedish Research Council (Swedish Research Council 2023-04654, VR 2018-00482, VR 2022-03845, VR 2022-04683, VR 2023-03403, VR grant 2021-03651), Knut and Alice Wallenberg Foundation (KAW 2018.0157, KAW 2018.0458, KAW 2019.0447, KAW 2022.0358); Switzerland: Swiss National Science Foundation (SNSF - PCEFP2\_194658); United Kingdom: Leverhulme Trust (Leverhulme Trust RPG-2020-004), Royal Society (NIF-R1-231091); United States of America: U.S. Department of Energy (ECA DE-AC02-76SF00515), Neubauer Family Foundation.

%
%


%
%
%
\printbibliography

\clearpage
 
\begin{flushleft}
\hypersetup{urlcolor=black}
{\Large The ATLAS Collaboration}

\bigskip

\AtlasOrcid[0000-0002-6665-4934]{G.~Aad}$^\textrm{\scriptsize 104}$,
\AtlasOrcid[0000-0001-7616-1554]{E.~Aakvaag}$^\textrm{\scriptsize 17}$,
\AtlasOrcid[0000-0002-5888-2734]{B.~Abbott}$^\textrm{\scriptsize 123}$,
\AtlasOrcid[0000-0002-0287-5869]{S.~Abdelhameed}$^\textrm{\scriptsize 119a}$,
\AtlasOrcid[0000-0002-1002-1652]{K.~Abeling}$^\textrm{\scriptsize 56}$,
\AtlasOrcid[0000-0001-5763-2760]{N.J.~Abicht}$^\textrm{\scriptsize 50}$,
\AtlasOrcid[0000-0002-8496-9294]{S.H.~Abidi}$^\textrm{\scriptsize 30}$,
\AtlasOrcid[0009-0003-6578-220X]{M.~Aboelela}$^\textrm{\scriptsize 45}$,
\AtlasOrcid[0000-0002-9987-2292]{A.~Aboulhorma}$^\textrm{\scriptsize 36e}$,
\AtlasOrcid[0000-0001-5329-6640]{H.~Abramowicz}$^\textrm{\scriptsize 154}$
% ...
\bigskip
\\

% ...
$^{17}$Department for Physics and Technology, University of Bergen, Bergen; Norway.\\
% ...
$^{30}$Physics Department, Brookhaven National Laboratory, Upton NY; United States of America.\\
% ...
$^{36}$$^{(a)}$Facult\'e des Sciences Ain Chock, Universit\'e Hassan II de Casablanca;$^{(b)}$Facult\'{e} des Sciences, Universit\'{e} Ibn-Tofail, K\'{e}nitra;$^{(c)}$Facult\'e des Sciences Semlalia, Universit\'e Cadi Ayyad, LPHEA-Marrakech;$^{(d)}$LPMR, Facult\'e des Sciences, Universit\'e Mohamed Premier, Oujda;$^{(e)}$Facult\'e des sciences, Universit\'e Mohammed V, Rabat;$^{(f)}$Institute of Applied Physics, Mohammed VI Polytechnic University, Ben Guerir; Morocco.\\
% ...
$^{45}$Physics Department, Southern Methodist University, Dallas TX; United States of America.\\
% ...
$^{50}$Fakult\"{a}t Physik , Technische Universit{\"a}t Dortmund, Dortmund; Germany.\\
% ...
$^{56}$II. Physikalisches Institut, Georg-August-Universit\"{a}t G\"ottingen, G\"ottingen; Germany.\\
% ...
$^{104}$CPPM, Aix-Marseille Universit\'e, CNRS/IN2P3, Marseille; France.\\
% ...
$^{119}$$^{(a)}$New York University Abu Dhabi, Abu Dhabi;$^{(b)}$United Arab Emirates University, Al Ain; United Arab Emirates.\\
% ...
$^{123}$Homer L. Dodge Department of Physics and Astronomy, University of Oklahoma, Norman OK; United States of America.\\
% ...
$^{154}$Raymond and Beverly Sackler School of Physics and Astronomy, Tel Aviv University, Tel Aviv; Israel.\\
% ...
$^{a}$ Also Affiliated with an institute covered by a cooperation agreement with CERN.\\
$^{b}$ Also at An-Najah National University, Nablus; Palestine.\\
$^{c}$ Also at Borough of Manhattan Community College, City University of New York, New York NY; United States of America.\\
$^{d}$ Also at Center for Interdisciplinary Research and Innovation (CIRI-AUTH), Thessaloniki; Greece.\\
$^{e}$ Also at Centro Studi e Ricerche Enrico Fermi; Italy.\\
$^{f}$ Also at CERN, Geneva; Switzerland.\\
% ...

\end{flushleft}


\end{document}